\documentclass[12pt]{report}
\begin{document}
\thispagestyle{empty}
\begin{center}
\LARGE{\bf{Calculation of Higher Mass-Dimensional \\ Effective
Lagrangians \\ in Quantum Field Theory}}

\vspace{1cm}

\large by

\vspace{1cm}

Joseph Ambrose G. Pagaran

\vspace{1.5cm}

A graduate thesis

\vspace{1cm}

in partial fulfillment of the requirements \\ for the degree of

\vspace{1cm}

{\bf Master of Science in Physics}

\vspace{1cm}

Submitted to the

\vspace{1cm}

Department of Physics\\ College of Science

De La Salle University\\

Taft Avenue, Manila \\Philippines

\vspace{1.5cm}

March 2003
\end{center}

\newpage

\vspace{0.5cm}
\begin{center}
\begin{Large}
\bf{APPROVAL SHEET}
\end{Large}
\end{center}

\vspace{0.8cm}

$\!\!\!\!\!\!\!\!\!$This thesis hereto entitled:

\vspace{0.2cm}
\begin{center}
\bf{``CALCULATION OF HIGHER MASS-DIMENSIONAL EFFECTIVE
LAGRANGIANS IN QUANTUM FIELD THEORY"}
\end{center}
\vspace{0.2cm}

$\!\!\!\!\!\!\!\!\!$prepared and submitted by \textbf{Joseph~Ambrose~G.~Pagaran} in partial fulfillment of the requirements for the degree of \textbf{Master of Science in Physics} has been examined and is recommended for acceptance and approval for ORAL EXAMINATION.

\vspace{1.5cm}

~~~~~~~~~~~~~~~~~~~~~~~~~~~~~~~~~~~~~~~~~~~~~~~~~~~~~~~~~~~~~~~~~~~~~~~~~~~~~~~~~~~(sgd)

~~~~~~~~~~~~~~~~~~~~~~~~~~~~~~~~~~~~~~~~~~~~~~~~~~~~~~~~~~~~~~~~~~~~~~~~~~Emmanuel T. Rodulfo, Ph.D.

\vspace{0.1cm}

~~~~~~~~~~~~~~~~~~~~~~~~~~~~~~~~~~~~~~~~~~~~~~~~~~~~~~~~~~~~~~~~~~~~~~~~~~~~~~~~~~Adviser

\vspace{0.8cm}

~~~~~Approved by Committee on Oral Examination with a grade of PASSED on March 1, 2003.

\vspace{0.3cm}

\begin{center}
(sgd)

Ermys~B.~Bornilla,~M.S.

Chair
\end{center}

\vspace{0.8cm}
~~~~~~~~~~~~(sgd)~~~~~~~~~~~~~~~~~~~~~~~~~~~~~~~~~~~~~~~~~~~~~~~~~~~~~~~~~~~~~~~~~(sgd)

Reuben~V.~Quiroga,~Ph.D.~~~~~~~~~~~~~~~~~~~~~~~~~~~~~~~~~~~~~~~~Jose~Perico~H.~Esguerra,~Ph.D.

~~~~~~~~~~~Member~~~~~~~~~~~~~~~~~~~~~~~~~~~~~~~~~~~~~~~~~~~~~~~~~~~~~~~~~~~~~Member

\vspace{0.8cm}

Accepted in partial fulfillment of the requirements for the degree of Master of Science in Physics.

\vspace{0.5cm}
~~~~~~~~~~~~~~~~~~~~~~~~~~~~~~~~~~~~~~~~~~~~~~~~~~~~~~~~~~~~~~~~~~~~~~~~~(sgd)

~~~~~~~~~~~~~~~~~~~~~~~~~~~~~~~~~~~~~~~~~~~~~~~~~~~~~~~~~~~~~Gerardo~C.~Janairo, Ph.D.

~~~~~~~~~~~~~~~~~~~~~~~~~~~~~~~~~~~~~~~~~~~~~~~~~~~~~~~~~~~~~~~~~~~~~~~~~Dean

~~~~~~~~~~~~~~~~~~~~~~~~~~~~~~~~~~~~~~~~~~~~~~~~~~~~~~~~~~~~~~~~~College of Science

\newpage
\pagenumbering{roman}
\setcounter{page}{3}

\abstract{A prescription for calculating low-energy one-loop higher-mass dimensional effective Lagrangians for non-Abelian field theories is constructed in the spirit of quasilocal background field method. Basis of Lorentz and gauge-invariant monomials of similar mass-dimensions acting as building blocks are matrix-multiplied in a specified order (usually dictated by a permutation of tensorial indices) generating the much needed invariants. The same set of building blocks is used to generate higher-order corrections for a specific mass-dimension. Though the gauge group, the spacetime dimensions, the order of corrections that can be included, and the mass-dimensions that can be formed are all kept arbitrary in the prescription, we constructed basis invariants from 3 up to 12 mass-dimensions to accommodate higher-order corrections up to fourth-order. With these basis, we pursued solving the zeroth-order corrections leading to invariants from 2 up to 16 mass-dimensions, for first-order from 4 up to 8 mass-dimensions, second and third order corrections from 6 up to 8 mass-dimensions. As a result, we  have reproduced the zeroth-order corrections showing dependence on the covariant derivative of the background matrix potential. Previous calculation was done up to 12 mass-dimensions but this dependence was not shown in closed form. For higher-order corrections, the case for 4 up to 6 mass-dimensions are also reproduced. Finally, we calculated the case for 8 mass-dimensions which is reduced only by exploiting the antisymmetry of the fieldstrength tensor and the freedom to throw away total derivatives.}

\newpage
\thispagestyle{empty}
\
\vspace{3.5in}

\begin{center}
  $
   To~Marius~Adrian.$\\
\end{center}

\newpage
\vspace{-0.4cm}
\begin{huge}\textbf{Acknowledgments}\end{huge}

\vspace{1cm}
Many people and institutions were involved in the completion of this MS thesis. First and foremost, I wish to express my sincerest gratitude to Associate Professor, Dr. Emmanuel T. Rodulfo for giving me the opportunity to work under his supervision at the Theoretical and Computational Physics Group, Department of Physics, De La Salle University. His guidance and mentorship, encouragement and support at all levels were invaluable.

My thanks also go to the members of my panel, Assistant Professor, Mr. Ermys Bornilla, Associate Professor, Dr. Reuben Quiroga and Assistant Professor, Dr. Jose Perico Esguerra of the National Institute of Physics, University of the Philippines - Diliman for reading previous drafts of this thesis. They have provided many valuable comments that improved the presentation and contents of the present work.

I also want to thank Associate Professor, Dr. Jerrold Garcia and Mr. William Yu of the Ateneo High Performance Computing Group of the Ateneo de Manila University and Associate Professor of Mathematics, Dr. Jose Tristan of the Mathematics Department of De La Salle University - Manila for allowing me the use of their licensed Mathematica. I want to mention Dr. Garcia's crucial role in introducing the linux-compiled version of Macsyma.

I thank the following Professors of Physics for their comments: Dr. Robert Delbourgo of the University of Tasmania, Hobart, Tasmania, and Dr. Valery Gusynin of the Bogolyubov Institute for Theoretical Physics, Kieve, Ukraine.

I also would like to express my gratitude to my friends and colleagues from the Physics Departments of the De La Salle University - Manila, the Ateneo de Manila University, and Mapua Institute of Technology for their support and encouragement. I thank Sytem Adminstrator, Mr. Clint Dominic Bennett, for sharing his technical expertise on networking and server usage.

My parents, Laurentino and Epifania, and my sisters Mary Ursulita, Laurentine, and Merriam, receive my deepest gratitude and love for their dedication and the many years of support during my undergraduate studies that provided the foundation for this work. I thank Ms. Chereque Clemente for her generosity and encouragement, and for proofreading this thesis.

Finally, I want to dedicate this thesis to my Lord. I am forever grateful to the Lord God Almighty for giving me the strength, patience, passion, and wisdom. It would not have been possible to complete this master's program without Him.

\vspace{0.75cm}
~~~~~~~~~~~~~~~~~~~~~~~~~~~~~~~~~~~~~~~~~~~~~~~~~~~~~~~~~~~~~~~~~~~~~~~~~~Joseph Ambrose G. Pagaran

\newpage
Typeset using \LaTeX2e by Joseph Ambrose G. Pagaran

Copyright \copyright 2003 by Joseph Ambrose G. Pagaran. All rights reserved.

\begin{footnotesize}
No part of this thesis may be reproduced, stored in a retrieval system, or transmitted, in any form or by any means, electronic, mechanical, photocopying, recording or otherwise, without the prior written permission of the copyright holder.
\end{footnotesize}
\vspace{0.25cm}

\textbf{Trademark~Notices}
\begin{footnotesize}
\begin{center}
\begin{enumerate}
\item[] Apple and Macintosh are registered trademarks of Apple Computer, Inc.
\item[] Acrobat and the Acrobat logo are trademarks of Adobe Systems Inc. or its subsidiaries and may be registered in certain jurisdictions.
\item[] GNU is a registered trademark of Free Software Foundation Inc.
\item[] Linux is a registered trademark of Linus Torvalds.
\item[] Mathematica are registered trademark of Wolfram Research, Inc.
\item[] MS-DOS, MS-Windows, Windows-NT, Windows 95, Windows 98 and MS-Excel are registered trademarks of Microsoft Corporation.
\item[] Notespad is  registered trademark of Bremer Corporation.
\item[] PostScript is a registered trademark of Adobe Systems Inc.
\item[] TeX is a trademark of the Americal Mathematical Society.
\item[] XWindows is a trademark of MIT.
\item[] UNIX is a registered trademark in the United States and other countries, licensed exclusively through The Open Group
\item[] All other trademarks and service marks referred to in this thesis are the property of their respective owners.
\end{enumerate}
\end{center}

\vspace{0.2cm}
The name "Macsyma" is a registered trademark of Macsyma Inc. in the U.S. and
other major countries.  Macsyma Inc. inherits an agreement from Symbolics
Inc. permitting the D.O.E. to use the name "D.O.E. Macsyma."  No other entity
has the right to use the name Macsyma or closely similar names in the
distribution of mathematical software.  Some parties who distribute
derivatives of M.I.T. Macsyma use names such as "Paramax" and "Aljabar;"
these names are not in conflict with the registered trademark "Macsyma."

The name "Maxima" is considered by Macsyma Inc. to be an infringement of
the Macsyma trademark, though Macsyma have refrained from legal action.  Macsyma request
that (1) distributors of such software alter the names of their software and (2)
information sources which list software packages not to list mathematical
software which uses this name.
\end{footnotesize}

\newpage
\textbf{\huge{Notations}}

\vspace{5mm}

\normalsize
 \[
 R_{\mu\nu.\rho}
  \equiv
  {\cal D}_{\rho} R_{\mu\nu}
\,\,\,\,\,\,\,\,\,\,\,\,\,\,\,\,\,\,\,\,\,\,\,\,\,\,\,\,\,\,\,\,\,\,\,\,
        R_{\mu\nu.\nu_1\nu_2\ldots\nu_q}
        \equiv
   {\cal D}^q_{\nu_1\nu_2\ldots\nu_q}
        R_{\mu\nu}
        ={\cal D}_{\nu_q}\ldots
        {\cal D}_{\nu_2}
        {\cal D}_{\nu_1}
        R_{\mu\nu}
 \]

\[
 \!\!\!\!\!\!
 \!\!\!\!\!\!\!\!\!\!\!\!\!\!\!\!\!\!\!\!\!\!\!\!\!\!\!\!\!\!\!\!\!\!\!\!
 \!\!\!\!\!\!\!\!\!\!\!\!\!\!\!
 {\cal D}^q_{\nu_1\nu_2\ldots\nu_q}
 \equiv
 {\cal D}_{\nu_q}\ldots
        {\cal D}_{\nu_2}
        {\cal D}_{\nu_1}
\!\!\!\!\!\!\!\!\!\!\!
\,\,\,\,\,\,\,\,\,\,\,\,\,\,\,\,\,\,\,\,\,\,\,\,\,\,\,\,\,\,\,\,\,\,\,\,
 \!\!\!\!\!\!\!\!\!\!\!\!\!\!\!\!\!\!\!\!\!\!\!\!\!\!\!\!\!\!\!\!\!\!\!\!
 \!\!\!\!\!\!\!\!\!\!\!\!\!\!\!\!\!\!\!\!\!\!\!\!\!\!\!\!\!\!\!\!\!\!\!\!
 \!\!\!\!\!\!\!\!\!\!\!\!\!\!\!\!\!\!\!\!\!\!\!\!\!\!\!\!\!\!\!\!\!\!\!\!
 \!\!\!\!\!\!\!\!\!\!\!\!\!\!\!\!\!\!\!\!\!\!\!\!\!\!\!\!\!\!\!\!
 \!\!\!\!\!\!\!\!\!\!\!\!\!\!\!\!\!
  Q_{,\mu}
  \equiv
  \partial_{\mu} Q
   =
  \frac{\partial Q}{\partial x_{\mu}}
\,\,\,\,\,\,\,\,\,\,\,\,
 \]

 \[
 \!\!\!\!\!\!
 \!\!\!\!\!\!\!\!\!\!\!\!\!\!\!\!\!\!\!\!\!\!\!\!\!\!\!\!\!\!\!\!\!\!\!\!
 \!\!\!\!\!\!\!\!\!\!\!\!\!\!\!
    (x-x')^q_{\nu_1\nu_2\ldots\nu_q}
    \equiv (x-x')_{\nu_1}(x-x')_{\nu_2}\ldots(x-x')_{\nu_q}
 \!\!\!\!\!\!\!\!\!\!\!\!\!\!\!\!\!\!\!\!\!\!\!\!\!\!\!\!\!\!\!\!\!\!\!\!
 \!\!\!\!\!\!\!\!\!\!\!\!\!\!\!\!\!\!\!\!\!\!\!\!\!\!\!\!\!\!\!\!\!\!\!\!
 \!\!\!\!\!\!\!\!\!\!\!\!\!\!\!\!\!\!\!\!\!\!\!\!\!\!\!\!\!\!\!\!\!\!\!\!
 \!\!\!\!\!\!\!\!\!\!\!\!\!\!\!\!\!\!\!\!\!\!\!\!\!\!\!\!\!\!\!\!\!\!\!\!
 \!\!\!\!\!\!\!\!\!\!\!\!\!\!\!\!\!\!\!\!\!\!\!\!\!\!\!\!
  T_{/\mu}
  \equiv
  \frac{\partial T}{\partial p_{\mu}}
\,\,\,\,\,\,\,\,\,\,\,\,
 \]

\begin{eqnarray*}
{\cal D}_{({\mu_\ell}{\nu_\ell}{\rho_\ell})}
&\equiv&
{\cal D}_{{\mu_\ell}\left\{{\nu_\ell}{\rho_\ell}\right\}}
+{\cal D}_{{\nu_\ell}\left\{{\rho_\ell}{\mu_\ell}\right\}}
+{\cal D}_{{\rho_\ell}\left\{{\mu_\ell}{\nu_\ell}\right\}}
\\
{\cal D}_{{\mu_\ell}\left\{{\nu_\ell}{\rho_\ell}\right\}}
&\equiv&
{\cal D}_{{\mu_\ell}{\nu_\ell}{\rho_\ell}}
+{\cal D}_{{\mu_\ell}{\rho_\ell}{\nu_\ell}}
\\
{\cal D}_{<{\tau_\ell}}\!Y_{{\mu_\ell}{\lambda_\ell}>}
&\equiv &
{\cal D}_{{\lambda_\ell}}\!Y_{{\mu_\ell}{\nu_\ell}} + {\cal D}_{{\nu_\ell}}\!Y_{{\mu_\ell}{\lambda_\ell}}
\end{eqnarray*}

 \[
 \!\!\!\!\!\!\!\!\!\!\!\!\!\!\!\!\!\!\!\!\!\!\!\!\!\!\!\!\!\!\!\!\!
 \!\!\!\!\!\!
(A \circ B)_{\mu_{1}\mu_{2} \ldots \mu_{p+q}} =
  \left\{
 \begin{array}{ll}
  A_{\mu_{1}\mu_{2} \ldots \mu_{p}}
  B_{\mu_{1}\mu_{2} \ldots \mu_{p}\mu_{p+1} \ldots \mu_{q}}
   & \mbox{if}\,\,\, p \leq q
 \\
  A_{\mu_{1}\mu_{2} \ldots \mu_{q}\mu_{q+1} \ldots \mu_{p}}
  B_{\mu_{1}\mu_{2} \ldots \mu_{q}}
   & \mbox{if}\,\,\, p \geq q
 \end{array}
  \right.
 \]

\[
  [m_1]\ldots[m_\ell]\equiv{\cal L}^{(1)[\sum_{w=0}^\ell m_w}_\ell
\]

\[
 c_2\equiv\frac{\hbar}{2(2\pi)^D}
\,\,\,\,\,\,\,\,\,\,\,\,\,\,\,\,\,\,\,\,
 c_4\equiv\frac{\hbar}{2(4\pi)^{D/2}}
\]

\[
(q_0,q_1,\ldots,q_{n-1}\mbox{;}A,B)
 \equiv
\prod^n_{\ell=0}
   \int^\infty_0 ds_\ell
   \frac{{s_0}^{q_0}\ldots {s_{n-1}}^{q_{n-1}}}{A\left(\sum^n_{r=0} s_r\right)^{B+D/2}}
\]

\[
\{{\mu}_{\ell},{\nu}_{\ell},{\rho}_{\ell},{\sigma}_{\ell},{\alpha}_{\ell},{\beta}_{\ell}\}\to\{{\mbox{a}}_{\ell},{\mbox{b}}_{\ell},{\mbox{c}}_{\ell},{\mbox{d}}_{\ell},{\mbox{e}}_{\ell},{\mbox{f}}_{\ell}\}
\]

\[
\{{\lambda}_{\ell},{\tau}_{\ell},{\kappa}_{\ell},{\eta}_{\ell},{\xi}_{\ell},{\varrho}_{\ell}\}\to\{{\mbox{l}}_{\ell},{\mbox{t}}_{\ell},{\mbox{k}}_{\ell},{\mbox{h}}_{\ell},{\mbox{x}}_{\ell},{\mbox{g}}_{\ell}\}
\]

\tableofcontents \thispagestyle{empty}
\setlength{\baselineskip}{25pt} \setcounter{page}{0}
\thispagestyle{empty}
\chapter{Introduction}
\pagenumbering{arabic}\setcounter{page}{1}
A physical phenomenon may involve arbitrary interactions including perhaps self-interactions. Such event can only be analyzed completely by introducing corrections in the equations that describe an ordinary interaction. Usually the included interaction terms depend on higher-order derivatives of the fields and appear in the effective Lagrangian.\footnote{For example in perturbation theory for gravity: $g_0+\delta g$, where $g_0$ is a background metric that is a solution of the field equations. The terms quadratic in $\delta g$ are regarded as the free action, the rest of the higher-order higher-derivative terms are the interactions (including possibly self-interactions).\cite{HwHt}} These effective Lagrangians containing higher-order derivative corrections arise naturally in higher-dimensional theories. Physics described by effective Lagrangians of this kind form a class of higher-derivative and higher dimensional theories.

Various areas of physics belong to this class. Each inclusion of higher-derivative term in one of these areas corresponds a reason. In general relativity for example,  higher-derivative terms of the metric\cite{Birr} appearing as curvature-squared term in the Lagrangian\cite{Stll,Boul} are added to a more standard lower-derivative theory as quantum corrections. Even for small coefficients, these terms can dominate. In another instance, they appear in brane theories in two classes\cite{Polc,John}. First, they appear as derivatives of the world-volume fields and second, as derivatives of the bulk fields to the world volume of the brane. The inclusion of higher-derivative terms occur in cosmic strings\cite{Curt,Maed}, in Dirac's relativistic model of the classical radiating electrons\cite{Dirc}, in quantum gravity \cite{Buch,Avrm}, in blackholes \cite{Mign}, and in quantum cosmology \cite{Mazz} to mention a few. In all instances, their presence dramatically recast the original lower-derivative theory into a new one no matter how small it may naively appear to perturb the original one.

Theories containing unconstrained higher-derivative corrections have very distinctive features. Their presence makes two seemingly identical theories very different\cite{Simn}, one a lower-derivative theory and the other the same theory with a higher derivative correction included. Mathematically, there is nothing inconsistent with these features. One feature is that they have more degrees of freedom\footnote{This is true at the classical level. At the quantum level, non-commuting variables in the lower-derivative theory, such as positions and velocities, become commuting in the higher-derivative theory.\cite{Mazz,Simn}} and they lack a lowest-energy bound than their lower-derivative counterparts.  More degrees of freedom\footnote{This means that more initial data are needed to solve the dynamical equations of motion.} might be more accurate physically as the most interesting new families of solutions appear progressively for each order of accommodated higher-derivative corrections.

Of course, we want to avoid the above problems. These are avoided if we consider nonlocal theories whose higher-order higher-derivative terms appear as a perturbative (Taylor-series) expansion about some small parameter.\footnote{This is small in the sense that $\delta g$ is kept smaller than $g_0$ locally. (See footnote 1 of this chapter). This is so if there are no large metric fluctuations below the Planck scale\cite{HwHt}, a necessity if perturbation is to make sense.}
They are not plagued with such problems as they contain implicit contraints keeping the number of degrees of freedom fixed and maintain a lowest-energy bound. Their relatives, the truncated expansions of some nonlocal (but otherwise well-behaved) theory, also enjoy the absence of such problems. Other theories, like treating the added higher-derivative terms as small, also avoid these problems. Through these implicit constraints, a lot of the unnecessary degrees of freedom are not counted consequently removing solutions that cannot be (Taylor-series) expanded about the small parameter. These imposed constraints do not correspond to higher-order terms with the higher powers of the expansion of the small parameter. Instead, each term in the series expansion contributes commensurately less with increasing order of degree. The series expansion can be implemented finitely or infinitely. While convergence\footnote{String perturbation theory expansion does not converge. They have to be augmented by non-perturbative objects, like $D$-branes.\cite{Polc}} is not an issue in the former, it is demanded in the latter that equations of motion converge.\footnote{The terms mentioned in footnote 2 include $(\nabla \delta g)^2$, multiplied by powers of $\delta g$. Actually the volume integral of such an interaction is not bounded by the free action and perturbation theory does not make sense. This makes perturbation for gravity (in the sense described in footnote 1) not renormalizable.\cite{tHVt} This is because $\delta g > g_0$ locally.}

Higher derivatives lead to ghosts\cite{HwHt}, states with negative norm. This means that the S matrix would not be unitary and there will be states with negative probabilities. Ref. \cite{HwHt} circumvented these features by considering fourth-order corrections. They have shown that perturbation theory for gravity in dimensions greater than two requires higher derivatives in the free action. Finally, higher-order Lagrangians appear in higher-dimensional theories such as Kaluza Klein and string theories. They are introduced for various reasons like spontaneous compactification from purely gravitational higher dimensional theories. Higher dimensional theories has been largely studied in cosmology\cite{Shaf,Maed2} and in Black holes\cite{Mign}.

\section{Background of the Study}
In the area of effective field theory, calculation of effective Lagrangian (with its fields allowed to vary arbitrarily) involves the construction of the Green function equation in momentum-space. One feature we choose to take advantage of in this method is that proper-time equations\footnote{We will interchangably refer proper-time equations with explicit solutions, unless otherwise indicated. Proper-time equations relate quantities to the dynamical properties of particles with space-time coordinates that depend upon a proper-time parameter. They depend on generalized non-Abelian fieldstrength tensor $Y$ and generalized background matrix potential $X$. We prefer to express them as a proper-time integration though they can be equivalently expressed in terms of a gamma function (See Appendix D.5).  This is an artifact of the proper-time method that isolates divergences in integrals with respect to the proper-time parameter, which is independent of the coordinate system and of the gauge. \cite{Schw}} can be obtained in the strong but slowly varying background fields.

The method of obtaining explicit solutions\footnote{Refer to previous footnote (No. 7).} was pioneered by Fock\cite{Fck} when he proposed a method of solving wave equations in external electromagnetic fields by an integral transform in the proper-time parameter. It was Schwinger\cite{Schw,Sch} who generalized the proper-time method and applied it to the covariant calculation of one-loop effective Lagrangian for constant electromagnetic fields. Then DeWitt\cite{DeWt} in the-so-called quasilocal background field method\cite{bfta}-\cite{bftz} reformulated Fock-Schwinger proper-time method in geometrical language. He applied it to the case of external gravitational field.

In this work, we will perform background-field-method-prescribed calculations which begin by imposing covariant restrictions on the background converting the nonlocal equation satisfied by the Green function equation to a quasilocal one. The covariant restrictions will assure us a fixed number of degrees of freedom and a maintained lowest-energy bound. The trend\footnote{Consider getting a sneak peek at Eqns (\ref{1stknd})-(\ref{3rdknd}).} in the calculation is towards an even more relaxed (in terms of covariant restrictions in the fields) version similar to Rodulfo\cite{Rodf-Diss} in application to gauge fields, similar to Brown and Duff\cite{Brw-Dff} in scalar fields. Our calculation goes beyond the covariant restriction imposed by Schwinger (who imposed first-order covariant restrictions on the field strength tensor), Brown and Duff (who imposed first as well as second-order covariant restrictions), and Rodulfo (who imposed third-order covariant restrictions). Our work aims to calculate explicitly the solution corresponding to the eighth-order partial differential Green function equation as obtained by Tiamzon\cite{Tiam,Tiam-Rodf}.

\section{Organization of Thesis}
This work is structured as follows: The fundamentals of the background field formalism and its elegance in accommodating higher-derivative corrections are highlighted first. An example is presented to lay ground for generalization. Choosing a constraint on the general form will give us a specific differential equation which this work aims to solve explicitly. This differential equation is presented in Chapter 2 in both space-time and momentum-space configurations. To solve the equation, Chapter 3 displays the Gaussian solution upon which we generate new families of solutions by letting the non-Gaussian sector operator act on the Gaussian part in momentum space. Though the solution presented in Chapter 3 is exact and closed-form, a Taylor-series-expansion (TSE) equivalent will be used instead to generate the basis invariants. These basis invariants grouped according to their total mass-dimensions are presented in Chapter 4. The zeroth-order one-loop effective Lagrangians are explicitly solved for mass-dimensions two up to sixteen including derivatives in the background matrix potential not previously shown. For higher-order corrections, another set of grouped basis invariants is presented in the latter part of Chapter 4. Prescriptions on how to use these grouped basis invariants are described for each mass-dimension, from mass-dimensions four up to twelve accommodating first-order up to fourth-order corrections. Full implementation of the prescriptions is illustrated explicitly both in the ungauged case up to ten mass-dimensions and in the usual case up to eight mass-dimensions. For the ungauged case, handling $p$, $X$, and $s$ integraions are presented step by step. These steps are skipped in the presentation of four up to eight mass-dimensional one-loop effective Lagrangians in the gauged case. The main calculation is presented in the latter part of Chapter 5. This is about the one-loop eight mass-dimensional one-loop Lagrangian. The repeated use of total derivatives, or equivalently integration by parts, cyclic matrix permutations, and antisymmetry of the field strength tensor is presented as the need arises. Use of Bianchi identities is yet to be implemented. Such further reduction is reserved in another occasion.

\chapter{Statement of the Problem}
\section{Background Field Method Fundamentals}
In this chapter, we review first DeWitt's\footnote{Bryce DeWitt of the University of Texas, Austin, USA invented background field method, and developed the methodology of ghost loops in gauge theory. His name is associated with the Wheeler-DeWitt equation, which provides the basis for most work on quantum cosmology, and with the Schwinger-DeWitt expansion, which is widely used in studying field theories in curved space-time and in string theory computations.} background field formalism\cite{DeWt,DWt-Ish} then we state the problem this thesis aims to achieve. That is, we present the relevant Green function equation for real boson fields and display its generalized form in terms of the generalized background gauge connection and the generalized matrix potential. Then in the next section, we impose (fourth-order) covariant restrictions on both the background gauge connection and the matrix potential to obtain the eighth-order partial differential equation in momentum-space. This is the section where the statement of the problem is formulated. The last two sections of this chapter will be devoted to the scope and delimitations, and the significance of this work.

To begin our review, consider the general form\cite{Brw-Dff,Rodf-Diss,tHft} of the bilinear Lagrangian\footnote{In the spirit of background field formalism, the bilinear Lagrangian appears as a second term of the power-series-expanded and quantum-field-variable-$\phi$-replaced-by-$\phi+A$ Lagrangian\cite{DeWt,tHft,Velt}
\begin{eqnarray*}
    {\cal L}(\phi+A)
    ={\cal L}(A)
    +\left.
        \frac{1}{2}
        \frac{\delta^2{\cal L}}{\delta\phi_i\delta\phi_j}
      \right|_{\phi=A}\phi_i\phi_j+\ldots,
\end{eqnarray*}
which is now a power series expanded in $\phi$ about $A$. This must represent the one-loop quantum corrections. This is upon
invoking the classical equation of motion on the background
\cite{GNW,Ichi}
\begin{eqnarray*}
    \left.
    \frac{\delta {\cal L}(\phi)}{\delta\phi_i}\right|_{\phi=A}
    =0,
\end{eqnarray*}} for real boson fields $\phi$ in $D$ dimensions.
\begin{eqnarray}\label{Lmnw}
    L=
    \frac{1}{2}M^{ij}          \phi^i          \phi^j
    +N^{ij}_{\mu}   \phi^i_{,\mu}   \phi^j
    +\frac{1}{2}W^{ij}_{\mu\nu}\phi^i_{,\mu}   \phi^j_{,\nu}.
\end{eqnarray}
Capitalized Roman Alphabets:
$W$, $N$, and $M$, are arbitrary external spacetime dependent
source functions evaluated at $\phi=A$. They may be chosen to have
the (anti)symmetry properties \cite{Rodf-Diss,tHft,Rodf}:
\begin{eqnarray}
    W^{ij}_{\mu\nu}&=&W^{ij}_{\nu\mu}=W^{ji}_{\mu\nu},
    \label{Wij}\\
    N^{ij}_{\mu}&=&-N^{ji}_{\mu},
    \label{Nij}\\ \label{Mij}
    M^{ij}&=&M^{ji}.
\end{eqnarray}
So that in flat Euclidean $D$-dimensional spacetime,
\begin{eqnarray}
    W^{ij}_{\mu\nu}&=&-\delta_{\mu\nu}\delta^{ij},
    \label{EWij}\\ \label{ED}
    \delta_{\mu\mu}&=&D,
    \\ \label{En}
    \delta^{ii}&=&d.
\end{eqnarray}
If one forms the tensor quantities
\begin{eqnarray}
    X&\equiv& M-N_\mu N_\mu \label{X}    \\
    Y_{\mu\nu}&\equiv&
     N_{\nu,\mu}- N_{\mu,\nu}
    +[N_\mu,N_\nu],\label{Y}
\end{eqnarray}
which together with $\phi$ transform according to
\begin{eqnarray}
    X&\longrightarrow & e^{\Lambda(x)}Xe^{-\Lambda(x)}
    \label{Xto}\\ \label{Yto}
    Y_{\mu\nu}&\longrightarrow &e^{\Lambda(x)}Y_{\mu\nu}e^{-\Lambda(x)}
    \\ \phi&\longrightarrow & e^{\Lambda(x)}\phi \label{pto}
\end{eqnarray}
for some arbitrary antisymmetric matrix $\Lambda^{ij}(x)$, then
the bilinear Lagrangian (\ref{Lmnw}) may be cast in the manifestly
gauge invariant form \cite{Rodf-Diss,Rodf}
\begin{eqnarray}\label{Lbi}
    {\cal L}^{(1)}=\frac{1}{2}\phi\left({\cal D}^2+X\right)\phi.
\end{eqnarray}

The Lagrangian induced by these one-loop effects, ${\cal L}^{(1)}$
in (\ref{Lbi}) is then given by the functional
integral over the quantum fields
\begin{eqnarray}\label{expL}
    \exp\int d^Dx\,\,{\cal L}^{(1)}
    =\int d[\phi]
    \exp\int d^Dx\,
    \frac{1}{2}\phi\left({\cal D}^2+X\right)\phi
\end{eqnarray}
subject to the condition that
\begin{eqnarray}
    {\cal L}^{(1)}\stackrel{A\to 0}{\longrightarrow}0.\,
\end{eqnarray}
Differentiating (\ref{expL}) with respect to $X$, one finds that
${\cal L}^{(1)}$ is determined by the coincidence limit of
two-point correlation function\footnote{Here, Tr denotes a trace over gauge indices (including possibly spinor indices) while tr is a Lorentz trace.}
\begin{eqnarray}\label{Corrf}
    \frac{\partial {\cal L}^{(1)}}{\partial X}
    =\frac{1}{2}\mbox{Tr}\left<\phi(x)\phi(x')\right>
\end{eqnarray}
where the (Euclidean) Green function
\begin{eqnarray}\label{Egf}
    G(x,x')\equiv\left<\phi^j(x)\phi^k(x')\right>
    =\frac{\int d[\phi]\,\phi^j(x)\phi^k(x')
    \exp\int d^Dx\,
    \frac{1}{2}\phi\left({\cal D}^2+X\right)\phi}
    {\int d[\phi]\,
    \exp\int d^Dx\,
    \frac{1}{2}\phi\left({\cal D}^2+X\right)\phi}
\end{eqnarray}
is the solution to the differential (Green function) equation
\begin{eqnarray}\label{Grnf}
    \left[\partial^2\!+\!X(x)\!+\!
        N_{\mu,\mu}
    \!+\!N_\mu(x)\partial_\mu\!+\!N_\mu(x)N_\mu(x)
    \right]
    \left<\phi(x)\phi(x')\right>
    \!=\!-\delta(x,x').
\end{eqnarray} 
Provided one can find some way of solving this non-local equation, the one-loop correction to the effective Lagrangian is given by\cite{DeWt}
\begin{eqnarray}
	L^{(1)}=\frac{1}{2}\mbox{Tr}\int dX \,\,\,G(x,x')
\end{eqnarray} 

The differential equation (\ref{Grnf}) has been generalized by
Tiamzon \cite{Tiam} and by Tiamzon and Rodulfo \cite{Tiam-Rodf} by
imposing covariant restrictions
\begin{eqnarray}\label{DnY}
    {\cal D}^nY=0,
    \;\;\;\;\;\;\;\;\;\;\;\;\;\;\;\;\;\;\;\;
    n\geq 1
\end{eqnarray}
on the background\---\---in the spirit of quasilocal
approximation. The case of $n=1$ was shown to be exactly soluble
in reference \cite{Brw-Dff}, while the case of $n=3$ was
considered in \cite{Rodf-Diss}.\footnote{The case when $n=4$ has
been derived in \cite{Tiam}. Their result will be be shown in the
next section. This was not solved explicitly. Its explicit
solution will be dealt with in this work.} The restriction
(\ref{DnY}) is the non-Abelian analogue to the one imposed by
Schwinger \cite{Schw} on the Maxwell field tensor,
$F_{\mu\nu,\rho}=0$ for constant external electromagnetic fields.
Such restriction is quasilocal and leads one to a closed form of
the one-loop effective Lagrangian in terms of invariants in even
powers of $Y_{\mu\nu}$.

Indeed, the non-trivial extension (\ref{DnY}) corresponds to the
general form of the background connection given by the finite sum
\cite{Tiam-Rodf}
\begin{eqnarray}\label{Nmu}
    N_\mu
    =e^{\Lambda(x)}\partial_\mu e^{-\Lambda(x)}
    +\sum^n_{q=1}
    \frac{(-1)^{1+q}}{(1+q)!}
    ({\cal D}^{q-1}Y(x))_\mu\circ (x-x')^q
\end{eqnarray}
as generalized by Tiamzon and Rodulfo in \cite{Tiam-Rodf} provided
(\ref{DnY}) is adhered to.

Employing the Fock-Schwinger gauge \cite{Flie,Flig,Shif},
\begin{eqnarray}\label{FSg}
    (x-x')\cdot N(x)=0
\end{eqnarray}
in $n$-repeated covariant differentiations
\begin{eqnarray}\label{ndFSg}
    (x-x')^{n+1}\circ{\cal D}^n N(x)=0,
\end{eqnarray}
one finds equivalently,
\begin{eqnarray}\label{Ngen}
    N_\mu(x)
    =\sum^{n-1}_{p=0}
    \frac{1+p}{(2+p)!}
    \left[{\cal D}^p Y(x')\circ (x-x')^{1+p}\right]_\mu.
\end{eqnarray}
Here, the pure gauge term in (\ref{Nmu}) is absorbed by the gauge
transformation
$    N_\mu\rightarrow
    e^{-\Lambda}
    \left(\partial_\mu+N_\mu\right)
    e^{\Lambda}
$
leaving only the invariant quantities as in (\ref{Ngen}). So that
in the completely non-local limit ($n\to\infty$) of the background
gauge connection, this may be expressed in a closed integral form
given by \cite{Pag-Rodf}
\begin{eqnarray}\label{Nint}
    N_\mu(x)=\int^1_0 d\alpha \,\alpha Y_{\mu}(\alpha x)(x-x')
\end{eqnarray}
where we have rescaled the manifold in accordance with
$x\rightarrow\alpha x$. Since this rescaling does not affect the
coefficients of expansion,
\begin{eqnarray}
    Y_{\nu\mu}(x')
    =e^{(x-x'){\cal D}_x}
    \left[Y_{\nu\mu}(x)\right]_{x=x'}
    \equiv
    e^{(x-x'){\cal D}_x}Y_{\nu\mu}(x').
\end{eqnarray}
We can then write $Y_{\mu\nu}$ as
\begin{eqnarray}
    Y_{\nu\mu}(\alpha x)
    =e^{\alpha(x-x'){\cal D}_{\alpha x}}
    \left[Y_{\nu\mu}(x')\right]
    =e^{\alpha(x-x')\partial_{\alpha x}}
    \left[Y_{\nu\mu}(x')\right].
\end{eqnarray}
Formally, this alternative non-local expression (\ref{Nint}) for
$N_\mu$ may be substituted into the Green function equation
(\ref{Grnf}).

Imposing similarly covariant restrictions on the matrix potential
\begin{eqnarray}\label{DnX}
    {\cal D}^lX=0,
    \;\;\;\;\;\;\;\;\;\;\;\;\;\;\;\;\;\;\;\;
    l\geq 1,
\end{eqnarray}
the matrix potential $X(x)$ has the general form
\cite{Tiam,Tiam-Rodf}
\begin{eqnarray}\label{Xgen}
    X(x)
    =\sum^{l-1}_{q=0}
    \frac{1}{q!}
    {\cal D}^q X(x')\circ (x-x')^q.
\end{eqnarray}

For any $n$ and $l$, the generalized differential equation
is given by
\begin{eqnarray}
    &&\!\!\!\!\!
    \left\{
    \partial^2+\sum^{l-1}_{q=0}\frac{1}{q!}{\cal D}^qX(x')\circ(x-x')^q
    -\sum^{n-1}_{q=0}\frac{1+q}{(2+q)!}
        \left[
        {\cal D}^{(q)}\cdot Y(x')
        \right]\circ(x-x')^q
    \right.
    \nonumber\\&&
    \;\;\;\;\;\;\;\;\;\;\;\;\;\;\;\;\;\;\;\;\;\,\;\;\;\;\;\;\;\;\;\;\;\;\;\;\;\;\;\;\;\;\;\;\;\;\;
    +\sum^{n-1}_{q=0}\frac{2(1+q)}{(2+q)!}
       \left[
    {\cal D}^{q}Y(x')\circ(x-x')^{1+q}
       \right]\cdot\partial
    \nonumber\\&&
    \!\!\!\!
    \left.
    +\sum^{n-1}_{r,q=0}\frac{(1+r)(1+q)}{(2+r)!(2+q)!}
        \left[
    {\cal D}^{r}Y(x')
        \circ(x-x')^{1+r}
    \right]\cdot
    \left[
    {\cal D}^{q}Y(x')
        \circ(x-x')^{1+q}
        \right]
    \right\}
    \nonumber\\&&
        \;\;\;\;\;\;\;\;\;\;\;\;\;\;\;\;\;\;\;\;\;\;\;\;\;\;\;\;\;\;\;\;\;\;\;\;\;\;\;\;\;\;\;\;\;\;\;\;\;\;\;\;\;\;\;\;\;
    \times\langle\phi(x)\phi(x')\rangle
    =-\delta(x,x').\label{qGfgen}
\end{eqnarray}
as it has been derived by Tiamzon\cite{Tiam} Tiamzon and
Rodulfo\cite{Tiam-Rodf} progressively by induction.

Upon transformation to a $D$-dimensional Euclidean momentum space
through
\begin{eqnarray}
    \langle\phi(x)\phi(x')\rangle
    =\int\frac{d^Dp}{(2\pi)^D}e^{ip(x-x')}G(p),
\end{eqnarray}
which effectively entails the replacements:
\begin{eqnarray}
    (x-x')\rightarrow-i\frac{\partial}{\partial p},
    \;\;\;\;\;\;\;\;\;\;
    \partial\rightarrow ip,
\end{eqnarray}

the general quasilocal Green function equation in momentum space
takes the form
\begin{eqnarray}
    &&\!\!\!\!\!\!\!\!\!\!\!
    \left\{
    \!-\!p^2 \!+\! \sum^{l-1}_{q=0}\frac{(-i)^q}{q!} {\cal D}^qX(x')\!\circ\!
    \left(\!\frac{\partial}{\partial p}\!\right)^q
    \!-\!\sum^{n-1}_{q=0}\frac{3(-i)^q(1\!+\!q)}{(2\!+\!q)!}
        \left[\!
        {\cal D}^{(q)}\cdot Y(x')
        \!\right]\circ\left(\!\frac{\partial}{\partial p}\!\right)^q
    \right.
    \nonumber\\
    &&\;
    +
    \sum^{n-1}_{r,q=0}\frac{(-i)^{r\!+\!q}(1\!+\!r)(1\!+\!q)}{(2\!+\!r)!(2\!+\!q)!}
        \left[\!
    {\cal D}^{r}Y(x')\!
        \circ\left(\!\frac{\partial}{\partial p}\!\right)^{1\!+\!r}\!
    \right]\!\cdot\!
    \left[\!
    {\cal D}^{q}Y(x')
        \!\circ\!\left(\!\frac{\partial}{\partial p}\!\right)^{1\!+\!q}\!
        \right]
    \nonumber\\
    &&\;
    \left.
    +\sum^{n-1}_{q=0}\frac{2(-i)^q(1+q)}{(2+q)!}p\cdot\!
       \left[
    {\cal D}^{q}Y(x')\circ\left(\frac{\partial}{\partial p}\right)^{1+q}
       \right]
    \right\}
     G(p)
    =-1.\label{qGfmom}
\end{eqnarray}

\section{Eighth-order Partial Differential Equation}
In this work, $n$ and $l$ are set to four in (\ref{DnY}) as
\begin{eqnarray}\label{D4Y}
   {\cal D}^4 Y=0
\end{eqnarray} 
and in
(\ref{DnX}) as well as in (\ref{qGfgen}), (\ref{qGfmom}), and in
(\ref{qDf}) as this will be appropriate in the calculation of higher mass-dimensional one-loop effective Lagrangian with higher-derivative corrections. As shown by Tiamzon and Rodulfo \cite{Tiam-Rodf}, when
$n=l=4$ in (\ref{DnY}) and (\ref{DnX}), the background gauge
connection takes the form
\begin{eqnarray}\label{N44}
    N_\mu
    \!&=&\!
     -\frac{1}{2}Y_{\mu\nu}(x')(x\!-\!x')_\nu
    \!-\frac{1}{3}
      Y_{\mu\nu.\rho}(x')(x\!-\!x')_{\rho\nu}^2
    \nonumber
    \\&&\!-\frac{1}{8}
        Y_{\mu\nu.\rho\sigma}(x')
        (x\!-\!x')^3_{\sigma\rho\nu}
    \!-\!\frac{1}{30}
        Y_{\mu\nu.\rho\sigma\kappa}(x')
        (x\!-\!x')^4_{\kappa\sigma\rho\nu}
\end{eqnarray}
where $Y$ (later also $X$) and their covariant derivatives are all
evaluated  at the fixed reference point $x'$.

Similarly, the matrix potential
takes the form
\begin{eqnarray}\label{X44}
    X(x)\!&=&\!X(x')
    \!+\!X_{.\rho}(x')(x\!-\!x')_\rho
    \!+\!\frac{1}{2}
        X_{.\rho\sigma}(x')(x\!-\!x')_{\rho\sigma}^2
    \nonumber
    \\&&\!+\!\frac{1}{6}X_{.\rho\sigma\kappa}(x')(x\!-\!x')_{\rho\sigma\kappa}^3.
\end{eqnarray}
Upon substitution of (\ref{N44}) and (\ref{X44}) into the Green
function equation (\ref{Grnf}), the following differential equation\footnote{In later chapters, we prefer to denote ${\cal D}^n_{\mu_1\ldots\mu_n}$ as ${\cal D}_{\mu_1\ldots\mu_n}$.}
\begin{eqnarray}
    \left\{
    \partial^2
        +X
        -Y_{{\mu}{\nu}}(x-x')_{\nu}\partial_{\mu}
        +\left(
     {\cal D}_{\mu}X
    +\frac{1}{3}{\cal D}_{\nu}Y_{{\mu}{\nu}}
        \right)
        (x-x')_{\mu}
    \right.
&&\nonumber\\
    -\frac{2}{3}{\cal D}_{\nu}Y_{{\rho}{\mu}}
        (x-x')^2_{{\mu}{\nu}}\partial_{\rho}
&&\nonumber\\
    \left[
    \frac{1}{2}{\cal D}^2_{{\nu}{\mu}}X
    +\frac{1}{4}Y_{{\rho}{\mu}}Y_{{\rho}{\nu}}
    -\frac{1}{8}
        \left(
    {\cal D}^2_{{\nu}{\rho}}Y_{{\rho}{\mu}}
    +{\cal D}^2_{{\rho}{\nu}}Y_{{\rho}{\mu}}
        \right)
    \right](x-x')^2_{{\mu}{\nu}}
&&\nonumber\\
    -\frac{1}{8}
    {\cal D}^2_{{\rho}{\nu}}Y_{{\sigma}{\mu}}
        (x-x')^3_{{\mu}{\nu}{\rho}}\partial_{\sigma}
&&\nonumber\\
    +\left[
    \frac{1}{6}
        {\cal D}^3_{{\mu}{\nu}{\rho}}X
    -\frac{1}{30}
        \left(
        {\cal D}^3_{{\rho}{\nu}{\sigma}}Y_{{\sigma}{\mu}}
        +{\cal D}^3_{{\rho}{\sigma}{\nu}}Y_{{\sigma}{\mu}}
        +{\cal D}^3_{{\sigma}{\rho}{\nu}}Y_{{\sigma}{\mu}}
        \right)
    \right.
&&\nonumber\\
    \left.
        +\frac{1}{6}
        \left(
            Y_{{\sigma}{\mu}}{\cal D}_{\rho}Y_{{\sigma}{\nu}}
            +({\cal D}_{\nu}Y_{{\sigma}{\rho}}
        \right)
    \right](x-x')^3_{{\mu}{\nu}{\rho}}
&&\nonumber\\
    -\frac{1}{15}{\cal D}^3_{{\nu}{\rho}{\sigma}}Y_{{\alpha}{\mu}}
        (x-x')^4_{{\mu}{\nu}{\rho}{\sigma}}\partial_{\alpha}
&&\nonumber\\
    +\left(
        \frac{1}{9}
            {\cal D}_{\nu}Y_{{\alpha}{\mu}}
            {\cal D}_{\sigma}Y_{{\alpha}{\rho}}
        +\frac{1}{8}
            Y_{{\alpha}{\mu}}
            {\cal D}^2_{{\sigma}{\rho}}Y_{{\alpha}{\nu}}
    \right)
        (x-x')^4_{{\mu}{\nu}{\rho}{\sigma}}
&&\nonumber\\
    \left(
        \frac{1}{30}
            Y_{{\beta}{\mu}}
            {\cal D}^3_{{\alpha}{\sigma}{\rho}}Y_{{\beta}{\nu}}
        +\frac{1}{12}
            {\cal D}_{\nu}Y_{{\beta}{\mu}}
            {\cal D}^2_{{\alpha}{\sigma}}Y_{{\beta}{\rho}}
    \right)
        (x-x')^5_{{\mu}{\nu}{\rho}{\sigma}{\alpha}}
&&\nonumber\\
    +\left(
        \frac{1}{45}
            {\cal D}_{\nu}Y_{{\gamma}{\mu}}
            {\cal D}^3_{{\beta}{\alpha}{\sigma}}Y_{{\gamma}{\rho}}
        +\frac{1}{64}
            {\cal D}^2_{{\rho}{\nu}}Y_{{\gamma}{\mu}}
            {\cal D}^2_{{\beta}{\alpha}}Y_{{\gamma}{\sigma}}
    \right)
        (x-x')^6_{{\mu}{\nu}{\rho}{\sigma}{\alpha}{\beta}}
&&\nonumber\\
    +\frac{1}{120}
        {\cal D}^2_{{\nu}{\rho}}Y_{{\xi}{\mu}}
        {\cal D}^3_{{\gamma}{\beta}{\alpha}}Y_{{\xi}{\sigma}}
            (x-x')^7_{{\mu}{\nu}{\rho}{\sigma}{\alpha}{\beta}{\gamma}}
&&\nonumber\\ \label{qGf4}
    \left.
    +\frac{1}{900}
        {\cal D}^3_{{\sigma}{\rho}{\nu}}Y_{{\zeta}{\mu}}
        {\cal D}^3_{{\xi}{\gamma}{\beta}}Y_{{\zeta}{\alpha}}
            (x-x')^8_{{\mu}{\nu}{\rho}{\sigma}{\alpha}{\beta}{\gamma}{\xi}}
    \right\}
    \left<\phi(x)\phi(x')\right>
    &\!\!\!\!=\!\!\!\!&-\delta(x,x').
\end{eqnarray}
is obtained, where $X$ and $Y$ and their covariant derivatives are all
evaluated at reference point $x'$. 
Fourier transforming (\ref{qGf4}) to a $D$-dimensional Euclidean
momentum space through
\begin{equation}
    \left<
        \phi(x)\phi(x')
    \right>
        =\int \frac{d^Dp}{(2\pi)^D}e^{ip\cdot (x-x')}G(p).
\end{equation}
This essentially entails the replacements:
\begin{eqnarray}
    (x-x')\to -i\frac{\partial}{\partial p},
    \,\,\,\, \partial \to ip
\end{eqnarray} 
The Fourier transformed Green function $G(p)$ correspondingly
satisfies an eighth-order partial differential equation in momentum
space
\begin{eqnarray}
\left\{ \nonumber
 -p^2 +X
 -Y_{{\mu}{\nu}}p_{\mu}\frac{\partial}{\partial p_{\nu}}
 -i({\cal D}_{\mu}X
   +{\cal D}_{\nu}Y_{{\mu}{\nu}})\frac{\partial}{\partial p_{\mu}}
 +\frac{2}{3}i{\cal D}_{\rho}Y_{{\mu}{\nu}}p_{\mu}
  \frac{\partial^2}{\partial p_{\nu}
                   \partial p_{\rho}}
 \right.
&&
\\ \nonumber
  -\left[
    \frac{1}{2}{\cal D}^2_{{\mu}{\nu}}X
  -\frac{1}{4}Y_{{\mu}{\rho}}Y_{{\rho}{\nu}}
  +\frac{3}{8}
      \left(
             {\cal D}^2_{{\nu}{\rho}} Y_{{\mu}{\rho}}
             - {\cal D}^2_{{\rho}{\nu}} Y_{{\rho}{\mu}}
      \right)
   \right]
\frac{\partial^2}{\partial p_{\mu}
                              \partial p_{\nu}}
&&
\\ \nonumber
+\frac{1}{4}{\cal D}^2_{{\sigma}{\rho}}Y_{{\mu}{\nu}}p_{\mu}
\frac{\partial^3}{\partial p_{\nu}
                              \partial p_{\rho}
                              \partial p_{\sigma}}
&&
\\ \nonumber
+\left[
  \frac{i}{6}{\cal D}^3_{{\rho}{\nu}{\mu}}X
 -\frac{i}{30}{\cal D}^3_{{\rho}{\nu}{\sigma}}Y_{{\sigma}{\mu}}
 -\frac{i}{10}\left({\cal D}^3_{{\rho}{\sigma}{\nu}}Y_{{\sigma}{\mu}}
                    +{\cal D}^3_{{\sigma}{\rho}{\nu}}Y_{{\sigma}{\mu}}
                       \right)
  \right.\,\,\,\,\,\,\,\,\,\,\,\,\,\,\,\,\,\,\,\,\,\,\,\,\,\,\,\,\,\,\,\,\,\,
&&
\\ \nonumber
  \left.
 +\frac{i}{6}\left(Y_{{\sigma}{\mu}}{\cal D}_{{\rho}}Y_{{\sigma}{\nu}}
                           +({\cal D}_{{\nu}}Y_{{\sigma}{\mu}})Y_{{\sigma}{\rho}}
                       \right)
  \right]
\frac{\partial^3}{\partial p_{\mu}
                              \partial p_{\nu}
                             \partial p_{\rho}}
&&
\\ \nonumber
+\left(
  {\cal D}^3_{{\sigma}{\rho}{\nu}}Y_{{\alpha}{\mu}}p_{\alpha}
  +\frac{1}{9}{\cal D}_{{\nu}}Y_{{\alpha}{\mu}}
                        {\cal D}_{{\sigma}}Y_{{\alpha}{\rho}}
  +\frac{1}{8}Y_{{\alpha}{\mu}}
                        {\cal D}^2_{{\sigma}{\rho}}Y_{{\alpha}{\nu}}
  \right)
\frac{\partial^4}{\partial p_{\mu}
                              \partial p_{\nu}
                             \partial p_{\rho}
                             \partial p_{\sigma}}
&&
\\ \nonumber
-\left(
     \frac{i}{30}Y_{{\beta}{\mu}} D^3_{{\alpha}{\sigma}{\rho}}Y_{{\beta}{\nu}}
   +\frac{i}{12}D_{{\nu}}Y_{{\beta}{\mu}} D^2_{{\alpha}{\sigma}}Y_{{\beta}{\rho}}
 \right)
\frac{\partial^5}{\partial p_{\mu}
                              \partial p_{\nu}
                              \partial p_{\rho}
                             \partial p_{\sigma}
                             \partial p_{\alpha}}
&&
\\ \nonumber
-\left(
     \frac{1}{45}{\cal D}_{{\nu}}Y_{{\gamma}{\mu}}
                           {\cal D}^3_{{\beta}{\alpha}{\sigma}}Y_{{\gamma}{\rho}}
 + \frac{1}{64}{\cal D}^2_{{\rho}{\nu}}Y_{{\gamma}{\mu}}
                           {\cal D}^2_{{\beta}{\alpha}}Y_{{\gamma}{\sigma}}
 \right)
\frac{\partial^6}{\partial p_{\mu}
                              \partial p_{\nu}
                              \partial p_{\rho}
                             \partial p_{\sigma}
                             \partial p_{\alpha}
                             \partial p_{\beta}}
&&
\\ \nonumber
    +\frac{i}{120}
        {\cal D}^2_{{\rho}{\nu}}Y_{{\xi}{\mu}}
        {\cal D}^3_{{\gamma}{\beta}{\alpha}}Y_{{\xi}{\sigma}}
\frac{\partial^7}{\partial p_{\mu}
                              \partial p_{\nu}
                              \partial p_{\rho}
                             \partial p_{\sigma}
                             \partial p_{\alpha}
                             \partial p_{\beta}
                             \partial p_{\gamma}}
&&
\\ \left.
    +\frac{1}{900}
        {\cal D}^3_{{\sigma}{\rho}{\nu}}Y_{{\zeta}{\mu}}
        {\cal D}^3_{{\xi}{\gamma}{\beta}}Y_{{\zeta}{\alpha}}
\frac{\partial^8}{\partial p_{\mu}
                              \partial p_{\nu}
                              \partial p_{\rho}
                             \partial p_{\sigma}
                             \partial p_{\alpha}
                             \partial p_{\beta}
                             \partial p_{\gamma}
                             \partial p_{\xi}}
                             \right\}G(p)=-1\nonumber\\ \label{8pde}
\end{eqnarray}

\section{General and Specific Objectives}
Our goal is to solve this eighth-order partial differential equation in momentum-space (\ref{8pde}) explicitly. Such task is equivalent to the calculation of higher mass-dimensional effective Lagrangians for 4 up to 26 mass-dimensions.\footnote{This is counting the total mass-dimensions of the product of field strength tensor (and its covariant derivatives) invariants. The scheme of identifying mass-dimensions of these quantities are described in detail in Section 4.1.} However, we limit our calculation up to 12 mass-dimensions. This corresponds to solution of (\ref{8pde}) which can lead to invariants of mass-dimension equal to or lower than twelve.

In our calculation, the use of a symbolic software\footnote{Thanks to Dr. Jose Tristan Reyes of the Math Department of De La Salle University,  and to Dr. Jerrold Garcia and Mr. William Yu of the Ateneo High Performance Computing Group of the Ateneo de Manila University for allowing me to use their licensed Mathematica software. The other software is Maxima 5.5,  for Windows (with enhancements by W. Schelter) and for Unix-Linux OS. Both were licensed under the GNU Public License. Thanks to Dr. Garcia for introducing me the linux-compiled version of Macsyma. Text-output handling were done with the aid of a spreadsheet software such as MS Excel. In this thesis, we will interchangeably  refer Macsyma with Maxima.} will be implemented for the first time. Consequently, the evaluation eases out complications introduced when handling tensors with a number of indices(either Lorentz or internal).

To verify whether our results agree with those found in literature, we want to compare only in the effective Lagrangians for four, and six mass-dimensions. That is, we implement standard simplification procedures\footnote{The simplification process includes\cite{Mue} the application of product rule, cyclic matrix permutations, integration by parts, Bianchi identites, and antisymmetry of $F_{\mu\nu}$ (in other instances perhaps mirror transformation).} to obtain results that completely agree\footnote{Effective Lagrangians with derivative corrections can be arbitrarily simplified using these identities and relations between $F$, $DF$, for our case including $D^2F$ and $D^3F$ relations. Different usage of such identities and relations can lead to seemingly different result once comparison is made with those found in literature\cite{vanN}.} with those found in literature\cite{Rodf-Diss,Avra,ven,Gilk,Fradk}. We implement such reduction scheme\cite{Mue} for the case of four and six mass-dimensions. For eight mass-dimensions, we limit our reduction process by applying only product rule, cyclic matrix permutations, integration by parts and the antisymmetry of $F_{\mu\nu}$ in the reduction process. Comparison to those found in literature for eight mass-dimensions can only be made once the use of Bianchi identities are applied.\footnote{The goal in the reduction process is to obtain a minimal set of linearly independent (gauge invariant) monomials.}  Results involving ten and twelve mass-dimensions will be presented just to showcase the facility of using symbolic software. We emphasize that the main result of this thesis will revolve around the one-loop effective Lagrangians for eight up to sixteen mass-dimensions in the zeroth-order corrections case (particularly the closed form expression of the ${\cal O}({\cal D}_\mu)$ terms) and up to twelve mass-dimension in the higher-order corrections case. Such reduction process for eight up to twelve mass-dimensions effective Lagrangians and obtain a minimal set of invariants will be reserved in another occasion.

The techniques of handling $n$-fold proper time integration (for $n=2,3$) will be presented and used extensively in our calculation. The case of handling two- up three-fold proper-time integrations were applied in Ref. \cite{Rodf-Diss}. But we present them in Appendix D in a manner that the case when $n>3$ can be obtained similarly and when the integrand is expressed in powers of the proper-time variables.

Also the algorithm used in our calculation can run in computers with limited memory capacity and with symbolic software that can handle at least algebraic manipulations. We treat the monomials as a textual ouput rather than their usual representation as tensor.\footnote{Tensor packages are not necessary. They become necessary when we treat monomials as tensors. See Chapter 6 for a possible recalculation of this work with the aid of tensor packages.} This is because an algebraic equivalent algorithm of (kronecker delta) index contraction is used. Also the algorithm avoids a polynomial expansion that usually leads to a very large number of terms.\footnote{See Section 6.3. Algorithm following direct polynomial expansion algorithm leads to ${10}^4$, ${10}^5$, ${10}^{10}$, and ${10}^{13}$ terms, when first- ($\ell=1$), second- ($\ell=2$), third- ($\ell=3$), fourth-order ($\ell=4$ in Eqn. (\ref{L1G}))  corrections are accommodated respectively.} An ordinary PC can handle ${10}^3$ terms. We device an algorithm that uses about $3,000$ terms but can accommodate higher-order corrections in the calculation of one-loop effective Lagrangians.

\section{Scope and Delimitations}
This work limits itself to the study of fields with a flat metric, eliminating thence the theories with curved metric. The extension to include gravity appears to be soluble for conformal gravity\cite{vanN}.

Calculations are done up to one loop. Dyson-Schwinger equation\cite{Itzy,Oko} extends beyond one-loop correction provided one can obtain an explicit solution to the equation of the appropriate Green function\cite{Brw-Dff}.

Prescriptions for calculating one-loop effective Lagrangians are made for mass-dimensions four up to twelve. The case of four and six mass-dimensions are worked out completely, eight mass-dimensions partially. Higher than eight mass-dimensions, results are presented in their unsimplified form. All of the above-mentioned processes are done with the aid of a symbolic manipulation software. Although we cannot claim that the calculation is totally automated, tasks like simplification procedures with the use of identities and relations (among $F^n,{\cal D}F^{n-1},\ldots,{\cal D}^{n-1}F$) are still done by hand. The calculation therefore can be considered quasi-automated.

Our results will purely belong to a mathematical physics domain. However, we refrain from showing rigorous proofs of convergence, of minimality of set, of linear independence of monomials, etc. Direct relevance and application to physics of the present work will not also be discussed. Such issues of relevance of the calculation are directed as a recommendation for future work\footnote{See Section 6.4}. Also, nothing whatsoever will be mentioned about the advantages and the reasons of using the softwares for certain portions, whole or part, in the implementation of the algorithm with a PC.

\chapter{Solution to the Eighth-order Differential Equation}
In this chapter, we briefly describe in both mathematical as well as algorithmic manner the procedure we will implement in solving the eighth-order differential equation (\ref{8pde}) in momentum space. We present the Gaussian solution and we reuse this six times\footnote{First in ${G_0}_{/{\mu}}$, second in ${G_0}_{/{\mu}{\nu}}$, up to sixth in ${G_0}_{/{\mu}{\nu}{\rho}{\sigma}{\alpha}{\beta}}$, partial momentum-space differentiations.} to solve the rest of the non-Gaussian sector. The solution will generate an expression in powers of momentum integration. The odd-powers of momentum integration vanish leaving the even-powers behind which are then explicitly integrated out.\footnote{The momentum integrations in $D$-dimensions are presented in the appendix. See Appendix C.} Then, the product of invariants, whose total mass-dimensionality is twelve or less is identified.\footnote{The following functions in MS Excel were used for the termwise determination and classification of monomials according to mass-dimensions: \texttt{IF}, \texttt{LEN}, and sort functions.} Then a series of simplifications follow: proper-time integrations\footnote{Handling two- up to three-fold proper-time integrations are presented in the appendix. See Appendix D.}, Bianchi identities\footnote{The use of such identities are exploited only for results relevant in mass-dimensions six or less.}, integration by parts (equivalently total derivatives), and relations between invariants and its covariant derivatives\footnote{Such relations are applied for results relevant in mass-dimensions six or less.}

\section{Gaussian Sector Solution}
Before solving the eighth-order differential equation in momentum space, we consider its quadratic form that leads to a Gaussian-like solution.

For the case when $n=1$ and $l=2$,
\begin{equation}
  Y_{\mu\nu.\rho} = 0, \;\;\;\;  X_{.\rho\sigma} = 0.
  \label{DYDXres}
\end{equation} 
(\ref{qGfmom}) may be written
as
\begin{eqnarray}\label{D0G0}
    \Delta_0(p)G_0(p)=-1,
\end{eqnarray} 
where
\begin{eqnarray}\label{D0p}
    \Delta_0(p)
    =-p^2+X-Y_{\mu\nu}p_\mu\frac{\partial}{\partial p_\nu}
    -iX_{.\mu}\frac{\partial}{\partial p_\mu}
    +\frac{1}{4}Y^2_{\mu\nu}\frac{\partial^2}{\partial p_\mu\partial
    p_\nu}.
\end{eqnarray} 
This case (\ref{D0G0}) was demonstrated to be exactly soluble in
\cite{Rodf-Diss} following Brown and Duff \cite{Brw-Dff}
yielding the result\footnote{The tensors are expressed in matrix form. For example,
\begin{eqnarray*}
\dot{X}\!\cdot\!(iY)^{-3}\left[\tan(iYs)-iYs\right]\!\cdot\!\dot{X}
&\equiv &
        ({\cal D}_\mu X) (iY_{\mu\nu})^{-3}\left[\tan(iY_{\nu\rho}\,s)-iY_{\nu\rho}\,s\right] ({\cal D}_\rho X)
\\
 2i\dot{X}\!\cdot\! Y^{-2}\left[1-\sec (iYs)\right]\!\cdot\! p
&\equiv &
        2i({\cal D}_\mu X)\,Y^{-2}_{\mu\nu}\left[\delta_{\nu\lambda}-\sec (iY_{\nu\lambda}\,s)\right] p_{\lambda}
\\
  \frac{1}{2}p\!\cdot\! Y^{-2}2iY^{-1}\tan(iYs)\!\cdot\! p
&\equiv &
        \frac{1}{2}p_\lambda\, (2iY^{-1}_{\lambda\mu} \tan(iY_{\mu\tau}\,s)\!\cdot\! p_\tau
\end{eqnarray*}
For the fieldstrength tensors, the following are some illustrations on how tensors with indices are equivalently expressed in matrix form:
\begin{eqnarray*}
 Y^{-1}_{\mu\nu}Y_{\mu\nu} = 1
\,\,\,\,\,\,\,\,\,
 Y_{\mu\nu}Y_{\nu\mu} \equiv Y^2
\,\,\,\,\,\,\,\,\,
 Y_{\mu\nu}Y_{\nu\rho}Y_{\rho\mu} Y_{\alpha\beta}Y_{\beta\alpha}\equiv Y^3 Y^2
\,\,\,\,\,\,\,\,\,
{\cal D}_\mu{\cal X}\equiv \dot{{\cal X}}
\end{eqnarray*}
}
\begin{eqnarray}\label{LG}
    G_0(p)
    \!\!\!\!&=&\!\!\!\!\int^\infty_0 ds\,\,\,
    \!\exp\!\left\{
    Xs
    \!+\!\frac{1}{2}\mbox{tr}\ln\sec(iYs)
    \!+\!\dot{X}\!\cdot\!(iY)^{-3}
    \left[\tan(iYs)-iYs\right]\!\cdot\!\dot{X}
    \right.
    \nonumber\\&&
    \left.
    \!+\!2i\dot{X}\!\cdot\! Y^{-2}\left[1-\sec (iYs)\right]\!\cdot\! p
    \!+\!\frac{1}{2}p\!\cdot\! 2iY^{-1}\tan(iYs)\!\cdot\! p
    \right\},
\end{eqnarray}
where tr is a Lorentz trace and $s$ is some proper-time integration variable. Later, it will be convenient to redefine
\begin{eqnarray}\label{Xredef}
   X\equiv -m^2+{\cal X}.
\end{eqnarray}
To recover the free Euclidean propagator,
\begin{eqnarray}
    \lim_{A\to 0} G_0(p)
    =\int^\infty_0 ds\,\,\,e^{-(p^2+m^2)s}=\frac{1}{p^2+m^2}
\end{eqnarray}
one simply switches the background off, that is, $N_\mu\to 0$ and
$X\to -m^2$. The general quasilocal Green function equation
(\ref{qGfmom}) then provides the unrestricted Green function $G$
that accommodates all covariant derivative corrections through the
perturbative expansion\cite{Rodf-Diss}
\begin{eqnarray}\label{G}
    G(p)=-(\Delta_0(p)
      +\Delta_1(p))^{-1}
        =G_0(p)\sum^\infty_{\ell=0}(\Delta_1(p)_{n,l} G_0(p))^\ell,
\end{eqnarray} 
where convergence is assured provided the backgrounds are strong
and slowly varying so that $\Delta_1\ll \Delta_0$. This expansion
(\ref{G}) solves perturbatively the general quasilocal Green function equation (in momentum space)
\begin{eqnarray}\label{DpGp}
  \Delta_1(p)_{n,l}G(p)=-1
\end{eqnarray} 
in terms of the known Green
function $G_0$. 
Here\cite{Tiam,Tiam-Rodf}
\begin{eqnarray}
    \Delta_1(p)_{n,l}
    \!\!\!&=&\!\!\! -p^2+\sum^{l-1}_{q=2}\frac{(-i)^q}{q!}
    {\cal D}^q X\!\circ\!\left(\frac{\partial}{\partial p}\right)^q
    \!+\!\sum^{n-1}_{q=1}\frac{3(-i)^{2+q}(1+q)}{(2+q)!}
    \left[{\cal D}^{(q)}\cdot Y\right]
    \!\circ\!\left(\frac{\partial}{\partial p}\right)^q
    \nonumber\\ \nonumber &&
    +\sum^{n-1}_{q=1}\frac{2(-i)^q(1+q)}{(2+q)!}
    p\cdot
    \left[
    {\cal D}^q Y
    \circ\left(\frac{\partial}{\partial p}\right)^{1+q}\right]
    \\  &&
    +\sum^{n-1}_{r=0}\sum^{n-1}_{q=0}
    (1-\delta_{r0}\delta_{q0})
    \frac{(-i)^{2+r+q}(1+r)(1+q)}{(2+r)!(2+q)!}
    \nonumber\\ &&
    \;\;\;\;\;\;\;\;\;\;\;\;\;\;\;
    \times\left[
    {\cal D}^r Y
    \circ\left(\frac{\partial}{\partial p}\right)^{1+r}\right]
    \cdot
    \left[
    {\cal D}^q Y
    \circ\left(\frac{\partial}{\partial
    p}\right)^{1+q}\right].\nonumber\\\label{qDf}
\end{eqnarray}
This is a general formula for any $n$ and $l$ (usually $l=n+1$ as exemplified below). For example,
\begin{eqnarray}
    \Delta_0(p)
        &=& \Delta_1(p)_{1,2}
\label{1stknd}
 \,\,\,\mathrm{i.e.}\,\,\,{\cal D}Y_{\mu\nu}=0\,\mathrm{and}\,{\cal D}^2\!{\cal X}=0
\\
    \Delta_1(p)_{\mathrm{Rodulfo}} &=& \Delta_1(p)_{2,3}-\Delta_0(p)
 \,\,\,\mathrm{i.e.}\,\,\,{\cal D}^2Y_{\mu\nu}=0\,\mathrm{and}\,{\cal D}^3\!{\cal X}=0
\label{2ndknd}
\\
   \Delta_1(p)_{\mathrm{Tiamzon}} &=& \Delta_1(p)_{3,4}-\Delta_0(p)
 \,\,\,\mathrm{i.e.}\,\,\,{\cal D}^3Y_{\mu\nu}=0\,\mathrm{and}\,{\cal D}^4\!{\cal X}=0
\label{3rdknd}
\end{eqnarray}
This is the operator in the Gaussian sector used in \cite{Brw-Dff}, the non-Gaussian sector used in \cite{Rodf-Diss}, and an improvement in \cite{Tiam}, respectively. This thesis uses the third kind (\ref{3rdknd}). In all cases, $\Delta_0(p)$ is given in (\ref{D0p}). From this time on, $\Delta_1(p)_{\mathrm{Tiamzon}}$ in (\ref{3rdknd}) will be referred to as as $\Delta_1(p)$.

The one-loop effective Lagrangian  calculated from the coincidence
limit of the two-point function in coordinate space (\ref{Corrf})
is given by
\begin{eqnarray}\label{L1G}
    {\cal L}^{(1)}
    &=&\frac{\hbar}{2(2\pi)^D}
    \mbox{Tr}
    \int dX\int d^Dp\,\,\,G_0(p)\sum^{\infty}_{\ell=0}
    \left(\Delta_1(p) G_0\right)^\ell
    \\&\equiv &{\cal L}^{(1)}_0+{\cal L}^{(1)}_1+{\cal
    L}^{(1)}_2+\ldots,
\end{eqnarray}
where
\begin{eqnarray}\label{L0L1L2}
  \begin{array}{c}
          {\cal L}^{(1)}_0
    = \frac{\hbar}{2(2\pi)^D}\mbox{Tr}\int dX\int d^Dp\,\,\,G_0(p)
    \\ \\
        {\cal L}^{(1)}_1
    = \frac{\hbar}{2(2\pi)^D}\mbox{Tr}\int dX
    \int d^Dp\,\,\,G_0(p)\Delta_1(p)G_0(p)
     \\ \\
    {\cal L}^{(1)}_2
    = \frac{\hbar}{2(2\pi)^D}\mbox{Tr}\int dX
    \int d^Dp\,\,\,G_0(p)\Delta_1(p)G_0(p)\Delta_1(p)G_0(p)\;\\ \vdots \\ \mbox{etc.}
  \end{array}
\end{eqnarray}
and Tr denotes a trace over gauge indices (including possibly spinor indices).
In this work, we will be working up to ${\cal L}^{(1)}_4$.

\section{Partial Momentum-Space Derivatives of $G_0(p)$}
It can be seen from (\ref{8pde}) with (\ref{G}) that the operator (\ref{qDf}) with the covariant restrictions (\ref{3rdknd}) is explicitly written as
\begin{eqnarray}
\nonumber
\Delta_1(p)\,\,\,\equiv\,\,\,
 -i(
   {\cal D}_{\nu}Y_{{\mu}{\nu}})\frac{\partial}{\partial p_{\mu}}
 +\frac{2}{3}i({\cal D}_{\rho}Y_{{\mu}{\nu}})p_{\mu}
  \frac{\partial^2}{\partial p_{\nu}
                   \partial p_{\rho}}
&&
\\ \nonumber
  -\left[
    \frac{1}{2}({\cal D}^2_{{\mu}{\nu}}X)
  +\frac{3}{8}
      \left(
             ({\cal D}^2_{{\nu}{\rho}} Y_{{\mu}{\rho}})
             - ({\cal D}^2_{{\rho}{\nu}} Y_{{\rho}{\mu}})
      \right)
   \right]
\frac{\partial^2}{\partial p_{\mu}
                              \partial p_{\nu}}
+\frac{1}{4}({\cal D}^2_{{\sigma}{\rho}}Y_{{\mu}{\nu}})p_{\mu}
\frac{\partial^3}{\partial p_{\nu}
                              \partial p_{\rho}
                              \partial p_{\sigma}}
&&
\\ \nonumber
+\left[
  \frac{i}{6}({\cal D}^3_{{\rho}{\nu}{\mu}}X)
 -\frac{i}{30}({\cal D}^3_{{\rho}{\nu}{\sigma}}Y_{{\sigma}{\mu}})
 -\frac{i}{10}\left(({\cal D}^3_{{\rho}{\sigma}{\nu}}Y_{{\sigma}{\mu}})
                    +({\cal D}^3_{{\sigma}{\rho}{\nu}}Y_{{\sigma}{\mu}})
                       \right)
  \right.\,\,\,\,\,\,\,\,\,\,\,\,\,\,\,\,\,\,\,\,\,\,\,\,\,\,\,\,\,\,\,\,\,\,
&&
\\ \nonumber
  \left.
 +\frac{i}{6}\left(Y_{{\sigma}{\mu}}({\cal D}_{{\rho}}Y_{{\sigma}{\nu}})
                           +({\cal D}_{{\nu}}Y_{{\sigma}{\mu}})Y_{{\sigma}{\rho}}
                       \right)
  \right]
\frac{\partial^3}{\partial p_{\mu}
                              \partial p_{\nu}
                             \partial p_{\rho}}
&&
\\ \nonumber
+\left(
  ({\cal D}^3_{{\sigma}{\rho}{\nu}}Y_{{\alpha}{\mu}})p_{\alpha}
  +\frac{1}{9}({\cal D}_{{\nu}}Y_{{\alpha}{\mu}})
                        ({\cal D}_{{\sigma}}Y_{{\alpha}{\rho}})
  +\frac{1}{8}Y_{{\alpha}{\mu}}
                        ({\cal D}^2_{{\sigma}{\rho}}Y_{{\alpha}{\nu}})
  \right)
\frac{\partial^4}{\partial p_{\mu}
                              \partial p_{\nu}
                             \partial p_{\rho}
                             \partial p_{\sigma}}
&&
\\ \nonumber
-\left(
     \frac{i}{30}Y_{{\beta}{\mu}} (D^3_{{\alpha}{\sigma}{\rho}}Y_{{\beta}{\nu}})
   +\frac{i}{12}(D_{{\nu}}Y_{{\beta}{\mu}}) (D^2_{{\alpha}{\sigma}}Y_{{\beta}{\rho}})
 \right)
\frac{\partial^5}{\partial p_{\mu}
                              \partial p_{\nu}
                              \partial p_{\rho}
                             \partial p_{\sigma}
                             \partial p_{\alpha}}
&&
\\ \nonumber
-\left(
     \frac{1}{45}({\cal D}_{{\nu}}Y_{{\gamma}{\mu}})
                           ({\cal D}^3_{{\beta}{\alpha}{\sigma}}Y_{{\gamma}{\rho}})
 + \frac{1}{64}({\cal D}^2_{{\rho}{\nu}}Y_{{\gamma}{\mu}})
                           ({\cal D}^2_{{\beta}{\alpha}}Y_{{\gamma}{\sigma}})
 \right)
\frac{\partial^6}{\partial p_{\mu}
                              \partial p_{\nu}
                              \partial p_{\rho}
                             \partial p_{\sigma}
                             \partial p_{\alpha}
                             \partial p_{\beta}}
&&
\\ \nonumber
    +\frac{i}{120}
        ({\cal D}^2_{{\rho}{\nu}}Y_{{\delta}{\mu}})
        ({\cal D}^3_{{\gamma}{\beta}{\alpha}}Y_{{\delta}{\sigma}})
\frac{\partial^7}{\partial p_{\mu}
                              \partial p_{\nu}
                              \partial p_{\rho}
                             \partial p_{\sigma}
                             \partial p_{\alpha}
                             \partial p_{\beta}
                             \partial p_{\gamma}}
&&
\\
    +\frac{1}{900}
        ({\cal D}^3_{{\sigma}{\rho}{\nu}}Y_{{\zeta}{\mu}})
        ({\cal D}^3_{{\xi}{\gamma}{\beta}}Y_{{\zeta}{\alpha}})
\frac{\partial^8}{\partial p_{\mu}
                              \partial p_{\nu}
                              \partial p_{\rho}
                             \partial p_{\sigma}
                             \partial p_{\alpha}
                             \partial p_{\beta}
                             \partial p_{\gamma}
                             \partial p_{\xi}}. \label{D1p8}
\end{eqnarray}
We intend to let this operator act on (\ref{LG}) with the prescription (\ref{G}) with $\ell=0$ up to $\ell=4$.

The last two terms of (\ref{8pde}) [correspondingly the last two terms in (\ref{D1p8})] will not contribute product of invariants (and its derivatives) of total mass-dimensions equal to twelve or less in our calcuation. In addition, terms containing ${\cal D}^3Y$ or ${\cal D}^4 X$ vanish too as covariantly restricted by Eqn. (\ref{3rdknd}). That is, we have the further-reduced $(\Delta_1G_0(p))^{\mathrm{red}}$ sixth-order in momentum-space derivatives:
\begin{eqnarray}
\nonumber
(\Delta_1G_0(p))^{\mathrm{red}} \!\!\!&\equiv &\!\!\!
 -i(
   {\cal D}_{\nu}Y_{{\mu}{\nu}})G_{0/\mu} 
 \!+\!\frac{2}{3}i({\cal D}_{\rho}Y_{{\mu}{\nu}})p_{\mu}
        G_{0/\nu\rho}  
\\ \nonumber
&&
  \!-\!\left[
    \frac{1}{2}({\cal D}^2_{{\mu}{\nu}}X)
  \!+\!\frac{3}{8}
      \left(
             ({\cal D}^2_{{\nu}{\rho}} Y_{{\mu}{\rho}})
             \!-\! ({\cal D}^2_{{\rho}{\nu}} Y_{{\rho}{\mu}})
      \right)
   \right]
       G_{0/\mu\nu}  
\!+\!\frac{1}{4}({\cal D}^2_{{\sigma}{\rho}}Y_{{\mu}{\nu}})p_{\mu}
       G_{0/\nu\rho\sigma}
\\ \nonumber
&&
\!+\!\left[
  \frac{i}{6}({\cal D}^3_{{\rho}{\nu}{\mu}}X)
 \!+\!\frac{i}{6}\left[Y_{{\sigma}{\mu}}({\cal D}_{{\rho}}Y_{{\sigma}{\nu}})
                           \!+\!({\cal D}_{{\nu}}Y_{{\sigma}{\mu}})Y_{{\sigma}{\rho}}
                       \right]
  \right]
       G_{0/\mu\nu\rho}
\\ \nonumber
&&
+\left[
  \frac{1}{9}({\cal D}_{{\nu}}Y_{{\alpha}{\mu}})
                        ({\cal D}_{{\sigma}}Y_{{\alpha}{\rho}})
  \!+\!\frac{1}{8}Y_{{\alpha}{\mu}}
                        ({\cal D}^2_{{\sigma}{\rho}}Y_{{\alpha}{\nu}})
  \right]
       G_{0/\mu\nu\rho\sigma}
\\ \nonumber
&&
   \!-\!\frac{i}{12}(D_{{\nu}}Y_{{\beta}{\mu}}) (D^2_{{\alpha}{\sigma}}Y_{{\beta}{\rho}})
       G_{0/\mu\nu\rho\sigma\alpha}
 \!-\frac{1}{64}({\cal D}^2_{{\rho}{\nu}}Y_{{\gamma}{\mu}})
                           ({\cal D}^2_{{\beta}{\alpha}}Y_{{\gamma}{\sigma}})
       G_{0/\mu\nu\rho\sigma\alpha\beta}
\nonumber\\
\!\!\!\!\!\!\!\!\!\!\!\!\!
\label{6DG}
\end{eqnarray}
That is, we have the ensuing task specifically in the operator (\ref{D1p8}) when acting on $G_0(p)$ is reduced to solving momentum-space partial derivatives up to order six only:
\begin{eqnarray}
   {G_0 }_{/{\mu}}
\,\,\,\,\,
   {G_0 }_{/{\mu}{\nu}}
\,\,\,\,\,
\ldots
\,\,\,\,\,
   {G_0 }_{/{\mu}{\nu}{\rho}{\sigma}{\alpha}{\beta}}.
\label{G0ps}
\end{eqnarray}

The sixth-order momentum-space partial differential equation given (\ref{6DG}) is good only for incorporating first-order correction as prescribed in the second equation of (\ref{L0L1L2}) specifically the
\begin{eqnarray}
   G_0(p)((\Delta_1(p) G_0(p))
\label{task1}
\end{eqnarray} 
part. Since we have the goal of incorporating up to fourth-order corrections prescribed by (\ref{L0L1L2}) from ${\cal L}^{(1)}_2$ to ${\cal L}^{(1)}_4$ specifically the portions:
\begin{eqnarray}
\label{task12}
   && G_0(p)(\Delta_1(p) G_0(p))_1(\Delta_1(p) G_0(p))_2
\\ && G_0(p)(\Delta_1(p) G_0(p))_1(\Delta_1(p) G_0(p))_2(\Delta_1(p) G_0(p))_3
\label{task13}
\\ && G_0(p)(\Delta_1(p) G_0(p))_1(\Delta_1(p) G_0(p))_2(\Delta_1(p) G_0(p))_3(\Delta_1(p) G_0(p))_4,
\label{task2}
\end{eqnarray}
it is therefore appropriate to introduce a subscript $\ell$ as in $(\Delta_1(p) G_0(p))_\ell$ to avoid repeating the indices more than two times. In this regard, we propose to implement the following system of replacements of variables in $G_0(p))$ as given in (\ref{D0p}) with  the following:
\begin{eqnarray}
 s \to s_\ell
\end{eqnarray}
for the proper-time variable in (\ref{LG}) and in (\ref{D1p8}), and system of index replacements
\begin{eqnarray} \label{gklat1}
\{{\mu},{\nu},{\rho},{\sigma},{\alpha},{\beta}\}
& \to &
\{{\mu}_{\ell},{\nu}_{\ell},{\rho}_{\ell},{\sigma}_{\ell},{\alpha}_{\ell},{\beta}_{\ell}\}
\\ \label{gklat2}
\{{\lambda},{\tau},{\kappa},{\eta},{\xi},{\varrho}\}
& \to &
\{{\lambda}_{\ell},{\tau}_{\ell},{\kappa}_{\ell},{\eta}_{\ell},{\xi}_{\ell},{\varrho}_{\ell}\}.
\end{eqnarray}
for the $X$ and $Y$, and for the momentum tensors, respectively. As an illustration, the product
\begin{eqnarray}
   G_0(p)((\Delta_1(p) G_0(p))_1((\Delta_1(p) G_0(p))_2
\end{eqnarray} 
requires the subscripts to be set as $\ell\!=\!1$ in $(\!(\Delta_1\!(p) G_0(p)\!)_1$ and $\ell\!=\!2$ in $(\!(\Delta_1\!(p) G_0\!(p)\!)_1$ with $\ell$-subscripted $(\Delta_1G_0(p))^{\mathrm{red}}_\ell$ sixth-order in momentum-space derivatives:
\begin{eqnarray}
\nonumber
\left(\Delta_1G_0(p)\right)^{\mathrm{red}}_\ell\!\!\!&\equiv &\!\!\!
 -i(
   {\cal D}_{\nu_\ell}Y_{{\mu_\ell}{\nu_\ell}})G_{0/\mu_\ell} 
 \!+\!\frac{2}{3}i({\cal D}_{\rho_\ell}Y_{{\mu_\ell}{\nu_\ell}})p_{\mu_\ell}
        G_{0/\nu_\ell\rho_\ell}  
\\ \nonumber
&&
  \!-\!\left[
    \frac{1}{2}({\cal D}^2_{{\mu_\ell}{\nu_\ell}}X)
  \!+\!\frac{3}{8}
      \left(
             ({\cal D}^2_{{\nu_\ell}{\rho_\ell}} Y_{{\mu_\ell}{\rho_\ell}})
             \!-\! ({\cal D}^2_{{\rho_\ell}{\nu_\ell}} Y_{{\rho_\ell}{\mu_\ell}})
      \right)
   \right]
       G_{0/\mu_\ell\nu_\ell}  
\\ \nonumber
&&
\!+\!\frac{1}{4}({\cal D}^2_{{\sigma_\ell}{\rho_\ell}}Y_{{\mu_\ell}{\nu_\ell}})p_{\mu_\ell}
       G_{0/\nu_\ell\rho_\ell\sigma_\ell}
%
%
\\ \nonumber
&&
\!+\!\left[
  \frac{i}{6}({\cal D}^3_{{\rho_\ell}{\nu_\ell}{\mu_\ell}}X)
 \!+\!\frac{i}{6}\left(Y_{{\sigma_\ell}{\mu_\ell}}({\cal D}_{{\rho_\ell}}Y_{{\sigma_\ell}{\nu_\ell}})
                           \!+\!({\cal D}_{{\nu_\ell}}Y_{{\sigma_\ell}{\mu_\ell}})Y_{{\sigma_\ell}{\rho_\ell}}
                       \right)
  \right]
       G_{0/\mu_\ell\nu_\ell\rho_\ell}
%
%
\\ \nonumber
&&
+\left[
  \frac{1}{9}({\cal D}_{{\nu_\ell}}Y_{{\alpha_\ell}{\mu_\ell}})
                        ({\cal D}_{{\sigma_\ell}}Y_{{\alpha_\ell}{\rho_\ell}})
  \!+\!\frac{1}{8}Y_{{\alpha_\ell}{\mu_\ell}}
                        ({\cal D}^2_{{\sigma_\ell}{\rho_\ell}}Y_{{\alpha_\ell}{\nu_\ell}})
  \right]
       G_{0/\mu_\ell\nu_\ell\rho_\ell\sigma_\ell}
%
%
\\ \nonumber
&&
   \!-\frac{i}{12}(D_{{\nu_\ell}}Y_{{\beta_\ell}{\mu_\ell}})(D^2_{{\alpha_\ell}{\sigma_\ell}}Y_{{\beta_\ell}{\rho_\ell}})
       G_{0/\mu_\ell\nu_\ell\rho_\ell\sigma_\ell\alpha_\ell}
%
%
\\ 
&&
 \!-\frac{1}{64}({\cal D}^2_{{\rho_\ell}{\nu_\ell}}Y_{{\gamma_\ell}{\mu_\ell}})
                           ({\cal D}^2_{{\beta_\ell}{\alpha_\ell}}Y_{{\gamma_\ell}{\sigma_\ell}})
       G_{0/\mu_\ell\nu_\ell\rho_\ell\sigma_\ell\alpha_\ell\beta_\ell}
%
%
%
%
\label{6DG_ell}
\end{eqnarray}

That is, (\ref{L0L1L2}) becomes
\begin{eqnarray}\label{L0L1L2_ell}
  \begin{array}{c}
          {\cal L}^{(1)}_0
    = \frac{\hbar}{2(2\pi)^D}\mbox{Tr}\int dX\int d^Dp\,\,\,G_0(p)
    \\ \\
        {\cal L}^{(1)}_1
    = \frac{\hbar}{2(2\pi)^D}\mbox{Tr}\int dX
    \int d^Dp\,\,\,G_0(p)\left(\Delta_1G_0(p)\right)^{\mathrm{red}}_{\ell=1}
     \\ \\
    {\cal L}^{(1)}_2
    = \frac{\hbar}{2(2\pi)^D}\mbox{Tr}\int dX
    \int d^Dp\,\,\,G_0(p)\left(\Delta_1G_0(p)\right)^{\mathrm{red}}_{\ell=1}\left(\Delta_1G_0(p)\right)^{\mathrm{red}}_{\ell=2}\;\\ \vdots \\ \mbox{etc.}
  \end{array}
\end{eqnarray}
We will be working up to $\left(\Delta_1G_0(p)\right)^{\mathrm{red}}_{\ell=4}$ for ${\cal L}^{(1)}_4$ only.

\section{$G_0(p)$ and $G_{0/\mu\nu\rho\ldots}(p)$}

Before setting $\ell$ subscripts to any positive integer $k$ and build a $(\Delta_1(p) G_0(p))_k$, let us first dissect the exactly soluble sector $G_0(p)$ as given in (\ref{LG}). One may rewrite (\ref{LG}) in terms of $P(s)$, $Q(s)$ and $R(s)$
\begin{eqnarray}\label{Gx}
    G_0 (p)
        =\int_0^\infty ds\,\,\,
        e^{Xs+P(s)+Q(s)\cdot p+\frac{1}{2}p\cdot R(s)\cdot p}.
\end{eqnarray}
where
\begin{eqnarray}
P(s) &=& \dot{X}\cdot Y^{-3}\left(Ys+i\tan iYs\right) \cdot \dot{X} + \frac{1}{2}\mathrm{tr}\ln \sec iYs
\label{Ps}
\\
Q(s) &=& 2i \dot{X}\cdot Y^{-2}\left( 1 - \sec iYs\right)
\label{Qs}
\\
R(s) & = & 2 i Y^{-1} \tan iYs
\label{Rs}
\end{eqnarray}
in matrix form.

Now we let $\Delta_1(p)$ given in (\ref{D1p8}) with ${\cal D}^4{\cal X} =0$ and ${\cal D}^3 Y =0$ act on (\ref{Gx}). This is just the reduced sixth-order partial differential equation in momentum-space given in (\ref{6DG}). We wish to compute for (\ref{G0ps}), the partial momentum derivatives, in terms of (\ref{Ps}), (\ref{Qs}), and (\ref{Rs}).
For purposes of generalization, we choose to let
\begin{eqnarray}
  \Theta &=& Xs+P+Q\cdot p+\frac{1}{2}p\cdot R \cdot p
\label{THETA}
\\
         &=& Xs + P + Q_\lambda p_\lambda + \frac{1}{2} p_\lambda R_{\lambda\tau} p_\tau
 \label{tensorform}
\end{eqnarray}
expressed first in matrix form (\ref{THETA}) then in tensorial form (\ref{tensorform}).
So that (\ref{Gx}) can be rewritten as
\begin{eqnarray}
  G_0(p) = \int_0^\infty ds\,\,\, e^{\Theta}.
\label{Gxrewrite}
\end{eqnarray}
The first derivative in momentum space\footnote{Maxima instruction to obtain $G_{\emptyset/\mu}(p)$ up to $G_{\emptyset/\mu\nu\rho\ldots\beta}(p)$ can be found in Appendix B.2.\\}, for example, is given by
\begin{eqnarray}
\frac{\partial G_0(p)}{\partial p_{{\nu}_{\ell}}}\equiv
G_{0/{\mu}_{\ell}}(p)
 =\int^\infty_0 ds_\ell\,\,\,e^{\Theta}
 \left[
 R_{{{\mu}_{\ell}}{{\lambda}_{\ell}}}p_{{{\lambda}_{\ell}}}+Q_{{{\mu}_{\ell}}}
 \right].
\label{Gpa}
\end{eqnarray}
We present below the first up to sixth partial derivative of the exactly soluble sector with respect to momentum-space variable $p$:
\begin{eqnarray}
G_{0/{\mu}_{\ell}{\nu}_{\ell}}(p)
&=&\int^\infty_0 ds\,\,\,e^{\Theta}
 \left[
  p_{{{\lambda}_{\ell}}} p_{{{\tau}_{\ell}}} R_{{{\mu}_{\ell}}{{\lambda}_{\ell}}} R_{{{\nu}_{\ell}}{{\tau}_{\ell}}}
\right.
\nonumber\\&&\,\,\,\,\,\,\,\,\,\,\,
\left.
 + p_{{{\lambda}_{\ell}}} (Q_{{{\mu}_{\ell}}} R_{{{\nu}_{\ell}}{{\lambda}_{\ell}}} + Q_{{{\nu}_{\ell}}} R_{{{\mu}_{\ell}}{{\lambda}_{\ell}}})
 + R_{{{\mu}_{\ell}}{{\nu}_{\ell}}} + Q_{{{\mu}_{\ell}}} Q_{{{\nu}_{\ell}}}
 \right]\label{Gpab}
\end{eqnarray}
\begin{eqnarray}
G_{0/{\mu}_{\ell}{\nu}_{\ell}{\rho}_{\ell}}(p)
&=&\int^\infty_0 ds\,\,\,e^{\Theta}
 \left[
 R_{{{\mu}_{\ell}}{{\lambda}_{\ell}}} R_{{{\nu}_{\ell}}{{\tau}_{\ell}}} R_{{{\rho}_{\ell}}{{\kappa}_{\ell}}}   p_{{{\lambda}_{\ell}}} p_{{{\tau}_{\ell}}} p_{{{\kappa}_{\ell}}}
\right.
\nonumber\\&&\,\,\,\,\,\,\,\,\,\,\,
 + ( Q_{{{\rho}_{\ell}}} R_{{{\mu}_{\ell}}{{\lambda}_{\ell}}} R_{{{\nu}_{\ell}}{{\tau}_{\ell}}}
    + Q_{{{\mu}_{\ell}}} R_{{{\nu}_{\ell}}{{\lambda}_{\ell}}} R_{{{\rho}_{\ell}}{{\tau}_{\ell}}}
    + Q_{{{\nu}_{\ell}}} R_{{{\mu}_{\ell}}{{\lambda}_{\ell}}} R_{{{\rho}_{\ell}}{{\tau}_{\ell}}}) p_{{{\lambda}_{\ell}}} p_{{{\tau}_{\ell}}}
\nonumber\\&&\!\!\!\!\!\!\!\!\!\!\!\!\!\!\!\!\!\!\!\!
 + (R_{{{\mu}_{\ell}}{{\nu}_{\ell}}} R_{{{\rho}_{\ell}}{{\lambda}_{\ell}}}
   + Q_{{{\mu}_{\ell}}} Q_{{{\nu}_{\ell}}} R_{{{\rho}_{\ell}}{{\lambda}_{\ell}}}
   + Q_{{{\mu}_{\ell}}} Q_{{{\rho}_{\ell}}} R_{{{\nu}_{\ell}}{{\lambda}_{\ell}}}
   + Q_{{{\nu}_{\ell}}} Q_{{{\rho}_{\ell}}} R_{{{\mu}_{\ell}}{{\lambda}_{\ell}}}
   +R_{{{\mu}_{\ell}}}c R_{{{\nu}_{\ell}}{{\lambda}_{\ell}}}
\nonumber\\&&\!\!\!\!\!\!\!\!\!\!
   \!\!\!+\! R_{{{\mu}_{\ell}}{{\lambda}_{\ell}}} R_{{{\nu}_{\ell}}{{\rho}_{\ell}}})   p_{{{\lambda}_{\ell}}}
 \!+\! Q_{{{\rho}_{\ell}}} R_{{{\mu}_{\ell}}{{\nu}_{\ell}}}
\left.
 \!+\! Q_{{{\mu}_{\ell}}} R_{{{\nu}_{\ell}}{{\rho}_{\ell}}}
 \!+\! Q_{{{\nu}_{\ell}}} R_{{{\mu}_{\ell}}{{\rho}_{\ell}}}
 \!+\! Q_{{{\mu}_{\ell}}} Q_{{{\nu}_{\ell}}} Q_{{{\rho}_{\ell}}}
 \right]
\label{Gpabc}
\end{eqnarray}
The rest of the momentum-space partial derivatives (fourth- up to sixth-order) of $G_{0}(p)$ are found in Appendix A. See Eqns. (\ref{Gpabcd})-(\ref{Gpabcdef}).

\section{$G_\emptyset(p)$ and $G_{\emptyset/\mu\nu\rho\ldots}(p)$}
After obtaining the general expressions (\ref{Gpa})-(\ref{Gpabc})  and (\ref{Gpabcd})-(\ref{Gpabcdef}), it is now appropriate to find the Taylor series expansion of (\ref{Qs}) and (\ref{Rs})
\begin{eqnarray}
Q(s) &=& 2i \dot{X}\cdot Y^{-2}\left( 1 - \sec iYs\right)
\nonumber\\
     &=& + i s^2 \dot{X}
         -\frac{5i}{12} s^4 \dot{X} Y^2
         +\frac{61i}{360} s^6 \dot{X} Y^4
         -\frac{277i}{4032} s^8 \dot{X} Y^8
         +\ldots
\label{taylQs}
\end{eqnarray}
\begin{eqnarray}
 R(s) = 2i\tan iYs
 =
 \!-2\,s
 \!+ \frac{2}{3} \,{s^3}\,{Y^2}
 \!- \frac{4}{15} \,{s^5}\,{Y^4}
 \!+ \frac{34}{315} \,{s^7}\,{Y^6}
 \!- \frac{124}{2835} \,{s^9}\,{Y^8}
 \!+ \ldots
\label{taylRs}
\end{eqnarray}
in matrix form. Considering the first term in (\ref{Qs}) and the first two terms in (\ref{taylRs}), (\ref{Gpa}), for example, becomes
\begin{eqnarray}
G_{\emptyset/{\mu}_{\ell}}(p)
 = \int^\infty_0 ds_\ell\,\,\,e^{\Theta_{\mathrm{TSE}}}
     \left(-2\,s_\ell\, \delta_{\mu_\ell\lambda_\ell}
   +i s^2 {\cal D}_{\mu_\ell} X
   + \frac{2}{3} \,{s_\ell}^3\,Y^2_{\mu_\ell\lambda_\ell}\right) p_{{{\lambda}_{\ell}}}
\label{Gpataylor}
\end{eqnarray}
with $\emptyset$ subscript in $G_{\emptyset}(p)$ reminding us that we are using the Taylor series expanded $G_0(p)$.\footnote{
Using the system of index replacements presented in (\ref{gklat1})-(\ref{gklat2}), in maxima code this is \texttt{Gpb} is the variable name assigned to $G_{0/\mu_\ell}(p)$. \texttt{Gpa:R\_azlz *p\_lz= -2 *sz *p\_lz + i*sz2 * D\_X +(2/3) *sz*sz*sz* Y2\_azlz *p\_lz} with z replaced subscript $\ell$. See Appendix B. \\}

In (\ref{Gpataylor}), (\ref{THETA}) with (\ref{Ps})-(\ref{Rs}) in Taylor-series expanded (TSE) form is
\begin{eqnarray}
 \Theta_{\mathrm{TSE}} = Xs
            -\frac{1}{4}s^2\, (\mathrm{tr}Y^2-4i\dot{X}\cdot p)
            +\frac{1}{2} p\cdot \left( -2\,s\,\mathbf{1} + \frac{2}{3} \,{s^3}\,{Y^2}\right)\cdot p.
\label{tse}
\end{eqnarray}
so that (\ref{Gxrewrite}) becomes
\begin{eqnarray}
  G_{\emptyset}(p) = \int^\infty_0 ds\,\,\, e^{\Theta_{\mathrm{TSE}}}.
\label{Gxrewritten}
\end{eqnarray}

This is because
\begin{eqnarray}
P(s) &=& \dot{X}\cdot Y^{-3}\left(Ys+i\tan iYs\right) \cdot \dot{X} + \frac{1}{2}\mathrm{tr}\ln \sec iYs
\nonumber\\
     &=& \dot{X}\cdot Y^{-3}
        \left[
          \frac{1}{3}s^3 Y^3
         -\frac{2}{15} s^5 Y^5
         +\frac{17}{315} s^7 Y^7
         -\frac{62}{2835} s^9 Y^9 +\ldots
        \right]\cdot \dot{X}
\nonumber\\&&
         -\frac{1}{4}s^2 \mathrm{tr}Y^2 +\frac{1}{24} s^4 \mathrm{tr}Y^4
         -\frac{1}{90} s^6 \mathrm{tr}Y^6 + \frac{17}{5040} s^8 \mathrm{tr}Y^8
         +\ldots
\label{taylPs}
\end{eqnarray}
$Q(s)$ from (\ref{taylQs}), and $R(s)$ from (\ref{taylRs}).

The following are the Taylor-series expanded equivalent of Eqns. (\ref{Gpa})-(\ref{Gpabc}), and (\ref{Gpabcd})-(\ref{Gpabcdef}):
\begin{eqnarray}
G_{\emptyset/{\mu}_{\ell}{\nu}_{\ell}}(p)
\!\!&=&\!\! \int^\infty_0 ds_\ell\,\,\,e^{\Theta_{\mathrm{TSE}}}
\left[\left(
 \frac{4}{9} {s_\ell}^6 Y^2_{{\mu_\ell}{\lambda_\ell}} Y^2_{{\nu_\ell}{\tau_\ell}}
 -\frac{4}{3} {s_\ell}^4 \delta_{{\mu_\ell}{\lambda_\ell}} Y^2_{{\nu_\ell}{\tau_\ell}}
 -\frac{4}{3} {s_\ell}^4 \delta_{{\nu_\ell}{\tau_\ell}} Y^2_{{\mu_\ell}{\lambda_\ell}}
\right.\right.
\nonumber\\&&
\left.
 +4 {s_\ell}^2 \delta_{{\mu_\ell}{\lambda_\ell}} \delta_{{\nu_\ell}{\tau_\ell}}
\right)p_{\lambda_\ell} p_{\tau_\ell}
 +\frac{2}{3} Y^2_{{\mu_\ell}{\nu_\ell}} {s_\ell}^3
 - 2\, \delta_{{\mu_\ell}{\nu_\ell}} {s_\ell}
- {\cal D}_{{\mu_\ell}}X {\cal D}_{{\nu_\ell}}X {s_\ell}^2
\nonumber\\&&
-2i {\cal D}_{{\mu_\ell}}X {d}_{{\nu_\ell}{\lambda_\ell}} p_{\lambda_\ell} {s_\ell}^2
-2i {d}_{{\mu_\ell}{\lambda_\ell}} {\cal D}_{{\nu_\ell}}X p_{\lambda_\ell} {s_\ell}^2
+\frac{2i}{3} {\cal D}_{{\nu_\ell}}X p_{\lambda_\ell} {s_\ell}^4 Y^2_{{\mu_\ell}{\lambda_\ell}}
\nonumber\\&&
\left.
+\frac{2i}{3} {\cal D}_{{\mu_\ell}}X p_{\lambda_\ell} {s_\ell}^4 Y^2_{{\nu_\ell}{\lambda_\ell}}
\right]\,\,\,\,\,\,\,\mathrm{etc.}
\label{Gpabtaylor}
\end{eqnarray}

Then (\ref{6DG_ell}) with subscript $0\to \emptyset$ and $\Theta \to \Theta_{\mathrm{TSE}}$ becomes
\begin{eqnarray}
\nonumber
\left(\Delta_1G_{\emptyset}(p)\right)^{\mathrm{red}}_\ell\!\!\!&\equiv &\!\!\!
 -i(
   {\cal D}_{\nu_\ell}Y_{{\mu_\ell}{\nu_\ell}})G_{\emptyset/\mu_\ell} 
 \!+\!\frac{2}{3}i({\cal D}_{\rho_\ell}Y_{{\mu_\ell}{\nu_\ell}})p_{\mu_\ell}
        G_{\emptyset/\nu_\ell\rho_\ell}  
\\ \nonumber
&&
  \!-\!\left[
    \frac{1}{2}({\cal D}^2_{{\mu_\ell}{\nu_\ell}}X)
  \!+\!\frac{3}{8}
      \left(
             ({\cal D}^2_{{\nu_\ell}{\rho_\ell}} Y_{{\mu_\ell}{\rho_\ell}})
             \!-\! ({\cal D}^2_{{\rho_\ell}{\nu_\ell}} Y_{{\rho_\ell}{\mu_\ell}})
      \right)
   \right]
       G_{\emptyset/\mu_\ell\nu_\ell}  
\\ \nonumber
&&
\!+\!\frac{1}{4}({\cal D}^2_{{\sigma_\ell}{\rho_\ell}}Y_{{\mu_\ell}{\nu_\ell}})p_{\mu_\ell}
       G_{\emptyset/\nu_\ell\rho_\ell\sigma_\ell}
%
%
\\ \nonumber
&&
\!+\!\left[
  \frac{i}{6}({\cal D}^3_{{\rho_\ell}{\nu_\ell}{\mu_\ell}}X)
 \!+\!\frac{i}{6}\left(Y_{{\sigma_\ell}{\mu_\ell}}({\cal D}_{{\rho_\ell}}Y_{{\sigma_\ell}{\nu_\ell}})
                           \!+\!({\cal D}_{{\nu_\ell}}Y_{{\sigma_\ell}{\mu_\ell}})Y_{{\sigma_\ell}{\rho_\ell}}
                       \right)
  \right]
       G_{\emptyset/\mu_\ell\nu_\ell\rho_\ell}
%
%
\\ \nonumber
&&
+\left[
  \frac{1}{9}({\cal D}_{{\nu_\ell}}Y_{{\alpha_\ell}{\mu_\ell}})
                        ({\cal D}_{{\sigma_\ell}}Y_{{\alpha_\ell}{\rho_\ell}})
  \!+\!\frac{1}{8}Y_{{\alpha_\ell}{\mu_\ell}}
                        ({\cal D}^2_{{\sigma_\ell}{\rho_\ell}}Y_{{\alpha_\ell}{\nu_\ell}})
  \right]
       G_{\emptyset/\mu_\ell\nu_\ell\rho_\ell\sigma_\ell}
%
%
\\ \nonumber
&&
   \!-\frac{i}{12}(D_{{\nu_\ell}}Y_{{\beta_\ell}{\mu_\ell}})(D^2_{{\alpha_\ell}{\sigma_\ell}}Y_{{\beta_\ell}{\rho_\ell}})
       G_{\emptyset/\mu_\ell\nu_\ell\rho_\ell\sigma_\ell\alpha_\ell}
%
%
\\ 
&&
 \!-\frac{1}{64}({\cal D}^2_{{\rho_\ell}{\nu_\ell}}Y_{{\gamma_\ell}{\mu_\ell}})
                           ({\cal D}^2_{{\beta_\ell}{\alpha_\ell}}Y_{{\gamma_\ell}{\sigma_\ell}})
       G_{\emptyset/\mu_\ell\nu_\ell\rho_\ell\sigma_\ell\alpha_\ell\beta_\ell}
%
%
%
%
\label{6DGTayl}
\end{eqnarray}
where we substitute $G_{\emptyset/{\mu}_{\ell}{\nu}_{\ell}{\rho}_{\ell}\ldots}(p)$'s, set $\ell=1$ up to $\ell=4$ in (\ref{L0L1L2_ell}) and
obtain
\begin{eqnarray}\label{L0L1L2_emptyset}
  \begin{array}{c}
          {\cal L}^{(1)}_0
    = \frac{\hbar}{2(2\pi)^D}\mbox{Tr}\int dX\int d^Dp\,\,\,G_{\emptyset}(p)
    \\ \\
        {\cal L}^{(1)}_1
    = \frac{\hbar}{2(2\pi)^D}\mbox{Tr}\int dX
    \int d^Dp\,\,\,G_{\emptyset}(p)\left(\Delta_1G_{\emptyset}(p)\right)^{\mathrm{red}}_{\ell=1}
     \\ \\
    {\cal L}^{(1)}_2
    = \frac{\hbar}{2(2\pi)^D}\mbox{Tr}\int dX
    \int d^Dp\,\,\,G_{\emptyset}(p)\left(\Delta_1G_{\emptyset}(p)\right)^{\mathrm{red}}_{\ell=1}\left(\Delta_1G_{\emptyset}(p)\right)^{\mathrm{red}}_{\ell=2}\;\\ \vdots \\ \mbox{etc.}
  \end{array}
\end{eqnarray}
up to $\left(\Delta_1G_{\emptyset}(p)\right)^{\mathrm{red}}_{\ell=4}$ for ${\cal L}^{(1)}_4$. This will be
our prescription for solving the zeroth-order up to fourth-order corrections of the one-loop effective Lagrangian ${\cal L}^{(1)}$ calculation.

\chapter{Higher Mass-Dimensional One-Loop Effective Lagrangians}


\section{Mass-Dimensional Term Hunting}
Our calculation can be reduced to an act of counting mass-dimensions. In this regard, we describe below a scheme on how mass-dimensions of certain quantities are identified.

From the scalar quantity (\ref{L1G}) the known (Gaussian) resolvent (\ref{LG}) quantity can be equivalently recasted as (\ref{Gx}). In either form, the quantity $G_0(p)$ has an overall zero mass-dimension. That is,
\begin{eqnarray}
   \label{dim-g1}
  \mbox{dim}(m^2s)&=&0
\\ \label{dim-g2}
  \mbox{dim}(Xs)&=&0
\\ \label{dim-g3}
  \mbox{dim}(P(s))&=&0
\\ \label{dim-g4}
  \mbox{dim}(Q(s)\cdot p)&=&0
\\ \label{dim-g5}
  \mbox{dim}(p\cdot R(s)\cdot p)&=&0.
\end{eqnarray}
From these, we identify the following mass-dimensions of the following quantities:
\begin{eqnarray}
  \label{dim-s}
  \mbox{dim}(s)&=&-2
\\ \label{dim-X}
  \mbox{dim}(X)&=&+2,
\end{eqnarray} 
respectively, from (\ref{dim-g1}) and (\ref{dim-g2}). At this point, there is an ambiguity in the mass-dimension of $Q(s)$ and $p$. As $Q(s)$ can have a $+1$ and $p$ a $-1$ mass-dimension, or vice-versa. In both cases, the overall mass-dimension is zero. The same thing can be said about $R(s)$ and $p^2$. Either $\mbox{dim}(R(s))=+2$ and $\mbox{dim}(p^2)=-2$, or vice-versa.

We consider, therefore, the functions $P(s)$, $Q(s)$, and $R(s)$ as given Eqns. (\ref{Ps}), (\ref{Qs}), and (\ref{Rs}), respectively in order to resolve the ambiguity identifying mass-dimensions of $Q(s)$ and $R(s)$.
First, we consider Eqn. (\ref{Ps}),
\begin{eqnarray*}
P(s) &=&
 +\frac{1}{3}s^3 {\cal D}X\cdot Y^{-3}Y^3 \cdot {\cal D}X
 -\frac{1}{4}s^2 \mathrm{tr}Y^2 
 + \ldots
\end{eqnarray*}
the first term in the Taylor series expansion of $\dot{X}\cdot Y^{-3}\left(Ys+i\tan iYs\right) \cdot \dot{X}$ and $\frac{1}{2}\mathrm{tr}\ln \sec iYs$, respectively.
Following (\ref{dim-g3}), each term must have zero mass-dimension. So that
\begin{eqnarray}
   \label{dim-D}
  \mbox{dim}({\cal D}_\mu)&=&+1
\\ \label{dim-Y}
  \mbox{dim}(Y_{\mu\nu})&=&+2,
\end{eqnarray}
Eqn. (\ref{Rs}),
\begin{eqnarray*}
  R(s)=2iY^{-1} \tan(iYs)
      =-2s+\frac{2}{3}Y^{2}s^{3}-\frac{4}{15}Y^{4}s^{5}+\ldots.
\end{eqnarray*}
\begin{eqnarray}
  \mbox{dim}(R(s))=-2
\end{eqnarray}
following (\ref{dim-s}) and (\ref{dim-Y}).
Hence,
\begin{eqnarray}
  \mbox{dim}(p)&=&+1
  \label{dim-p}
\\
  \mbox{dim}(Q(s))&=&-1.
\end{eqnarray}
The latter can be seen alternatively from Eqn. (\ref{Qs})
\begin{eqnarray*}
  Q(s)=2i{\cal D}X \cdot Y^{-2}(1 - \sec{iYs} ),
\end{eqnarray*}
following (\ref{dim-X}) and (\ref{dim-D}).

\section{Basis Invariants for $0$th-Order Corrections}

After having identified the mass-dimensions of $s$ in (\ref{dim-s}), $X$ in (\ref{dim-X}), ${\cal D}_\mu$ in (\ref{dim-D}), $Y_{\mu\nu}$ in (\ref{dim-Y}), and $p_\lambda$ in (\ref{dim-p}), we are now ready to build the basis upon which we matrix multiply them and come up with invariants whose total mass-dimensions is twelve or less.

There are two kinds of basis we will construct. We will construct basis for zeroth-order corrections and another for higher-order corrections. In this chapter, we will construct the basis invariants and solve for the mass-dimensional one-loop effective Lagrangians accommodating  zeroth-order corrections. Also in this chapter, we will construct the basis invariants needed and the prescriptions on how these basis will be used for accommodating higher-order corrections in the calculation of one-loop effective Lagrangians. However, we choose to delay until the chapter to implement the prescriptions described in the latter part of this chapter.


To obtain the zeroth-order one-loop effective Lagrangian ${\cal L}^{(1)}_0$ indicated in the first equation of (\ref{L0L1L2}), we must perform the $p$-integration which can be handled by \cite{Brw-Dff,Rodf-Diss}
\begin{eqnarray}
\int d^Dp e^{Q\cdotp+\frac{1}{2}p\cdot R\cdot p} = \pi^{D/2}
 \mathrm{exp} \left[-\frac{1}{2} Q\cdot R^{-1}\cdot Q -\mathrm{tr}\frac{1}{2}\ln \left(-\frac{R}{2s}\right)\right].
\end{eqnarray}
We then perform the $X$-integration and obtain the closed form of the zeroth-order one-loop effective Lagrangian
\begin{eqnarray}
 {\cal L}^{(1)}_0
 = \frac{\hbar}{2 (4\pi)^{D/2}}
   \int^\infty_0\!\!\! \frac{ds}{s^{1+D/2}}
 \times
  \left[
 e^{Xs+P(s)-\frac{1}{2}Q(s)\cdot R^{-1}(s)\cdot Q(s)
   -\frac{1}{2}\mathrm{tr} \ln \left(-\frac{R(s)}{2s}\right)}
   -e^{X_0s}
  \right]
\label{Lzeroth}
\end{eqnarray}
where $X_0$ represents the zero reference of the background potential $X$.
Equivalently, Rodulfo\cite{Brw-Dff,Rodf-Diss} simplified (\ref{Lzeroth}) as\footnote{The $X$ and $p$ integration has already been performed. The overall sign $+$ is for bosons, $-$ for fermions.}
\begin{eqnarray}
  {\cal L}^{(1)}_0
&=&\pm\frac{\hbar}{2(4\pi)^{D/2}}
\mathrm{Tr}
 \int^\infty_0 ds\,s^{-1-D/2}e^{-m^2s}
\nonumber\\&&\times
 \left[\right.
e^{+s{\cal X}}
e^{-\frac{1}{2}\mathrm{tr}\ln (iYs)^{-1}\sin iYs}
e^{+\dot{{\cal X}}\cdot (iY)^{-3} \left(2\tan \frac{iYs}{2} - iYs\right)\cdot\dot{{\cal X}}}
%
%
-e^{{\cal X}_0s}
\left. \right]\label{Ltaylor}
\end{eqnarray}
This will be used in the extraction of zeroth-order corrections by Taylor-series\footnote{Taylor-series expansions were obtained using Mathematica. \\} expanding the separate exponential functions.

In the following Taylor series expansion, we choose to represent in matrix form. But we restore the tensorial form as we present the one-loop effective Lagrangians containing zeroth-order corrections two up to sixteen mass-dimensional invariants.

For the first term in (\ref{Ltaylor}),
\begin{eqnarray}
 e^{{\cal X}s}=1 \!+\! s\,{\cal X} \!+\! {\frac{{s^2}\,{{\cal X}^2}}{2}} \!+\!
  {\frac{{s^3}\,{{\cal X}^3}}{6}} \!+\!
  {\frac{{s^4}\,{{\cal X}^4}}{24}} \!+\!
  {\frac{{s^5}\,{{\cal X}^5}}{120}} \!+\!
  {\frac{{s^6}\,{{\cal X}^6}}{720}} \!+\!
  {\frac{{s^7}\,{{\cal X}^7}}{5040}} \!+\!
  {\frac{{s^8}\,{{\cal X}^8}}{40320}} \!+\!
  \ldots.
\end{eqnarray}
\newpage
There are seven terms (upon collecting only the invariants whose mass-dimensions is equal to 16 or less)
\begin{small}
\begin{eqnarray}
a[0] &=& +1
\\
a[2] &=& +{\cal X} s
\label{a2}\\
a[4] &=& +\frac{1}{2} {\cal X}^2 s^2
\label{a4}\\
a[6] &=& +\frac{1}{6} {\cal X}^3 s^3
\label{a6}\\
a[8] &=& +\frac{1}{24} {\cal X}^4 s^4
\label{a8}\\
a[10] &=& +\frac{1}{120} {\cal X}^5 s^5            
\label{a10}\\
a[12] &=& +\frac{1}{720} {\cal X}^6 s^6            
\label{a12}\\
a[14] &=& +\frac{1}{5040} {\cal X}^7 s^7
\label{a14}\\
a[16] &=& +\frac{1}{40320} {\cal X}^8 s^8
\label{a16}
\end{eqnarray}
\end{small}

For the second term in (\ref{Ltaylor}),
\begin{eqnarray}
 e^{-\frac{1}{2}\mathrm{tr}\ln (iYs)^{-1}\sin iYs}=e^{\,+\,\frac{(iYs)^2}{12}\,+\,\frac{(iYs)^4}{360}\,+\,\frac{(iYs)^6}{5670}
\,+\, {\frac{{(iYs)^8}}{75600}} \,+\,
  {\frac{{(iYs)^{10}}}{935550}} \,+\,\ldots}.\label{2ndLtaylor}
\end{eqnarray}
There are four terms. That is after using the series expansion
\begin{eqnarray}
\ln (x)^{-1}\sin x = -\frac{x^2}{6}-\frac{x^4}{180}-\frac{x^6}{2835}-\frac{x^8}{37800}-\frac{x^{10}}{467775}-\ldots
\end{eqnarray}
For the first term $e^{+\frac{(iYs)^2}{12}}$ in (\ref{2ndLtaylor}),
\begin{small}
\begin{eqnarray}
b[0]&=& +1 
\\
b[4]&=& -\frac{1}{12} Y^2 s^2
\label{b4}\\
b[8]&=& +\frac{1}{288} Y^2 Y^2 s^4
\label{b8}\\
b[12]&=& -\frac{1}{10368} Y^2 Y^2 Y^2 s^6
\label{b12}\\
b[16]&=& +\frac{1}{497664} Y^2 Y^2 Y^2 Y^2 s^8
\label{b16}
\end{eqnarray}
\end{small}
\newpage
This is after using the power series expansion:
\begin{eqnarray}
e^{+\frac{x^2}{12}}=1 + {\frac{{x^2}}{12}} + {\frac{{x^4}}{288}} +
  {\frac{{x^6}}{10368}} + {\frac{{x^8}}{497664}}+
  {\frac{{x^{10}}}{29859840}} + \ldots.
\end{eqnarray}
For the second term $e^{+\frac{(iYs)^4}{360}}$ in (\ref{2ndLtaylor})
\begin{eqnarray}
c[0]&=& +1                     
\\
c[8]&=& +\frac{1}{360} Y^4 s^4
\label{c8}
\\
c[16]&=& +\frac{1}{259200} Y^4 Y^4 s^8
\label{c16}
\end{eqnarray}
This is after using the power series expansion:
\begin{eqnarray}
e^{+\frac{x^4}{360}}=1 + {\frac{{x^4}}{360}} + {\frac{{x^8}}{259200}} +
  {\frac{{x^{12}}}{279936000}} +  \ldots.
\end{eqnarray}
For the third term $e^{+\frac{(iYs)^6}{5670}}$ in (\ref{2ndLtaylor})
\begin{eqnarray}
d[0]&=& +1                         
\\
d[12]&=& -\frac{1}{5670} Y^6 s^6
\label{d12}
\end{eqnarray}
This is after using the power series expansion:
\begin{eqnarray}
e^{+\frac{x^6}{5670}}=1 + {\frac{{x^6}}{5670}} +
  {\frac{{x^{12}}}{64297800}} +
  {\frac{{x^{18}}}{1093705578000}} + \ldots.
\end{eqnarray}
For the fourth term $e^{+{\frac{{(iYs)^8}}{75600}}}$
\begin{eqnarray}
e[0] &=& +1
\\
e[16] &=& +\frac{1}{75600} Y^8 s^8
\label{e16}
\end{eqnarray}
This is after using the power series expansion:
\begin{eqnarray}
e^{+\frac{x^8}{75600}}=1 + {\frac{{x^8}}{75600}} +
  {\frac{{x^{16}}}{11430720000}} + \ldots.
\end{eqnarray}

\newpage
For third term in (\ref{Ltaylor})
\begin{eqnarray}
e^{\dot{{\cal X}}\cdot (iY)^{-3}\left(2\tan \frac{iYs}{2} - iYs\right)\cdot \dot{{\cal X}}}
=e^{\dot{{\cal X}}\cdot (iY)^{-3}\left({\frac{{(iYs)^3}}{12}} + {\frac{{(iYs)^5}}{120}} +
  {\frac{17\,{(iYs)^7}}{20160}} +
  {\frac{31\,{(iYs)^9}}{362880}} +
\ldots
\right)\cdot \dot{{\cal X}}}\label{3rdLtaylor}
\end{eqnarray}
This is after using the series expansion
\begin{eqnarray}
2 \tan \frac{x}{2} - x= {\frac{{x^3}}{12}} + {\frac{{x^5}}{120}} +
  {\frac{17\,{x^7}}{20160}} +
  {\frac{31\,{x^9}}{362880}} +
  {\frac{691\,{x^{11}}}{79833600}} + \ldots
\end{eqnarray}
There are two terms to consider. The first term in ({\ref{3rdLtaylor}}):
\begin{eqnarray}
e^{+{\frac{{1}}{12}}\dot{{\cal X}}\cdot (iY)^{-3}  (iYs)^3\cdot \dot{{\cal X}}}
=e^{+{\frac{{1}}{12}}\dot{{\cal X}}\cdot Y^{-3}  Y^3\cdot \dot{{\cal X}} s^3}
\end{eqnarray}
using
\begin{eqnarray}
 e^{\frac{x}{12}}=
1 + {\frac{x}{12}} + {\frac{{x^2}}{288}} +
  {\frac{{x^3}}{10368}} + {\frac{{x^4}}{497664}} + \ldots
\end{eqnarray}
\begin{eqnarray}
f[0] &=& +1
\\
f[6] &=& +\frac{1}{12}\dot{{\cal X}}\cdot Y^{-3}  Y^3\cdot \dot{{\cal X}}s^3
\label{f6}\\
f[12] &=& +\frac{{1}}{288}\left(\dot{{\cal X}}\cdot Y^{-3}  Y^3\cdot \dot{{\cal X}}\right)^2 s^6
\label{f12}
\end{eqnarray}
The second term in ({\ref{3rdLtaylor}}):
\begin{eqnarray}
e^{+{\frac{{1}}{120}}\dot{{\cal X}}\cdot (iY)^{-3} (iYs)^5\cdot \dot{{\cal X}}}
=e^{-{\frac{{1}}{120}}\dot{{\cal X}}\cdot Y^{-3} Y^5\cdot \dot{{\cal X}}s^5}
\end{eqnarray}
using
\begin{eqnarray}
 e^{-x/120}=
1 - {\frac{x}{120}} + {\frac{{x^2}}{28800}} -
  {\frac{{x^3}}{10368000}} + \ldots
\end{eqnarray}
\begin{eqnarray}
g[0] &=& +1
\\
g[10] &=& - {\frac{1}{120}} \dot{{\cal X}}\cdot Y^{-3} Y^5\cdot \dot{{\cal X}}s^5
\label{g10}
\end{eqnarray}
%
%
%
%

For the constant of $X$-integration, the last term in (\ref{Ltaylor}):
\begin{eqnarray}
 e^{-{\cal X}_0s}=1 - s\,{\cal X} + {\frac{{s^2}\,{{{\cal X}_0}^2}}{2}} -
  {\frac{{s^3}\,{{{\cal X}_0}^3}}{6}} +
  {\frac{{s^4}\,{{{\cal X}_0}^4}}{24}} -
  {\frac{{s^5}\,{{{\cal X}_0}^5}}{120}} +
  {\frac{{s^6}\,{{{\cal X}_0}^6}}{720}} -
  {\frac{{s^7}\,{{{\cal X}_0}^7}}{5040}} + \ldots.
\end{eqnarray}
There are seven terms
\begin{small}
\begin{eqnarray}
h[0] &=& -1
\\
h[2] &=& -{{\cal X}_0} s                               
\label{h2}\\
h[4] &=& -\frac{1}{2} {{\cal X}_0}^2 s^2               
\label{h4}\\
h[6] &=& -\frac{1}{6} {{\cal X}_0}^3 s^3               
\label{h6}\\
h[8] &=& -\frac{1}{24} {{\cal X}_0}^4 s^4              
\label{h8}\\
h[10] &=& -\frac{1}{120} {{\cal X}_0}^5 s^5            
\label{h10}\\
h[12] &=& -\frac{1}{720} {{\cal X}_0}^6 s^6            
\label{h12}\\
h[14] &=& -\frac{1}{5040} {{\cal X}_0}^7 s^7
\label{h14}\\
h[16] &=& -\frac{1}{40320} {{\cal X}_0}^8 s^8
\label{h16}
\end{eqnarray}
\end{small}

\subsection{${\cal L}^{(1)}_0$: $0$th-order Mass-Dimensional Lagrangians}
After having gathered the terms of interest that resulted in the Taylor series expansion, we now group the terms into their common mass-dimensions (as indicated in their names and the numbers inside the $[\,]$ notation).

\begin{large}
\textbf{Two Mass-Dimensions}
\end{large}

The two mass-dimensional one-loop effective Lagrangian
\begin{eqnarray}\label{0[2]}
 {\cal L}^{(1)[2]}_0
  \!\!&=&\!\! \frac{\hbar}{2(4\pi)^{D/2}}
\int^\infty_0 \frac{ds\,e^{-m^2s}}{s^{-0+D/2}}
   \mbox{Tr}
\left(
   {\cal {\cal X}}-{\cal {\cal X}}_0
\right)
\end{eqnarray}
The terms contained here are contributed from (\ref{a2}) and (\ref{h2}), respectively. Eqn. (\ref{0[2]}) is exactly Eqn. (2.71) of \cite{Rodf-Diss}.

\begin{large}
\textbf{Four Mass-Dimensions}
\end{large}

That is, the four mass-dimensional one-loop effective Lagrangian
has the zeroth-order correction:
\begin{eqnarray}\label{0[4]}
 {\cal L}^{(1)[4]}_0
  \!\!&=&\!\! \frac{\hbar}{2(4\pi)^{D/2}}
\int^\infty_0 \frac{ds\,e^{-m^2s}}{s^{-1+D/2}}
   \mbox{Tr}
\left[
   \frac{1}{2}({\cal X}^2-{{\cal X}_0}^2)
   +\frac{1}{12}Y_{\mu\nu}Y_{\mu\nu}
\right]
\end{eqnarray}
The terms contained here are contributed from (\ref{a4}), (\ref{h4}) and (\ref{b4}), respectively.
Eqn. (\ref{0[4]}) is exactly Eqn. (2.72) of \cite{Rodf-Diss}.

\begin{large}
\textbf{Six Mass-Dimensions}
\end{large}

The six mass-dimensional one-loop effective Lagrangian
 ~has the zeroth-order correction:
\begin{eqnarray}
 {\cal L}^{(1)[6]}_0
  \!\!&=&\!\! \frac{\hbar}{2(4\pi)^{D/2}}
\int^\infty_0 \frac{ds\,e^{-m^2s}}{s^{-2+D/2}}
   \mbox{Tr}
\left[
   \frac{1}{6}({\cal X}^3-{{\cal X}_0}^3)
   +\frac{1}{12}{\cal X}Y_{\mu\nu}Y_{\mu\nu}
   +\frac{1}{12}({\cal D}_\mu{\cal X})({\cal D}_\mu{\cal X})
\right].
\nonumber\\
\label{0[6]}
\end{eqnarray}
The terms contained here are contributed from (\ref{a6}), (\ref{h6}), the products 
\begin{eqnarray}
a[2]b[4] &=& -\frac{1}{12} {\cal X} Y^2 s^3 
\end{eqnarray}
and (\ref{f6}), respectively. Eqn. (\ref{0[6]}) is exactly Eqn. (2.73) of \cite{Rodf-Diss}.

\begin{large}
\textbf{Eight Mass-Dimensions}
\end{large}

The eight mass-dimensional one-loop effective Lagrangian
has the zeroth-order correction:
\begin{eqnarray}
 {\cal L}^{(1)[8]}_0
  \!\!&=&\!\! \frac{\hbar}{2(4\pi)^{D/2}}
\int^\infty_0 \frac{ds\,e^{-m^2s}}{s^{-3+D/2}}
   \mbox{Tr}
\left[
   \frac{1}{24}({\cal X}^4-{{\cal X}_0}^4)
   +\frac{1}{288}Y_{\mu\nu}Y_{\mu\nu}Y_{\rho\sigma}Y_{\rho\sigma}
\right.
\nonumber\\&&\label{0[8]}
\left.
   +\frac{1}{360}Y_{\mu\nu}Y_{\nu\rho}Y_{\rho\sigma}Y_{\sigma\mu}
   +\frac{1}{24}{\cal X}^2Y_{\mu\nu}Y_{\mu\nu}
   +\frac{1}{12}{\cal X}({\cal D}_\mu{\cal X})({\cal D}_\mu{\cal X})
\right]
\end{eqnarray}
The terms contained here are contributed from (\ref{a8}), (\ref{h8}), (\ref{b8}) and products:
\begin{eqnarray}
a[4]b[4] &=& -\frac{1}{24} {\cal X}^2 Y^2 s^4 
\\
a[2]f[6] &=& +\frac{1}{12} {\cal X} \dot{{\cal X}}\cdot Y^{-3}  Y^3\cdot \dot{{\cal X}} s^4,
\end{eqnarray}
respectively. Eqn. (\ref{0[8]}) is Eqn. (2.74) of \cite{Rodf-Diss} in the long-wavelength limit (${\cal D}_\mu \to 0$). In Eqn. (\ref{0[8]}), we have lifted this limit and showed explicitly the complete zeroth-order correction with eight mass-dimension. What is new here is the inclusion of the ${\cal O}({\cal D}_\mu) $ term.

\begin{large}
\textbf{Ten Mass-Dimensions}
\end{large}

The ten mass-dimensional one-loop effective Lagrangian
~has the zeroth-order correction:
\begin{eqnarray}
 {\cal L}^{(1)[10]}_0
  \!\!&=&\!\! \frac{\hbar}{2(4\pi)^{D/2}}
\int^\infty_0 \frac{ds\,e^{-m^2s}}{s^{-4+D/2}}
   \mbox{Tr}
\left[
   \frac{1}{120}({\cal X}^5-{{\cal X}_0}^5)
   +\frac{1}{288}{\cal X}Y_{\mu\nu}Y_{\mu\nu}Y_{\rho\sigma}Y_{\rho\sigma}
\right.
\nonumber\\&&
   +\frac{1}{360}{\cal X}Y_{\mu\nu}Y_{\nu\rho}Y_{\rho\sigma}Y_{\sigma\mu}
   +\frac{1}{72}{\cal X}^3Y_{\mu\nu}Y_{\mu\nu}
   +\frac{1}{120}{\cal X}({\cal D}_\mu{\cal X})Y_{\mu\nu}Y_{\mu\nu}({\cal D}_\mu{\cal X})
\nonumber\\&&
   +\frac{1}{24}{\cal X}^2({\cal D}_\mu{\cal X})({\cal D}_\mu{\cal X})
\label{0[10]}
\left.
   +\frac{1}{144} Y_{\mu\nu}Y_{\mu\nu} ({\cal D}_\mu{\cal X})({\cal D}_\mu{\cal X})
\right].
\end{eqnarray}
The terms contained here are contributed from (\ref{a10}), (\ref{h10}), products of Eqns. (\ref{a2})$\times$(\ref{b8}), (\ref{a2})$\times$(\ref{c8}), (\ref{a6})$\times$(\ref{b4}):
\begin{eqnarray}
a[2]b[8] &=& +\frac{1}{288} {\cal X} Y^2 Y^2 s^5  
\\
a[2]c[8] &=& +\frac{1}{360} {\cal X} Y^4 s^5  
\\
a[6]b[4] &=& -\frac{1}{72} {\cal X}^3 Y^2 s^5,  
\end{eqnarray}
Eqn. (\ref{g10}), products: 
\begin{eqnarray}
a[4]f[6] &=& +\frac{1}{24} {\cal X}^2 \dot{{\cal X}}\cdot Y^{-3}  Y^3\cdot \dot{{\cal X}}s^5
\\
b[4]f[6] &=& -\frac{1}{144} Y^2 \dot{{\cal X}}\cdot Y^{-3}  Y^3\cdot \dot{{\cal X}}s^5,
\end{eqnarray}
respectively. Eqn. (\ref{0[10]}) is Eqn. (2.75) of \cite{Rodf-Diss} in the long-wavelength limit (${\cal D}_\mu \to 0$). In Eqn. (\ref{0[10]}), we have lifted this limit and showed explicitly including the ${\cal O}({\cal D}_\mu) $ terms. This completes the zeroth-order correction containing invariants whose mass-dimension is ten.

\begin{large}
\textbf{Twelve Mass-Dimensions}
\end{large}

The twelve mass-dimensional one-loop effective Lagrangian
~has the zeroth-order correction:
\begin{eqnarray}
 {\cal L}^{(1)[12]}_0
  \!\!&=&\!\! \frac{\hbar}{2(4\pi)^{D/2}}
\int^\infty_0 \frac{ds\,e^{-m^2s}}{s^{-5+D/2}}
   \mbox{Tr}
\left[
   \frac{1}{720}({\cal X}^6-{{\cal X}_0}^6)
\right.
\nonumber\\&&
   +\frac{1}{10368}Y_{\mu\nu}Y_{\mu\nu}Y_{\rho\sigma}Y_{\rho\sigma}Y_{\alpha\beta}Y_{\alpha\beta}
   -\frac{1}{5670}Y_{\mu\nu}Y_{\nu\rho}Y_{\rho\sigma}Y_{\sigma\alpha}Y_{\alpha\beta}Y_{\beta\gamma}
\nonumber\\&&
   +\frac{1}{576}{\cal X}^2Y_{\mu\nu}Y_{\mu\nu}Y_{\rho\sigma}Y_{\rho\sigma}
   +\frac{1}{4320}Y_{\mu\nu}Y_{\mu\nu}Y_{\rho\sigma}Y_{\sigma\alpha}Y_{\alpha\beta}Y_{\beta\rho}
   +\frac{1}{720}{\cal X}^2Y_{\mu\nu}Y_{\nu\rho}Y_{\rho\sigma}Y_{\sigma\mu}
\nonumber\\&&
   +\frac{1}{288} {\cal X}^4 Y_{\mu\nu}Y_{\mu\nu}
   +\frac{1}{288}({\cal D}_\mu{\cal X})({\cal D}_\mu{\cal X})({\cal D}_\nu{\cal X})({\cal D}_\nu{\cal X})
   +\frac{1}{72}{\cal X}^3({\cal D}_\mu{\cal X})({\cal D}_\mu{\cal X})
\nonumber\\&&
\left.
   +\frac{11}{720}{\cal X} ({\cal D}_\mu{\cal X}) Y_{\mu\nu}Y_{\mu\nu}Y_{\rho\sigma}Y_{\rho\sigma} ({\cal D}_\mu{\cal X})
\right]
\label{0[12]}
\end{eqnarray}
The terms contained here are contributed from (\ref{a12}), (\ref{h12}), (\ref{b12}), (\ref{d12}), products:
\begin{small}
\begin{eqnarray}
a[4]b[8] &=& +\frac{1}{576} {\cal X}^2 Y^2 Y^2 s^6    
\\
b[4]c[8] &=& -\frac{1}{4320} Y^2 Y^4 s^6   
\\
a[4]c[8] &=& +\frac{1}{720} {\cal X}^2 Y^4 s^6  
\\
a[8]b[4] &=& -\frac{1}{288} {\cal X}^4 Y^2 s^6  
\end{eqnarray}
\end{small}
Eqn. (\ref{f12}), products:
\begin{small}
\begin{eqnarray}
a[6]f[6] &=& +\frac{1}{72} {\cal X}^3 \dot{{\cal X}}\cdot Y^{-3}  Y^3\cdot \dot{{\cal X}} s^6
\\
a[2]g[10] &=& - {\frac{1}{120}}{\cal X} \dot{{\cal X}}\cdot Y^{-3} Y^5\cdot \dot{{\cal X}}s^6
\\
a[2]b[4]f[6] &=& -\frac{1}{144} {\cal X} Y^2 \dot{{\cal X}}\cdot Y^{-3}  Y^3\cdot \dot{{\cal X}}s^6
\end{eqnarray}
\end{small}
respectively. The last two terms are combined resulting to the last term in (\ref{0[12]}). Eqn. (\ref{0[12]}) is Eqn. (2.75) of \cite{Rodf-Diss} in the long-wavelength limit (${\cal D}_\mu \to 0$). In Eqn. (\ref{0[12]}) we have lifted this limit and included the ${\cal O}({\cal D}_\mu) $ terms. This is the closed form of the zeroth-order correction that contains twelve mass-dimensional invariants. Again like the eight, and ten-mass dimensional zeroth-order contributions we have explicitly displayed the ${\cal O}({\cal D}_\mu) $ terms.

\begin{large}
\textbf{Fourteen Mass-Dimensions}
\end{large}

The fourteen mass-dimensional one-loop effective Lagrangian
~has the zeroth-order correction:
\begin{eqnarray}
 {\cal L}^{(1)[14]}_0
  \!\!&=&\!\! \frac{\hbar}{2(4\pi)^{D/2}}
\int^\infty_0 \frac{ds\,e^{-m^2s}}{s^{-6+D/2}}
   \mbox{Tr}
\left[
  \frac{1}{5040} \left( {\cal X}^7-{{\cal X}_0}^7
                  \right)
  -\frac{1}{2160} {\cal X}^3 Y_{\mu\nu}Y_{\nu\rho}Y_{\rho\sigma}Y_{\sigma\mu}
\right.
\nonumber\\ && \!\!\!\!\!\!\!\!\!
  +\frac{1}{1440} {\cal X}^5 Y_{\mu\nu}Y_{\mu\nu}
  -\frac{1}{2160} {\cal X}^3 Y_{\mu\nu}Y_{\nu\rho}Y_{\rho\sigma}Y_{\sigma\mu}
  +\frac{1}{1728} {\cal X}^3 Y_{\mu\nu}Y_{\mu\nu} Y_{\rho\sigma}Y_{\rho\sigma}
\nonumber\\ && \!\!\!\!\!\!\!\!\!
  +\frac{1}{10368} {\cal X}  Y_{\mu\nu}Y_{\mu\nu} Y_{\rho\sigma}Y_{\rho\sigma} Y_{\alpha\beta}Y_{\alpha\beta}
  +\frac{1}{4320} {\cal X} Y_{\mu\nu}Y_{\mu\nu} Y_{\rho\sigma}Y_{\sigma\alpha}Y_{\alpha\beta}Y_{\beta\rho}
\nonumber\\ && \!\!\!\!\!\!\!\!\!
  -\frac{1}{5670} {\cal X}  Y_{\mu\nu}Y_{\nu\rho}Y_{\rho\sigma}Y_{\sigma\alpha}Y_{\alpha\beta}Y_{\beta\mu}
  +\frac{1}{288} {\cal X}^4 ({\cal D}_\mu{\cal X}) ({\cal D}_\mu{\cal X})
\nonumber\\ && \!\!\!\!\!\!\!\!\!
  - {\frac{1}{1440}} ({\cal D}_\mu{\cal X}) Y_{\mu\nu}Y_{\nu\rho} ({\cal D}_\rho{\cal X}) Y_{\sigma\alpha}Y_{\sigma\alpha}
  +\frac{1}{3456} ({\cal D}_\mu{\cal X})({\cal D}_\mu{\cal X}) Y_{\rho\sigma}Y_{\rho\sigma} Y_{\alpha\beta}Y_{\alpha\beta}
\nonumber\\ && \!\!\!\!\!\!\!\!\!
  -\frac{1}{1440} ({\cal D}_\mu{\cal X}) Y_{\mu\nu}Y_{\nu\rho} ({\cal D}_{\rho}{\cal X}) Y_{\sigma\alpha}Y_{\sigma\alpha}
  +\frac{1}{3456} ({\cal D}_\mu{\cal X}) ({\cal D}_\mu{\cal X}) Y_{\mu\nu}Y_{\mu\nu} Y_{\rho\sigma}Y_{\rho\sigma}
\nonumber\\ && \!\!\!\!\!\!\!\!\!
  +\frac{1}{4320} ({\cal D}_\mu{\cal X})({\cal D}_\mu{\cal X}) Y_{\mu\nu}Y_{\nu\rho}Y_{\rho\sigma}Y_{\sigma\mu}
  -\frac{1}{240} {\cal X}^2 ({\cal D}_\mu{\cal X}) Y_{\mu\nu}Y_{\nu\rho} ({\cal D}_\rho{\cal X})
\nonumber\\ && \!\!\!\!\!\!\!\!\!
\left.
  +\frac{1}{288} {\cal X}^2 ({\cal D}_\mu{\cal X})({\cal D}_\mu{\cal X}) Y_{\mu\nu}Y_{\mu\nu}
  +\frac{1}{288} {\cal X} ({\cal D}_\mu{\cal X})({\cal D}_\mu{\cal X})({\cal D}_\nu{\cal X})({\cal D}_\nu{\cal X})
\right]
\nonumber\\ \label{0[14]}
\end{eqnarray}
from 
(\ref{a14}),
(\ref{h14}), and the products:
\begin{eqnarray}
%
%
a[10]b[4] &=& -\frac{1}{1440} {\cal X}^5 Y^2 s^7
\label{a10b4}\\             
a[6]c[8] &=& -\frac{1}{2160} {\cal X}^3 Y^4  s^7
\label{a6c8}\\              
a[6]b[8] &=& +\frac{1}{1728} {\cal X}^3 Y^2 Y^2 s^7
\label{a6b8}\\              
a[2]b[12] &=& -\frac{1}{10368} {\cal X}  Y^2 Y^2 Y^2 s^7
\label{a2b12}\\               
a[2]b[4]c[8] &=& -\frac{1}{4320} {\cal X} Y^2 Y^4 s^7
\label{a2b4c8}\\              
a[2]d[12] &=& -\frac{1}{5670} {\cal X}  Y^6 s^7
\label{a2d12}\\               
a[8]f[6] &=& +\frac{1}{288} {\cal X}^4 \dot{{\cal X}}\cdot Y^{-3}  Y^3\cdot \dot{{\cal X}}s^7
\label{a8f6}\\             
b[4]g[10] &=& + {\frac{1}{1440}} \dot{{\cal X}}\cdot Y^{-3} Y^5\cdot \dot{{\cal X}} Y^2 s^7
\label{b4g10}\\             
b[8]f[6] &=& +\frac{1}{3456} \dot{{\cal X}}\cdot Y^{-3}  Y^3\cdot \dot{{\cal X}} Y^2 Y^2 s^7
\label{b8f6}\\              
c[8]f[6]&=& +\frac{1}{4320} \dot{{\cal X}}\cdot Y^{-3}  Y^3\cdot \dot{{\cal X}} Y^4 s^7
\label{c8f6}\\             
a[4]g[10] &=& - {\frac{1}{240}} {\cal X}^2 \dot{{\cal X}}\cdot Y^{-3} Y^5\cdot \dot{{\cal X}}s^7
\label{a4g10}\\              
a[4]b[4]f[6] &=& -\frac{1}{288} {\cal X}^2 \dot{{\cal X}}\cdot Y^{-3}  Y^3\cdot \dot{{\cal X}} Y^2 s^7
\label{a2b4f6}\\              
a[2]f[12] &=& +\frac{{1}}{288}{\cal X}\left(\dot{{\cal X}}\cdot Y^{-3}  Y^3\cdot \dot{{\cal X}}\right)^2 s^7,
\label{a2f12}
\end{eqnarray}
respectively.

\begin{large}
\textbf{Sixteen Mass-Dimensions}
\end{large}

The sixteen mass-dimensional one-loop effective Lagrangian
~has the zeroth-order correction:
\begin{eqnarray}
 {\cal L}^{(1)[16]}_0
  \!\!&=&\!\! \frac{\hbar}{2(4\pi)^{D/2}}
\int^\infty_0 \frac{ds\,e^{-m^2s}}{s^{-7+D/2}}
   \mbox{Tr}
\left[\frac{1}{40320}\left(
             {\cal X}^8 - {{\cal X}_0}^8
                     \right)
+\frac{1}{8640} {\cal X}^6  Y_{\mu\nu}Y_{\mu\nu}
\right.
\nonumber\\ && \!\!\!\!\!\!\!\!\!
+\frac{1}{6912} {\cal X}^4  Y_{\mu\nu}Y_{\mu\nu} Y_{\rho\sigma}Y_{\rho\sigma}
+\frac{1}{20736} {\cal X}^2  Y_{\mu\nu}Y_{\mu\nu} Y_{\rho\sigma}Y_{\rho\sigma} Y_{\alpha\beta}Y_{\alpha\beta}
\nonumber\\ && \!\!\!\!\!\!\!\!\!
+\frac{1}{8640}  {\cal X}^2  Y_{\mu\nu}Y_{\mu\nu} Y_{\rho\sigma}Y_{\sigma\alpha}Y_{\alpha\beta}Y_{\beta\rho}
+\frac{1}{8640} {\cal X}^4 Y_{\mu\nu}Y_{\nu\rho}Y_{\rho\sigma}Y_{\sigma\mu}
\nonumber\\ && \!\!\!\!\!\!\!\!\!
-\frac{1}{11340} {\cal X}^2 Y_{\mu\nu}Y_{\nu\rho}Y_{\rho\sigma}Y_{\sigma\alpha}Y_{\alpha\beta}Y_{\beta\mu}
+\frac{1}{497664} Y_{\mu\nu}Y_{\mu\nu} Y_{\rho\sigma}Y_{\rho\sigma} Y_{\alpha\beta}Y_{\alpha\beta} Y_{\gamma\delta}Y_{\gamma\delta}
\nonumber\\ && \!\!\!\!\!\!\!\!\!
+\frac{1}{103680} Y_{\mu\nu}Y_{\nu\rho}Y_{\rho\sigma}Y_{\sigma\mu} Y_{\alpha\beta}Y_{\alpha\beta} Y_{\gamma\delta}Y_{\gamma\delta}
+\frac{1}{259200} Y_{\mu\nu}Y_{\nu\rho}Y_{\rho\sigma}Y_{\sigma\mu} Y_{\alpha\beta}Y_{\beta\gamma}Y_{\gamma\delta}Y_{\delta\alpha}
\nonumber\\ && \!\!\!\!\!\!\!\!\!
-\frac{1}{68040}  Y_{\mu\nu}Y_{\nu\rho}Y_{\rho\sigma}Y_{\sigma\alpha}Y_{\alpha\beta}Y_{\beta\mu} Y_{\gamma\delta}Y_{\gamma\delta}
+\frac{1}{75600} Y_{\mu\nu}Y_{\nu\rho}Y_{\rho\sigma}Y_{\sigma\alpha}Y_{\alpha\beta}Y_{\beta\gamma}Y_{\gamma\delta}Y_{\delta\mu}
\nonumber\\ && \!\!\!\!\!\!\!\!\!
+\frac{1}{1440}  {\cal X}^5 ({\cal D}_\mu{\cal X})({\cal D}_\mu{\cal X})
- {\frac{1}{720}} {\cal X}^3 ({\cal D}_\mu{\cal X}) Y_{\mu\nu}Y_{\nu\sigma} ({\cal D}_\sigma{\cal X})
\nonumber\\ && \!\!\!\!\!\!\!\!\!
+\frac{1}{864} {\cal X}^3 ({\cal D}_\mu{\cal X})({\cal D}_\mu{\cal X}) Y_{\nu\rho}Y_{\nu\rho}
+\frac{1}{1440} {\cal X} ({\cal D}_\mu{\cal X}) Y_{\mu\nu}Y_{\nu\rho} ({\cal D}_\rho{\cal X}) Y_{\sigma\alpha}Y_{\sigma\alpha}
\nonumber\\ && \!\!\!\!\!\!\!\!\!
+\frac{{1}}{576}{\cal X}^2 ({\cal D}_\mu{\cal X}) ({\cal D}_\mu{\cal X}) ({\cal D}_\nu{\cal X})({\cal D}_\nu{\cal X})
+\frac{{1}}{3456}({\cal D}_\mu{\cal X})({\cal D}_\mu{\cal X})({\cal D}_\nu{\cal X})({\cal D}_\nu{\cal X}) Y_{\rho\sigma}Y_{\rho\sigma}
\nonumber\\ && \!\!\!\!\!\!\!\!\!
\left.
-\frac{1}{1440}({\cal D}_\mu{\cal X})({\cal D}_\mu{\cal X})({\cal D}_\nu{\cal X}) Y_{\nu\rho}Y_{\rho\sigma}({\cal D}_\sigma{\cal X})
\right] \label{0[16]}
\end{eqnarray}
from 
(\ref{a16}),
(\ref{h16})
the products:
\begin{eqnarray}
%
a[12]b[4] &=& -\frac{1}{8640} {\cal X}^6  Y^2 s^8
\label{a12b4}\\              
a[8]b[8] &=& +\frac{1}{6912} {\cal X}^4  Y^2 Y^2 s^8
\label{a8b8}\\              
a[4]b[12] &=& -\frac{1}{20736} {\cal X}^2  Y^2 Y^2 Y^2 s^8
\label{a4b12}\\               
a[4]b[4]c[8] &=& -\frac{1}{8640}  {\cal X}^2  Y^2  Y^4 s^8
\label{a4b4c8}\\                 
a[8]c[8] &=& +\frac{1}{8640} {\cal X}^4 Y^4 s^8
\label{a8c8}\\              
a[4]d[12] &=& -\frac{1}{11340} {\cal X}^2 Y^6 s^8,
\label{a4d12}\\               
\end{eqnarray}
(\ref{b16}),
\begin{eqnarray}
b[8]c[8] &=& +\frac{1}{103680} Y^4 Y^2 Y^2 s^8
\label{b8c8}\\                
\end{eqnarray}
(\ref{c16}),
\begin{eqnarray}
b[4]d[12]&=& +\frac{1}{68040}  Y^6 Y^2 s^8
\label{b4d12}\\              
\end{eqnarray}
(\ref{e16}), and the products:
\begin{eqnarray}
a[10]f[6] &=& +\frac{1}{1440}  {\cal X}^5 \dot{{\cal X}}\cdot Y^{-3}  Y^3\cdot \dot{{\cal X}}s^8
\label{a10f6}\\             
a[6]g[10] &=&  - {\frac{1}{720}} {\cal X}^3 \dot{{\cal X}}\cdot Y^{-3} Y^5\cdot \dot{{\cal X}}s^8
\label{a6g10}\\                 
a[6]b[4]f[6] &=& -\frac{1}{864} {\cal X}^3 \dot{{\cal X}}\cdot Y^{-3}  Y^3\cdot \dot{{\cal X}} Y^2 s^8
\label{a6b4f6}\\                 
a[2]b[4]g[10] &=& -\frac{1}{1440} {\cal X} \dot{{\cal X}}\cdot Y^{-3} Y^5\cdot \dot{{\cal X}} Y^2 s^8
\label{a2b4g10}\\                
a[4]f[12] &=& +\frac{{1}}{576}{\cal X}^2\left(\dot{{\cal X}}\cdot Y^{-3}  Y^3\cdot \dot{{\cal X}}\right)^2 s^8
\label{a4f12}\\               
b[4]f[12] &=& -\frac{{1}}{3456}\left(\dot{{\cal X}}\cdot Y^{-3}  Y^3\cdot \dot{{\cal X}}\right)^2 Y^2 s^8
\label{b4f12}\\                
f[6]g[10] &=& -\frac{1}{1440}\dot{{\cal X}}\cdot Y^{-3}  Y^3\cdot \dot{{\cal X}} \dot{{\cal X}}\cdot Y^{-3} Y^5\cdot \dot{{\cal X}}s^8
\label{f6g10}
\end{eqnarray}

\section{Basis Invariants for Higher-Order Corrections}

The building block of the prescription described in (\ref{L0L1L2_emptyset}) is $\left(\Delta_1G_\emptyset(p)\right)^{\mathrm{red}}_\ell$. In this section, we expand it\footnote{There are 2,974 terms.} and choose to collect\footnote{We used MS Excel spreadsheet software to sort the generated invariants according to increasing string length.}
the terms of similar mass-dimensions\footnote{In the 2,974-term collection, mass-dimensions range from 3 up to 26.}, ranging from [3] up to [12]. With the notation
\begin{eqnarray}
  [\ell]\equiv(\Delta_1G_{\emptyset_{\mathrm{t}}}(p))^{\mathrm{red}}_{\ell}
\label{ellll}
\end{eqnarray}
denoting the common mass-dimension of the collected terms.

A replacement $G_{\emptyset}\to G_{\emptyset_{\mathrm{t}}}$ (with $G_{\emptyset}$ given in (\ref{Gxrewritten})) will be implemented:
\begin{eqnarray}
  G_{\emptyset}(p)\to G_{\emptyset_{\mathrm{t}}}(p) \equiv \int^\infty_0 ds_0\,\,\, e^{\Theta_{\mathrm{TSE}_\mathrm{t}}}= \int^\infty_0 ds_0\,\,\, e^{({\cal X}-m^2-p^2)s_0}.
\end{eqnarray}
with $\Theta_{\mathrm{TSE}}$ now truncated\footnote{This is upon considering only the first term of Taylor expanded $R(s)$ given in (\ref{taylRs}) with $P(s)$ and $Q(s)$ completely disregarded.} $\Theta_{\mathrm{TSE}_\mathrm{t}}$ and $X$ redefined as $X \to {\cal X} -m^2$. Note that a subscript $0$ is instituted on the proper-time parameter $s$. This is to distinguish higher-fold proper-time integrations with respect to $s_1,s_2,\ldots$ that may become relevant in higher-order calculations.

We will group terms coming from different order of corrections that will give a total mass-dimension of twelve or less. Note that the collection includes odd mass-dimensions. We know that such odd mass-dimensions vanish owing to their odd powers in $p$-integration. We consider them here as they do not vanish in the higher order expansion for example, when they are multiplied to any odd mass-dimension collection of invariants.

With the knowledge on the mass-dimensions discussed in the earlier part of this chapter, we enumerate below the result of collecting invariants containing similar total mass-dimensions and then contracting kronecker delta indices:\footnote{Refer to Appendix B for mass-dimensional basis with uncontracted kronecker delta indices}

%
%
%
\begin{small}
\begin{eqnarray}
[3]_\ell=+\int^\infty_0 d_{s_\ell}
    \frac{2i}{3}
      \left[ {\cal D}_{\mu_\ell}\!Y_{{\lambda_\ell}{\mu_\ell}}\,p\,{s_\ell}
            +4 {\cal D}_{\kappa_\ell}\!Y_{{\lambda_\ell}{\tau_\ell}}\,p^3\,{s_\ell}^2\right]
      e^{({\cal X}-m^2-p^2)s_\ell}
\label{thr}
\end{eqnarray}
\end{small}
The first term in (\ref{thr}) agrees with Eqn. 2.94 of \cite{Rodf-Diss}.  Ref. \cite{Rodf-Diss} disregarded the last term which we choose to display instead. Here, $p^n$ means $p_{\lambda_\ell}p_{\tau_\ell}\ldots$ whose subscripts are identified from the unpaired tensorial indices. 

\begin{small}
\begin{eqnarray}
[4]_\ell\!\!\!&=&\!\!\!
\int^\infty_0 d_{s_\ell}
\left\{
+ {s_\ell} {\cal D}_{{\mu_\ell}{\mu_\ell}}{\cal X}
+ {p}^2 {s_\ell}^2
   \left[- 2 {\cal D}_{{\lambda_\ell}{\tau_\ell}}{\cal X}
    + {\cal D}_{{\tau_\ell}{\mu_\ell}}\!Y_{{\lambda_\ell}{\mu_\ell}}
    + {\cal D}_{{\mu_\ell}{\tau_\ell}}\!Y_{{\lambda_\ell}{\mu_\ell}}
    + {\cal D}_{{\mu_\ell}{\mu_\ell}}\!Y_{{\lambda_\ell}{\tau_\ell}}
\right.\right.
\nonumber\\&&\,\,\,\,\,\,\,\,\,\,\,\,\,\,\,\,\,\,\,\,
\left.\left.
    + \frac{3}{2}\left({\cal D}_{{\mu_\ell}{\tau_\ell}}\!Y_{{\mu_\ell}{\lambda_\ell}}
                  - {\cal D}_{{\tau_\ell}{\mu_\ell}}\!Y_{{\lambda_\ell}{\mu_\ell}}\right)
   \right]
- 2 {p}^4 {s_\ell}^3 {\cal D}_{{\eta_\ell}{\kappa_\ell}}\!Y_{{\lambda_\ell}{\tau_\ell}}
\right\}
      e^{({\cal {\cal X}}-m^2-p^2)s_\ell}
\label{for-v1}
\end{eqnarray}
\end{small}

The first and second terms in (\ref{for-v1}) agree with Eqn. 2.92 and (first term) Eqn. 2.95  of \cite{Rodf-Diss}, respectively. Knowing that

\begin{small}
\begin{eqnarray*}
{\cal D}_{{\tau_\ell}{\mu_\ell}}\!Y_{{\lambda_\ell}{\mu_\ell}}
- \frac{3}{2}{\cal D}_{{\tau_\ell}{\mu_\ell}}\!Y_{{\lambda_\ell}{\mu_\ell}}
&=&- \frac{1}{2}{\cal D}_{{\tau_\ell}{\mu_\ell}}\!Y_{{\lambda_\ell}{\mu_\ell}}
\nonumber\\
{\cal D}_{{\mu_\ell}{\tau_\ell}}\!Y_{{\lambda_\ell}{\mu_\ell}}
+ \frac{3}{2}{\cal D}_{{\mu_\ell}{\tau_\ell}}\!Y_{{\lambda_\ell}{\mu_\ell}}
=
{\cal D}_{{\mu_\ell}{\tau_\ell}}\!Y_{{\lambda_\ell}{\mu_\ell}}
- \frac{3}{2}{\cal D}_{{\mu_\ell}{\tau_\ell}}\!Y_{{\lambda_\ell}{\mu_\ell}}
&=&
+ \frac{1}{2}{\cal D}_{{\mu_\ell}{\tau_\ell}}\!Y_{{\lambda_\ell}{\mu_\ell}}
\end{eqnarray*}
\end{small}

Although the term ${\cal D}_{{\mu_\ell}{\mu_\ell}}\!Y_{{\lambda_\ell}{\tau_\ell}}$ vanishes after performing the $p^2$ momentum integration in the calculation of ${\cal L}^{(1)[4]}_1$, it will not vanish in ${\cal L}^{(1)[8]}_2$, etc. These simplifications would make the $p^2$ term in (\ref{for-v1}) agree with (second term) Eqn 2.95  of \cite{Rodf-Diss}. Again Ref. \cite{Rodf-Diss} disregarded the last term which we choose to display instead, including terms that will become relevant in higher mass-dimensions. Hence, we have the simplified version of (\ref{for-v1})
\begin{eqnarray}
[4]_\ell\!\!\!&=&\!\!\!
\int^\infty_0 d_{s_\ell}
\left\{
+ {s_\ell} {\cal D}^2{\cal X}
+ {p}^2 {s_\ell}^2
   \left[- 2 {\cal D}_{{\lambda_\ell}{\tau_\ell}}{\cal X}
    + \frac{1}{2}
      {\cal D}_{[{\mu_\ell}{\tau_\ell}]}\!Y_{{\lambda_\ell}{\mu_\ell}}
+{\cal D}^2Y_{{\lambda_\ell}{\tau_\ell}}
\right.
\right]
\nonumber\\&&\,\,\,\,\,\,\,\,\,\,\,\,\,\,\,\,\,\,\,\,
\left.
- 2 {p}^4 {s_\ell}^3 {\cal D}_{{\eta_\ell}{\kappa_\ell}}\!Y_{{\lambda_\ell}{\tau_\ell}}
\right\}
      e^{({\cal {\cal X}}-m^2-p^2)s_\ell}
\label{for}
\end{eqnarray}
where
\begin{eqnarray}
      {\cal D}_{[{\mu_\ell}{\tau_\ell}]}\!Y_{{\lambda_\ell}{\mu_\ell}}
&\equiv&
      {\cal D}_{{\mu_\ell}{\tau_\ell}}\!Y_{{\lambda_\ell}{\mu_\ell}}
    - {\cal D}_{{\tau_\ell}{\mu_\ell}}\!Y_{{\lambda_\ell}{\mu_\ell}}
\\
{\cal D}^2 
&\equiv &{\cal D}_{{\mu_\ell}{\mu_\ell}}
\end{eqnarray}

The $p^4$ term will become relevant in $[8]$, $[10]$, and $[12]$ mass-dimensions appearing in the higher-order corrections. 
\begin{eqnarray}
[5]_\ell\!\!\!&=&\!\!\!
\int^\infty_0 d_{s_\ell}
\left\{
+\frac{2i}{3}\,{p}\,{s_\ell}^2
(
 {\cal D}_{({\lambda_\ell}{\mu_\ell}{\mu_\ell})}{\cal X}
\right.
 + 2 
    {\cal D}_{<{\lambda_\ell}}\!Y_{{\mu_\ell}{\nu_\ell}>} 
     Y_{{\mu_\ell}{\nu_\ell}}
)
\nonumber\\&&\,\,\,\,\,\,\,\,\,\,\,\,\,\,\,\,\,\,\,\,
\left.
-\frac{4i}{3}\,{p}^3\,{s_\ell}^3
(
 {\cal D}_{{\kappa_\ell}{\tau_\ell}{\lambda_\ell}}{\cal X}
 + ({\cal D}_{{\tau_\ell}}\!Y_{{\mu_\ell}{\lambda_\ell}}) Y_{{\mu_\ell}{\kappa_\ell}}
 + ({\cal D}_{{\kappa_\ell}}\!Y_{{\mu_\ell}{\tau_\ell}}) Y_{{\mu_\ell}{\lambda_\ell}}
)
\right\}
      e^{({\cal X}-m^2-p^2)s_\ell}
\nonumber\\
\label{fiv}
\end{eqnarray}
where
\begin{eqnarray}
{\cal D}_{({\mu_\ell}{\nu_\ell}{\rho_\ell})}{\cal X}
&\equiv&
{\cal D}_{{\mu_\ell}\left\{{\nu_\ell}{\rho_\ell}\right\}}{\cal X}
+{\cal D}_{{\nu_\ell}\left\{{\rho_\ell}{\mu_\ell}\right\}}{\cal X}
+{\cal D}_{{\rho_\ell}\left\{{\mu_\ell}{\nu_\ell}\right\}}{\cal X}
\\
{\cal D}_{{\mu_\ell}\left\{{\nu_\ell}{\rho_\ell}\right\}}{\cal X}
&\equiv&
{\cal D}_{{\mu_\ell}{\nu_\ell}{\rho_\ell}}{\cal X}
+{\cal D}_{{\mu_\ell}{\rho_\ell}{\nu_\ell}}{\cal X}
\\
{\cal D}_{<{\tau_\ell}}\!Y_{{\mu_\ell}{\lambda_\ell}>}=
&\equiv &
{\cal D}_{{\lambda_\ell}}\!Y_{{\mu_\ell}{\nu_\ell}} + {\cal D}_{{\nu_\ell}}\!Y_{{\mu_\ell}{\lambda_\ell}}
\end{eqnarray}
%
\begin{small}
\begin{eqnarray}
[6]_\ell\!\!\!&=&\!\!\!
\int^\infty_0 d_{s_\ell}
\left\{
+\frac{1}{2}\,{s_\ell}^2 Y_{{\mu_\ell}{\nu_\ell}}
(
 {\cal D}^2
Y_{{\mu_\ell}{\nu_\ell}}
 + {\cal D}_{\left\{
             {\rho_\ell}{\nu_\ell}\right\}}\!Y_{{\mu_\ell}{\rho_\ell}}
)
\right.
\nonumber\\&&\,\,\,\,\,\,\,\,\,\,\,\,\,\,\,\,\,\,\,\,
+\frac{4}{9}\,{s_\ell}^2
(
 ({\cal D}_{{\nu_\ell}}\!Y_{{\mu_\ell}{\nu_\ell}})({\cal D}_{\rho_\ell}Y_{{\mu_\ell}{\rho_\ell}})
 + ({\cal D}_{\rho_\ell}Y_{{\mu_\ell}{\nu_\ell}}) ({\cal D}_{<\rho_\ell}Y_{{\mu_\ell}{\nu_\ell}>})
)
\nonumber\\&&\,\,\,\,\,\,\,\,\,\,\,\,\,\,\,\,\,\,\,\,
+\frac{4}{3} \,p^2 \,{s_\ell}^3 ({\cal D}_{{\mu_\ell}}{\cal X})
  (
    (
  {\cal D}_{<{\tau_\ell}}Y_{{\lambda_\ell}{\mu_\ell}>}
    )+({\cal D}_{{\nu_\ell}}Y_{{\mu_\ell}{\nu_\ell}})
  )
\nonumber\\&&\,\,\,\,\,\,\,\,\,\,\,\,\,\,\,\,\,\,\,\,
-\,{p}^2\,{s_\ell}^3
(
  (
  {\cal D}_{{\tau_\ell}{\lambda_\ell}}\!Y_{{\mu_\ell}{\nu_\ell}}
 + {\cal D}_{{\tau_\ell}{\nu_\ell}}\!Y_{{\mu_\ell}{\lambda_\ell}}
 + {\cal D}_{{\nu_\ell}{\lambda_\ell}}\!Y_{{\mu_\ell}{\tau_\ell}}
  )
   Y_{{\mu_\ell}{\nu_\ell}}
\nonumber\\&&\,\,\,\,\,\,\,\,\,\,\,\,\,\,\,\,\,\,\,\,\,\,\,\,\,\,\,\,\,\,\,\,\,\,\,\,\,\,\,\,
 +
  (
  {\cal D}_{\left\{{\tau_\ell}{\nu_\ell}\right\}}\!Y_{{\mu_\ell}{\nu_\ell}}
 + {\cal D}^2
                                 Y_{{\mu_\ell}{\tau_\ell}}
  ) Y_{{\mu_\ell}{\lambda_\ell}}
)
\nonumber\\&&\,\,\,\,\,\,\,\,\,\,\,\,\,\,\,\,\,\,\,\,
-\frac{8}{9}\,{p}^2\,{s_\ell}^3
(2
 (
  ({\cal D}_{{\nu_\ell}}\!Y_{{\mu_\ell}{\nu_\ell}}) ({\cal D}_{{\tau_\ell}}\!Y_{{\mu_\ell}{\lambda_\ell}}) + ({\cal D}_{{\nu_\ell}}\!Y_{{\mu_\ell}{\lambda_\ell}}) ({\cal D}_{{\tau_\ell}}\!Y_{{\mu_\ell}{\nu_\ell}})
 )
\nonumber\\&&\,\,\,\,\,\,\,\,\,\,\,\,\,\,\,\,\,\,\,\,\,\,\,\,\,\,\,\,\,\,\,\,\,\,\,\,\,\,\,\,
 + ({\cal D}_{{\lambda_\ell}}\!Y_{{\mu_\ell}{\nu_\ell}}) ({\cal D}_{{\tau_\ell}}\!Y_{{\mu_\ell}{\nu_\ell}})  + ({\cal D}_{{\nu_\ell}}\!Y_{{\mu_\ell}{\lambda_\ell}}) ({\cal D}_{{\nu_\ell}}\!Y_{{\mu_\ell}{\tau_\ell}})
)
\nonumber\\&&\,\,\,\,\,\,\,\,\,\,\,\,\,\,\,\,\,\,\,\,
\left.
+ 2\,{p}^4\,{s_\ell}^4 ({\cal D}_{{\eta_\ell}{\kappa_\ell}}\!Y_{{\mu_\ell}{\tau_\ell}}) Y_{{\mu_\ell}{\lambda_\ell}}
+\frac{16}{9}\,{p}^4\,{s_\ell}^4 ({\cal D}_{{\tau_\ell}}\!Y_{{\mu_\ell}{\lambda_\ell}}) ({\cal D}_{\eta_\ell}Y_{{\mu_\ell}{\kappa_\ell}})
\right\}
      e^{({\cal X}-m^2-p^2)s_\ell}
\nonumber\\
\label{six}
\end{eqnarray}
\end{small}
%
%
\begin{small}
\begin{eqnarray}
[7]_\ell\!\!\!&=&\!\!\!
\int^\infty_0 d_{s_\ell}
\left\{
+\frac{2i}{3}\,{p}\,{s_\ell}^2
(
 {\cal D}_{({\lambda_\ell}{\mu_\ell}{\mu_\ell})}{\cal X}
 + 2 ({\cal D}_{{\nu_\ell}}\!Y_{{\mu_\ell}{\nu_\ell}}) Y_{{\mu_\ell}{\lambda_\ell}}
\right.
 +
 2 (
    {\cal D}_{<{\lambda_\ell}}\!Y_{{\mu_\ell}{\nu_\ell}>}
   ) Y_{{\mu_\ell}{\nu_\ell}}
)
\nonumber\\&&\,\,\,\,\,\,\,\,\,\,\,\,\,\,\,\,\,\,\,\,
-\frac{i}{2} \,p \,{s_\ell}^3 {\cal D}_{{\mu_\ell}}{\cal X}
 (
  {\cal D}^2
            Y_{{\lambda_\ell}{\mu_\ell}}
+{\cal D}_{\left\{{\mu_\ell}{\nu_\ell}\right\}}Y_{{\lambda_\ell}{\nu_\ell}}
  )
\nonumber\\&&\,\,\,\,\,\,\,\,\,\,\,\,\,\,\,\,\,\,\,\,
+\frac{3i}{4} \,p \,{s_\ell}^3 {\cal D}_{{\mu_\ell}}{\cal X}
 (
      {\cal D}_{\left\{{\lambda_\ell}{\nu_\ell}\right\}}Y_{{\mu_\ell}{\nu_\ell}}
     +{\cal D}_{\left\{{\mu_\ell}{\nu_\ell}\right\}}Y_{{\lambda_\ell}{\nu_\ell}}
  +(4/3)
     {\cal D}_{\left\{{\mu_\ell}{\lambda_\ell}\right\}}{\cal X}
  )
\nonumber\\&&\,\,\,\,\,\,\,\,\,\,\,\,\,\,\,\,\,\,\,\,
+i \,p^3 \,{s_\ell}^4 {\cal D}_{{\mu_\ell}}{\cal X}
  (
     {\cal D}_{{\kappa_\ell}{\tau_\ell}}Y_{{\lambda_\ell}{\mu_\ell}}
    + {\cal D}_{{\kappa_\ell}{\mu_\ell}}Y_{{\lambda_\ell}{\tau_\ell}}+{\cal D}_{{\mu_\ell}{\kappa_\ell}}Y_{{\lambda_\ell}{\tau_\ell}}
   )
\nonumber\\&&\,\,\,\,\,\,\,\,\,\,\,\,\,\,\,\,\,\,\,\,
\left.
-\frac{4i}{3}\,{p}^3\,{s_\ell}^3
(
 {\cal D}_{{\kappa_\ell}{\tau_\ell}{\lambda_\ell}}{\cal X} + ({\cal D}_{{\tau_\ell}}\!Y_{{\mu_\ell}{\lambda_\ell}}) Y_{{\mu_\ell}{\kappa_\ell}} + ({\cal D}_{{\kappa_\ell}}\!Y_{{\mu_\ell}{\tau_\ell}}) Y_{{\mu_\ell}{\lambda_\ell}}
)
\right\}
      e^{({\cal X}-m^2-p^2)s_\ell}
\nonumber\\
\label{svn}
\end{eqnarray}
\end{small}
\begin{small}
\begin{eqnarray}
[8]_\ell\!\!\!&=&\!\!\!
\int^\infty_0 d_{s_\ell}
\left\{
\frac{-1}{3} \,{s_\ell}^3 ({\cal D}_{{\mu_\ell}{\nu_\ell}}{\cal X}) Y^2_{{\mu_\ell}{\nu_\ell}}
\!+\!\frac{1}{4} \,{s_\ell}^3 Y^2_{{\mu_\ell}{\nu_\ell}}
  (
   {\cal D}_{\left\{{\rho_\ell}{\nu_\ell}\right\}}\!Y_{{\rho_\ell}{\mu_\ell}}
   )
\right.
\!+\!\frac{1}{3} \,{s_\ell}^3 ({\cal D}_{{\mu_\ell}}{\cal X})
  (
     {\cal D}_{({\nu_\ell}{\nu_\ell}{\mu_\ell})}{\cal X}
  )
\nonumber\\&&\,\,\,\,\,\,\,\,\,\,\,\,\,\,\,\,\,\,\,\,
+\frac{1}{3} \,{s_\ell}^3 {\cal D}_{{\mu_\ell}}{\cal X}
  (
     2 ({\cal D}_{{\rho_\ell}}Y_{{\nu_\ell}{\rho_\ell}}) Y_{{\nu_\ell}{\mu_\ell}}
    +2 ({\cal D}_{<{\rho_\ell}}Y_{{\nu_\ell}{\mu_\ell}>}
	) Y_{{\nu_\ell}{\rho_\ell}}
   )
\nonumber\\&&\,\,\,\,\,\,\,\,\,\,\,\,\,\,\,\,\,\,\,\,
+\frac{1}{8} \,{s_\ell}^3
   (
    {\cal D}_{{\rho_\ell}{\nu_\ell}}\!Y_{{\mu_\ell}{\nu_\ell}}
     (
      {\cal D}^2
		Y_{{\mu_\ell}{\rho_\ell}}
 + {\cal D}_{\{{\sigma_\ell}{\rho_\ell}\}}\!Y_{{\mu_\ell}{\sigma_\ell}}
      )
\nonumber\\&&\,\,\,\,\,\,\,\,\,\,\,\,\,\,\,\,\,\,\,\,\,\,\,\,\,\,\,\,\,\,\,\,\,\,\,\,\,\,\,\,
        + {\cal D}_{{\nu_\ell}{\rho_\ell}}\!Y_{{\mu_\ell}{\nu_\ell}}
     (
       {\cal D}^2
		Y_{{\mu_\ell}{\rho_\ell}}
 + {\cal D}_{\{{\sigma_\ell}{\rho_\ell}\}}\!Y_{{\mu_\ell}{\sigma_\ell}}
      )
\nonumber\\&&\,\,\,\,\,\,\,\,\,\,\,\,\,\,\,\,\,\,\,\,\,\,\,\,\,\,\,\,\,\,\,\,\,\,\,\,\,\,\,\,
        + {\cal D}^2
     			Y_{{\mu_\ell}{\nu_\ell}}
      (
       {\cal D}^2
		Y_{{\mu_\ell}{\nu_\ell}}
 + {\cal D}_{\{{\sigma_\ell}{\nu_\ell}\}}\!Y_{{\mu_\ell}{\sigma_\ell}}
      )
\nonumber\\&&\,\,\,\,\,\,\,\,\,\,\,\,\,\,\,\,\,\,\,\,\,\,\,\,\,\,\,\,\,\,\,\,\,\,\,\,\,\,\,\,
        + {\cal D}_{{\sigma_\ell}{\rho_\ell}}\!Y_{{\mu_\ell}{\nu_\ell}}
      (
        (
         {\cal D}_{\{{\sigma_\ell}{\rho_\ell}\}}\!Y_{{\mu_\ell}{\nu_\ell}}
        )
          +
       (
        {\cal D}_{\{{\sigma_\ell}{\nu_\ell}\}}\!Y_{{\mu_\ell}{\rho_\ell}}
       )
 +
       (
        {\cal D}_{\{{\rho_\ell}{\nu_\ell}\}}\!Y_{{\mu_\ell}{\sigma_\ell}}
       )
      )
   )
\nonumber\\&&\,\,\,\,\,\,\,\,\,\,\,\,\,\,\,\,\,\,\,\,
-\frac{2}{3} \,p^2 \,{s_\ell}^4 {\cal D}_{{\mu_\ell}}{\cal X}
  (
     (
         {\cal D}_{{\tau_\ell}{\lambda_\ell}{\mu_\ell}}{\cal X}
	+{\cal D}_{\{{\tau_\ell}{\mu_\ell}\}{\lambda_\ell}}{\cal X}
      )
       +2 ({\cal D}_{{\tau_\ell}}Y_{{\nu_\ell}{\lambda_\ell}}) Y_{{\nu_\ell}{\mu_\ell}}
\nonumber\\&&\,\,\,\,\,\,\,\,\,\,\,\,\,\,\,\,\,\,\,\,\,\,\,\,\,\,\,\,\,\,\,\,\,\,\,\,\,\,\,\,
       +
     (
       {\cal D}_{{\lambda_\ell}}Y_{{\nu_\ell}{\mu_\ell}}
	+{\cal D}_{{\mu_\ell}}Y_{{\nu_\ell}{\lambda_\ell}}
      ) Y_{{\nu_\ell}{\tau_\ell}}
       +({\cal D}_{{\tau_\ell}}Y_{{\nu_\ell}{\mu_\ell}}
	+{\cal D}_{{\mu_\ell}}Y_{{\nu_\ell}{\tau_\ell}}) Y_{{\nu_\ell}{\lambda_\ell}}
   )
\nonumber\\&&\,\,\,\,\,\,\,\,\,\,\,\,\,\,\,\,\,\,\,\,
+\frac{2}{3} \,p^2 \,{s_\ell}^4
      (
       ({\cal D}_{{\lambda_\ell}{\mu_\ell}}{\cal X}) Y^2_{{\mu_\ell}{\tau_\ell}}
      + ({\cal D}_{{\mu_\ell}{\tau_\ell}}{\cal X}) Y^2_{{\mu_\ell}{\lambda_\ell}}
      )
\nonumber\\&&\,\,\,\,\,\,\,\,\,\,\,\,\,\,\,\,\,\,\,\,
+\frac{1}{2} \,p^2 \,{s_\ell}^4
    (
      (
       {\cal D}_{\{{\nu_\ell}{\mu_\ell}\}}\!Y_{{\lambda_\ell}{\mu_\ell}}
      ) Y^2_{{\nu_\ell}{\tau_\ell}} +
      (
       {\cal D}_{\{{\tau_\ell}{\nu_\ell}\}}\!Y_{{\mu_\ell}{\nu_\ell}}
      ) Y^2_{{\mu_\ell}{\lambda_\ell}}
    )
\nonumber\\&&\,\,\,\,\,\,\,\,\,\,\,\,\,\,\,\,\,\,\,\,
-\frac{1}{3} \,p^2 \,{s_\ell}^4
    (
      (
      {\cal D}_{\{{\rho_\ell}{\nu_\ell}\}}\!Y_{{\mu_\ell}{\nu_\ell}}
	 + {\cal D}^2
			Y_{{\mu_\ell}{\rho_\ell}}
      ) Y^2_{{\rho_\ell}{\lambda_\ell}}
\nonumber\\&&\,\,\,\,\,\,\,\,\,\,\,\,\,\,\,\,\,\,\,\,\,\,\,\,\,\,\,\,\,\,\,\,\,\,\,\,\,\,\,\,
              +
      (
      {\cal D}_{{\rho_\ell}{\nu_\ell}}\!Y_{{\mu_\ell}{\lambda_\ell}}
 + {\cal D}_{\{{\rho_\ell}{\lambda_\ell}\}}\!Y_{{\mu_\ell}{\nu_\ell}}
      ) Y^2_{{\nu_\ell}{\rho_\ell}}
    )
\nonumber\\&&\,\,\,\,\,\,\,\,\,\,\,\,\,\,\,\,\,\,\,\,
-\frac{1}{4} \,p^2 \,{s_\ell}^4
  (
   +(
     {\cal D}_{{\mu_\ell}{\lambda_\ell}}\!Y_{{\nu_\ell}{\tau_\ell}}
 + {\cal D}_{{\tau_\ell}{\lambda_\ell}}\!Y_{{\nu_\ell}{\mu_\ell}}
    ) {\cal D}^2
			Y_{{\nu_\ell}{\mu_\ell}}
\nonumber\\&&\,\,\,\,\,\,\,\,\,\,\,\,\,\,\,\,\,\,\,\,\,\,\,\,\,\,\,\,\,\,\,\,\,\,\,\,\,\,\,\,
   +(
     {\cal D}^2
		Y_{{\nu_\ell}{\mu_\ell}}
 + {\cal D}_{{\rho_\ell}{\mu_\ell}}\!Y_{{\nu_\ell}{\rho_\ell}}
    ) {\cal D}_{{\tau_\ell}{\mu_\ell}}\!Y_{{\nu_\ell}{\lambda_\ell}}
\nonumber\\&&\,\,\,\,\,\,\,\,\,\,\,\,\,\,\,\,\,\,\,\,\,\,\,\,\,\,\,\,\,\,\,\,\,\,\,\,\,\,\,\,
   +(
     {\cal D}_{\{{\lambda_\ell}{\rho_\ell}\}}\!Y_{{\nu_\ell}{\mu_\ell}}
     ) {\cal D}_{{\tau_\ell}{\mu_\ell}}\!Y_{{\nu_\ell}{\rho_\ell}}
   +(
     {\cal D}_{\{{\rho_\ell}{\tau_\ell}\}}\!Y_{{\nu_\ell}{\mu_\ell}}
     ) {\cal D}_{{\rho_\ell}{\mu_\ell}}\!Y_{{\nu_\ell}{\lambda_\ell}}
\nonumber\\&&\,\,\,\,\,\,\,\,\,\,\,\,\,\,\,\,\,\,\,\,\,\,\,\,\,\,\,\,\,\,\,\,\,\,\,\,\,\,\,\,
   +(
     {\cal D}_{{\lambda_\ell}{\rho_\ell}}\!Y_{{\nu_\ell}{\mu_\ell}} + {\cal D}_{{\rho_\ell}{\mu_\ell}}\!Y_{{\nu_\ell}{\lambda_\ell}}
    ) {\cal D}_{{\mu_\ell}{\rho_\ell}}\!Y_{{\nu_\ell}{\tau_\ell}}
\nonumber\\&&\,\,\,\,\,\,\,\,\,\,\,\,\,\,\,\,\,\,\,\,\,\,\,\,\,\,\,\,\,\,\,\,\,\,\,\,\,\,\,\,
   +(
     {\cal D}_{{\mu_\ell}{\lambda_\ell}}\!Y_{{\nu_\ell}{\rho_\ell}} + {\cal D}_{{\mu_\ell}{\rho_\ell}}\!Y_{{\nu_\ell}{\lambda_\ell}}
    ) {\cal D}_{{\rho_\ell}{\tau_\ell}}\!Y_{{\nu_\ell}{\mu_\ell}}
   +(
     {\cal D}_{\{{\tau_\ell}{\rho_\ell}\}}\!Y_{{\nu_\ell}{\lambda_\ell}}
    ) {\cal D}^2
		Y_{{\nu_\ell}{\rho_\ell}}
\nonumber\\&&\,\,\,\,\,\,\,\,\,\,\,\,\,\,\,\,\,\,\,\,\,\,\,\,\,\,\,\,\,\,\,\,\,\,\,\,\,\,\,\,
   +(
     2{\cal D}_{{\rho_\ell}{\mu_\ell}}\!Y_{{\nu_\ell}{\mu_\ell}}
 + {\cal D}_{{\mu_\ell}{\rho_\ell}}\!Y_{{\nu_\ell}{\mu_\ell}}
    ) {\cal D}_{{\tau_\ell}{\rho_\ell}}\!Y_{{\nu_\ell}{\lambda_\ell}}
\nonumber\\&&\,\,\,\,\,\,\,\,\,\,\,\,\,\,\,\,\,\,\,\,\,\,\,\,\,\,\,\,\,\,\,\,\,\,\,\,\,\,\,\,
   +(
     {\cal D}^2
		Y_{{\nu_\ell}{\mu_\ell}}
 + {\cal D}_{\{{\rho_\ell}{\mu_\ell}\}}\!Y_{{\nu_\ell}{\rho_\ell}}
    ) {\cal D}_{{\lambda_\ell}{\tau_\ell}}\!Y_{{\nu_\ell}{\mu_\ell}}
\nonumber\\&&\,\,\,\,\,\,\,\,\,\,\,\,\,\,\,\,\,\,\,\,\,\,\,\,\,\,\,\,\,\,\,\,\,\,\,\,\,\,\,\,
   +(
     {\cal D}_{\{{\tau_\ell}{\rho_\ell}\}}\!Y_{{\nu_\ell}{\rho_\ell}}
 + {\cal D}^2
		Y_{{\nu_\ell}{\tau_\ell}}
    ) {\cal D}^2
		Y_{{\nu_\ell}{\lambda_\ell}}
\nonumber\\&&\,\,\,\,\,\,\,\,\,\,\,\,\,\,\,\,\,\,\,\,\,\,\,\,\,\,\,\,\,\,\,\,\,\,\,\,\,\,\,\,
   +(
     {\cal D}_{\{{\rho_\ell}{\mu_\ell}\}}\!Y_{{\nu_\ell}{\lambda_\ell}}
 + {\cal D}_{{\mu_\ell}{\lambda_\ell}}\!Y_{{\nu_\ell}{\rho_\ell}}
     ) {\cal D}_{{\tau_\ell}{\rho_\ell}}\!Y_{{\nu_\ell}{\mu_\ell}}
\nonumber\\&&\,\,\,\,\,\,\,\,\,\,\,\,\,\,\,\,\,\,\,\,\,\,\,\,\,\,\,\,\,\,\,\,\,\,\,\,\,\,\,\,
   +(
     {\cal D}_{\{{\tau_\ell}{\mu_\ell}\}}\!Y_{{\nu_\ell}{\rho_\ell}}
 + {\cal D}_{{\rho_\ell}{\mu_\ell}}\!Y_{{\nu_\ell}{\tau_\ell}}
    ) {\cal D}_{{\rho_\ell}{\mu_\ell}}\!Y_{{\nu_\ell}{\lambda_\ell}}
\nonumber\\&&\,\,\,\,\,\,\,\,\,\,\,\,\,\,\,\,\,\,\,\,\,\,\,\,\,\,\,\,\,\,\,\,\,\,\,\,\,\,\,\,
   +(
     {\cal D}_{\{{\lambda_\ell}{\rho_\ell}\}}\!Y_{{\nu_\ell}{\mu_\ell}}
 ({\cal D}_{\{{\tau_\ell}{\rho_\ell}\}}\!Y_{{\nu_\ell}{\mu_\ell}}
    )
\nonumber\\&&\,\,\,\,\,\,\,\,\,\,\,\,\,\,\,\,\,\,\,\,\,\,\,\,\,\,\,\,\,\,\,\,\,\,\,\,\,\,\,\,
   +(
     {\cal D}_{\{{\rho_\ell}{\mu_\ell}\}}\!Y_{{\nu_\ell}{\mu_\ell}}
	)
({\cal D}_{{\tau_\ell}{\lambda_\ell}}\!Y_{{\nu_\ell}{\rho_\ell}} + {\cal D}_{{\rho_\ell}{\tau_\ell}}\!Y_{{\nu_\ell}{\lambda_\ell}} + {\cal D}_{{\rho_\ell}{\lambda_\ell}}\!Y_{{\nu_\ell}{\tau_\ell}}
    )
\nonumber\\&&\,\,\,\,\,\,\,\,\,\,\,\,\,\,\,\,\,\,\,\,\,\,\,\,\,\,\,\,\,\,\,\,\,\,\,\,\,\,\,\,
   +(
    {\cal D}_{\{{\lambda_\ell}{\mu_\ell}\}}\!Y_{{\nu_\ell}{\mu_\ell}}
 ({\cal D}_{\{{\tau_\ell}{\rho_\ell}\}}\!Y_{{\nu_\ell}{\rho_\ell}}
 + {\cal D}^2
		Y_{{\nu_\ell}{\tau_\ell}}
    )
  )
\nonumber\\&&\,\,\,\,\,\,\,\,\,\,\,\,\,\,\,\,\,\,\,\,
+\frac{2}{3} \,p^4 \,{s_\ell}^5
   (
    ({\cal D}_{{\mu_\ell}{\kappa_\ell}}\!Y_{{\lambda_\ell}{\tau_\ell}}) Y^2_{{\mu_\ell}{\eta_\ell}} + ({\cal D}_{{\eta_\ell}{\mu_\ell}}\!Y_{{\lambda_\ell}{\tau_\ell}}) Y^2_{{\mu_\ell}{\kappa_\ell}} + ({\cal D}_{{\eta_\ell}{\kappa_\ell}}\!Y_{{\lambda_\ell}{\mu_\ell}}) Y^2_{{\mu_\ell}{\tau_\ell}}
   )
\nonumber\\&&\,\,\,\,\,\,\,\,\,\,\,\,\,\,\,\,\,\,\,\,
+\frac{1}{2} \,p^4 \,{s_\ell}^5
  (
   (
    {\cal D}_{{\eta_\ell}{\mu_\ell}}\!Y_{{\nu_\ell}{\lambda_\ell}}
 + {\cal D}_{{\mu_\ell}{\lambda_\ell}}\!Y_{{\nu_\ell}{\eta_\ell}}
   ) {\cal D}_{{\kappa_\ell}{\tau_\ell}}\!Y_{{\nu_\ell}{\mu_\ell}}
\nonumber\\&&\,\,\,\,\,\,\,\,\,\,\,\,\,\,\,\,\,\,\,\,\,\,\,\,\,\,\,\,\,\,\,\,\,\,\,\,\,\,\,\,
 + (
    {\cal D}_{{\eta_\ell}{\mu_\ell}}\!Y_{{\nu_\ell}{\tau_\ell}}
 + {\cal D}_{{\nu_\ell}{\tau_\ell}}\!Y_{{\mu_\ell}{\eta_\ell}}
   ) {\cal D}_{{\kappa_\ell}{\mu_\ell}}\!Y_{{\nu_\ell}{\lambda_\ell}}
\nonumber\\&&\,\,\,\,\,\,\,\,\,\,\,\,\,\,\,\,\,\,\,\,\,\,\,\,\,\,\,\,\,\,\,\,\,\,\,\,\,\,\,\,
 + (
    {\cal D}_{\{{\eta_\ell}{\mu_\ell}\}}\!Y_{{\nu_\ell}{\kappa_\ell}}
   ) {\cal D}_{{\mu_\ell}{\tau_\ell}}\!Y_{{\nu_\ell}{\lambda_\ell}}
\nonumber\\&&\,\,\,\,\,\,\,\,\,\,\,\,\,\,\,\,\,\,\,\,\,\,\,\,\,\,\,\,\,\,\,\,\,\,\,\,\,\,\,\,
 + (
    {\cal D}_{\{{\tau_\ell}{\mu_\ell}\}}\!Y_{{\nu_\ell}{\lambda_\ell}}
 + {\cal D}_{{\lambda_\ell}{\tau_\ell}}\!Y_{{\nu_\ell}{\mu_\ell}}
   ) {\cal D}_{{\eta_\ell}{\kappa_\ell}}\!Y_{{\nu_\ell}{\mu_\ell}}
\nonumber\\&&\,\,\,\,\,\,\,\,\,\,\,\,\,\,\,\,\,\,\,\,\,\,\,\,\,\,\,\,\,\,\,\,\,\,\,\,\,\,\,\,
 + (
    {\cal D}_{\{{\lambda_\ell}{\mu_\ell}\}}\!Y_{{\nu_\ell}{\mu_\ell}}
 + {\cal D}^2
	Y_{{\nu_\ell}{\lambda_\ell}}
   ) {\cal D}_{{\eta_\ell}{\kappa_\ell}}\!Y_{{\nu_\ell}{\tau_\ell}}
\nonumber\\&&\,\,\,\,\,\,\,\,\,\,\,\,\,\,\,\,\,\,\,\,\,\,\,\,\,\,\,\,\,\,\,\,\,\,\,\,\,\,\,\,
 + (
    {\cal D}^2
		Y_{{\nu_\ell}{\eta_\ell}}
 + {\cal D}_{\{{\mu_\ell}{\eta_\ell}\}}\!Y_{{\nu_\ell}{\mu_\ell}}
   ) {\cal D}_{{\kappa_\ell}{\tau_\ell}}\!Y_{{\nu_\ell}{\lambda_\ell}}
  )
\nonumber\\&&\,\,\,\,\,\,\,\,\,\,\,\,\,\,\,\,\,\,\,\,
\left.
- \,p^6 \,{s_\ell}^6 ({\cal D}_{{\kappa_\ell}{\tau_\ell}}\!Y_{{\mu_\ell}{\lambda_\ell}}) ({\cal D}_{{\varrho_\ell}{\xi_\ell}}\!Y_{{\mu_\ell}{\eta_\ell}})
\right\}
      e^{({\cal X}-m^2-p^2)s_\ell}
\label{eit}
\end{eqnarray}
\end{small}
\begin{small}
\begin{eqnarray}
[9]_\ell\!\!\!&=&\!\!\!\!\!\!
\int^\infty_0\!\! d_{s_\ell}\!
\left\{
\!\frac{-2i}{9}\,{p} \,{s_\ell}^4
  (
   (
    {\cal D}_{\{{\mu_\ell}{\nu_\ell}\}{\nu_\ell}}{\cal X}
 \! + \! {\cal D}_{{\nu_\ell}{\nu_\ell}{\mu_\ell}}{\cal X}
\!\! + \! 2 ({\cal D}_{{\rho_\ell}}\!Y_{{\nu_\ell}{\rho_\ell}})Y_{{\nu_\ell}{\mu_\ell}}
\right.
\! \!               + \!
    (
    2 {\cal D}_{<{\mu_\ell}}\!Y_{{\nu_\ell}{\rho_\ell}>}
    ) Y_{{\nu_\ell}{\rho_\ell}}
   ) Y^2_{{\mu_\ell}{\lambda_\ell}}
\nonumber\\&&\,\,\,\,\,\,\,\,\,\,\,\,\,\,\,\,\,\,\,\,\,\,\,\,\,\,\,\,\,\,\,\,\,\,\,
              +
    (
     {\cal D}_{{\nu_\ell}{\mu_\ell}{\lambda_\ell}}{\cal X}
 + {\cal D}_{\{{\nu_\ell}{\lambda_\ell}\}{\mu_\ell}}{\cal X}
 + ({\cal D}_{<{\nu_\ell}}\!Y_{{\rho_\ell}{\lambda_\ell}>}
    ) Y_{{\rho_\ell}{\mu_\ell}}
\nonumber\\&&\,\,\,\,\,\,\,\,\,\,\,\,\,\,\,\,\,\,\,\,\,\,\,\,\,\,\,\,\,\,\,\,\,\,\,
                +
    (
     {\cal D}_{<{\mu_\ell}}\!Y_{{\rho_\ell}{\lambda_\ell}>}
    ) \!Y_{{\rho_\ell}{\nu_\ell}} + 2 ({\cal D}_{{\nu_\ell}}\!Y_{{\rho_\ell}{\mu_\ell}}) Y_{{\rho_\ell}{\lambda_\ell}}
   ) Y^2_{{\mu_\ell}{\nu_\ell}}
  )
\nonumber\\&&\,\,\,\,\,\,\,\,\,\,\,\,\,\,\,\,\,\,\,\,
-\frac{2i}{3} p \,{s_\ell}^4 ({\cal D}_{{\mu_\ell}}{\cal X}) ({\cal D}_{{\nu_\ell}}{\cal X}) ({\cal D}_{{\nu_\ell}}Y_{{\lambda_\ell}{\mu_\ell}})
\nonumber\\&&\,\,\,\,\,\,\,\,\,\,\,\,\,\,\,\,\,\,\,\,
+\frac{i}{2} p \,{s_\ell}^4 {\cal D}_{{\mu_\ell}}{\cal X}
    (
        ({\cal D}^2
		Y_{{\nu_\ell}{\lambda_\ell}}
+{\cal D}_{\{{\rho_\ell}{\lambda_\ell}\}}Y_{{\nu_\ell}{\rho_\ell}}
) Y_{{\nu_\ell}{\mu_\ell}}
\nonumber\\&&\,\,\,\,\,\,\,\,\,\,\,\,\,\,\,\,\,\,\,\,\,\,\,\,\,\,\,\,\,\,\,\,\,\,\,
	+
        (
             ({\cal D}_{\{{\lambda_\ell}{\mu_\ell}\}}Y_{{\nu_\ell}{\rho_\ell}}
		)
        +
            ({\cal D}_{\{{\lambda_\ell}{\rho_\ell}\}}Y_{{\nu_\ell}{\mu_\ell}}
		)
        +   ({\cal D}_{\{{\rho_\ell}{\mu_\ell}\}}Y_{{\nu_\ell}{\lambda_\ell}}
		)
        ) Y_{{\nu_\ell}{\rho_\ell}}
\nonumber\\&&\,\,\,\,\,\,\,\,\,\,\,\,\,\,\,\,\,\,\,\,\,\,\,\,\,\,\,\,\,\,\,\,\,\,\,
	+   ({\cal D}^2
			Y_{{\nu_\ell}{\mu_\ell}}
	+{\cal D}_{\{{\rho_\ell}{\mu_\ell}\}}Y_{{\nu_\ell}{\rho_\ell}}
		) Y_{{\nu_\ell}{\lambda_\ell}}
     )
\nonumber\\&&\,\,\,\,\,\,\,\,\,\,\,\,\,\,\,\,\,\,\,\,
+\frac{4i}{9} p \,{s_\ell}^4 {\cal D}_{{\mu_\ell}}{\cal X}
       (2
            ({\cal D}_{<{\lambda_\ell}}Y_{{\nu_\ell}{\mu_\ell}>}
		) ({\cal D}_{{\rho_\ell}}Y_{{\nu_\ell}{\rho_\ell}})
      +2    ({\cal D}_{<{\rho_\ell}}Y_{{\nu_\ell}{\mu_\ell}>}
		)
	({\cal D}_{{\rho_\ell}}Y_{{\nu_\ell}{\lambda_\ell}})
\nonumber\\&&\,\,\,\,\,\,\,\,\,\,\,\,\,\,\,\,\,\,\,\,\,\,\,\,\,\,\,\,\,\,\,\,\,\,\,
             +2 ({\cal D}_{<{\rho_\ell}}Y_{{\nu_\ell}{\mu_\ell}>}
		)
		({\cal D}_{{\lambda_\ell}}Y_{{\nu_\ell}{\rho_\ell}})
        )
\nonumber\\&&\,\,\,\,\,\,\,\,\,\,\,\,\,\,\,\,\,\,\,\,
-i \,p^3 \,{s_\ell}^5 {\cal D}_{{\mu_\ell}}{\cal X}
        (
            ({\cal D}_{{\kappa_\ell}{\tau_\ell}}Y_{{\nu_\ell}{\lambda_\ell}}) Y_{{\nu_\ell}{\mu_\ell}}
           +({\cal D}_{{\kappa_\ell}{\tau_\ell}}Y_{{\nu_\ell}{\mu_\ell}}
	+{\cal D}_{\{{\kappa_\ell}{\mu_\ell}\}}Y_{{\nu_\ell}{\tau_\ell}}
		) Y_{{\nu_\ell}{\lambda_\ell}}
         )
\nonumber\\&&\,\,\,\,\,\,\,\,\,\,\,\,\,\,\,\,\,\,\,\,
-\frac{8i}{9} \,p^3 \,{s_\ell}^5 {\cal D}_{{\mu_\ell}}{\cal X}
        (
            ({\cal D}_{<{\lambda_\ell}}Y_{{\nu_\ell}{\mu_\ell}>}
		) {\cal D}_{{\kappa_\ell}}Y_{{\nu_\ell}{\tau_\ell}}
            +{\cal D}_{{\tau_\ell}}Y_{{\nu_\ell}{\lambda_\ell}}
		({\cal D}_{<{\kappa_\ell}}Y_{{\nu_\ell}{\mu_\ell}>}
            )
         )
\nonumber\\&&\,\,\,\,\,\,\,\,\,\,\,\,\,\,\,\,\,\,\,\,
+\frac{4i}{9}\,{p}^3 \,{s_\ell}^5
  (
   (
     {\cal D}_{{\mu_\ell}{\tau_\ell}{\lambda_\ell}}{\cal X} +
    ({\cal D}_{{\tau_\ell}}\!Y_{{\nu_\ell}{\lambda_\ell}}) Y_{{\nu_\ell}{\mu_\ell}} + ({\cal D}_{{\mu_\ell}}\!Y_{{\nu_\ell}{\tau_\ell}})Y_{{\nu_\ell}{\lambda_\ell}}
   ) Y^2_{{\mu_\ell}{\kappa_\ell}}
\nonumber\\&&\,\,\,\,\,\,\,\,\,\,\,\,\,\,\,\,\,\,\,\,\,\,\,\,\,\,\,\,\,\,\,\,\,\,\,
              +
    (
     {\cal D}_{{\kappa_\ell}{\mu_\ell}{\lambda_\ell}}{\cal X} + ({\cal D}_{{\mu_\ell}}\!Y_{{\nu_\ell}{\lambda_\ell}})Y_{{\nu_\ell}{\kappa_\ell}} + ({\cal D}_{{\kappa_\ell}}\!Y_{{\nu_\ell}{\mu_\ell}})Y_{{\nu_\ell}{\lambda_\ell}}
    ) Y^2_{{\mu_\ell}{\tau_\ell}}
\nonumber\\&&\,\,\,\,\,\,\,\,\,\,\,\,\,\,\,\,\,\,\,\,\,\,\,\,\,\,\,\,\,\,\,\,\,\,\,
\left.
              +
   (
     {\cal D}_{{\kappa_\ell}{\tau_\ell}{\mu_\ell}}{\cal X} + ({\cal D}_{{\tau_\ell}}\!Y_{{\nu_\ell}{\mu_\ell}})Y_{{\nu_\ell}{\kappa_\ell}} + ({\cal D}_{{\kappa_\ell}}\!Y_{{\nu_\ell}{\tau_\ell}})Y_{{\nu_\ell}{\mu_\ell}}
    ) Y^2_{{\mu_\ell}{\lambda_\ell}}
  )
\right\}
      e^{({\cal X}-m^2-p^2)s_\ell}
\nonumber\\
\label{nin}
\end{eqnarray}
\end{small}
\begin{small}
\begin{eqnarray}
[10]_\ell\!\!\!&=&\!\!\!
\int^\infty_0 d_{s_\ell}
\left\{
-\frac{1}{6} \,{s_\ell}^4
  (
    (
     {\cal D}_{{\sigma_\ell}{\rho_\ell}}\!Y_{{\mu_\ell}{\nu_\ell}} + {\cal D}_{{\sigma_\ell}{\nu_\ell}}\!Y_{{\mu_\ell}{\rho_\ell}} + {\cal D}_{{\nu_\ell}{\sigma_\ell}}\!Y_{{\mu_\ell}{\rho_\ell}}
     ) Y^2_{{\rho_\ell}{\sigma_\ell}}
\right.
\nonumber\\&&\,\,\,\,\,\,\,\,\,\,\,\,\,\,\,\,\,\,\,\,\,\,\,\,\,\,\,\,\,\,\,\,\,\,\,
           +
    (
     {\cal D}_{{\sigma_\ell}{\rho_\ell}}\!Y_{{\mu_\ell}{\rho_\ell}} + {\cal D}_{{\rho_\ell}{\sigma_\ell}}\!Y_{{\mu_\ell}{\rho_\ell}} + {\cal D}_{{\rho_\ell}{\rho_\ell}}\!Y_{{\mu_\ell}{\sigma_\ell}}
    ) Y^2_{{\nu_\ell}{\sigma_\ell}}
  ) \!Y_{{\mu_\ell}{\nu_\ell}}
\nonumber\\&&\,\,\,\,\,\,\,\,\,\,\,\,\,\,\,\,\,\,\,\,
+\frac{1}{2} \,{s_\ell}^4 ({\cal D}_{{\mu_\ell}}{\cal X})({\cal D}_{{\nu_\ell}}{\cal X})
        ({\cal D}_{{\mu_\ell}{\nu_\ell}}{\cal X}
	+(3/4) ({\cal D}_{\{{\nu_\ell}{\rho_\ell}\}}Y_{{\mu_\ell}{\rho_\ell}}
		)
\nonumber\\&&\,\,\,\,\,\,\,\,\,\,\,\,\,\,\,\,\,\,\,\,
+\frac{1}{3} \,{s_\ell}^4 {\cal D}_{{\mu_\ell}}{\cal X}
        (
           ({\cal D}_{<{\rho_\ell}}Y_{{\nu_\ell}{\mu_\ell}>}
		)
           ({\cal D}_{\{{\rho_\ell}{\sigma_\ell}\}}Y_{{\nu_\ell}{\sigma_\ell}}
	+{\cal D}^2
		Y_{{\nu_\ell}{\rho_\ell}})
\nonumber\\&&\,\,\,\,\,\,\,\,\,\,\,\,\,\,\,\,\,\,\,\,\,\,\,\,\,\,\,\,\,\,\,\,\,\,\,
	+{\cal D}_{{\rho_\ell}}Y_{{\nu_\ell}{\rho_\ell}}
           ({\cal D}^2
			Y_{{\nu_\ell}{\mu_\ell}}
	+{\cal D}_{\{{\sigma_\ell}{\mu_\ell}\}}Y_{{\nu_\ell}{\sigma_\ell}}
			)
\nonumber\\&&\,\,\,\,\,\,\,\,\,\,\,\,\,\,\,\,\,\,\,\,\,\,\,\,\,\,\,\,\,\,\,\,\,\,\,
	+{\cal D}_{{\rho_\ell}}Y_{{\nu_\ell}{\sigma_\ell}}
           (
              ({\cal D}_{\{{\rho_\ell}{\mu_\ell}\}}Y_{{\nu_\ell}{\sigma_\ell}}
		)
             +({\cal D}_{{\rho_\ell}{\sigma_\ell}}Y_{{\nu_\ell}{\mu_\ell}}
	+{\cal D}_{{\sigma_\ell}{\rho_\ell}}Y_{{\nu_\ell}{\mu_\ell}})
\nonumber\\&&\,\,\,\,\,\,\,\,\,\,\,\,\,\,\,\,\,\,\,\,\,\,\,\,\,\,\,\,\,\,\,\,\,\,\,
	+({\cal D}_{\{{\mu_\ell}{\sigma_\ell}\}}Y_{{\nu_\ell}{\rho_\ell}}
		)
           )
        )
\nonumber\\&&\,\,\,\,\,\,\,\,\,\,\,\,\,\,\,\,\,\,\,\,
-\frac{4}{27} \,{s_\ell}^4
  (
    2 ({\cal D}_{{\sigma_\ell}}\!Y_{{\rho_\ell}{\sigma_\ell}})
    (
     {\cal D}_{{\nu_\ell}}\!Y_{{\rho_\ell}{\mu_\ell}}
    )
 +
    (
     {\cal D}_{{\mu_\ell}}\!Y_{{\rho_\ell}{\sigma_\ell}} + {\cal D}_{{\sigma_\ell}}\!Y_{{\rho_\ell}{\mu_\ell}}
     ) {\cal D}_{{\nu_\ell}}\!Y_{{\rho_\ell}{\sigma_\ell}}
\nonumber\\&&\,\,\,\,\,\,\,\,\,\,\,\,\,\,\,\,\,\,\,\,\,\,\,\,\,\,\,\,\,\,\,\,\,\,\,
           +
     (
      {\cal D}_{{\mu_\ell}}\!Y_{{\rho_\ell}{\sigma_\ell}} + {\cal D}_{{\sigma_\ell}}\!Y_{{\rho_\ell}{\mu_\ell}}
     ) {\cal D}_{{\sigma_\ell}}\!Y_{{\rho_\ell}{\nu_\ell}}
   ) Y^2_{{\mu_\ell}{\nu_\ell}}
\nonumber\\&&\,\,\,\,\,\,\,\,\,\,\,\,\,\,\,\,\,\,\,\,
+\frac{1}{2} \,p^2 \,{s_\ell}^5 ({\cal D}_{{\mu_\ell}}{\cal X})({\cal D}_{{\nu_\ell}}{\cal X})
        ({\cal D}_{\{{\tau_\ell}{\nu_\ell}\}}Y_{{\lambda_\ell}{\mu_\ell}}
	+ {\cal D}_{{\nu_\ell}{\mu_\ell}}Y_{{\lambda_\ell}{\tau_\ell}})
\nonumber\\&&\,\,\,\,\,\,\,\,\,\,\,\,\,\,\,\,\,\,\,\,
-\frac{2}{3} \,p^2 \,{s_\ell}^5 {\cal D}_{{\mu_\ell}}{\cal X}
       (
          ({\cal D}_{<{\lambda_\ell}}Y_{{\nu_\ell}{\mu_\ell}>}
			)
          ({\cal D}^2
			Y_{{\nu_\ell}{\tau_\ell}}
		+{\cal D}_{\{{\rho_\ell}{\tau_\ell}\}}Y_{{\nu_\ell}{\rho_\ell}}
			)
\nonumber\\&&\,\,\,\,\,\,\,\,\,\,\,\,\,\,\,\,\,\,\,\,\,\,\,\,\,\,\,\,\,\,\,\,\,\,\,
	 +({\cal D}_{<{\rho_\ell}}Y_{{\nu_\ell}{\mu_\ell}>}
			)
          ({\cal D}_{{\rho_\ell}{\lambda_\ell}}Y_{{\nu_\ell}{\tau_\ell}}
		+{\cal D}_{{\tau_\ell}{\rho_\ell}}Y_{{\nu_\ell}{\lambda_\ell}}
		+{\cal D}_{{\tau_\ell}{\lambda_\ell}}Y_{{\nu_\ell}{\rho_\ell}})
\nonumber\\&&\,\,\,\,\,\,\,\,\,\,\,\,\,\,\,\,\,\,\,\,\,\,\,\,\,\,\,\,\,\,\,\,\,\,\,
	 +({\cal D}_{<{\lambda_\ell}}Y_{{\nu_\ell}{\rho_\ell}>}
			)
          ({\cal D}_{\{{\tau_\ell}{\mu_\ell}\}}Y_{{\nu_\ell}{\rho_\ell}}
		+{\cal D}_{{\tau_\ell}{\rho_\ell}}Y_{{\nu_\ell}{\mu_\ell}})
\nonumber\\&&\,\,\,\,\,\,\,\,\,\,\,\,\,\,\,\,\,\,\,\,\,\,\,\,\,\,\,\,\,\,\,\,\,\,\,
	 +{\cal D}_{{\rho_\ell}}Y_{{\nu_\ell}{\rho_\ell}}
          ({\cal D}_{{\tau_\ell}{\lambda_\ell}}Y_{{\nu_\ell}{\mu_\ell}}
		+{\cal D}_{\{{\tau_\ell}{\mu_\ell}\}}Y_{{\nu_\ell}{\lambda_\ell}}
			)
\nonumber\\&&\,\,\,\,\,\,\,\,\,\,\,\,\,\,\,\,\,\,\,\,\,\,\,\,\,\,\,\,\,\,\,\,\,\,\,
	 +{\cal D}_{{\rho_\ell}}Y_{{\nu_\ell}{\lambda_\ell}}
          ({\cal D}_{\{{\rho_\ell}{\mu_\ell}\}}Y_{{\nu_\ell}{\tau_\ell}}
		+{\cal D}_{{\rho_\ell}{\tau_\ell}}Y_{{\nu_\ell}{\mu_\ell}})
\nonumber\\&&\,\,\,\,\,\,\,\,\,\,\,\,\,\,\,\,\,\,\,\,\,\,\,\,\,\,\,\,\,\,\,\,\,\,\,
	 + ({\cal D}_{{\tau_\ell}}Y_{{\nu_\ell}{\rho_\ell}})
	({\cal D}_{\{{\mu_\ell}{\rho_\ell}\}}Y_{{\nu_\ell}{\lambda_\ell}}
		+{\cal D}_{{\rho_\ell}{\lambda_\ell}}Y_{{\nu_\ell}{\mu_\ell}}
			)
\nonumber\\&&\,\,\,\,\,\,\,\,\,\,\,\,\,\,\,\,\,\,\,\,\,\,\,\,\,\,\,\,\,\,\,\,\,\,\,
	 + ({\cal D}_{{\tau_\ell}}Y_{{\nu_\ell}{\lambda_\ell}})
		({\cal D}^2
			Y_{{\nu_\ell}{\mu_\ell}}
		+{\cal D}_{\{{\rho_\ell}{\mu_\ell}\}}Y_{{\nu_\ell}{\rho_\ell}}
			)
       )
\nonumber\\&&\,\,\,\,\,\,\,\,\,\,\,\,\,\,\,\,\,\,\,\,
+\frac{1}{3}\,{p}^2 \,{s_\ell}^5
  (
    (
      (
       {\cal D}_{{\rho_\ell}{\lambda_\ell}}\!Y_{{\mu_\ell}{\nu_\ell}} + {\cal D}_{{\rho_\ell}{\nu_\ell}}\!Y_{{\mu_\ell}{\lambda_\ell}} + {\cal D}_{{\nu_\ell}{\lambda_\ell}}\!Y_{{\mu_\ell}{\rho_\ell}}
      ) Y^2_{{\rho_\ell}{\tau_\ell}}
\nonumber\\&&\,\,\,\,\,\,\,\,\,\,\,\,\,\,\,\,\,\,\,\,\,\,\,\,\,\,\,\,\,\,\,\,\,\,\,
              +
      (
       {\cal D}_{{\tau_\ell}{\rho_\ell}}\!Y_{{\mu_\ell}{\nu_\ell}} + {\cal D}_{{\tau_\ell}{\nu_\ell}}\!Y_{{\mu_\ell}{\rho_\ell}} + {\cal D}_{{\nu_\ell}{\rho_\ell}}\!Y_{{\mu_\ell}{\tau_\ell}}
      ) Y^2_{{\rho_\ell}{\lambda_\ell}}
\nonumber\\&&\,\,\,\,\,\,\,\,\,\,\,\,\,\,\,\,\,\,\,\,\,\,\,\,\,\,\,\,\,\,\,\,\,\,\,
              +
      (
       {\cal D}_{{\tau_\ell}{\rho_\ell}}\!Y_{{\mu_\ell}{\rho_\ell}} + {\cal D}_{{\rho_\ell}{\tau_\ell}}\!Y_{{\mu_\ell}{\rho_\ell}} + {\cal D}_{{\rho_\ell}{\rho_\ell}}\!Y_{{\mu_\ell}{\tau_\ell}}
      ) Y^2_{{\nu_\ell}{\lambda_\ell}}
\nonumber\\&&\,\,\,\,\,\,\,\,\,\,\,\,\,\,\,\,\,\,\,\,\,\,\,\,\,\,\,\,\,\,\,\,\,\,\,
              +
      (
       {\cal D}_{{\tau_\ell}{\rho_\ell}}\!Y_{{\mu_\ell}{\lambda_\ell}} + {\cal D}_{{\rho_\ell}{\lambda_\ell}}\!Y_{{\mu_\ell}{\tau_\ell}} + {\cal D}_{{\tau_\ell}{\lambda_\ell}}\!Y_{{\mu_\ell}{\rho_\ell}}
       ) Y^2_{{\nu_\ell}{\rho_\ell}}
    ) \!Y_{{\mu_\ell}{\nu_\ell}}
\nonumber\\&&\,\,\,\,\,\,\,\,\,\,\,\,\,\,\,\,\,\,\,\,\,\,\,\,\,\,\,\,\,\,\,\,\,\,\,
              +
   (
     (
      {\cal D}_{{\rho_\ell}{\nu_\ell}}\!Y_{{\mu_\ell}{\nu_\ell}} + {\cal D}_{{\nu_\ell}{\rho_\ell}}\!Y_{{\mu_\ell}{\nu_\ell}} + {\cal D}_{{\nu_\ell}{\nu_\ell}}\!Y_{{\mu_\ell}{\rho_\ell}}
     ) Y^2_{{\rho_\ell}{\tau_\ell}}
\nonumber\\&&\,\,\,\,\,\,\,\,\,\,\,\,\,\,\,\,\,\,\,\,\,\,\,\,\,\,\,\,\,\,\,\,\,\,\,
                +
     (
      {\cal D}_{{\nu_\ell}{\rho_\ell}}\!Y_{{\mu_\ell}{\tau_\ell}} + {\cal D}_{{\nu_\ell}{\tau_\ell}}\!Y_{{\mu_\ell}{\rho_\ell}} + {\cal D}_{{\tau_\ell}{\nu_\ell}}\!Y_{{\mu_\ell}{\rho_\ell}}
     ) Y^2_{{\rho_\ell}{\nu_\ell}}
   ) Y_{{\mu_\ell}{\lambda_\ell}}
  )
\nonumber\\&&\,\,\,\,\,\,\,\,\,\,\,\,\,\,\,\,\,\,\,\,
+\frac{8}{27}\,{p}^2 \,{s_\ell}^5
 (
     (
         2 ({\cal D}_{{\nu_\ell}}\!Y_{{\rho_\ell}{\nu_\ell}}) ({\cal D}_{{\mu_\ell}}\!Y_{{\rho_\ell}{\lambda_\ell}})  + ({\cal D}_{{\mu_\ell}}\!Y_{{\rho_\ell}{\nu_\ell}}) ({\cal D}_{{\lambda_\ell}}\!Y_{{\rho_\ell}{\nu_\ell}})
\nonumber\\&&\,\,\,\,\,\,\,\,\,\,\,\,\,\,\,\,\,\,\,\,\,\,\,\,\,\,\,\,\,\,\,\,\,\,\,
         + 2
        (
          {\cal D}_{{\nu_\ell}}\!Y_{{\rho_\ell}{\mu_\ell}} + {\cal D}_{{\mu_\ell}}\!Y_{{\rho_\ell}{\nu_\ell}}
         ) {\cal D}_{{\nu_\ell}}\!Y_{{\rho_\ell}{\lambda_\ell}}
     ) Y^2_{{\mu_\ell}{\tau_\ell}}
\nonumber\\&&\,\,\,\,\,\,\,\,\,\,\,\,\,\,\,\,\,\,\,\,\,\,\,\,\,\,\,\,\,\,\,\,\,\,\,
              +
     (
       2 ({\cal D}_{{\nu_\ell}}\!Y_{{\rho_\ell}{\nu_\ell}}) ({\cal D}_{{\tau_\ell}}\!Y_{{\rho_\ell}{\mu_\ell}})  + ({\cal D}_{{\nu_\ell}}\!Y_{{\rho_\ell}{\mu_\ell}}) ({\cal D}_{{\nu_\ell}}\!Y_{{\rho_\ell}{\tau_\ell}})
\nonumber\\&&\,\,\,\,\,\,\,\,\,\,\,\,\,\,\,\,\,\,\,\,\,\,\,\,\,\,\,\,\,\,\,\,\,\,\,
                +
        (
            {\cal D}_{{\mu_\ell}}\!Y_{{\rho_\ell}{\nu_\ell}} + 2 {\cal D}_{{\nu_\ell}}\!Y_{{\rho_\ell}{\mu_\ell}}
         ) {\cal D}_{{\tau_\ell}}\!Y_{{\rho_\ell}{\nu_\ell}}
      ) Y^2_{{\mu_\ell}{\lambda_\ell}}
\nonumber\\&&\,\,\,\,\,\,\,\,\,\,\,\,\,\,\,\,\,\,\,\,\,\,\,\,\,\,\,\,\,\,\,\,\,\,\,
              +
      (
         2 ({\cal D}_{{\nu_\ell}}\!Y_{{\rho_\ell}{\mu_\ell}}) ({\cal D}_{{\tau_\ell}}\!Y_{{\rho_\ell}{\lambda_\ell}}) + ({\cal D}_{{\tau_\ell}}\!Y_{{\rho_\ell}{\mu_\ell}}) ({\cal D}_{{\nu_\ell}}\!Y_{{\rho_\ell}{\lambda_\ell}})
\nonumber\\&&\,\,\,\,\,\,\,\,\,\,\,\,\,\,\,\,\,\,\,\,\,\,\,\,\,\,\,\,\,\,\,\,\,\,\,
                + {\cal D}_{{\lambda_\ell}}\!Y_{{\rho_\ell}{\mu_\ell}} {\cal D}_{{\tau_\ell}}\!Y_{{\rho_\ell}{\nu_\ell}} + {\cal D}_{{\mu_\ell}}\!Y_{{\rho_\ell}{\lambda_\ell}}
         (
            {\cal D}_{{\nu_\ell}}\!Y_{{\rho_\ell}{\tau_\ell}} + {\cal D}_{{\tau_\ell}}\!Y_{{\rho_\ell}{\nu_\ell}}
         )
      ) Y^2_{{\mu_\ell}{\nu_\ell}}
 )
\nonumber\\&&\,\,\,\,\,\,\,\,\,\,\,\,\,\,\,\,\,\,\,\,
\frac{4}{3} \,p^4 \,{s_\ell}^6 ({\cal D}_{{\mu_\ell}}{\cal X})
       ( ({\cal D}_{{\lambda_\ell}}Y_{{\nu_\ell}{\mu_\ell}})
		 ({\cal D}_{{\eta_\ell}{\kappa_\ell}}Y_{{\nu_\ell}{\tau_\ell}})
		+({\cal D}_{{\mu_\ell}}Y_{{\nu_\ell}{\lambda_\ell}})({\cal D}_{{\eta_\ell}{\kappa_\ell}}Y_{{\nu_\ell}{\tau_\ell}})
\nonumber\\&&\,\,\,\,\,\,\,\,\,\,\,\,\,\,\,\,\,\,\,\,\,\,\,\,\,\,\,\,\,\,\,\,\,\,\,
	+{\cal D}_{{\tau_\ell}}Y_{{\nu_\ell}{\lambda_\ell}}
         ({\cal D}_{{\eta_\ell}{\kappa_\ell}}Y_{{\nu_\ell}{\mu_\ell}}
	+{\cal D}_{{\mu_\ell}{\eta_\ell}}Y_{{\nu_\ell}{\kappa_\ell}}
		)
       )
\nonumber\\&&\,\,\,\,\,\,\,\,\,\,\,\,\,\,\,\,\,\,\,\,
-\frac{2}{3}\,{p}^4 \,{s_\ell}^6
 (
    (
      ({\cal D}_{{\nu_\ell}{\kappa_\ell}}\!Y_{{\mu_\ell}{\tau_\ell}}) Y^2_{{\nu_\ell}{\eta_\ell}}
    + ({\cal D}_{{\eta_\ell}{\nu_\ell}}\!Y_{{\mu_\ell}{\tau_\ell}}) Y^2_{{\nu_\ell}{\kappa_\ell}}
\nonumber\\&&\,\,\,\,\,\,\,\,\,\,\,\,\,\,\,\,\,\,\,\,\,\,\,\,\,\,\,\,\,\,\,\,\,\,\,
    + ({\cal D}_{{\eta_\ell}{\kappa_\ell}}\!Y_{{\mu_\ell}{\nu_\ell}}) Y^2_{{\nu_\ell}{\tau_\ell}}
    ) \!Y_{{\mu_\ell}{\lambda_\ell}}
              + ({\cal D}_{{\eta_\ell}{\kappa_\ell}}\!Y_{{\mu_\ell}{\tau_\ell}}) Y^2_{{\nu_\ell}{\lambda_\ell}} Y_{{\mu_\ell}{\nu_\ell}}
 )
\nonumber\\&&\,\,\,\,\,\,\,\,\,\,\,\,\,\,\,\,\,\,\,\,
-\frac{16}{27}\,{p}^4 \,{s_\ell}^6
 (
   {\cal D}_{{\tau_\ell}}\!Y_{{\nu_\ell}{\lambda_\ell}}
   (
     ({\cal D}_{{\mu_\ell}}\!Y_{{\nu_\ell}{\kappa_\ell}}) Y^2_{{\mu_\ell}{\eta_\ell}} + ({\cal D}_{{\eta_\ell}}\!Y_{{\nu_\ell}{\mu_\ell}}) Y^2_{{\mu_\ell}{\kappa_\ell}}
   )
\nonumber\\&&\,\,\,\,\,\,\,\,\,\,\,\,\,\,\,\,\,\,\,\,\,\,\,\,\,\,\,\,\,\,\,\,\,\,\,
\left.
               +
   (
     ({\cal D}_{{\mu_\ell}}\!Y_{{\nu_\ell}{\lambda_\ell}}) Y^2_{{\mu_\ell}{\tau_\ell}} + ({\cal D}_{{\tau_\ell}}\!Y_{{\nu_\ell}{\mu_\ell}}) Y^2_{{\mu_\ell}{\lambda_\ell}}
   ) {\cal D}_{{\eta_\ell}}\!Y_{{\nu_\ell}{\kappa_\ell}}
 )
\right\}
      e^{({\cal X}-m^2-p^2)s_\ell}
\nonumber\\
\label{ten}
\end{eqnarray}
\end{small}
\begin{small}

\end{small}

Note from above that $[11]$ and $[13]$ up to $[28]$ are excluded in the list as they will not contribute in any sense in our calculation.

\subsection{${\cal L}^{(1)}_1$: $1$st-order Mass-Dimensional Lagrangians}

Using the compact notation (\ref{ellll}),\footnote{This corresponds to changing $\ell$ in the subscript of tensor indices as 1 if the term is contributed by a first order correction, 2 by a second order correction, and so on. In our presentation, left-most factor in $[\,\,][\,\,]\ldots[\,\,]$ denotes the first order corrrection and the right-most the highest order in the degree of corrections considered. The number of  $[\,\,]$ factors denotes the order of corrections. Adding all numbers indicated in each $[\,\,]$ should sum up to the common mass-dimension of the collected terms.} we now explicitly present the prescription described in (\ref{L0L1L2_emptyset}) in terms of the mass-dimensional basis $[\,\ell\,]$ given in (\ref{thr})-(\ref{twl}):

\begin{eqnarray}
{\cal L}^{(1)[4]}_1
    = \frac{\hbar}{2(2\pi)^D}\mbox{Tr}\int dX
    \int d^Dp 
    \,\,\, G_{\emptyset_{\mathrm{t}}}[4]
\label{1[4]}
\end{eqnarray}
with $[4]$ given in (\ref{for}).

\begin{eqnarray}
{\cal L}^{(1)[6]}_1
    = \frac{\hbar}{2(2\pi)^D}\mbox{Tr}\int dX
    \int d^Dp 
    \,\,\, G_{\emptyset_{\mathrm{t}}}[6]
\label{1[6]}
\end{eqnarray}
with $[6]$ given in (\ref{six}).

\begin{eqnarray}
{\cal L}^{(1)[8]}_1
    = \frac{\hbar}{2(2\pi)^D}\mbox{Tr}\int dX
    \int d^Dp 
    \,\,\, G_{\emptyset_{\mathrm{t}}}[8]
\label{1[8]}
\end{eqnarray}
with $[8]$ given in (\ref{eit}).

\begin{eqnarray}
{\cal L}^{(1)[10]}_1
    = \frac{\hbar}{2(2\pi)^D}\mbox{Tr}\int dX
    \int d^Dp 
    \,\,\, G_{\emptyset_{\mathrm{t}}}[10]
\label{1[10]}
\end{eqnarray}
with $[10]$ given in (\ref{ten}).

\begin{eqnarray}
{\cal L}^{(1)[12]}_1
    = \frac{\hbar}{2(2\pi)^D}\mbox{Tr}\int dX
    \int d^Dp 
    \,\,\, G_{\emptyset_{\mathrm{t}}}[12]
\label{1[12]}
\end{eqnarray}
with $[12]$ given in (\ref{twl}).

This is after performing the prescription
\begin{eqnarray}
{\cal L}^{(1)}_1
    = \frac{\hbar}{2(2\pi)^D}\mbox{Tr}\int dX
    \int d^Dp\,\,\,G_{\emptyset_{\mathrm{t}}}(p)\left(\Delta_1G_{\emptyset}(p)\right)^{\mathrm{red}}_{\ell=1}.
\end{eqnarray}
Notice that there are no first-order corrections that can be incorporated in the calculation of one loop effective Lagrangian in two-mass dimensions.

\subsection{${\cal L}^{(1)}_2$: $2$nd-order Mass-Dimensional Lagrangians}

\begin{eqnarray}
{\cal L}^{(1)[6]}_2
    = \frac{\hbar}{2(2\pi)^D}\mbox{Tr}\int dX
    \int d^Dp 
    \,\,\, G_{\emptyset_{\mathrm{t}}}[3][3]
\label{2[6]}
\end{eqnarray}
with $[3]$ given in (\ref{thr}).

\begin{eqnarray}
{\cal L}^{(1)[8]}_2
    = \frac{\hbar}{2(2\pi)^D}\mbox{Tr}\int dX
    \int d^Dp 
    \,\,\, G_{\emptyset_{\mathrm{t}}}\left([3][5]+[4][4]+[5][3]\right)
\label{2[8]}
\end{eqnarray}
with $[3]$, $[4]$, and $[5]$ given in (\ref{thr}), (\ref{for}), and (\ref{fiv}), respectively. Here, $[3][5]$ is just a swapping of indices $[5][3]$. That is, the subscript $\ell=1 \Leftrightarrow \ell=2$ is swapped. As an illustration, $[3][5]$ means $[3]$, set $\ell=1$ and $[5]$ set $\ell=2$. The other way around, $[5][3]$ means $[5]$, set $\ell=1$ and $[3]$ set $\ell=2$.

\begin{eqnarray}
{\cal L}^{(1)[10]}_2
    \!\!=\!\! \frac{\hbar}{2(2\pi)^D}\!\mbox{Tr}\!\int\!\!dX\!\!\!
    \int\!\! d^Dp 
    \,\,\, G_{\emptyset_{\mathrm{t}}}\!\!\left([3][7]+[4][6]+[5][5]+[6][4]+[7][3]\right)
\label{2[10]}
\end{eqnarray}
with $[4]$, $[5]$, and $[6]$ given in (\ref{for}), (\ref{fiv}), and (\ref{six}), respectively.

\begin{eqnarray}
{\cal L}^{(1)[12]}_2
\!\!&=&\!\! \frac{\hbar}{2(2\pi)^D}\mbox{Tr}\int dX
    \int d^Dp 
\nonumber\\
    &&\!\!\!\!\!\!\!\!\!\!\!\!\!\!\!\!\!\!\!\!\!\!\!
    G_{\emptyset_{\mathrm{t}}}\!\left([3][9]+[4][8]+[5][7]+[6][6]+[7][5]+[8][4]+[9][3]\right)
\label{2[12]}
\end{eqnarray}
with $[4]$, $[5]$, and $[6]$ given in (\ref{for}), (\ref{fiv}), and (\ref{six}), respectively.

This is after performing the prescription
\begin{eqnarray}
{\cal L}^{(1)}_2
    = \frac{\hbar}{2(2\pi)^D}\mbox{Tr}\int dX
    \int d^Dp\,\,\,G_{\emptyset_{\mathrm{t}}}(p)\left(\Delta_1G_{\emptyset}(p)\right)^{\mathrm{red}}_{\ell=1}
                                   \left(\Delta_1G_{\emptyset}(p)\right)^{\mathrm{red}}_{\ell=2}.
\end{eqnarray}
Notice that there are no second-order corrections that can be incorporated in the calculation of one loop effective Lagrangian in two and four mass-dimensions.

\subsection{${\cal L}^{(1)}_3$: $3$rd-order Mass-Dimensional Lagrangians}

\begin{eqnarray}
{\cal L}^{(1)[10]}_3
    = \frac{\hbar}{2(2\pi)^D}\mbox{Tr}\int\!\!dX\!\!
    \int \!\!d^Dp\!\! \int^\infty_0\!\! ds\,\,\, G_{\emptyset_{\mathrm{t}}}\left([3][3][4]+[3][4][3]+[4][3][3]\right)
\label{3[10]}
\end{eqnarray}
with $[3]$, and $[4]$ given in (\ref{thr}), and (\ref{for}), respectively. Here, $[3][3][4]$ is just a swapping of indices with  $[3][4][3]$ and similarly with  $[4][3][3]$. As an illustration, $[3][3][4]$ means leftmost $[3]$, set $\ell=1$ mid $[3]$ set $\ell=2$ and rightmost $[4]$ set $\ell=3$. The leftmost object is always set $\ell=1$, the second object set $\ell=2$, the rightmost set $\ell=3$.

\begin{eqnarray}
\!\!\!\!\!\!\!\!\!\!\!\!\!\!\!\!\!\!
{\cal L}^{(1)[12]}_3
    \!\!\!\!&=&\!\!\!\!\frac{\hbar}{2(2\pi)^D}\!\mbox{Tr}\int\!\!dX\!\!
    \int\!\!d^Dp 
    \,\,\,
\nonumber\\&&\!\!\!\!\!\!\!\!\!\!\!\!\!\!\!\!\!\!\!\!\!\!\!\!
G_{\emptyset_{\mathrm{t}}}\!\!\left([3][4][5]\!+\![3][5][4]\!+\![4][5][3]\!+\![4][3][5]\!+\![5][3][4]\!+\![5][4][3]\!+\![4][4][4]\right)
\label{3[12]}
\end{eqnarray}
with $[3]$, $[4]$, and $[5]$ given in (\ref{thr}), (\ref{for}), and (\ref{fiv}), respectively.

This is after performing the prescription
\begin{eqnarray}
{\cal L}^{(1)}_3
\!\!=\!\!\frac{\hbar}{2(2\pi)^D}\mbox{Tr}\!\int\!\!dX\!
    \int\!\!d^Dp\,\,\,G_{\emptyset_{\mathrm{t}}}(p)
                                   \left(\Delta_1G_{\emptyset}(p)\right)^{\mathrm{red}}_{\ell=1}
                                   \left(\Delta_1G_{\emptyset}(p)\right)^{\mathrm{red}}_{\ell=2}
                                   \left(\Delta_1G_{\emptyset}(p)\right)^{\mathrm{red}}_{\ell=3}.
\nonumber\\
\end{eqnarray}
Notice that there are no third-order corrections that can be incorporated in the calculation of one-loop effective Lagrangian in two, four, six, and eight mass-dimensions.

\subsection{${\cal L}^{(1)}_4$: $4$th-order Mass-Dimensional Lagrangians}

\begin{eqnarray}
{\cal L}^{(1)[12]}_4
    \!\!=\!\! \frac{\hbar}{2(2\pi)^D}\mbox{Tr}\int dX
    \int d^Dp 
    \,\,\, G_{\emptyset_{\mathrm{t}}}[3][3][3][3]
\label{4[12]}
\end{eqnarray}
with $[3]$ given in (\ref{thr}).

This is after performing the prescription
\begin{eqnarray}
{\cal L}^{(1)}_4
\!\!\!\!&=&\!\!\!\!\frac{\hbar}{2(2\pi)^D}\!\mbox{Tr}\!\int\!\!dX\!
    \int\!\!d^Dp\,\,\,
\nonumber\\&&
\!\!\!\!\!\!\!\!\!\!\!\!\!\!\!\!\!
G_{\emptyset_{\mathrm{t}}}(p)\left(\Delta_1G_{\emptyset}(p)\right)^{\mathrm{red}}_{\ell=1}
                                   \left(\Delta_1G_{\emptyset}(p)\right)^{\mathrm{red}}_{\ell=2}
                                   \left(\Delta_1G_{\emptyset}(p)\right)^{\mathrm{red}}_{\ell=3}
                                   \left(\Delta_1G_{\emptyset}(p)\right)^{\mathrm{red}}_{\ell=4}.
\end{eqnarray}
Notice that there are no third-order corrections that can be incorporated in the calculation of one loop effective Lagrangian in two, four, six, eight, and ten mass-dimensions.



\chapter{One-Loop Lagrangians with Higher-Order Corrections}
Consider the ungauged case when
\begin{eqnarray}
Y\to 0
\end{eqnarray}
in Eqns. (\ref{thr})-(\ref{twl}). Later we will relax this limit but for purposes of illustration, we present the case of pure $X$ and its covariant derivatives of $X$. That is, we illustrate how straightforward our calculation will be in determining the higher-order corrections. Also, we show step by step how the ensuing integrations: $p$ (momentum), $X$ (background potential), and $s$ (proper-time integration) are performed.

\section{Pure $X$ Mass-Dimensional Lagrangians}

\subsection{Pure $X$: Two Mass-Dimensions}


\begin{eqnarray}
 {\cal L}^{(1)[2]}_{Y\to 0}
 = {\cal L}^{(1)[2]}_{0\,\,Y\to 0} + {\cal L}^{(1)[2]}_{1\,\,Y\to 0}+\ldots
\end{eqnarray}
where ${\cal L}^{(1)[2]}_{k\,\,Y\to 0}$ for $k>0$ does not exist. For the case when $k=0$
from (\ref{0[2]}), it is
\begin{eqnarray}
 {\cal L}^{(1)[2]}_{0\,Y\to 0}
  \!\!&=&\!\! \frac{\hbar}{2(4\pi)^{D/2}}
\int^\infty_0 \frac{ds\,e^{-m^2s}}{s^{-0+D/2}}
   \mbox{Tr}
\left(
   {\cal {\cal X}}-{\cal {\cal X}}_0
\right)
\end{eqnarray}

\subsection{Pure $X$: Four Mass-Dimensions}


\begin{eqnarray}
 {\cal L}^{(1)[4]}_{Y\to 0}
 = {\cal L}^{(1)[4]}_{0\,\,Y\to 0} + {\cal L}^{(1)[4]}_{1\,\,Y\to 0}+\ldots
\end{eqnarray}
where from (\ref{0[4]})
\begin{eqnarray}
 {\cal L}^{(1)[4]}_{0\,Y\to 0}
  \!\!&=&\!\! \frac{\hbar}{2(4\pi)^{D/2}}
\int^\infty_0 \frac{ds\,e^{-m^2s}}{s^{-1+D/2}}
   \mbox{Tr}
\left[
   \frac{1}{2}({\cal X}^2-{{\cal X}_0}^2)
\right].
\end{eqnarray}
and
\begin{eqnarray}
\!\!\!\!\!
{\cal L}^{(1)[4]}_{1\, Y\to 0}
    \!\!&=&\!\! \frac{\hbar}{2(2\pi)^D}\!\!
\!\int^\infty_0\!\!\!ds\frac{e^{-m^2s}}{s^{-1+D/2}}
 \frac{1}{6}\mbox{Tr}
 {\cal D}_{{\mu}{\mu}}{\cal X}
\label{Y01[4]}
\end{eqnarray}

The following is the explicit derivation of ${\cal L}^{(1)[4]}_{1\, Y\to 0}$:

To compute for ${\cal L}^{(1)[4]}_{1\,\,Y\to 0}$, consider from the prescription (\ref{1[4]}) with $[4]$ given in (\ref{for}) but in the limit $Y\to 0$.
\begin{eqnarray}
{\cal L}^{(1)[4]}_{1\, Y\to 0}
    \!\!\!\!&=&\!\!\!\! \frac{\hbar}{2(2\pi)^D}\mbox{Tr}\!\!
    \int^\infty_0\!\! d{s_1}    \int^\infty_0\!\! d{s_0}
    \!\!\int dX\!\!
\nonumber\\&&\,\,\,\,\,\,\,\,\,\,\,\,\,\,\,\,\,\,\,\,
\times
    \int d^Dp\,\,\,
      e^{(X
         -p^2)s_0}
\left\{
 {s_0} {\cal D}_{{\mu}{\mu}}X
 - 2{p}^2 {s_0}^2
   {\cal D}_{{\lambda}{\tau}}X
\right\}
      e^{(X
         -p^2)s_1}
\end{eqnarray}
expressed in terms of $X$. The indices of the momentum tensor can be identified from the unpaired indices of ${\cal D}$. That is, $p^2 =p_\lambda\, p_\tau $ for this instance.

Factoring out the $X$ and $p^2$ in the exponents
\begin{eqnarray}
{\cal L}^{(1)[4]}_{1\, Y\to 0}
    \!\!\!\!&=&\!\!\!\! \frac{\hbar}{2(2\pi)^D}\mbox{Tr}\!\!
    \int^\infty_0\!\! d{s_1}    \int^\infty_0\!\! d{s_0}
    \!\!\int dX\,\,\,
      e^{(s_1
         +s_0)X}
\nonumber\\&&\,\,\,\,\,\,\,\,\,\,\,\,\,\,\,\,\,\,\,\,
\times
    \int d^Dp\,\,\,
\left\{
 {s_0} {\cal D}_{{\mu}{\mu}}X
 - 2{p}^2 {s_0}^2
   {\cal D}_{{\lambda}{\tau}}X
\right\}
      e^{-p^2(s_1
         +s_0)}
\end{eqnarray}

Performing $p^0$ and $p^2$ momentum integration using the formulas (\ref{pdzer}) and (\ref{pdtwo}), respectively.
\begin{eqnarray}
{\cal L}^{(1)[4]}_{1\, Y\to 0}
    \!\!\!\!&=&\!\!\!\! \frac{\hbar}{2(2\pi)^D}\mbox{Tr}\!\!
    \int^\infty_0\!\! d{s_1}    \int^\infty_0\!\! d{s_0}
    \!\!\int dX\,\,\,
      e^{X(s_0
         +s_1)}
\nonumber\\&&\,\,\,\,\,\,\,\,\,\,\,\,\,\,\,\,\,\,\,\,
\times
\left\{
 (1,0;1,0)
 {\cal D}_{{\mu}{\mu}}X
 - 2(2,0;2,1)
  \delta_{\lambda\tau}
   {\cal D}_{{\lambda}{\tau}}X
\right\}
\end{eqnarray}
where
\begin{eqnarray}
 (1,0;1,0)&\equiv &\frac{\pi^{D/2}}{2^1(s_0+s_1)^{0+D/2}} s_0\,e^{X(s_0+s_1)}
\\
 (2,0;2,0)&\equiv &\frac{\pi^{D/2}}{2^2(s_0+s_1)^{1+D/2}} {s_0}^2 e^{X(s_0+s_1)}
\end{eqnarray}

Index contraction\footnote{Because indices are dummy they can be renamed at will.} and performing $X$-integration using
\begin{eqnarray}\label{Xinteg}
 \int dX \,\, e^{X(s_0+s_1)}
   &=& \frac{e^{X(s_0+s_1)}}{s_0+s_1}, \,\,\,\,\,\, \mbox{where}\,\, X\equiv -m^2 + {\cal X}\nonumber
\\ &=& \frac{e^{-m^2(s_0+s_1)}}{s_0+s_1}\left[1+ {\cal X}(s_0+s_1)+\frac{1}{2}{\cal X}^2(s_0+s_1)^2+\ldots\right]
\end{eqnarray}
 and considering\footnote{Because ${\cal X}(s_0+s_1)$ term will contribute in the six mass-dimensions and $\frac{1}{2!}{\cal X}^2(s_0+s_1)$ term will contribute in the eight mass-dimensions and so on.}
\begin{eqnarray}
 \left[\underbrace{1}_{\Uparrow}+{\cal X}(s_1+s_0)+\frac{1}{2}{\cal X}^2(s_1+s_0)^2+\ldots\right]
\label{eX1st}
\end{eqnarray}
(the first term indicated by an arrow) we have
\begin{eqnarray}
{\cal L}^{(1)[4]}_{1\, Y\to 0}
    \!\!\!\!&=&\!\!\!\! \frac{\hbar}{2(2\pi)^D}\mbox{Tr}\!\!
    \int^\infty_0\!\! d{s_1}    \int^\infty_0\!\! d{s_0}
\left\{
 (1,0;1,1)
 - 2(2,0;2,2)
   %
\right\}
      {\cal D}_{{\mu}{\mu}}{\cal X}
      e^{-p^2(s_0
         +s_1)}
\nonumber\\
\end{eqnarray}

Performing proper-time integrations\footnote{See Appendix D for a detailed derivation.}
\begin{footnotesize}
\begin{eqnarray}
(1,0;1,1)
\!\!\!\!&\equiv &\!\!\!\!
 \int^\infty_0\!\!\!ds_1\!\!
 \int^\infty_0\!\!\! ds_0\,
  \frac{s_0 e^{-m^2(s_0+s_1)}}{1(s_0+s_1)^{1+D/2}}
 \!=\!\frac{1}{2}\!\int^\infty_0\!\!\!ds\frac{e^{-m^2s}}{s^{-1+D/2}}
 \!=\!\frac{1}{2}\frac{\Gamma[2-D/2]}{m^{4-D/2}}
\\
(2,0;2,2)
\!\!\!\!&\equiv &\!\!\!\!
 \int^\infty_0\!\!\!ds_1\!\!
 \int^\infty_0\!\!\!ds_0\,
  \frac{{s_0}^2 e^{-m^2(s_0+s_1)}}{2(s_0+s_1)^{2+D/2}}
  \!=\!\frac{1}{2}\cdot\frac{1}{3}\!\int^\infty_0\!\!\!ds\frac{e^{-m^2s}}{s^{-1+D/2}}
 \!=\!\frac{1}{6}\frac{\Gamma[2-D/2]}{m^{4-D/2}}
\end{eqnarray}
\end{footnotesize}
we have
\begin{eqnarray}
{\cal L}^{(1)[4]}_{1\, Y\to 0}
    \!\!&=&\!\! \frac{\hbar}{2(2\pi)^D}\!\!
\!\int^\infty_0\!\!\!ds\frac{e^{-m^2s}}{s^{-1+D/2}}
\mbox{Tr}
\left\{
  \left(\frac{1}{2}-\frac{1}{3}\right)
 {\cal D}_{{\mu}{\mu}}{\cal X}
\right\}
\end{eqnarray}
thereby obtaining (\ref{Y01[4]}). This agrees with Eqn (2.108) of \cite{Rodf-Diss}.

\begin{large}
\textbf{Higher Mass-Dimensions Contributed by $[4]$}
\end{large}

\textbf{$[4]{\cal X}$: Six Mass-Dimensions}.

The contributions of $[4]$ to higher mass-dimensions as one moves to higher order terms in the expansion
\begin{eqnarray}
 \left[1+\underbrace{{\cal X}(s_1+s_0)}_{\Uparrow}+\frac{1}{2}(s_1+s_0)^2+\ldots\right]
\label{eX2nd}
\end{eqnarray}
(the second term indicated by an arrow)
\begin{eqnarray}
{\cal L}^{(1)[6]}_{1\,\,\,[4]{\cal X}\,\,\,Y\to 0}
 \!\!&=&\!\!\pi^{D/2}\,\,
  \int^\infty_0\!\!\!ds_1
  \int^\infty_0\!\!\!ds_0
\nonumber\\&&
\left\{
  (2,0\mathrm{;}\,1,0)
 -2(3,0\mathrm{;}\,2,1)
\right\}{\cal X}D_{\mu\mu}{\cal X}
\label{L1[4]X}
\end{eqnarray}
Because the proper-time integrations
\begin{footnotesize}
\begin{eqnarray}
(1,0;1,0)
\!\!\!\!&\equiv &\!\!\!\!
 \int^\infty_0\!\!\!ds_1\!\!
 \int^\infty_0\!\!\! ds_0\,
  \frac{s_0 e^{-m^2(s_0+s_1)}}{1(s_0+s_1)^{0+D/2}}
 =\frac{1}{2}\int^\infty_0\!\!\!ds\frac{e^{-m^2s}}{s^{-2+D/2}}
 =\frac{1}{2}\frac{\Gamma[3-D/2]}{m^{6-D/2}}
\\
(2,0;2,1)
\!\!\!\!&\equiv &\!\!\!\!
 \int^\infty_0\!\!\!ds_1\!\!
 \int^\infty_0\!\!\!ds_0\,
  \frac{{s_0}^2 e^{-m^2(s_0+s_1)}}{2(s_0+s_1)^{1+D/2}}
 =\frac{1}{2}\!\cdot\!\frac{1}{3}\int^\infty_0\!\!\!ds\frac{e^{-m^2s}}{s^{-2+D/2}}
 =\frac{1}{6}\frac{\Gamma[3-D/2]}{m^{6-D/2}}
\end{eqnarray}
\end{footnotesize}
yeild the same prefactor as $(1,0;1,1)$ and (2,0;2,2), respectively, the first-order correction of mass-dimensions six contributed by $[4]{\cal X}$ is given by the Lagrangian
\begin{eqnarray}
{\cal L}^{(1)[6]}_{1\, Y\to 0}
  \!\!&=&\!\frac{\hbar}{2(4\pi)^{D/2}}
\int^\infty_0 \frac{e^{-m^2s}ds}{s^{-2+D/2}}\,\,\,
   \frac{1}{6}  \mathrm{Tr}{\cal X}D_{{\mu}{\mu}}{\cal X}.
\label{1[4]X}
\end{eqnarray}

\textbf{$[4]{\cal X}^2$: Eight Mass-Dimensions}.
With four and six mass-dimensions invariants agreeing with the results obtained by Rodulfo, this work showcases the calculation for the eight mass-dimensions.
Consider the contribution of $[4]$ to higher mass-dimensions as one moves to higher order terms in the expansion
\begin{eqnarray}
 \left[1+{\cal X}(s_1+s_0)+\underbrace{\frac{1}{2}{\cal X}^2(s_1+s_0)^2}_{\Uparrow}+\frac{1}{6}(s_1+s_0)^3+\ldots\right]
\label{eX3rd}
\end{eqnarray}
(the third term indicated by an arrow)

In similar manner, the first order correction of mass-dimensions eight contributed by $[4]{\cal X}^2$ is given by the Lagrangian
\begin{eqnarray}
{\cal L}^{(1)[8]}_{1\,\,\,[4]{\cal X}^2\,\,\,Y\to 0}
  \!\!&=&\!\frac{\hbar}{2(4\pi)^{D/2}}
\int^\infty_0 \frac{e^{-m^2s}ds}{s^{-3+D/2}}\,\,\,
   \frac{1}{12}  \mathrm{Tr}{\cal X}^2D_{{\mu}{\mu}}{\cal X}
\label{1XX[4]}
\end{eqnarray}

Contributions to higher mass-dimensions can be obtained in a similar fashion
\begin{eqnarray}
{\cal L}^{(1)[10]}_{1\,\,\,[4]{\cal X}^3\,\,\,Y\to 0}
  \!\!&=&\!\frac{\hbar}{2(4\pi)^{D/2}}
\int^\infty_0 \frac{e^{-m^2s}ds}{s^{-4+D/2}}\,\,\,
   \frac{1}{36}  \mathrm{Tr}{\cal X}^3D_{{\mu}{\mu}}{\cal X}
\end{eqnarray}
\begin{eqnarray}
{\cal L}^{(1)[12]}_{1\,\,\,[4]{\cal X}^4\,\,\,Y\to 0}
  \!\!&=&\!\frac{\hbar}{2(4\pi)^{D/2}}
\int^\infty_0 \frac{e^{-m^2s}ds}{s^{-5+D/2}}\,\,\,
   \frac{1}{144}  \mathrm{Tr}{\cal X}^4D_{{\mu}{\mu}}{\cal X}
\end{eqnarray}
\begin{eqnarray}
{\cal L}^{(1)[14]}_{1\,\,\,[4]{\cal X}^5\,\,\,Y\to 0}
  \!\!&=&\!\frac{\hbar}{2(4\pi)^{D/2}}
\int^\infty_0 \frac{e^{-m^2s}ds}{s^{-6+D/2}}\,\,\,
   \frac{1}{720}  \mathrm{Tr}{\cal X}^5D_{{\mu}{\mu}}{\cal X}
\end{eqnarray}
\begin{eqnarray}
{\cal L}^{(1)[16]}_{1\,\,\,[4]{\cal X}^6\,\,\,Y\to 0}
  \!\!&=&\!\frac{\hbar}{2(4\pi)^{D/2}}
\int^\infty_0 \frac{e^{-m^2s}ds}{s^{-7+D/2}}\,\,\,
   \frac{1}{4320}  \mathrm{Tr}{\cal X}^6D_{{\mu}{\mu}}{\cal X}
\end{eqnarray}
with the prefactors and exponents of ${\cal X}$ dictated by the coefficients and exponents of a Taylor series of an exponential function.

\subsection{Pure $X$: Six Mass-Dimensions}

From (\ref{six}) prescribed by (\ref{1[6]}) but in the limit $Y \to 0$
\begin{eqnarray}
 {\cal L}^{(1)[6]}_{1\,\,\,Y\to 0} = 0.
\end{eqnarray}
There are no first-order corrections.
Also from (\ref{thr}) prescribed by  (\ref{2[6]}) in the same limit $Y \to 0$
\begin{eqnarray}
 {\cal L}^{(1)[6]}_{2\,\,\,Y\to 0} = 0.
\end{eqnarray}
The one-loop pure $X$ six mass-dimensional effective Lagrangian is
\begin{eqnarray}
 {\cal L}^{(1)[6]}_{Y\to 0}
  \!\!&=&\!\! \frac{\hbar}{2(4\pi)^{D/2}}
\int^\infty_0 \frac{ds\,e^{-m^2s}}{s^{-2+D/2}}
   \mbox{Tr}
\left[
   \frac{1}{6}({\cal X}^3-{{\cal X}_0}^3)
   +\frac{1}{12}({\cal D}_\mu{\cal X})({\cal D}_\mu{\cal X})
   +\frac{1}{6}{\cal X}{\cal D}_{\mu\mu}{\cal X}
\right].
\nonumber\\
\end{eqnarray}
which is contributed mainly from (\ref{1[4]X}) and ${\cal L}^{(1)[6]}_{0\,\,\,Y\to 0}$.

Using the relation:
\begin{eqnarray}
{\cal X} ({\cal D}_{\mu\mu}{\cal X})
 = -({\cal D}_{\mu}{\cal X})({\cal D}_{\mu}{\cal X}) + {\cal D}_{\mu}\left(({\cal D}_{\mu}{\cal X}){\cal X}\right)
\end{eqnarray}
\begin{eqnarray}
 {\cal L}^{(1)[6]}_{Y\to 0}
  \!\!&=&\!\! \frac{\hbar}{2(4\pi)^{D/2}}
\int^\infty_0 \frac{ds\,e^{-m^2s}}{s^{-2+D/2}}
   \mbox{Tr}
\left[
   \frac{1}{6}({\cal X}^3-{{\cal X}_0}^3)
   -\frac{1}{12}({\cal D}_\mu{\cal X})({\cal D}_\mu{\cal X})
\right].
\end{eqnarray}

\subsection{Pure $X$: Eight Mass-Dimensions}

The prescription (\ref{1[8]}) with (\ref{eit}) in the limit $Y \to 0$
\begin{eqnarray}
 {\cal L}^{(1)[8]}_{1\,\,\,Y\to 0}
\!\!\!\!&=&\!\!\!\!
 \frac{\hbar}{2(2\pi)^D}\mbox{Tr}\!\!
    \int^\infty_0\!\! d{s_1}    \int^\infty_0\!\! d{s_0}
    \!\!\int dX\!\!
    \int d^Dp\,\,\,
      e^{(X-p^2)s_1}
\left\{
\frac{1}{3} \,{s_0}^3 ({\cal D}_{{\mu}}{X})
  (
     {\cal D}_{({\nu}{\nu}{\mu})}{X}
   )
\right.
\nonumber\\&&\!\!\!\!\!\!\!\!\!\!\!\!\!\!
\left.
-\frac{2}{3} \,p^2 \,{s_0}^4 ({\cal D}_{{\mu}}{X})
  (
         {\cal D}_{{\tau}{\lambda}{\mu}}{X}
+{\cal D}_{\{{\tau}{\mu}\}{\lambda}}{X}
   )
\right\}
      e^{(X-p^2)s_0}
\end{eqnarray}
Performing the $p$-integration, $X$-integration (after redefining $X\to {\cal X}-m^2$ and considering (\ref{eX1st})) and renaming dummy indices, we have
\begin{eqnarray}
 {\cal L}^{(1)[8]}_{1\,\,\,Y\to 0}
\!\!\!\!&=&\!\!\!\!
 \frac{\hbar}{2(2\pi)^D}\mbox{Tr}\!\!
    \int^\infty_0\!\! d{s_1}    \int^\infty_0\!\! d{s_0}\,e^{-m^2(s_1+s_0)}\,
\left\{
\frac{1}{3}\,(3,0;1,1) ({\cal D}_{{\mu}}{\cal X})
  (
     {\cal D}_{({\nu}{\nu}{\mu})}{\cal X}
   )
\right.
\nonumber\\&&\!\!\!\!\!\!\!\!\!\!\!\!\!\!
\left.
-\frac{2}{3}\,(4,0;2,2) ({\cal D}_{{\mu}}{\cal X})
  (
         {\cal D}_{{\nu}{\nu}{\mu}}{\cal X}
+{\cal D}_{\{{\nu}{\mu}\}{\nu}}{\cal X}
   )
\right\}
\end{eqnarray}
Performing two-fold proper-time integration:
\begin{eqnarray}
 {\cal L}^{(1)[8]}_{1\,\,\,Y\to 0}
  \!\!&=&\!\! \frac{\hbar}{2(4\pi)^{D/2}}
\int^\infty_0 \frac{ds\,e^{-m^2s}}{s^{-2+D/2}}
   \mbox{Tr}
\left\{
\frac{1}{3}\cdot\frac{1}{4} ({\cal D}_{{\mu}}{\cal X})
  (
     {\cal D}_{({\nu}{\nu}{\mu})}{\cal X}
   )
\right.
\nonumber\\&&\!\!\!\!\!\!\!\!\!\!\!\!\!\!
\left.
-\frac{2}{3}\cdot\frac{1}{10} ({\cal D}_{{\mu}}{\cal X})
  (
         {\cal D}_{{\nu}{\nu}{\mu}}{\cal X}
+{\cal D}_{\{{\nu}{\mu}\}{\nu}}{\cal X}
   )
\right\}
\end{eqnarray}
Because
\begin{eqnarray}
     {\cal D}_{({\nu}{\nu}{\mu})}{\cal X}
 ={\cal D}_{{\nu}{\nu}{\mu}}{\cal X}
	+{\cal D}_{{\mu}{\nu}{\nu}}{\cal X}
	+{\cal D}_{{\nu}{\mu}{\nu}}{\cal X}
\end{eqnarray}
so that
\begin{eqnarray*}
     ({\cal D}_{{\mu}}{\cal X})({\cal D}_{({\nu}{\nu}{\mu})}{\cal X})
 &=&({\cal D}_{{\mu}}{\cal X})({\cal D}_{{\nu}{\nu}{\mu}}{\cal X})
	+({\cal D}_{{\mu}}{\cal X})({\cal D}_{{\mu}{\nu}{\nu}}{\cal X})
	+({\cal D}_{{\mu}}{\cal X})({\cal D}_{{\nu}{\mu}{\nu}}{\cal X})
\\
 &=&-({\cal D}_{{\mu}{\mu}}{\cal X})({\cal D}_{{\nu}{\nu}}{\cal X})
	-({\cal D}_{{\mu}{\nu}}{\cal X})({\cal D}_{{\mu}{\nu}}{\cal X})
	-({\cal D}_{{\mu}{\nu}}{\cal X})({\cal D}_{{\nu}{\mu}}{\cal X})
\end{eqnarray*}
and
\begin{eqnarray*}
         ({\cal D}_{{\mu}}{\cal X})({\cal D}_{\{{\nu}{\mu}\}{\nu}}{\cal X})
&=&({\cal D}_{{\mu}}{\cal X})({\cal D}_{{\nu}{\mu}{\nu}}{\cal X})
 +({\cal D}_{{\mu}}{\cal X})({\cal D}_{{\mu}{\nu}{\nu}}{\cal X})
\\
&=&-({\cal D}_{{\mu}{\nu}}{\cal X})({\cal D}_{{\nu}{\mu}}{\cal X})
 -({\cal D}_{{\mu}{\nu}}{\cal X})({\cal D}_{{\mu}{\nu}}{\cal X})
\end{eqnarray*}
after applying integration by parts (or equivalently applying an appropriate total derivative relation).
Hence, we have
\begin{eqnarray}
 {\cal L}^{(1)[8]}_{1\,\,\,Y\to 0}
  \!\!&=&\!\! \frac{\hbar}{2(4\pi)^{D/2}}
\int^\infty_0 \frac{ds\,e^{-m^2s}}{s^{-2+D/2}}
   \mbox{Tr}
\left\{
 -\frac{1}{60} ({\cal D}_{{\mu}{\mu}}{\cal X})({\cal D}_{{\nu}{\nu}}{\cal X})
%
\right.
\nonumber\\&&\,\,\,\,\,\,\,\,\,\,\,\,\,\,\,\,\,\,\,\,\,\,\,\,\,\,\,\,\,\,\,\,\,\,\,\,\,\,\,\,\,\,\,\,\,\,\,\,\,\,\,\,\,\,\,\,\,\,\,\,\,\,\,\,\,\,
\left.
-\frac{1}{60} ({\cal D}_{{\mu}{\nu}}{\cal X})
	({\cal D}_{{\mu}{\nu}}{\cal X}+{\cal D}_{{\nu}{\mu}}{\cal X})
\right\}
\label{X1[8]}
\end{eqnarray}

The second-order correction in the pure $X$ case is:
\begin{eqnarray}
 {\cal L}^{(1)[8]}_{2\,\,\,Y\to 0} \!\!\!\!&=&\!\!\!\!
 \frac{\hbar}{2(2\pi)^D}\mbox{Tr}\!\!
    \int^\infty_0\!\! d{s_2}  \int^\infty_0\!\! d{s_1}    \int^\infty_0\!\! d{s_0}
    \!\!\int dX\!\!
    \int d^Dp\,\,\,
      e^{(X-p^2)s_2} e^{(X-p^2)s_1}      e^{(X-p^2)s_0}
\nonumber\\&&\!\!\!\!\!\!\!\!\!\!\!\!\!\!\!\!\!
\times
\left\{
 ({\cal D}_{{\mu_1}{\mu_1}}{X}) ({\cal D}_{{\mu_2}{\mu_2}}{X}) \,{s_0}\, {s_1}
-2 ({\cal D}_{{\mu_2}{\mu_2}}{X}) ({\cal D}_{{\lambda_1}{\tau_1}}{X}) \,p^2\, {s_0}^2\, {s_1}
\right.
\nonumber\\&&\!\!\!\!\!\!\!\!\!\!\!\!\!\!
\left.
-2 ({\cal D}_{{\mu_1}{\mu_1}}{X}) ({\cal D}_{{\lambda_2}{\tau_2}}{X}) \,p^2 {s_0}\, {s_1}^2
+4 ({\cal D}_{{\lambda_1}{\tau_1}}{X}) ({\cal D}_{{\lambda_2}{\tau_2}}{X}) \,p^4\, {s_0}^2\, {s_1}^2
\right\}
\end{eqnarray}
as prescribed in (\ref{2[8]}) with (\ref{for}) with $\ell=1$ and $\ell=2$.

Performing the $p$-integration, $X$-integration (after redefining $X\to {\cal X}-m^2$ and considering (\ref{eX1st})) and renaming dummy indices, we have
\begin{eqnarray}
 {\cal L}^{(1)[8]}_{2\,\,\,Y\to 0} \!\!\!\!&=&\!\!\!\!
 \frac{\hbar}{2(2\pi)^D}\mbox{Tr}\!\!
    \int^\infty_0\!\! d{s_2}  \int^\infty_0\!\! d{s_1}    \int^\infty_0\!\! d{s_0}
\left\{
 (1,1,0;1,1)({\cal D}_{{\mu}{\mu}}{\cal X}) ({\cal D}_{{\nu}{\nu}}{\cal X})
\right.
\nonumber\\&&\!\!\!\!\!\!\!\!\!\!\!\!\!\!\!\!\!
-2 (2,1,0;2,2)({\cal D}_{{\mu}{\mu}}{\cal X}) ({\cal D}_{{\nu}{\nu}}{\cal X})
-2 (1,2,0;2,2)({\cal D}_{{\mu}{\mu}}{\cal X}) ({\cal D}_{{\nu}{\nu}}{\cal X})
\nonumber\\&&\!\!\!\!\!\!\!\!\!\!\!\!\!\!
\left.
+4 (2,2,0;4,3)
 \left(
({\cal D}_{{\mu}{\mu}}{\cal X}) ({\cal D}_{{\nu}{\nu}}{\cal X})
+({\cal D}_{{\mu}{\nu}}{\cal X}) ({\cal D}_{{\mu}{\nu}}{\cal X})
+({\cal D}_{{\mu}{\nu}}{\cal X}) ({\cal D}_{{\nu}{\mu}}{\cal X})
 \right)
\right\}
\nonumber\\
\end{eqnarray}

Performing three-fold proper-time integration:
\begin{eqnarray}
 {\cal L}^{(1)[8]}_{2\,\,\,Y\to 0}
  \!\!&=&\!\! \frac{\hbar}{2(4\pi)^{D/2}}
\int^\infty_0 \frac{ds\,e^{-m^2s}}{s^{-2+D/2}}
\left\{
 \frac{1}{24}({\cal D}_{{\mu}{\mu}}{\cal X}) ({\cal D}_{{\nu}{\nu}}{\cal X})
\right.
\nonumber\\&&\!\!\!\!\!\!\!\!\!\!\!\!\!\!\!\!\!
-2\cdot\frac{1}{120}
\left( ({\cal D}_{{\mu}{\mu}}{\cal X}) ({\cal D}_{{\nu}{\nu}}{\cal X})
      +({\cal D}_{{\mu}{\mu}}{\cal X}) ({\cal D}_{{\nu}{\nu}}{\cal X})
\right)
\nonumber\\&&\!\!\!\!\!\!\!\!\!\!\!\!\!\!
\left.
+4\cdot \frac{1}{720}
 \left(
({\cal D}_{{\mu}{\mu}}{\cal X}) ({\cal D}_{{\nu}{\nu}}{\cal X})
+({\cal D}_{{\mu}{\nu}}{\cal X}) ({\cal D}_{{\mu}{\nu}}{\cal X})
+({\cal D}_{{\mu}{\nu}}{\cal X}) ({\cal D}_{{\nu}{\mu}}{\cal X})
 \right)
\right\}
\nonumber\\
\end{eqnarray}

Simplifying:
\begin{eqnarray}
 {\cal L}^{(1)[8]}_{2\,\,\,Y\to 0}
  \!\!&=&\!\! \frac{\hbar}{2(4\pi)^{D/2}}
\int^\infty_0 \frac{ds\,e^{-m^2s}}{s^{-2+D/2}}
\left\{
 \frac{1}{72}({\cal D}_{{\mu}{\mu}}{\cal X}) ({\cal D}_{{\nu}{\nu}}{\cal X})
\right.
\nonumber\\&&\,\,\,\,\,\,\,\,\,\,\,\,\,\,\,\,\,\,\,\,\,\,\,\,\,\,\,\,\,\,\,\,\,\,\,\,\,\,\,\,\,\,\,\,\,\,\,\,\,\,\,\,\,\,\,\,\,\,\,\,\,\,\,\,\,
\left.
-\frac{1}{180}({\cal D}_{{\mu}{\nu}}{\cal X})
\left(({\cal D}_{{\mu}{\nu}}{\cal X})
+({\cal D}_{{\nu}{\mu}}{\cal X})
\right)
\right\}
\nonumber\\
\label{X2[8]}
\end{eqnarray}
Combining (\ref{X1[8]}) and (\ref{X2[8]}), we have
\begin{eqnarray}
 {\cal L}^{(1)[8]}_{Y\to 0}
  \!\!&=&\!\! \frac{\hbar}{2(4\pi)^{D/2}}
\int^\infty_0 \frac{ds\,e^{-m^2s}}{s^{-2+D/2}}
\left\{
 -\frac{1}{360}({\cal D}_{{\mu}{\mu}}{\cal X}) ({\cal D}_{{\nu}{\nu}}{\cal X})
\right.
\nonumber\\&&\,\,\,\,\,\,\,\,\,\,\,\,\,\,\,\,\,\,\,\,\,\,\,\,\,\,\,\,\,\,\,\,\,\,\,\,\,\,\,\,\,\,\,\,\,\,\,\,\,\,\,\,\,\,\,\,\,\,\,\,\,\,\,\,\,
\left.
-\frac{1}{45}({\cal D}_{{\mu}{\nu}}{\cal X})
\left(({\cal D}_{{\mu}{\nu}}{\cal X})
+({\cal D}_{{\nu}{\mu}}{\cal X})
\right)
\right\}
\end{eqnarray}
Adding (\ref{0[8]}) and (\ref{1XX[4]})
\begin{eqnarray}
 {\cal L}^{(1)[8]}_{Y\to 0}
  \!\!&=&\!\! \frac{\hbar}{2(4\pi)^{D/2}}
\int^\infty_0 \frac{ds\,e^{-m^2s}}{s^{-2+D/2}}
\left\{
 \frac{1}{24}\left({\cal X}-{\cal X}_0\right)
 +\frac{1}{12}{\cal X}({\cal D}_{{\mu}}{\cal X}) ({\cal D}_{{\mu}}{\cal X})
\right.
\nonumber\\&&\,\,\,\,\,\,\,\,\,\,\,\,\,\,\,\,\,\,\,\,\,\,\,\,\,\,\,\,\,\,\,\,\,\,\,\,\,\,\,\,\,\,\,\,\,\,\,\,\,\,\,\,\,\,\,\,\,\,\,\,\,\,\,\,\,
 +\frac{1}{12}{\cal X}^2({\cal D}_{{\mu}{\mu}}{\cal X})
 -\frac{1}{360}({\cal D}_{{\mu}{\mu}}{\cal X}) ({\cal D}_{{\nu}{\nu}}{\cal X})
\nonumber\\&&\,\,\,\,\,\,\,\,\,\,\,\,\,\,\,\,\,\,\,\,\,\,\,\,\,\,\,\,\,\,\,\,\,\,\,\,\,\,\,\,\,\,\,\,\,\,\,\,\,\,\,\,\,\,\,\,\,\,\,\,\,\,\,\,\,
\left.
 -\frac{1}{45}({\cal D}_{{\mu}{\nu}}{\cal X})
\left(({\cal D}_{{\mu}{\nu}}{\cal X})
+({\cal D}_{{\nu}{\mu}}{\cal X})
\right)
\right\}
\end{eqnarray}

Because
\begin{eqnarray}
 {\cal D}_\mu \left({\cal X}^2({\cal D}_{{\mu}}{\cal X})\right)
 = 2{\cal X}({\cal D}_{{\mu}}{\cal X})({\cal D}_{{\mu}}{\cal X})+{\cal X}^2({\cal D}_{{\mu}{\mu}}{\cal X})
\end{eqnarray}
and if the trace of tensors that are self-contracted vanish, we have
\begin{eqnarray}
 {\cal L}^{(1)[8]}_{Y\to 0}
  \!\!&=&\!\! \frac{\hbar}{2(4\pi)^{D/2}}
\int^\infty_0 \frac{ds\,e^{-m^2s}}{s^{-2+D/2}}
\left\{
 \frac{1}{24}\left({\cal X}-{\cal X}_0\right)
 -\frac{1}{12}{\cal X}({\cal D}_{{\mu}}{\cal X}) ({\cal D}_{{\mu}}{\cal X})
\right.
\nonumber\\&&\,\,\,\,\,\,\,\,\,\,\,\,\,\,\,\,\,\,\,\,\,\,\,\,\,\,\,\,\,\,\,\,\,\,\,\,\,\,\,\,\,\,\,\,\,\,\,\,\,\,\,\,\,\,\,\,\,\,\,\,\,\,\,\,\,
\left.
 -\frac{2}{45}({\cal D}_{{\mu}{\nu}}{\cal X})
 ({\cal D}_{{\nu}{\mu}}{\cal X})
\right\}
\label{LX[8]}
\end{eqnarray}
Using the relation for reordering of derivatives within a factor:
\begin{eqnarray}
  {\cal D}_{\mu\nu}{\cal A}&=&{\cal D}_{\nu\mu}{\cal A} -i \left[Y_{\mu\nu},{\cal A}\right]
\end{eqnarray}
for any tensor $\cal A$. This relationship is used for the simplification of the last term in (\ref{LX[8]}) but in the limit $Y \to 0$.

\section{Explicit Calculation of ${\cal L}^{(1)[4,6,8]}_{1,2}$}

Having shown in the previous section the step by step calculation of ${\cal L}^{(1)[4,6,8]}$ for the ungauged case, we choose to skip most of the steps and present only the order by order corrections for each mass-dimension.

\subsection{Four Mass-Dimensions}
\begin{large}
\textbf{First-Order Corrections: Four Mass-Dimensions}
\end{large}

Using the prescription (\ref{1[4]}) with basis invariants (\ref{for})

\begin{eqnarray}
 {\cal L}^{(1)[4]}_{1}
  \!\!&=&\!\! \frac{\hbar}{2(4\pi)^{D/2}}
\int^\infty_0 \frac{ds\,e^{-m^2s}}{s^{-1+D/2}} \frac{1}{6}\mbox{Tr}{\cal D}_{\mu\mu}{\cal X}
\label{L14}
\end{eqnarray}

The simplifications are as follows: $(1,0;1,1)=\frac{1}{2}$, $(2,0;2,2)=\frac{1}{6}$, trace of self-contracted tensors vanish, and $Y_{\mu\nu} = -Y_{\nu\mu}$

Since there are no second-order corrections, we have the complete four mass-dimensional one-loop effective Lagrangian
\begin{eqnarray}\label{L4}
 {\cal L}^{(1)[4]}
  \!\!&=&\!\! \frac{\hbar}{2(4\pi)^{D/2}}
\int^\infty_0 \frac{ds\,e^{-m^2s}}{s^{-1+D/2}}
   \mbox{Tr}
\left[
   \frac{1}{2}({\cal X}^2-{{\cal X}_0}^2)
   +\frac{1}{12}Y_{\mu\nu}Y_{\mu\nu}
   +\frac{1}{6}{\cal D}_{\mu\mu}{\cal X}
\right]
\label{L1[4]done}
\end{eqnarray}
This is upon incorporating the results contributed from the zeroth-order corrections (\ref{0[4]}). This result is in agreement with that of Refs. \cite{Rodf-Diss,Avra,Gilk,Fradk}

\subsection{Six Mass-Dimensions}
\begin{large}
\textbf{First-Order Corrections: Six Mass-Dimensions}
\end{large}

Using the prescription (\ref{1[6]}) with basis invariants (\ref{six})
we have
\begin{eqnarray}
 {\cal L}^{(1)[6]}_{1}
  \!\!&=&\!\! \frac{\hbar}{2(4\pi)^{D/2}}
\int^\infty_0 \frac{ds\,e^{-m^2s}}{s^{-2+D/2}}
  \left(\frac{-1}{540}\right)\mbox{Tr}
\left(
 ({\cal D}_{\rho}Y_{\mu\nu}) ({\cal D}_{\rho}Y_{\mu\nu})
+({\cal D}_{\nu}Y_{\mu\nu}) ({\cal D}_{\rho}Y_{\mu\rho})
\right.
\nonumber\\&&\,\,\,\,\,\,\,\,\,\,\,\,\,\,\,\,\,\,\,\,\,\,\,\,\,\,\,\,\,\,\,\,\,\,\,\,\,\,\,\,\,\,\,\,\,\,\,\,\,\,\,\,\,\,\,\,\,\,\,\,\,\,\,\,\,\,\,\,\,\,\,\,\,\,\,\,\,\,\,\,
\left.
+({\cal D}_{\rho}Y_{\mu\nu}) ({\cal D}_{\nu}Y_{\mu\rho})
\right)
\end{eqnarray}
and using the identity
\begin{eqnarray}
({\cal D}_{\rho}Y_{\mu\nu}) ({\cal D}_{\nu}Y_{\mu\rho})
=\frac{1}{2}({\cal D}_{\rho}Y_{\mu\nu}) ({\cal D}_{\rho}Y_{\mu\nu})
\end{eqnarray}
we have the first-order correction for six mass-dimensions
\begin{eqnarray}
 {\cal L}^{(1)[6]}_{1}
  \!\!&=&\!\! \frac{\hbar}{2(4\pi)^{D/2}}
\int^\infty_0 \frac{ds\,e^{-m^2s}}{s^{-2+D/2}}
  \left(\frac{-1}{540}\right)\mbox{Tr}
\left(
 \frac{3}{2}({\cal D}_{\rho}Y_{\mu\nu}) ({\cal D}_{\rho}Y_{\mu\nu})
+({\cal D}_{\nu}Y_{\mu\nu})({\cal D}_{\rho}Y_{\mu\rho})
\right)
\nonumber\\
\end{eqnarray}
Further simplification can be implemented:
\begin{eqnarray}
({\cal D}_{\rho}Y_{\mu\nu}) ({\cal D}_{\rho}Y_{\mu\nu})
= 2({\cal D}_{\nu}Y_{\mu\nu})({\cal D}_{\rho}Y_{\mu\rho})
-4Y_{\mu\nu}Y_{\nu\rho}Y_{\rho\mu} + \mathrm{tot.div.}
\end{eqnarray}
so that we have the further simplified version of the first order correction:
\begin{eqnarray}
 {\cal L}^{(1)[6]}_{1}
  \!\!&=&\!\! \frac{\hbar}{2(4\pi)^{D/2}}
\int^\infty_0 \frac{ds\,e^{-m^2s}}{s^{-2+D/2}}
  \left(\frac{-1}{270}\right)\mbox{Tr} 
\left(
 2({\cal D}_{\nu}Y_{\mu\nu})({\cal D}_{\rho}Y_{\mu\rho})
-3 Y_{\mu\nu}Y_{\nu\rho}Y_{\rho\mu}
\right)
\nonumber\\
\label{L16}
\end{eqnarray}

The proper-time integrations used are $(2,0;1,1)=\frac{1}{3}$, $(3,0;2,2)=\frac{1}{8}$, and \\ $(4,0;4,3)=\frac{1}{20}$.
Although there is a new term in (\ref{six})
\begin{eqnarray}
\frac{4}{3} p^2 {s_0}^3
{\cal D}_{\mu}{X} ({\cal D}_{\tau}Y_{{\lambda}{\mu}}+{\cal D}_{\mu}Y_{{\lambda}{\tau}}+ {\cal D}_{\nu}Y_{{\mu}{\nu}})
\end{eqnarray}
not included in \cite{Rodf-Diss}. In the simplification process such term vanishes.
We also used the following:
\begin{eqnarray}
 Y_{\mu\nu} {\cal D}_{\rho\nu}Y_{\mu\rho} = - {\cal D}_{\nu}Y_{\mu\nu} {\cal D}_{\rho}Y_{\mu\rho} + \mathrm{tot.der.}
\end{eqnarray}

\begin{large}
\textbf{Second-Order Corrections: Six Mass-Dimensions}
\end{large}

Using the prescription (\ref{2[6]}) with basis invariants (\ref{thr}) with $\ell=1$ and $\ell=2$
\begin{eqnarray}
 {\cal L}^{(1)[6]}_{2}
  \!\!&=&\!\! \frac{\hbar}{2(4\pi)^{D/2}}
\int^\infty_0 \frac{ds\,e^{-m^2s}}{s^{-2+D/2}}
  \left(\frac{-1}{108}\right)\mbox{Tr}
 ({\cal D}_{\nu}Y_{\mu\nu})({\cal D}_{\rho}Y_{\mu\rho})
\label{L26}
\end{eqnarray}
combining (\ref{L4}) (but considering (\ref{eX1st})), (\ref{L16}) and (\ref{L26}) we have
\begin{eqnarray}
 {\cal L}^{(1)[6]}
  \!\!&=&\!\! \frac{\hbar}{2(4\pi)^{D/2}}
\int^\infty_0 \frac{ds\,e^{-m^2s}}{s^{-2+D/2}}
   \mbox{Tr}
\left[
   \frac{1}{6}({\cal X}^3-{{\cal X}_0}^3)
   +\frac{1}{12}{\cal X}Y_{\mu\nu}Y_{\mu\nu}
   +\frac{1}{12}({\cal D}_\mu{\cal X})({\cal D}_\mu{\cal X})
\right.
\nonumber\\&&\,\,\,\,\,\,\,\,\,\,\,\,\,\,\,\,\,\,\,\,\,\,\,\,\,\,\,\,\,\,\,\,\,\,\,\,\,\,\,\,\,\,\,\,\,\,\,\,\,\,\,\,\,
\left.
   +\frac{1}{6}{\cal X}({\cal D}_{\mu\mu}{\cal X})
   -\frac{1}{60} ({\cal D}_{\nu}Y_{\mu\nu})({\cal D}_{\rho}Y_{\mu\rho})
   +\frac{1}{90} Y_{\mu\nu}Y_{\nu\rho}Y_{\rho\mu}
\right].
\nonumber\\
\end{eqnarray}
Further simplifications:
\begin{eqnarray}
 {\cal L}^{(1)[6]}
  \!\!&=&\!\! \frac{\hbar}{2(4\pi)^{D/2}}
\int^\infty_0 \frac{ds\,e^{-m^2s}}{s^{-2+D/2}}
   \mbox{Tr}
\left[
   \frac{1}{6}({\cal X}^3-{{\cal X}_0}^3)
   +\frac{1}{12}{\cal X}Y_{\mu\nu}Y_{\mu\nu}
   -\frac{1}{12}({\cal D}_\mu{\cal X})({\cal D}_\mu{\cal X})
\right.
\nonumber\\&&\,\,\,\,\,\,\,\,\,\,\,\,\,\,\,\,\,\,\,\,\,\,\,\,\,\,\,\,\,\,\,\,\,\,\,\,\,\,\,\,\,\,\,\,\,\,\,\,\,\,\,\,\,\,\,\,\,\,\,\,\,\,\,\,\,\,\,\,
\left.
   -\frac{1}{60} ({\cal D}_{\nu}Y_{\mu\nu})({\cal D}_{\rho}Y_{\mu\rho})
   +\frac{1}{90} Y_{\mu\nu}Y_{\nu\rho}Y_{\rho\mu}
\right].
\label{L6}
\end{eqnarray}
This result is in agreement with \cite{Rodf-Diss,ven}

\subsection{Eight Mass-Dimensions}
\begin{large}
\textbf{First-Order Corrections: Eight Mass-Dimensions}
\end{large}

Using the prescription (\ref{1[8]}) with basis invariants (\ref{eit}) and implementing the algorithm presented in Appendix E, the 154-term result can be reduced to the following:
\begin{eqnarray}
 {\cal L}^{(1)[8]}_{1}
  \!\!\!&=&\!\!\! \frac{\hbar}{2(4\pi)^{D/2}}
\int^\infty_0\!\!\frac{ds\,e^{-m^2s}}{s^{-3+D/2}}
   \mbox{Tr}
\left\{
\!\frac{1}{30} ({\cal D}_{{\mu}}{\cal X}) ({\cal D}_{{\rho}}Y_{{\nu}{\rho}}) Y_{{\nu}{\mu}}
\right.
\!+\!\frac{1}{30} ({\cal D}_{{\mu}}{\cal X}) ({\cal D}_{{\mu}}Y_{{\nu}{\rho}}) Y_{{\nu}{\rho}}
\nonumber\\&&\!\!\!\!\!\!\!\!\!\!\!\!\!\!\!\!
+\frac{1}{30} ({\cal D}_{{\mu}}{\cal X}) ({\cal D}_{{\rho}}Y_{{\nu}{\mu}}) Y_{{\nu}{\rho}}
+\frac{1}{20} ({\cal D}_{{\mu}{\nu}}{\cal X}) Y^2_{{\mu}{\nu}}
+\frac{1}{36} ({\cal D}_{{\mu}{\nu}}Y_{{\rho}{\rho}}) Y^2_{{\mu}{\nu}}
+\frac{2}{45} ({\cal D}_{{\mu}{\rho}}Y_{{\nu}{\rho}}) Y^2_{{\mu}{\nu}}
\nonumber\\&&\!\!\!\!\!\!\!\!\!\!\!\!\!\!\!\!
+\frac{1}{36} ({\cal D}_{{\mu}{\rho}}Y_{{\rho}{\nu}}) Y^2_{{\mu}{\nu}}
-\frac{1}{80} ({\cal D}_{{\nu}{\rho}}Y_{{\mu}{\rho}}) Y^2_{{\mu}{\nu}}
-\frac{1}{180} ({\cal D}_{{\nu}{\rho}}Y_{{\rho}{\mu}}) Y^2_{{\mu}{\nu}}
-\frac{1}{180} ({\cal D}_{{\rho}{\mu}}Y_{{\nu}{\rho}}) Y^2_{{\mu}{\nu}}
\nonumber\\&&\!\!\!\!\!\!\!\!\!\!\!\!\!\!\!\!
-\frac{1}{45} ({\cal D}_{{\rho}{\mu}}Y_{{\rho}{\nu}}) Y^2_{{\mu}{\nu}}
+\frac{1}{144} ({\cal D}_{{\rho}{\nu}}Y_{{\rho}{\mu}}) Y^2_{{\mu}{\nu}}
-\frac{1}{180} ({\cal D}_{{\rho}{\rho}}Y_{{\nu}{\mu}}) Y^2_{{\mu}{\nu}}
-\frac{1}{60} ({\cal D}_{{\mu}{\nu}}{\cal X}) ({\cal D}_{{\mu}{\nu}}{\cal X})
\nonumber\\&&\!\!\!\!\!\!\!\!\!\!\!\!\!\!\!\!
-\frac{1}{60} ({\cal D}_{{\mu}{\nu}}{\cal X}) ({\cal D}_{{\nu}{\mu}}{\cal X})
-\frac{1}{60} ({\cal D}_{{\mu}{\mu}}{\cal X}) ({\cal D}_{{\nu}{\nu}}{\cal X})
+\frac{1}{1120} ({\cal D}_{{\rho}{\nu}}Y_{{\mu}{\nu}}) ({\cal D}_{{\rho}{\sigma}}Y_{{\mu}{\sigma}})
\nonumber\\&&\!\!\!\!\!\!\!\!\!\!\!\!\!\!\!\!
+\frac{1}{1120} ({\cal D}_{{\rho}{\nu}}Y_{{\mu}{\nu}}) ({\cal D}_{{\sigma}{\sigma}}Y_{{\mu}{\rho}})
+\frac{1}{1120} ({\cal D}_{{\rho}{\rho}}Y_{{\mu}{\nu}}) ({\cal D}_{{\sigma}{\sigma}}Y_{{\mu}{\nu}})
-\frac{1}{120} ({\cal D}_{{\nu}{\sigma}}Y_{{\mu}{\rho}}) ({\cal D}_{{\rho}{\sigma}}Y_{{\mu}{\nu}})
\nonumber\\&&\!\!\!\!\!\!\!\!\!\!\!\!\!\!\!\!
-\frac{1}{20} ({\cal D}_{{\nu}{\sigma}}Y_{{\mu}{\nu}}) ({\cal D}_{{\sigma}{\rho}}Y_{{\mu}{\rho}})
-\frac{1}{20} ({\cal D}_{{\rho}{\sigma}}Y_{{\mu}{\rho}}) ({\cal D}_{{\sigma}{\nu}}Y_{{\mu}{\nu}})
-\frac{1}{240} ({\cal D}_{{\nu}{\rho}}Y_{{\mu}{\sigma}}) ({\cal D}_{{\rho}{\sigma}}Y_{{\mu}{\nu}})
\nonumber\\&&\!\!\!\!\!\!\!\!\!\!\!\!\!\!\!\!
-\frac{1}{40} ({\cal D}_{{\nu}{\rho}}Y_{{\mu}{\rho}}) ({\cal D}_{{\sigma}{\sigma}}Y_{{\mu}{\nu}})
-\frac{1}{40} ({\cal D}_{{\nu}{\sigma}}Y_{{\mu}{\nu}}) ({\cal D}_{{\rho}{\rho}}Y_{{\mu}{\sigma}})
-\frac{1}{40} ({\cal D}_{{\nu}{\sigma}}Y_{{\mu}{\nu}}) ({\cal D}_{{\rho}{\sigma}}Y_{{\mu}{\rho}})
\nonumber\\&&\!\!\!\!\!\!\!\!\!\!\!\!\!\!\!\!
-\frac{1}{40} ({\cal D}_{{\rho}{\nu}}Y_{{\mu}{\rho}}) ({\cal D}_{{\sigma}{\sigma}}Y_{{\mu}{\nu}})
+\frac{1}{48} ({\cal D}_{{\nu}{\mu}}Y_{{\nu}{\rho}}) ({\cal D}_{{\nu}{\rho}}Y_{{\mu}{\nu}})
+\frac{1}{48} ({\cal D}_{{\rho}{\nu}}Y_{{\mu}{\nu}}) ({\cal D}_{{\sigma}{\mu}}Y_{{\rho}{\sigma}})
\nonumber\\&&\!\!\!\!\!\!\!\!\!\!\!\!\!\!\!\!
+\frac{1}{48} ({\cal D}_{{\rho}{\sigma}}Y_{{\mu}{\nu}}) ({\cal D}_{{\sigma}{\mu}}Y_{{\rho}{\nu}})
+\frac{1}{480} ({\cal D}_{{\rho}{\sigma}}Y_{{\mu}{\nu}}) ({\cal D}_{{\rho}{\sigma}}Y_{{\mu}{\nu}})
-\frac{1}{840} ({\cal D}_{{\sigma}{\rho}}Y_{{\mu}{\nu}}) ({\cal D}_{{\sigma}{\rho}}Y_{{\mu}{\nu}})
\nonumber\\&&\!\!\!\!\!\!\!\!\!\!\!\!\!\!\!\!
+\frac{11}{480} ({\cal D}_{{\rho}{\sigma}}Y_{{\mu}{\nu}}) ({\cal D}_{{\sigma}{\nu}}Y_{{\mu}{\rho}})
+\frac{29}{1120} ({\cal D}_{{\nu}{\rho}}Y_{{\mu}{\nu}}) ({\cal D}_{{\rho}{\sigma}}Y_{{\mu}{\sigma}})
+\frac{29}{1120} ({\cal D}_{{\nu}{\rho}}Y_{{\mu}{\nu}}) ({\cal D}_{{\sigma}{\rho}}Y_{{\mu}{\sigma}})
\nonumber\\&&\!\!\!\!\!\!\!\!\!\!\!\!\!\!\!\!
+\frac{29}{1120} ({\cal D}_{{\nu}{\rho}}Y_{{\mu}{\nu}}) ({\cal D}_{{\sigma}{\sigma}}Y_{{\mu}{\rho}})
+\frac{29}{1120} ({\cal D}_{{\nu}{\sigma}}Y_{{\mu}{\sigma}}) ({\cal D}_{{\rho}{\rho}}Y_{{\mu}{\nu}})
+\frac{29}{1120} ({\cal D}_{{\rho}{\rho}}Y_{{\mu}{\nu}}) ({\cal D}_{{\sigma}{\nu}}Y_{{\mu}{\sigma}})
\nonumber\\&&\!\!\!\!\!\!\!\!\!\!\!\!\!\!\!\!
+\frac{31}{3360} ({\cal D}_{{\nu}{\rho}}Y_{{\mu}{\sigma}}) ({\cal D}_{{\sigma}{\rho}}Y_{{\mu}{\nu}})
+\frac{31}{3360} ({\cal D}_{{\rho}{\nu}}Y_{{\mu}{\sigma}}) ({\cal D}_{{\sigma}{\rho}}Y_{{\mu}{\nu}})
+\frac{37}{672} ({\cal D}_{{\rho}{\nu}}Y_{{\mu}{\nu}}) ({\cal D}_{{\sigma}{\rho}}Y_{{\mu}{\sigma}})
\nonumber\\&&\!\!\!\!\!\!\!\!\!\!\!\!\!\!\!\!
-\frac{43}{840} ({\cal D}_{{\sigma}{\nu}}Y_{{\mu}{\rho}}) ({\cal D}_{{\sigma}{\rho}}Y_{{\mu}{\nu}})
+\frac{5}{96} ({\cal D}_{{\rho}{\nu}}Y_{{\mu}{\sigma}}) ({\cal D}_{{\rho}{\sigma}}Y_{{\mu}{\nu}})
-\frac{67}{3360} ({\cal D}_{{\rho}{\sigma}}Y_{{\mu}{\nu}}) ({\cal D}_{{\sigma}{\rho}}Y_{{\mu}{\nu}})
\nonumber\\&&\!\!\!\!\!\!\!\!\!\!\!\!\!\!\!\!
\left.
-\frac{79}{1680} ({\cal D}_{{\nu}{\sigma}}Y_{{\mu}{\rho}}) ({\cal D}_{{\sigma}{\rho}}Y_{{\mu}{\nu}})
\right\}
\label{L18}
\end{eqnarray}

\begin{large}
\textbf{Second-Order Corrections: Eight Mass-Dimensions}
\end{large}

Using the prescription (\ref{2[8]}) with basis invariants (\ref{thr}) and (\ref{fiv}) and implementing the algorithm presented in Appendix E, the ($2\times 78$)-term result can be reduced to the following:

\begin{eqnarray}
 {\cal L}^{(1)[8]}_{2\,\,\,\,[3][5]+[5][3]}
  \!\!\!&=&\!\!\! \frac{\hbar}{2(4\pi)^{D/2}}
\int^\infty_0 \frac{ds\,e^{-m^2s}}{s^{-3+D/2}}
   \mbox{Tr}
\left\{
\frac{2}{945}\!({\cal D}_{{\mu}{\nu}{\rho}}{\cal X})\!({\cal D}_{{\rho}}Y_{{\mu}{\nu}})
\!+\!\frac{2}{945} ({\cal D}_{{\mu}{\rho}{\nu}}{\cal X}) ({\cal D}_{{\rho}}Y_{{\mu}{\nu}})
\right.
\nonumber\\&&\!\!\!\!\!\!\!
+\frac{2}{945} ({\cal D}_{{\nu}{\mu}{\rho}}{\cal X}) ({\cal D}_{{\rho}}Y_{{\mu}{\nu}})
+\frac{2}{945} ({\cal D}_{{\nu}{\rho}{\mu}}{\cal X}) ({\cal D}_{{\rho}}Y_{{\mu}{\nu}})
+\frac{2}{945} ({\cal D}_{{\rho}{\mu}{\nu}}{\cal X}) ({\cal D}_{{\rho}}Y_{{\mu}{\nu}})
\nonumber\\&&\!\!\!\!\!\!\!
+\frac{2}{945} ({\cal D}_{{\rho}{\nu}{\mu}}{\cal X}) ({\cal D}_{{\rho}}Y_{{\mu}{\nu}})
-\frac{8}{2835} ({\cal D}_{{\mu}}Y_{{\mu}{\nu}}) ({\cal D}_{{\nu}{\rho}{\rho}}{\cal X})
-\frac{8}{2835} ({\cal D}_{{\mu}}Y_{{\mu}{\nu}}) ({\cal D}_{{\rho}{\nu}{\rho}}{\cal X})
\nonumber\\&&\!\!\!\!\!\!\!
-\!\frac{8}{2835}\! ({\cal D}_{{\mu}}Y_{{\mu}{\nu}}) ({\cal D}_{{\rho}{\rho}{\nu}}{\cal X})
-\!\frac{37}{5670}\! ({\cal D}_{{\mu}{\rho}{\rho}}{\cal X}) ({\cal D}_{{\nu}}Y_{{\mu}{\nu}})
-\!\frac{37}{5670}\! ({\cal D}_{{\nu}}Y_{{\mu}{\nu}}) ({\cal D}_{{\rho}{\mu}{\rho}}{\cal X})
\nonumber\\&&\!\!\!\!\!\!\!
-\frac{37}{5670} ({\cal D}_{{\nu}}Y_{{\mu}{\nu}}) ({\cal D}_{{\rho}{\rho}{\mu}}{\cal X})
+\frac{4}{945} ({\cal D}_{{\nu}}Y_{{\mu}{\sigma}}) ({\cal D}_{{\rho}}Y_{{\rho}{\sigma}}) Y_{{\mu}{\nu}}
\nonumber\\&&\!\!\!\!\!\!\!
+\frac{4}{945} ({\cal D}_{{\nu}}Y_{{\rho}{\sigma}}) ({\cal D}_{{\rho}}Y_{{\mu}{\sigma}}) Y_{{\mu}{\nu}}
+\frac{4}{945} ({\cal D}_{{\nu}}Y_{{\rho}{\sigma}}) ({\cal D}_{{\sigma}}Y_{{\mu}{\rho}}) Y_{{\mu}{\nu}}
\nonumber\\&&\!\!\!\!\!\!\!
+\frac{4}{945} ({\cal D}_{{\rho}}Y_{{\mu}{\sigma}}) ({\cal D}_{{\rho}}Y_{{\nu}{\sigma}}) Y_{{\mu}{\nu}}
+\frac{4}{945} ({\cal D}_{{\rho}}Y_{{\mu}{\sigma}}) ({\cal D}_{{\rho}}Y_{{\sigma}{\nu}}) Y_{{\mu}{\nu}}
\nonumber\\&&\!\!\!\!\!\!\!
+\frac{4}{945} ({\cal D}_{{\rho}}Y_{{\nu}{\sigma}}) ({\cal D}_{{\sigma}}Y_{{\mu}{\rho}}) Y_{{\mu}{\nu}}
+\frac{4}{945} ({\cal D}_{{\rho}}Y_{{\rho}{\nu}}) ({\cal D}_{{\sigma}}Y_{{\mu}{\sigma}}) Y_{{\mu}{\nu}}
\nonumber\\&&\!\!\!\!\!\!\!
+\frac{4}{945} ({\cal D}_{{\rho}}Y_{{\sigma}{\nu}}) ({\cal D}_{{\sigma}}Y_{{\mu}{\rho}}) Y_{{\mu}{\nu}}
-\frac{1}{315} ({\cal D}_{{\nu}}Y_{{\mu}{\sigma}}) ({\cal D}_{{\rho}}Y_{{\sigma}{\rho}}) Y_{{\mu}{\nu}}
\nonumber\\&&\!\!\!\!\!\!\!
-\frac{1}{315} ({\cal D}_{{\rho}}Y_{{\nu}{\rho}}) ({\cal D}_{{\sigma}}Y_{{\mu}{\sigma}}) Y_{{\mu}{\nu}}
-\frac{4}{405} ({\cal D}_{{\nu}}Y_{{\mu}{\rho}}) ({\cal D}_{{\sigma}}Y_{{\rho}{\sigma}}) Y_{{\mu}{\nu}}
\nonumber\\&&\!\!\!\!\!\!\!
-\frac{4}{405} ({\cal D}_{{\nu}}Y_{{\mu}{\rho}}) ({\cal D}_{{\sigma}}Y_{{\sigma}{\rho}}) Y_{{\mu}{\nu}}
-\frac{4}{405} ({\cal D}_{{\rho}}Y_{{\mu}{\rho}}) ({\cal D}_{{\sigma}}Y_{{\nu}{\sigma}}) Y_{{\mu}{\nu}}
\nonumber\\&&\!\!\!\!\!\!\!
-\frac{4}{405} ({\cal D}_{{\rho}}Y_{{\mu}{\rho}}) ({\cal D}_{{\sigma}}Y_{{\sigma}{\nu}}) Y_{{\mu}{\nu}}
-\frac{16}{2835} ({\cal D}_{{\rho}}Y_{{\rho}{\sigma}}) ({\cal D}_{{\sigma}}Y_{{\mu}{\nu}}) Y_{{\mu}{\nu}}
\nonumber\\&&\!\!\!\!\!\!\!
\left.
-\frac{37}{2835} ({\cal D}_{{\rho}}Y_{{\sigma}{\rho}}) ({\cal D}_{{\sigma}}Y_{{\mu}{\nu}}) Y_{{\mu}{\nu}}
\right\}
\label{L28_35+53}
\end{eqnarray}

With basis invariants (\ref{for}) and implementing the algorithm presented in Appendix E yields the 285-term result which can be reduced to the following:

\begin{eqnarray}
 {\cal L}^{(1)[8]}_{2\,\,\,\,[4][4]}
  \!\!\!&=&\!\!\! \frac{\hbar}{2(4\pi)^{D/2}}
\int^\infty_0 \frac{ds\,e^{-m^2s}}{s^{-3+D/2}}
   \mbox{Tr}
\left\{
\frac{1}{72} ({\cal D}_{{\mu}{\mu}}X) ({\cal D}_{{\nu}{\nu}}X)
+\frac{1}{180} ({\cal D}_{{\mu}{\nu}}X) ({\cal D}_{{\mu}{\nu}}X)
\right.
\nonumber\\&&\!\!\!\!\!\!\!\!\!\!\!\!\!\!\!\!
+\frac{1}{180} ({\cal D}_{{\mu}{\nu}}X) ({\cal D}_{{\nu}{\mu}}X)
+\frac{13}{2520} ({\cal D}_{{\mu}{\nu}}X) ({\cal D}_{{\mu}{\rho}}Y_{{\nu}{\rho}})
-\frac{1}{336} ({\cal D}_{{\mu}{\nu}}X) ({\cal D}_{{\mu}{\rho}}Y_{{\rho}{\nu}})
\nonumber\\&&\!\!\!\!\!\!\!\!\!\!\!\!\!\!\!\!
+\frac{13}{2520} ({\cal D}_{{\mu}{\nu}}X) ({\cal D}_{{\nu}{\rho}}Y_{{\mu}{\rho}})
+\frac{2}{315} ({\cal D}_{{\mu}{\mu}}X) ({\cal D}_{{\nu}{\rho}}Y_{{\nu}{\rho}})
+\frac{1}{420} ({\cal D}_{{\mu}{\nu}}X) ({\cal D}_{{\nu}{\rho}}Y_{{\rho}{\mu}})
\nonumber\\&&\!\!\!\!\!\!\!\!\!\!\!\!\!\!\!\!
+\frac{2}{315} ({\cal D}_{{\mu}{\mu}}X) ({\cal D}_{{\nu}{\rho}}Y_{{\rho}{\nu}})
-\frac{1}{315} ({\cal D}_{{\mu}{\nu}}X) ({\cal D}_{{\rho}{\mu}}Y_{{\nu}{\rho}})
-\frac{19}{1680} ({\cal D}_{{\mu}{\nu}}X) ({\cal D}_{{\rho}{\mu}}Y_{{\rho}{\nu}})
\nonumber\\&&\!\!\!\!\!\!\!\!\!\!\!\!\!\!\!\!
-\frac{1}{315} ({\cal D}_{{\mu}{\nu}}X) ({\cal D}_{{\rho}{\nu}}Y_{{\mu}{\rho}})
-\frac{1}{168} ({\cal D}_{{\mu}{\nu}}X) ({\cal D}_{{\rho}{\nu}}Y_{{\rho}{\mu}})
-\frac{43}{5040} ({\cal D}_{{\mu}{\nu}}X) ({\cal D}_{{\rho}{\rho}}Y_{{\mu}{\nu}})
\nonumber\\&&\!\!\!\!\!\!\!\!\!\!\!\!\!\!\!\!
-\frac{1}{315} ({\cal D}_{{\mu}{\nu}}X) ({\cal D}_{{\rho}{\rho}}Y_{{\nu}{\mu}})
+\frac{1}{672} ({\cal D}_{{\mu}{\rho}}Y_{{\mu}{\nu}}) ({\cal D}_{{\nu}{\sigma}}Y_{{\rho}{\sigma}})
+\frac{1}{672} ({\cal D}_{{\mu}{\sigma}}Y_{{\mu}{\nu}}) ({\cal D}_{{\sigma}{\rho}}Y_{{\nu}{\rho}})
\nonumber\\&&\!\!\!\!\!\!\!\!\!\!\!\!\!\!\!\!
+\frac{1}{840} ({\cal D}_{{\mu}{\rho}}Y_{{\rho}{\sigma}}) ({\cal D}_{{\sigma}{\nu}}Y_{{\mu}{\nu}})
+\frac{1}{840} ({\cal D}_{{\sigma}{\nu}}Y_{{\mu}{\nu}}) ({\cal D}_{{\sigma}{\rho}}Y_{{\rho}{\mu}})
+\frac{1}{1120} ({\cal D}_{{\mu}{\rho}}Y_{{\mu}{\nu}}) ({\cal D}_{{\sigma}{\nu}}Y_{{\rho}{\sigma}})
\nonumber\\&&\!\!\!\!\!\!\!\!\!\!\!\!\!\!\!\!
+\frac{1}{1120} ({\cal D}_{{\mu}{\rho}}Y_{{\mu}{\nu}}) ({\cal D}_{{\sigma}{\rho}}Y_{{\nu}{\sigma}})
+\frac{1}{1120} ({\cal D}_{{\mu}{\rho}}Y_{{\mu}{\nu}}) ({\cal D}_{{\sigma}{\sigma}}Y_{{\nu}{\rho}})
+\frac{1}{1120} ({\cal D}_{{\mu}{\rho}}Y_{{\mu}{\nu}}) ({\cal D}_{{\sigma}{\sigma}}Y_{{\rho}{\nu}})
\nonumber\\&&\!\!\!\!\!\!\!\!\!\!\!\!\!\!\!\!
+\frac{1}{1260} ({\cal D}_{{\mu}{\rho}}Y_{{\nu}{\rho}}) ({\cal D}_{{\sigma}{\sigma}}Y_{{\mu}{\nu}})
+\frac{1}{1260} ({\cal D}_{{\mu}{\sigma}}Y_{{\rho}{\sigma}}) ({\cal D}_{{\nu}{\rho}}Y_{{\mu}{\nu}})
+\frac{1}{1260} ({\cal D}_{{\nu}{\rho}}Y_{{\mu}{\rho}}) ({\cal D}_{{\sigma}{\sigma}}Y_{{\mu}{\nu}})
\nonumber\\&&\!\!\!\!\!\!\!\!\!\!\!\!\!\!\!\!
+\frac{1}{1260} ({\cal D}_{{\nu}{\sigma}}Y_{{\mu}{\nu}}) ({\cal D}_{{\sigma}{\rho}}Y_{{\mu}{\rho}})
-\frac{1}{210} ({\cal D}_{{\rho}{\nu}}Y_{{\mu}{\nu}}) ({\cal D}_{{\sigma}{\mu}}Y_{{\sigma}{\rho}})
-\frac{1}{210} ({\cal D}_{{\rho}{\nu}}Y_{{\mu}{\nu}}) ({\cal D}_{{\sigma}{\rho}}Y_{{\sigma}{\mu}})
\nonumber\\&&\!\!\!\!\!\!\!\!\!\!\!\!\!\!\!\!
+\frac{1}{3360} ({\cal D}_{{\nu}{\rho}}Y_{{\mu}{\nu}}) ({\cal D}_{{\sigma}{\mu}}Y_{{\sigma}{\rho}})
+\frac{1}{3360} ({\cal D}_{{\nu}{\rho}}Y_{{\mu}{\nu}}) ({\cal D}_{{\sigma}{\rho}}Y_{{\sigma}{\mu}})
+\frac{1}{3360} ({\cal D}_{{\rho}{\rho}}Y_{{\mu}{\nu}}) ({\cal D}_{{\sigma}{\mu}}Y_{{\sigma}{\nu}})
\nonumber\\&&\!\!\!\!\!\!\!\!\!\!\!\!\!\!\!\!
+\frac{1}{3360} ({\cal D}_{{\rho}{\rho}}Y_{{\mu}{\nu}}) ({\cal D}_{{\sigma}{\nu}}Y_{{\sigma}{\mu}})
-\frac{1}{360} ({\cal D}_{{\rho}{\nu}}Y_{{\mu}{\nu}}) ({\cal D}_{{\sigma}{\mu}}Y_{{\rho}{\sigma}})
-\frac{1}{360} ({\cal D}_{{\rho}{\nu}}Y_{{\mu}{\nu}}) ({\cal D}_{{\sigma}{\rho}}Y_{{\mu}{\sigma}})
\nonumber\\&&\!\!\!\!\!\!\!\!\!\!\!\!\!\!\!\!
-\frac{1}{360} ({\cal D}_{{\rho}{\nu}}Y_{{\mu}{\nu}}) ({\cal D}_{{\sigma}{\sigma}}Y_{{\mu}{\rho}})
-\frac{1}{360} ({\cal D}_{{\rho}{\nu}}Y_{{\mu}{\nu}}) ({\cal D}_{{\sigma}{\sigma}}Y_{{\rho}{\mu}})
+\frac{1}{5040} ({\cal D}_{{\nu}{\rho}}Y_{{\mu}{\nu}}) ({\cal D}_{{\sigma}{\mu}}Y_{{\rho}{\sigma}})
\nonumber\\&&\!\!\!\!\!\!\!\!\!\!\!\!\!\!\!\!
+\frac{1}{5040} ({\cal D}_{{\nu}{\rho}}Y_{{\mu}{\nu}}) ({\cal D}_{{\sigma}{\rho}}Y_{{\mu}{\sigma}})
+\frac{1}{5040} ({\cal D}_{{\nu}{\rho}}Y_{{\mu}{\nu}}) ({\cal D}_{{\sigma}{\sigma}}Y_{{\mu}{\rho}})
+\frac{1}{5040} ({\cal D}_{{\nu}{\rho}}Y_{{\mu}{\nu}}) ({\cal D}_{{\sigma}{\sigma}}Y_{{\rho}{\mu}})
\nonumber\\&&\!\!\!\!\!\!\!\!\!\!\!\!\!\!\!\!
+\frac{1}{5040} ({\cal D}_{{\rho}{\rho}}Y_{{\mu}{\nu}}) ({\cal D}_{{\sigma}{\mu}}Y_{{\nu}{\sigma}})
+\frac{1}{5040} ({\cal D}_{{\rho}{\rho}}Y_{{\mu}{\nu}}) ({\cal D}_{{\sigma}{\nu}}Y_{{\mu}{\sigma}})
+\frac{1}{5040} ({\cal D}_{{\rho}{\rho}}Y_{{\mu}{\nu}}) ({\cal D}_{{\sigma}{\sigma}}Y_{{\mu}{\nu}})
\nonumber\\&&\!\!\!\!\!\!\!\!\!\!\!\!\!\!\!\!
+\frac{1}{5040} ({\cal D}_{{\rho}{\rho}}Y_{{\mu}{\nu}}) ({\cal D}_{{\sigma}{\sigma}}Y_{{\nu}{\mu}})
-\frac{1}{560} ({\cal D}_{{\rho}{\mu}}Y_{{\mu}{\nu}}) ({\cal D}_{{\sigma}{\nu}}Y_{{\sigma}{\rho}})
-\frac{1}{560} ({\cal D}_{{\rho}{\mu}}Y_{{\mu}{\nu}}) ({\cal D}_{{\sigma}{\rho}}Y_{{\sigma}{\nu}})
\nonumber\\&&\!\!\!\!\!\!\!\!\!\!\!\!\!\!\!\!
-\frac{1}{840} ({\cal D}_{{\rho}{\mu}}Y_{{\mu}{\nu}}) ({\cal D}_{{\sigma}{\nu}}Y_{{\rho}{\sigma}})
-\frac{1}{840} ({\cal D}_{{\rho}{\mu}}Y_{{\mu}{\nu}}) ({\cal D}_{{\sigma}{\rho}}Y_{{\nu}{\sigma}})
-\frac{1}{840} ({\cal D}_{{\rho}{\mu}}Y_{{\mu}{\nu}}) ({\cal D}_{{\sigma}{\sigma}}Y_{{\nu}{\rho}})
\nonumber\\&&\!\!\!\!\!\!\!\!\!\!\!\!\!\!\!\!
+\frac{3}{2240} ({\cal D}_{{\mu}{\rho}}Y_{{\mu}{\nu}}) ({\cal D}_{{\sigma}{\nu}}Y_{{\sigma}{\rho}})
+\frac{3}{2240} ({\cal D}_{{\mu}{\rho}}Y_{{\mu}{\nu}}) ({\cal D}_{{\sigma}{\rho}}Y_{{\sigma}{\nu}})
-\frac{1}{1680} ({\cal D}_{{\nu}{\sigma}}Y_{{\rho}{\sigma}}) ({\cal D}_{{\rho}{\mu}}Y_{{\mu}{\nu}})
\nonumber\\&&\!\!\!\!\!\!\!\!\!\!\!\!\!\!\!\!
-\frac{1}{1680} ({\cal D}_{{\sigma}{\mu}}Y_{{\mu}{\nu}}) ({\cal D}_{{\sigma}{\rho}}Y_{{\nu}{\rho}})
-\frac{5}{8064} ({\cal D}_{{\mu}{\nu}}Y_{{\mu}{\nu}}) ({\cal D}_{{\sigma}{\rho}}Y_{{\rho}{\sigma}})
-\frac{1}{840} ({\cal D}_{{\rho}{\mu}}Y_{{\mu}{\nu}}) ({\cal D}_{{\sigma}{\sigma}}Y_{{\rho}{\nu}})
\nonumber\\&&\!\!\!\!\!\!\!\!\!\!\!\!\!\!\!\!
\!+\!\frac{11}{40320} ({\cal D}_{{\mu}{\nu}}Y_{{\mu}{\nu}}) ({\cal D}_{{\sigma}{\rho}}Y_{{\sigma}{\rho}})
\!+\!\frac{11}{40320} ({\cal D}_{{\nu}{\mu}}Y_{{\mu}{\nu}}) ({\cal D}_{{\sigma}{\rho}}Y_{{\rho}{\sigma}})
\!+\!\frac{19}{20160} ({\cal D}_{{\mu}{\sigma}}Y_{{\rho}{\sigma}}) ({\cal D}_{{\rho}{\nu}}Y_{{\mu}{\nu}})
\nonumber\\&&\!\!\!\!\!\!\!\!\!\!\!\!\!\!\!\!
\left.
\!+\!\frac{19}{20160} ({\cal D}_{{\sigma}{\nu}}Y_{{\mu}{\nu}}) ({\cal D}_{{\sigma}{\rho}}Y_{{\mu}{\rho}})
\!+\!\frac{47}{40320} ({\cal D}_{{\nu}{\mu}}Y_{{\mu}{\nu}}) ({\cal D}_{{\sigma}{\rho}}Y_{{\sigma}{\rho}})
\right\}
\label{L28_44}
\end{eqnarray}

\begin{large}
\textbf{Lower Mass-Dimensional Invariant Contributions}
\end{large}

We have already made a point in Chapter 4 that $X$-integration depending on whether we are considering (\ref{eX1st}),
(\ref{eX2nd}), or (\ref{eX3rd}) could contribute same invariants but differs by a $\frac{1}{n!}{\cal X}^n$ factor.

In this regard (\ref{L14}) result contributes in the correction of eight mass-dimensional one-loop effective Lagrangian as
\begin{eqnarray}
 {\cal L}^{(1)[8]}_{1\,\,{\cal X}^2[4]}
  \!\!&=&\!\! \frac{\hbar}{2(4\pi)^{D/2}}
\int^\infty_0 \frac{ds\,e^{-m^2s}}{s^{-3+D/2}} \frac{1}{12}\mbox{Tr}{\cal X}^2{\cal D}_{\mu\mu}{\cal X}
\label{L14XX}
\end{eqnarray}
if $X$-integration is performed with the third therm in the expansion (\ref{eX3rd}).

In similar manner (\ref{L16}) contributes as
\begin{eqnarray}
 {\cal L}^{(1)[8]}_{1\,\,{\cal X}[6]}
  \!\!&=&\!\! \frac{\hbar}{2(4\pi)^{D/2}}
\int^\infty_0 \frac{ds\,e^{-m^2s}}{s^{-2+D/2}}
  \left(\frac{-1}{270}\right)\mbox{Tr} {\cal X}
\left\{
 2({\cal D}_{\nu}Y_{\mu\nu})({\cal D}_{\rho}Y_{\mu\rho})
-3 Y_{\mu\nu}Y_{\nu\rho}Y_{\rho\mu}
\right\}
\nonumber\\
\label{L16X}
\end{eqnarray}
and in the same way as (\ref{L26})
\begin{eqnarray}
 {\cal L}^{(1)[8]}_{2\,\,{\cal X}[6]}
  \!\!&=&\!\! \frac{\hbar}{2(4\pi)^{D/2}}
\int^\infty_0 \frac{ds\,e^{-m^2s}}{s^{-2+D/2}}
  \left(\frac{-1}{108}\right)\mbox{Tr}{\cal X}
 ({\cal D}_{\nu}Y_{\mu\nu})({\cal D}_{\rho}Y_{\mu\rho})
\label{L26X}
\end{eqnarray}

We now combine the results (\ref{0[8]}), (\ref{L18}), (\ref{L28_35+53}) (\ref{L28_44}), (\ref{L14XX}), (\ref{L16X}) and (\ref{L26X})
\begin{eqnarray}
 {\cal L}^{(1)[8]}
 =
 {\cal L}^{(1)[8]}_{0}
 +{\cal L}^{(1)[8]}_{1}
 +{\cal L}^{(1)[8]}_{1\,\,{\cal X}^2[4]}
 +{\cal L}^{(1)[8]}_{1\,\,{\cal X}[6]}
 +{\cal L}^{(1)[8]}_{2}
 +{\cal L}^{(1)[8]}_{2\,\,{\cal X}[6]}
\label{L18s}
\end{eqnarray}
This is an expression containing 132 terms.

Manipulating using the properties:
Cyclic matrix permutations
\begin{eqnarray}
  \mathrm{tr}(XYZ) = \mathrm{tr}(YZX)
\end{eqnarray}
integration by parts
\begin{eqnarray}
  \int dx\,\,\mathrm{tr}(D_\mu X_\mu Y) = - \int dx \,\,\mathrm{tr}(X_\mu D_\mu Y)
\end{eqnarray}
and antisymmetry of $Y_{\mu\nu}$
\begin{eqnarray}
 	Y_{\mu\nu} = -Y_{\nu\mu}
\end{eqnarray}
we have the following one-loop effective Lagrangian with eight mass-dimensional invariants:
\begin{small}
\begin{eqnarray}
 {\cal L}^{(1)[8]}
  \!\!\!\!&=&\!\!\!\! \frac{\hbar}{2(4\pi)^{D/2}}
\int^\infty_0 \frac{ds\,e^{-m^2s}}{s^{-2+D/2}}
\left\{
+\frac{1}{24}\left({\cal X}^4-{{\cal X}_0}^4\right)
\right.
+\frac{1}{90} {\cal X} Y_{\mu\nu} Y_{\nu\rho} Y_{\rho\mu}
\nonumber\\&&
+\frac{1}{288} Y_{\mu\nu} Y_{\mu\nu} Y_{\rho\sigma} Y_{\rho\sigma}
+\frac{1}{360} Y_{\mu\nu} Y_{\nu\rho} Y_{\rho\sigma} Y_{\sigma\mu}
+\frac{1}{24} {\cal X}^2 Y_{\mu\nu} Y_{\mu\nu}
-\frac{1}{12} {\cal X} ({\cal D}_{\mu}{\cal X})({\cal D}_{\mu}{\cal X})
\nonumber\\&&
-\frac{1}{60} {\cal X} ({\cal D}_{\nu}Y_{\mu\nu})({\cal D}_{\rho}Y_{\mu\rho})
+\frac{1}{30}({\cal D}_{\mu}{\cal X})
  \left( ({\cal D}_{<\mu}Y_{\nu\rho>})
        Y_{\nu\rho}
         +({\cal D}_{\rho}Y_{\nu\rho})Y_{\nu\mu}\right)
\nonumber\\&&
+Y_{\mu\nu}
\left[
 \frac{1}{135}
 \left(
   ({\cal D}_{<\nu}Y_{\mu\sigma >}
    ) ({\cal D}_{\rho}Y_{\rho\sigma})+({\cal D}_{\rho}Y_{\rho\nu})({\cal D}_{\sigma}Y_{\mu\sigma})
 \right)
 +\frac{4}{945}
   ({\cal D}_{\nu}Y_{\rho\sigma})({\cal D}_{<\rho}Y_{\mu\sigma >}
            )
\right]
\nonumber\\&&
+Y^2_{\mu\nu}
 \left[
 \frac{1}{60}{\cal D}_{\{\rho\mu\}}Y_{\nu\rho} 
  -\frac{1}{144} {\cal D}_{\{\nu\rho\}}Y_{\mu\rho} 
  -\frac{1}{180} {\cal D}_{\rho\rho}Y_{\nu\mu}
  +\frac{1}{20} {\cal D}_{\mu\nu}{\cal X}
 \right]
\nonumber\\&&
-\frac{1}{90}
 \left[
 ({\cal D}_{\mu\nu}{\cal X})({\cal D}_{\{\mu\nu\}}{\cal X})
 +\frac{1}{4} ({\cal D}_{\mu\mu}{\cal X})({\cal D}_{\nu\nu}{\cal X})
 \right]
\nonumber\\&&
{\cal D}_{\mu\nu}{\cal X}
\left[
 +\frac{11}{2520} {\cal D}_{\rho\nu}Y_{\mu\rho}
 +\frac{37}{3240} {\cal D}_{\nu\rho}Y_{\mu\rho}
 +\frac{179}{15120} {\cal D}_{\rho\mu}Y_{\nu\rho}
 +\frac{41}{5040} {\cal D}_{\mu\rho}Y_{\nu\rho}
\right]
\nonumber\\ &&
+\frac{8}{2835} ({\cal D}_{\mu\mu}{\cal X})({\cal D}_{\nu\rho}Y_{\nu\rho})
-\frac{377}{5040} ({\cal D}_{{\rho}{\sigma}}Y_{{\mu}{\rho}}) ({\cal D}_{{\sigma}{\nu}}Y_{{\mu}{\nu}})
-\frac{167}{6720} ({\cal D}_{{\rho}{\sigma}}Y_{{\mu}{\rho}}) ({\cal D}_{{\nu}{\sigma}}Y_{{\mu}{\nu}})
\nonumber\\ &&
-\frac{27}{1120} ({\cal D}_{{\rho}{\nu}}Y_{{\mu}{\rho}}) ({\cal D}_{{\sigma}{\sigma}}Y_{{\mu}{\nu}})
-\frac{29}{1260} ({\cal D}_{{\rho}{\nu}}Y_{{\mu}{\nu}}) ({\cal D}_{{\sigma}{\mu}}Y_{{\sigma}{\rho}})
+\frac{1}{2880} ({\cal D}_{{\nu}{\sigma}}Y_{{\rho}{\sigma}}) ({\cal D}_{{\mu}{\rho}}Y_{{\mu}{\nu}})
\nonumber\\ &&
-\frac{1}{120} ({\cal D}_{{\nu}{\sigma}}Y_{{\mu}{\rho}}) ({\cal D}_{{\rho}{\sigma}}Y_{{\mu}{\nu}})
+\frac{3}{160} ({\cal D}_{{\nu}{\sigma}}Y_{{\mu}{\rho}}) ({\cal D}_{{\sigma}{\rho}}Y_{{\mu}{\nu}})
-\frac{557}{10080} ({\cal D}_{{\rho}{\nu}}Y_{{\mu}{\nu}}) ({\cal D}_{{\sigma}{\rho}}Y_{{\sigma}{\mu}})
\nonumber\\ &&
-\frac{1}{3360} ({\cal D}_{{\rho}{\nu}}Y_{{\mu}{\nu}}) ({\cal D}_{{\rho}{\sigma}}Y_{{\sigma}{\mu}})
+\frac{37}{10080} ({\cal D}_{{\rho}{\nu}}Y_{{\mu}{\nu}}) ({\cal D}_{{\sigma}{\sigma}}Y_{{\mu}{\rho}})
+\frac{1}{672} ({\cal D}_{{\mu}{\rho}}Y_{{\mu}{\nu}}) ({\cal D}_{{\nu}{\sigma}}Y_{{\rho}{\sigma}})
\nonumber\\ &&
-\frac{1}{2880} ({\cal D}_{{\mu}{\rho}}Y_{{\mu}{\nu}}) ({\cal D}_{{\sigma}{\nu}}Y_{{\rho}{\sigma}})
+\frac{1}{1120} ({\cal D}_{{\mu}{\nu}}Y_{{\rho}{\sigma}}) ({\cal D}_{{\mu}{\nu}}Y_{{\rho}{\sigma}})
-\frac{1}{10080} ({\cal D}_{{\rho}{\rho}}Y_{{\mu}{\nu}}) ({\cal D}_{{\sigma}{\mu}}Y_{{\nu}{\sigma}})
\nonumber\\ &&
-\frac{79}{1680} ({\cal D}_{{\mu}{\nu}}Y_{{\rho}{\sigma}}) ({\cal D}_{{\nu}{\sigma}}Y_{{\rho}{\mu}})
+\frac{31}{3360} ({\cal D}_{{\mu}{\nu}}Y_{{\rho}{\sigma}}) ({\cal D}_{{\sigma}{\nu}}Y_{{\rho}{\mu}})
+\frac{1}{1120} ({\cal D}_{{\rho}{\rho}}Y_{{\mu}{\nu}}) ({\cal D}_{{\sigma}{\sigma}}Y_{{\mu}{\nu}})
\nonumber\\ &&
+\frac{1}{48} ({\cal D}_{{\rho}{\sigma}}Y_{{\mu}{\nu}}) ({\cal D}_{{\sigma}{\mu}}Y_{{\rho}{\nu}})
+\frac{31}{3360} ({\cal D}_{{\rho}{\nu}}Y_{{\mu}{\sigma}}) ({\cal D}_{{\sigma}{\rho}}Y_{{\mu}{\nu}})
-\frac{67}{3360} ({\cal D}_{{\rho}{\sigma}}Y_{{\mu}{\nu}}) ({\cal D}_{{\sigma}{\rho}}Y_{{\mu}{\nu}})
\nonumber\\ &&
+\frac{1}{48} ({\cal D}_{{\sigma}{\mu}}Y_{{\nu}{\rho}}) ({\cal D}_{{\nu}{\rho}}Y_{{\mu}{\sigma}})
+\frac{1}{3360} ({\cal D}_{{\rho}{\rho}}Y_{{\mu}{\nu}}) ({\cal D}_{{\sigma}{\nu}}Y_{{\sigma}{\mu}})
+\frac{263}{10080} ({\cal D}_{{\rho}{\rho}}Y_{{\mu}{\nu}}) ({\cal D}_{{\sigma}{\nu}}Y_{{\mu}{\sigma}})
\nonumber\\ &&
\left.
+\frac{19}{20160} ({\cal D}_{{\sigma}{\nu}}Y_{{\mu}{\nu}}) ({\cal D}_{{\sigma}{\rho}}Y_{{\mu}{\rho}})
+\frac{263}{10080} ({\cal D}_{{\nu}{\rho}}Y_{{\mu}{\nu}}) ({\cal D}_{{\sigma}{\rho}}Y_{{\mu}{\sigma}})
\right\}
\label{L8}
\end{eqnarray}
\end{small}
Eqn. (\ref{L8}) is a 54-term reduced form of 132-term Eqn. (\ref{L18s}).
This result awaits the application of the following identities\cite{ven}
\begin{eqnarray}
({\cal D}^2{\cal X}) Y^2_{\mu\nu}
  \!\!+\!\!2 ({\cal D}_\mu{\cal X}) ({\cal D}_\mu Y_{\nu\rho})Y_{\nu\rho}&\!\!\!\!=\!\!\!\!&0
\label{id-frst}
\\
({\cal D}_{\nu\mu} {\cal X})Y_{\mu\rho}Y_{\nu\rho}
  \!\!+\!\!({\cal D}_\mu {\cal X})({\cal D}_\nu Y_{\nu\rho})Y_{\mu\rho}
  \!\!+\!\!({\cal D}_\mu {\cal X})({\cal D}_\nu Y_{\rho\mu})Y_{\nu\rho}&\!\!\!\!=\!\!\!\!&0
\\
({\cal D}_\mu {\cal X})({\cal D}_\nu Y_{\nu\rho})Y_{\mu\rho}
  \!\!+\!\!{\cal X} ({\cal D}_\mu Y_{\mu\nu})^2
  \!\!+\!\!{\cal X} Y_{\mu\nu} ({\cal D}_{\rho\mu}Y_{\rho\nu})&\!\!\!\!=\!\!\!\!&0
\\
 ({\cal D}_\mu{\cal X}) ({\cal D}_\mu Y_{\nu\rho})Y_{\nu\rho}
  \!\!+\!\!{\cal X}({\cal D}_{\mu}Y_{\nu\rho})^2
  \!\!+\!\!{\cal X} Y_{\mu\nu} ({\cal D}^2Y_{\mu\nu})&\!\!\!\!=\!\!\!\!&0
\\
({\cal D}_\mu{\cal X}) ({\cal D}_\mu Y_{\nu\rho})Y_{\nu\rho}
  \!\!+\!\!2 ({\cal D}_\mu {\cal X})({\cal D}_\nu Y_{\rho\mu})Y_{\nu\rho}&\!\!\!\!=\!\!\!\!&0
\\
{\cal X}({\cal D}_{\mu}Y_{\nu\rho})^2
  \!\!+\!\!2 {\cal X} ({\cal D}_\mu Y_{\nu\rho})({\cal D}_\nu Y_{\rho\mu})&\!\!\!\!=\!\!\!\!&0
\\
2 {\cal X} Y_{\mu\nu} ({\cal D}_{\rho\mu}Y_{\rho\nu})
   +2 {\cal X}Y_{\mu\nu}Y_{\nu\rho}Y_{\rho\mu}
&\!\!\!\!=\!\!\!\!&{\cal X} Y_{\mu\nu} ({\cal D}^2Y_{\mu\nu})
\label{id-last}
\end{eqnarray}
and ordering of derivatives within a factor can be changed by continued application\cite{Mue} of
\begin{eqnarray}
	D_{\mu\nu}X = D_{\nu\mu} - i[Y_{\mu\nu},X]
\end{eqnarray}
for a further reduction.

\chapter{Summary and Future Work}
In this thesis, we have developed a prescription that can calculate one-loop effective Lagrangians for various mass-dimensions and for different orders of correction. We have worked out the calculation of ${\cal L}^{1[2]}$, ${\cal L}^{1[4]}$, and ${\cal L}^{1[6]}$ completely; ${\cal L}^{1[8]}$ partially; ${\cal L}^{1[10]}$ and ${\cal L}^{1[12]}$ in their unsimplified form. For appropriate mass-dimensions, first- up to fourth-order corrections were incorporated. The results obtained in the present work for ${\cal L}^{1[2]}$, ${\cal L}^{1[4]}$, and ${\cal L}^{1[6]}$ have been compared to those found in literature and were found to be in complete agreement. ${\cal L}^{1[8]}$ remains to be simplified further and reduced to a 17-term expression before comparison can be made.

\section{Effective Lagrangians}

The following is a list of the one-loop effective Lagrangians explicitly calculated in a quasi-automated manner incorporating zeroth-order corrections.

\vspace{0.5cm}
\begin{minipage}{5.5in}
\begin{tabular}[b]{c|c|c}
\hline
Mass-Dim 
& Name & Eqn. No.
\\ \hline\hline
Two      
& ${\cal L}^{1[2]}_0$ & (\ref{0[2]}) \\ \hline
Four     
& ${\cal L}^{1[4]}_0$ & (\ref{0[4]}) \\ \hline
Six     
& ${\cal L}^{1[6]}_0$ & (\ref{0[6]}) \\ \hline
Eight 
& ${\cal L}^{1[8]}_0$ & (\ref{0[8]}) \\ \hline
Ten 
& ${\cal L}^{1[10]}_0$ & (\ref{0[10]}) \\ \hline
Twelve 
& ${\cal L}^{1[12]}_0$ & (\ref{0[12]}) \\ \hline
Forteen 
& ${\cal L}^{1[14]}_0$ & (\ref{0[14]}) \\ \hline
Sixteen 
& ${\cal L}^{1[16]}_0$ & (\ref{0[16]}) \\ \hline
\end{tabular}
\end{minipage}

\vspace{1cm}
For higher order corrections, the following were calculated.

\vspace{0.5cm}
\begin{tabular}[b]{c|c|c|l}
\hline
Mass-Dim  & Name & Eqn. No. & Comments
\\ \hline\hline
Two  & ${\cal L}^{1[2]}_1$ &  & None \\ \hline
Four  & ${\cal L}^{1[4]}_1$ & (\ref{1[4]}) & ${\cal L}^{1[4]}={\cal L}^{1[4]}_0+{\cal L}^{1[4]}_1$ See (\ref{L1[4]done}).\\ \hline
Six   & ${\cal L}^{1[6]}_1$ & (\ref{1[6]}) & \\ 
      & ${\cal L}^{1[6]}_2$ & (\ref{2[6]}) & ${\cal L}^{1[6]}={\cal L}^{1[6]}_0+{\cal L}^{1[6]}_1+{\cal L}^{1[6]}_{1\,[4]{\cal X}}$ See (\ref{L6}).\\ 
      & ${\cal L}^{1[6]}_{1\,[4]{\cal X}}$ & (\ref{L1[4]X}) & From [4] upon considering (\ref{eX2nd})\\ \hline
Eight & ${\cal L}^{1[8]}_1$ & (\ref{1[8]}) & ${\cal L}^{1[8]}={\cal L}^{1[8]}_0+{\cal L}^{1[8]}_1$ See (\ref{L8}).\\ 
      & ${\cal L}^{1[8]}_2$ & (\ref{2[8]}) & Eqn. (\ref{L8}) awaits further simplification.\\ 
      & ${\cal L}^{1[8]}_{1\,[6]{\cal X}}$ & (\ref{L16X}) & From [6] upon considering (\ref{eX2nd})\\ 
      & ${\cal L}^{1[8]}_{2\,[6]{\cal X}}$ & (\ref{L26X}) & From [6] upon considering (\ref{eX2nd})\\ 
      & ${\cal L}^{1[8]}_{1\,[4]{\cal X}^2}$ & (\ref{L14XX}) & From [4] upon considering (\ref{eX3rd})\\ \hline
Ten & ${\cal L}^{1[10]}_1$ & (\ref{1[10]}) & prescription~only\\ 
      & ${\cal L}^{1[10]}_2$ & (\ref{2[10]}) & prescription~only\\ 
      & ${\cal L}^{1[10]}_3$ & (\ref{3[10]}) & prescription~only\\ \hline
Twelve & ${\cal L}^{1[12]}_1$ & (\ref{1[12]}) & prescription~only\\ 
       & ${\cal L}^{1[12]}_2$ & (\ref{2[12]}) & prescription~only\\ 
       & ${\cal L}^{1[12]}_3$ & (\ref{3[12]}) & prescription~only\\ 
       & ${\cal L}^{1[12]}_4$ & (\ref{4[12]}) & prescription~only\\ \hline
\end{tabular}

Apart from the above contributions, the following should be included for ten
\begin{eqnarray}
{\cal L}^{1[10]}_{1\,[8]{\cal X}}+{\cal L}^{1[10]}_{2\,[8]{\cal X}}
 + {\cal L}^{1[10]}_{1\,[6]{\cal X}^2}+{\cal L}^{1[10]}_{2\,[6]{\cal X}^2}
 + {\cal L}^{1[10]}_{1\,[4]{\cal X}^3}
\end{eqnarray}
and twelve
\begin{eqnarray}
{\cal L}^{1[12]}_{1\,[10]{\cal X}}+{\cal L}^{1[12]}_{2\,[10]{\cal X}}+{\cal L}^{1[12]}_{3\,[10]{\cal X}}
 + {\cal L}^{1[12]}_{1\,[8]{\cal X}^2}+{\cal L}^{1[12]}_{2\,[8]{\cal X}^2}
 + {\cal L}^{1[12]}_{1\,[6]{\cal X}^3}+ {\cal L}^{1[12]}_{2\,[6]{\cal X}^3}
 + {\cal L}^{1[12]}_{1\,[4]{\cal X}^4}
\end{eqnarray}
mass-dimensional one-loop effective Lagrangians. They are all contributed mainly from lower mass-dimensions.

The results for ${\cal L}^{1[2]}$, ${\cal L}^{1[4]}$, and ${\cal L}^{1[6]}$ are in complete agreement with those found in literature. See Refs. \cite{Rodf-Diss,Avra,ven,Gilk,Fradk}.

\section{Closed form expressions.}

Three closed form expressions have been obtained.
\begin{enumerate}
 \item[(i)] partial momentum-space derivatives of $G_0(p)$: first- up to sixth-order
 \item[(ii)] $p^8$ integration formula in both arbitrary and flat metric
 \item[(iii)] two- and three-fold proper-time integration formula
\end{enumerate}

We have calculated the closed form formula for handling $p^8$ integral in both flat and arbitrary space-time metric. See Eqn. (\ref{pdeight}) and Appendix C.6, respectively.

We have also developed a more general way of handling two- and three-fold proper-time integrals for various form of integrand. Refer to Chapter 5 or Appendix D on how these different forms were handled. Furthermore, we have also devised a compact notation
\begin{eqnarray*}
(q_0,q_1,\ldots,q_{n-1}\mbox{;}A,B)
 \equiv
\prod^n_{\ell=0}
   \int^\infty_0 ds_\ell
   \frac{{s_0}^{q_0}\ldots {s_{n-1}}^{q_{n-1}}}{A\left(\sum^n_{r=0} s_r\right)^{B+D/2}}
\end{eqnarray*}
as introduced in (\ref{psupp}). The use of such notation are illustrated in detail in Chapter 5.

Knowing the Gaussian sector, we have expressed the non-Gaussian sector of the Green function equation by letting the sixth-order operator act on the Gaussian sector. In so doing, the general closed forms of the partial momentum-space derivatives of $G_0(p)$ have been obtained in terms of $P(s)$, $Q(s)$, and $R(s)$. See Appendix A.

\section{Algorithm}
The following summarizes the procedural programming (advantages) used in this thesis, further proving the possibility of completely automating DeWitt's background field formalism.

\begin{large}
\textbf{Collect then Matrix Multiply}
\end{large}

There are two avenues to choose from when handling (\ref{L1G})
\begin{eqnarray*}
{\cal L}^{(1)}
    &=&\frac{\hbar}{2(2\pi)^D}
    \mbox{Tr}
    \int dX\int d^Dp\,\,\,G_0(p)\sum^{\infty}_{\ell=0}
    \left(\Delta_1(p) G_0\right)^\ell
\end{eqnarray*}
particularly the $\left(\Delta_1(p) G_0\right)^\ell$ term:
\begin{enumerate}
\item[1.)] Multiply the polynomial $\Delta_1(p) G_0$ by itself $\ell=1$, up to $\ell=4$ times consecutively. Then collect terms of similar mass-dimensions. Refer to (\ref{task1})-(\ref{task2}).
\item[2.)] Collect terms of similar mass-dimensions in $\Delta_1(p) G_0$ into basis then matrix multiply them to form the much needed products (the gauge invariant monomials) for a mass-dimension of interest.
\end{enumerate}

We have tried both avenues.

For the first avenue, the following results:

\vspace{0.5cm}
\begin{minipage}{5.4in}
\begin{tabular}[b]{ccccc}
 & $\ell=1$ & $\ell=2$ & $\ell=3$ & $\ell=4$ \\ \hline\hline
No.~of~terms\footnote{
 The numbers above are yet to be mutliplied to a 4-term $G_0(p)$ (in terms of $P$, $Q$, and $R$) as given in (\ref{THETA}) (or equivalently in (\ref{tensorform})). The number of terms will increase once TSE $P$, $Q$, and $R$ are specified.} & 2,974 & 8,844,676\footnote{We have only tried up to the case when $\ell\!=\!2$ and it took about 3 days to finish on a Pentium 133 PC not to mention the time it may take to sort, identify, and collect monomials of similar mass-dimensions.} & \small{2.630406642$\times$${10}^{10}$} & \small{7.822829355$\times$${10}^{13}$}
\\ \hline
\end{tabular}
\end{minipage}
\vspace{0.5cm}

So we reverse the process and the second avenue was conceptualized and was further developed. The method of obtaining immediately monomials of specific mass-dimensions and order of corrections can be observed following the notation introduced in (\ref{ellll})
\begin{eqnarray*}
  [\ell]\equiv(\Delta_1G_{\emptyset_{\mathrm{t}}}(p))^{\mathrm{red}}_{\ell}
\end{eqnarray*}
The following tabulation depicts how manageable the second avenue is.

For $\ell=1$

\begin{minipage}{5.2in}
\begin{tabular}[b]{cccccccccc}
\hline
 & $[3]$ & $[4]$ & $[5]$ & $[6]$ & $[7]$ & $[8]$ & $[9]$ & $[10]$ & $[12]$ \\ \hline \vspace{0.1cm}
No.~of~terms\footnote{The numbers in the table represent the $\Delta_1(p) G_0$ in the unsimplified version (contraction of indices has not been performed yet.).} & 3 & 10 & 12 & 23 & 42 & 112 &  42 & 104 & 334
\\ \hline
\end{tabular}
\end{minipage}

\newpage
\vspace{1cm}
The following table further depicts the ease in forming invariants in the mass-dimension and order of correction of interest if the second avenue were used.

\vspace{0.5cm}
\begin{center}
\begin{minipage}{5.2in}
\begin{tabular}[b]{c|c|r|c}
 $\ell$ & matrix mult. & No. of terms & mass-dim. \\
\hline\hline\hline
1  &  [4] &  10 & Four\\ \hline
         1 &   [6] &  23 & Six \\ \hline
         1 &   [8] &  112 & Eight \\ \hline
         1 &   [10] &  104 & Ten \\ \hline
         1 &   [12] &  334 & Twelve \\ \hline\hline
2  &  [3][3] &  9 & Six    \\ \hline
         2 &   [3][5] &  36 & Eight \\
         2 &   [4][4] & 100 & Eight \\
         2 &   [5][3] &  36 & Eight \\ \hline
        2  &   [3][7] &  126 & Ten \\
        2  &   [4][6] &  230 & Ten \\
        2  &   [5][5] &  144 & Ten \\
        2  &   [6][4] &  230 & Ten \\
        2  &   [7][3] &  126 & Ten \\ \hline
        2  &   [3][9] &  126 & Twelve \\
        2  &   [4][8] &  1,120 & Twelve \\
        2  &   [5][7] &  504 & Twelve \\
        2  &   [6][6] &  529 & Twelve \\
        2  &   [7][5] &  504 & Twelve \\
        2  &   [8][4] &  1,120 & Twelve \\
        2  &   [9][3] &  126 & Twelve \\ \hline\hline
3  &   [3][3][4] & 90 & Ten \\
        3  &   [3][4][3] & 90 & Ten \\
        3  &   [4][3][3] & 90 & Ten \\ \hline
        3  &   [3][4][5] & 360 & Twelve \\
        3  &   [3][5][4] & 360 & Twelve \\
        3  &   [4][5][3] & 360 & Twelve \\
        3  &   [4][3][5] & 360 & Twelve \\
        3  &   [5][3][4] & 360 & Twelve \\
        3  &   [5][4][3] & 360 & Twelve \\
        3  &   [4][4][4] & 1,000 & Twelve \\ \hline\hline
4 &  [4][4][4][4] &  10,000 & Twelve \\ \hline
\end{tabular}
\end{minipage}
\end{center}

\vspace{1cm}
\begin{large}
\textbf{System of Index Replacements}
\end{large}

\begin{enumerate}
\item[] We have introduced a system of temporary index replacements from Greek to Latin when symbolic computation is to be performed and then back to Greek when results are to be presented in \LaTeXe format.
\end{enumerate}
Below is the above-mentioned system of index replacements as given in (\ref{gklat1})-(\ref{gklat2}).
\begin{eqnarray*}
\{{\mu}_{\ell},{\nu}_{\ell},{\rho}_{\ell},{\sigma}_{\ell},{\alpha}_{\ell},{\beta}_{\ell}\}&\to&\{{\mbox{a}}_{\ell},{\mbox{b}}_{\ell},{\mbox{c}}_{\ell},{\mbox{d}}_{\ell},{\mbox{e}}_{\ell},{\mbox{f}}_{\ell}\}
\\
\{{\lambda}_{\ell},{\tau}_{\ell},{\kappa}_{\ell},{\eta}_{\ell},{\xi}_{\ell},{\varrho}_{\ell}\}&\to&\{{\mbox{l}}_{\ell},{\mbox{t}}_{\ell},{\mbox{k}}_{\ell},{\mbox{h}}_{\ell},{\mbox{x}}_{\ell},{\mbox{g}}_{\ell}\}
\end{eqnarray*}

\begin{large}
\textbf{Interface of Softwares}
\end{large}

\begin{enumerate}
\item[] We have used and interfaced three symbolic handling softwares, namely: Mathematica, Macsyma, and MS Excel for pattern matching, procedural programming, and data base pivoting needs of the calculation.
\end{enumerate}

\begin{large}
\textbf{Kronecker delta Index Contraction}
\end{large}

Since the monomials are treated as text output rather their usual tensor representation, we have devised an alternative algorithm to contract indices as dictated by even powers of $p$ integration in flat space. Instead of using symbolic softwares' tensor package to contract indices, we have used Mathematica and MS Excel to handle such task. Below is an illustration how the task is done.

For example, in the contraction of indices of the monomial (one of the $2,974$ terms generated by Macsyma):
\begin{eqnarray*}
+\frac{1}{13440} \left({\cal D}_{{\eta_1}{\kappa_1}}Y_{{\lambda_1}{\tau_1}}\right)
                 \left({\cal D}_{{\eta_2}{\kappa_2}}Y_{{\lambda_2}{\tau_2}}\right)
\delta_{{\lambda_1}{\eta_1}} \delta_{{\lambda_2}{\eta_2}} \delta_{{\tau_1}{\kappa_1}} \delta_{{\tau_2}{\kappa_2}}
\end{eqnarray*}
can be rewritten into a textual output following the system of index replacements:

\begin{verbatim}
+(1/13440) D_h1k1Y_l1t1 D_h2k2Y_l2t2 d_l1h1 d_l2h2 d_t1k1 d_t2k2
\end{verbatim}

Using MS Excel and Notespad to isolate $D^2Y~D^2Y$ and kronecker delta from its indices, , we have the following (in Mathematica format)

\begin{verbatim}
{h1,k1,l1,t1,h2,k2,l2,t2}/.{l1->h1,l2->h2,t1->k1,t2->k2}
\end{verbatim}

This is after manipulating the textual output using any text editor (for our case Notespad). If we enter the above in Mathematica, we obtain the output

\begin{verbatim}
{h1,k1,h1,k1,h2,k2,h2,k2}
\end{verbatim}

After a series of manipulation, we have the resulting (index-contracted) monomial:
\begin{eqnarray*}
+\frac{1}{13440} \left({\cal D}_{{\eta_1}{\kappa_1}}Y_{{\eta_1}{\kappa_1}}\right)
                 \left({\cal D}_{{\eta_2}{\kappa_2}}Y_{{\eta_2}{\kappa_2}}\right)
\end{eqnarray*}

\section{Recommendations}
\begin{enumerate}
\item The results presented here particularly the one-loop eight mass-dimensional effective Lagrangian await further simplification. While the antisymmetry of the fieldstrength tensor and the freedom to throw away total derivatives allowed us to obtain the most we can do to obtain the present simplified form of ${\cal L}^{(1)[8]}$, the identities (\ref{id-frst})-(\ref{id-last})
will hopefully reduce the present form (\ref{L8}) into 17 linearly independent invariant monomials as they have been obtained by van de Ven\cite{ven} via the heat kernel approach.
\item The prescriptions described in Chapter 4 to obtain ${\cal L}^{(1)[10]}$ and ${\cal L}^{(1)[12]}$ await similar reduction process. If such reduced Lagrangians are to be consistent with Ref. \cite{Mue} monomial counting for a specific mass-dimensions, there should be 79 linearly independent invariant monomials for ${\cal L}^{(1)[10]}$ as it has been worked completely in Ref. \cite{ven} and 554 such invariant monomials for ${\cal L}^{(1)[12]}$ as it has been counted by Ref. \cite{Mue}. Both effective Lagrangians remain to be worked out explicitly in the spirit of background field formalism.
\item The following options are left arbitrary throughout our calculation. Fixing them can be considered an extension to our present work.
\subitem Gauge group. This would facilitate in the reduction process and express one-loop effective Lagrangians in terms of linearly independent invariant monomials. Particularly, the minimal set upon which the reduction scheme will be based would be chosen by group representation methods rather than by hand.
\subitem Mass-Dimensions (higher than 12). This means accommodating higher order corrections. Mathematically, this is equivalent to handling higher powers of momentum and higher folds of proper-time integrations.
\subitem Further relaxation on the covariant restrictions in the fieldstrength tensor ($n>4$) and in the background matrix potential ($l>4$)
\item The following options can be changed.
\subitem Euclidean spacetime metric to a conformally flat metric, or in general curved spacetime metric that may be relevant in quantum gravity and early universe theories.

Also, the assumption of strong and slowly varying background fields as indicated after Eqn. \ref{G} can be changed to weak but rapidly changing one. Calculating mass-dimensional one-loop effective Lagrangians in this line of assumption remains an open problem.
\item The most interesting extension of this thesis is the application of the results to a specific field theory.
This is done simply by determining the relevant background connection $N_\mu$ so that the following quantities can be defined:
\begin{eqnarray*}
X&\equiv& M-N_\mu N_\mu     \\
    Y_{\mu\nu}&\equiv&
     N_{\nu,\mu}- N_{\mu,\nu}
    +[N_\mu,N_\nu]
\end{eqnarray*}
as these are given in (\ref{X}), and (\ref{Y})
and the covariant derivative
\begin{eqnarray}
 {\cal D}_\mu T &\equiv & \partial_\mu T + \left[N_\mu,T\right]
\\
 {\cal D}_\mu \phi &\equiv & \partial_\mu \phi + N_\mu \phi
\end{eqnarray}
for any tensor $T$, and scalar $\phi$.
One may alternatively determine these quantities if the relevant second order operator
\begin{eqnarray}
 \Delta \equiv (\partial + N_\mu)^2 + X
\end{eqnarray}
is identified. This is the operator part of (\ref{Grnf}). Here, $X = -m^2 +{\cal X}$.

Showing only the relevant second order operator\cite{Rodf-Diss,Tiam} for the following field theories:
\subitem \textbf{Pure Yang-Mills theory}
\begin{eqnarray}
 (\Delta^{\mathrm{vector}})^{ab}_{\alpha\beta}
   &=&
  \left(\partial_\mu\delta^{ab}\delta_{\alpha\beta}
        - g f^{abc} A^c_\mu \delta_{\alpha\beta} - 2g f^{abc} F^c_{\alpha\beta}
  \right)
\\
 (\Delta^{\mathrm{ghost}})^{ab}
  &=&
 \left(\partial_\mu \delta^{ab} - g f^{abc} A^c_\mu \right)
\end{eqnarray}
These are the vector and ghost operators, where
\begin{eqnarray}
 F^a_{\mu\nu} = \partial_\mu A^a_\nu - \partial_\nu A^a_\mu + g f^{abc} A^b_\mu A^c_\nu
\end{eqnarray}
\subitem \textbf{Scalar bosons in a Yang-Mills background}
\begin{eqnarray}
 \Delta^{\mathrm{boson}}
 \equiv
 (\partial_\mu - ig T^a_s A^a_\mu)^2 +m^2_s
\end{eqnarray}
with
\begin{eqnarray}
 \mathrm{Tr}\left(T^a_sT^b_s\right) = T_s \delta^{ab}
\end{eqnarray}
to normalize the generators of the fundamental representation.
\subitem \textbf{Dirac fermions in a Yang-Mills background}
\begin{eqnarray}
 \Delta^{\mathrm{fermion}}
 \equiv
 \gamma_\mu(\partial_\mu - ig T^a_f A^a_\mu)^2 + i m_f \mathbf{1}
\end{eqnarray}
where
\begin{eqnarray}
 \left\{\gamma_\mu,\gamma_\nu\right\} &=& 2 \delta_{\mu\nu} \mathbf{1}
\\
 \sigma_{\mu\nu} &=& \frac{i}{2}\left[\gamma_\mu,\gamma_\nu,\right]
\\
\mathrm{Tr}\mathbf{1} &=& 2^{D/2}
\end{eqnarray}
with
\begin{eqnarray}
 \mathrm{Tr}\left(T^a_fT^b_f\right) = T_f \delta^{ab}
\end{eqnarray}
to normalize the generators of the fundamental representation.
Here, $\gamma_\mu$ represents a $2^{[D/2]}\times 2^{[D/2]}$ Dirac matrix normalized as above.

See Refs. \cite{Rodf-Diss,Tiam} for details on the results of this thesis can be applied to Yang-Mills theory interacting with matter.
\item All calculations described since then are all up to one-loop. One may consider calculating higher loop effective Lagrangians.
\item All covariant derivatives are of integer type. One may consider calculating with fractional derivatives.
\end{enumerate}

\newpage
\addcontentsline{toc}{chapter}{Bibliography}

\newpage
\begin{appendix}
\addcontentsline{toc}{chapter}{Appendix}
\chapter{Partial Momentum-space Derivatives of $G_0(p)$}

\section{$G_{0/\mu\nu\rho\ldots}(p)$ in terms of $P(s)$, $Q(s)$ and $R(s)$}

We have the following momentum-space derivatives of $G_0(p)$ from first up to sixth order with $\Theta$ given in (\ref{THETA}):

\begin{scriptsize}

\begin{eqnarray}
G_{0/{\mu}_{\ell}{\nu}_{\ell}{\rho}_{\ell}{\sigma}_{\ell}}(p)
&=&\int^\infty_0 ds\,\,\,e^{\Theta}
 \left[
 R_{{{\mu}_{\ell}}{{\lambda}_{\ell}}} R_{{{\nu}_{\ell}}{{\tau}_{\ell}}} R_{{{\rho}_{\ell}}{{\kappa}_{\ell}}} R_{{{\sigma}_{\ell}}{{\eta}_{\ell}}} p_{{{\lambda}_{\ell}}} p_{{{\tau}_{\ell}}} p_{{{\kappa}_{\ell}}} p_{{\eta}_{\ell}}
\right.
 + (Q_{{{\mu}_{\ell}}} R_{{{\nu}_{\ell}}{{\lambda}_{\ell}}} R_{{{\rho}_{\ell}}{{\tau}_{\ell}}} R_{{\sigma}{{\kappa}_{\ell}}}
\nonumber\\&&\!\!\!\!\!\!\!\!\!\!\!\!\!\!\!\!\!\!\!\!\!\!\!\!\!\!\!\!\!\!\!\!\!\!\!\!\!\!\!\!\!\!\!\!\!\!\!\!\!\!\!\!\!\!\!\!\!\!\!\!\!\!\!\!\!
  + Q_{{{\nu}_{\ell}}} R_{{{\mu}_{\ell}}{{\lambda}_{\ell}}} R_{{{\rho}_{\ell}}{{\tau}_{\ell}}} R_{{\sigma}{{\kappa}_{\ell}}}
  + Q_{{{\rho}_{\ell}}} R_{{{\mu}_{\ell}}{{\lambda}_{\ell}}} R_{{{\nu}_{\ell}}{{\tau}_{\ell}}} R_{{\sigma}{{\kappa}_{\ell}}}
  + Q_{{{\sigma}_{\ell}}} R_{{{\mu}_{\ell}}{{\lambda}_{\ell}}} R_{{{\nu}_{\ell}}{{\tau}_{\ell}}} R_{{{\rho}_{\ell}}{{\kappa}_{\ell}}}) p_{{{\lambda}_{\ell}}} p_{{{\tau}_{\ell}}} p_{{{\kappa}_{\ell}}}
\nonumber\\&&\!\!\!\!\!\!\!\!\!\!\!\!\!\!\!\!\!\!\!\!\!\!\!\!\!\!\!\!\!\!\!\!\!\!\!\!\!\!\!\!\!\!\!\!\!\!\!\!\!\!\!\!\!\!\!\!\!\!\!\!\!\!\!\!\!
 + (R_{{{\mu}_{\ell}}{{\nu}_{\ell}}} R_{{{\rho}_{\ell}}{{\lambda}_{\ell}}} R_{{{\sigma}_{\ell}}{{\tau}_{\ell}}}
    + R_{{{\mu}_{\ell}}{{\rho}_{\ell}}} R_{{{\nu}_{\ell}}{{\lambda}_{\ell}}} R_{{{\sigma}_{\ell}}{{\tau}_{\ell}}}
    + R_{{{\mu}_{\ell}}{{\lambda}_{\ell}}} R_{{{\nu}_{\ell}}{{\rho}_{\ell}}} R_{{{\sigma}_{\ell}}{{\tau}_{\ell}}}
    + R_{{{\mu}_{\ell}}{{\lambda}_{\ell}}} R_{{{\nu}_{\ell}}{{\sigma}_{\ell}}} R_{{{\rho}_{\ell}}{{\tau}_{\ell}}}
\nonumber\\&&\!\!\!\!\!\!\!\!\!\!\!\!\!\!\!\!\!\!\!\!\!\!\!\!\!\!\!\!\!\!\!\!\!\!\!\!\!\!\!\!\!\!\!\!\!\!\!\!\!\!\!\!\!\!\!\!\!\!\!\!\!\!\!\!\!
    + R_{{{\mu}_{\ell}}}d R_{{{\nu}_{\ell}}{{\tau}_{\ell}}} R_{{{\rho}_{\ell}}{{\lambda}_{\ell}}}
    + R_{{{\mu}_{\ell}}{{\lambda}_{\ell}}} R_{{{\nu}_{\ell}}{{\tau}_{\ell}}} R_{{{\rho}_{\ell}}{{\sigma}_{\ell}}}
    + Q_{{{\mu}_{\ell}}} Q_{{{\rho}_{\ell}}} R_{{{\nu}_{\ell}}{{\lambda}_{\ell}}} R_{{{\sigma}_{\ell}}{{\tau}_{\ell}}}
    + Q_{{{\mu}_{\ell}}} Q_{{{\nu}_{\ell}}} R_{{{\rho}_{\ell}}{{\lambda}_{\ell}}} R_{{{\sigma}_{\ell}}{{\tau}_{\ell}}}
\nonumber\\&&\!\!\!\!\!\!\!\!\!\!\!\!\!\!\!\!\!\!\!\!\!\!\!\!\!\!\!\!\!\!\!\!\!\!\!\!\!\!\!\!\!\!\!\!\!\!\!\!\!\!\!\!\!\!\!\!\!\!\!\!\!\!\!\!\!
    + Q_{{{\nu}_{\ell}}} Q_{{{\rho}_{\ell}}} R_{{{\mu}_{\ell}}{{\lambda}_{\ell}}} R_{{{\sigma}_{\ell}}{{\tau}_{\ell}}}
    + Q_{{{\mu}_{\ell}}} Q_{{{\sigma}_{\ell}}} R_{{{\nu}_{\ell}}{{\lambda}_{\ell}}} R_{{{\rho}_{\ell}}{{\tau}_{\ell}}}
    + Q_{{{\nu}_{\ell}}} Q_{{{\sigma}_{\ell}}} R_{{{\mu}_{\ell}}{{\lambda}_{\ell}}} R_{{{\rho}_{\ell}}{{\tau}_{\ell}}}
    + Q_{{{\rho}_{\ell}}} Q_{{{\sigma}_{\ell}}} R_{{{\mu}_{\ell}}{{\lambda}_{\ell}}} R_{{{\nu}_{\ell}}{{\tau}_{\ell}}}) p_{{{\lambda}_{\ell}}} p_{{{\tau}_{\ell}}}
\nonumber\\&&\!\!\!\!\!\!\!\!\!\!\!\!\!\!\!\!\!\!\!\!\!\!\!\!\!\!\!\!\!\!\!\!\!\!\!\!\!\!\!\!\!\!\!\!\!\!\!\!\!\!\!\!\!\!\!\!\!\!\!\!\!\!\!\!\!
 + ( Q_{{{\mu}_{\ell}}} R_{{{\nu}_{\ell}}{{\rho}_{\ell}}} R_{{{\sigma}_{\ell}}{{\lambda}_{\ell}}}
    + Q_{{{\nu}_{\ell}}} R_{{{\mu}_{\ell}}{{\rho}_{\ell}}} R_{{{\sigma}_{\ell}}{{\lambda}_{\ell}}}
    + Q_{{{\rho}_{\ell}}} R_{{{\mu}_{\ell}}{{\nu}_{\ell}}} R_{{{\sigma}_{\ell}}{{\lambda}_{\ell}}}
    + Q_{{{\sigma}_{\ell}}} R_{{{\mu}_{\ell}}{{\lambda}_{\ell}}} R_{{{\nu}_{\ell}}{{\rho}_{\ell}}}
\nonumber\\&&\!\!\!\!\!\!\!\!\!\!\!\!\!\!\!\!\!\!\!\!\!\!\!\!\!\!\!\!\!\!\!\!\!\!\!\!\!\!\!\!\!\!\!\!\!\!\!\!\!\!\!\!\!\!\!\!\!\!\!\!\!\!\!\!\!
    + Q_{{{\sigma}_{\ell}}} R_{{{\mu}_{\ell}}{{\nu}_{\ell}}} R_{{{\rho}_{\ell}}{{\lambda}_{\ell}}}
    + Q_{{{\sigma}_{\ell}}} R_{{{\mu}_{\ell}}{{\rho}_{\ell}}} R_{{{\nu}_{\ell}}{{\lambda}_{\ell}}}
    + Q_{{{\mu}_{\ell}}} R_{{{\nu}_{\ell}}{{\sigma}_{\ell}}} R_{{{\rho}_{\ell}}{{\lambda}_{\ell}}}
    + Q_{{{\nu}_{\ell}}} R_{{{\mu}_{\ell}}{{\sigma}_{\ell}}} R_{{{\rho}_{\ell}}{{\lambda}_{\ell}}}
\nonumber\\&&\!\!\!\!\!\!\!\!\!\!\!\!\!\!\!\!\!\!\!\!\!\!\!\!\!\!\!\!\!\!\!\!\!\!\!\!\!\!\!\!\!\!\!\!\!\!\!\!\!\!\!\!\!\!\!\!\!\!\!\!\!\!\!\!\!
    + Q_{{{\mu}_{\ell}}} R_{{{\nu}_{\ell}}{{\lambda}_{\ell}}} R_{{{\rho}_{\ell}}{{\sigma}_{\ell}}}
    + Q_{{{\nu}_{\ell}}} R_{{{\mu}_{\ell}}{{\lambda}_{\ell}}} R_{{{\rho}_{\ell}}{{\sigma}_{\ell}}}
    + Q_{{{\rho}_{\ell}}} R_{{{\mu}_{\ell}}{{\sigma}_{\ell}}}R_{{{\nu}_{\ell}}{{\lambda}_{\ell}}}
    + Q_{{{\rho}_{\ell}}} R_{{{\mu}_{\ell}}{{\lambda}_{\ell}}} R_{{{\nu}_{\ell}}{{\sigma}_{\ell}}}
\nonumber\\&&\!\!\!\!\!\!\!\!\!\!\!\!\!\!\!\!\!\!\!\!\!\!\!\!\!\!\!\!\!\!\!\!\!\!\!\!\!\!\!\!\!\!\!\!\!\!\!\!\!\!\!\!\!\!\!\!\!\!\!\!\!\!\!\!\!
    + Q_{{{\mu}_{\ell}}} Q_{{{\nu}_{\ell}}} Q_{{{\rho}_{\ell}}} R_{{{\sigma}_{\ell}}{{\lambda}_{\ell}}}
    + Q_{{{\mu}_{\ell}}} Q_{{{\nu}_{\ell}}} Q_{{{\sigma}_{\ell}}} R_{{{\rho}_{\ell}}{{\lambda}_{\ell}}}
    + Q_{{{\mu}_{\ell}}} Q_{{{\rho}_{\ell}}} Q_{{{\sigma}_{\ell}}} R_{{{\nu}_{\ell}}{{\lambda}_{\ell}}}
    + Q_{{{\nu}_{\ell}}} Q_{{{\rho}_{\ell}}} Q_{{{\sigma}_{\ell}}} R_{{{\mu}_{\ell}}{{\lambda}_{\ell}}}) p_{{{\lambda}_{\ell}}}
\nonumber\\&&\!\!\!\!\!\!\!\!\!\!\!\!\!\!\!\!\!\!\!\!\!\!\!\!\!\!\!\!\!\!\!\!\!\!\!\!\!\!\!\!\!\!\!\!\!\!\!\!\!\!\!\!\!\!\!\!\!\!\!\!\!\!\!\!\!
 + R_{{{\mu}_{\ell}}{{\nu}_{\ell}}} R_{{{\rho}_{\ell}}{{\sigma}_{\ell}}}
 + R_{{{\mu}_{\ell}}{{\rho}_{\ell}}} R_{{{\nu}_{\ell}}{{\sigma}_{\ell}}}
 + R_{{{\mu}_{\ell}}{{\sigma}_{\ell}}} R_{{{\nu}_{\ell}}{{\rho}_{\ell}}}
 + Q_{{{\mu}_{\ell}}} Q_{{{\nu}_{\ell}}} R_{{{\rho}_{\ell}}{{\sigma}_{\ell}}}
 + Q_{{{\mu}_{\ell}}} Q_{{{\rho}_{\ell}}} R_{{{\nu}_{\ell}}{{\sigma}_{\ell}}}
 + Q_{{{\mu}_{\ell}}} Q_{{{\sigma}_{\ell}}} R_{{{\nu}_{\ell}}{{\rho}_{\ell}}}
\nonumber\\&&\!\!\!\!\!\!\!\!\!\!\!\!\!\!\!\!\!\!\!\!\!\!\!\!\!\!\!\!\!\!\!\!\!\!\!\!\!\!\!\!\!\!\!\!\!\!\!\!\!\!\!\!\!\!\!\!\!\!\!\!\!\!\!\!\!
 + Q_{{{\nu}_{\ell}}} Q_{{{\rho}_{\ell}}} R_{{{\mu}_{\ell}}{{\sigma}_{\ell}}}
 + Q_{{{\nu}_{\ell}}} Q_{{{\sigma}_{\ell}}} R_{{{\mu}_{\ell}}{{\rho}_{\ell}}}
\left.
 + Q_{{{\rho}_{\ell}}} Q_{{{\sigma}_{\ell}}} R_{{{\mu}_{\ell}}{{\nu}_{\ell}}}
 + Q_{{{\mu}_{\ell}}} Q_{{{\nu}_{\ell}}} Q_{{{\rho}_{\ell}}} Q_{{{\sigma}_{\ell}}}
 \right]\label{Gpabcd}
\end{eqnarray}



\end{scriptsize}

\section{List of TSE $Q(s)$ and $R(s)$ in Maxima Code}

The following are the complete list of $R$'s and $Q$'s needed to solve for (\ref{G0ps}) or in terms of $R$ in Eqns. (\ref{Gpa})-(\ref{Gpabc}), and (\ref{Gpabcd})-(\ref{Gpabcdef})

\begin{scriptsize}
\begin{verbatim}
R_ab:-2*s*d_ab+Y2_ab*s*s*s*2/3; R_ac:-2*s*d_ac+Y2_ac*s*s*s*2/3; R_ad:-2*s*d_ad+Y2_ad*s*s*s*2/3;
R_ae:-2*s*d_ae+Y2_ae*s*s*s*2/3; R_af:-2*s*d_af+Y2_af*s*s*s*2/3; R_al:-2*s*d_al+Y2_al*s*s*s*2/3;
/**/
R_bc:-2*s*d_bc+Y2_bc*s*s*s*2/3; R_bd:-2*s*d_bd+Y2_bd*s*s*s*2/3; R_bl:-2*s*d_bl+Y2_bl*s*s*s*2/3;
R_bt:-2*s*d_bt+Y2_bt*s*s*s*2/3; R_be:-2*s*d_be+Y2_be*s*s*s*2/3;
R_bf:-2*s*d_bf+Y2_bf*s*s*s*2/3;
/**/
R_cd:-2*s*d_cd+Y2_cd*s*s*s*2/3; R_ck:-2*s*d_ck+Y2_ck*s*s*s*2/3; R_ce:-2*s*d_ce+Y2_ce*s*s*s*2/3;
R_cf:-2*s*d_cf+Y2_cf*s*s*s*2/3; R_cl:-2*s*d_cl+Y2_cl*s*s*s*2/3; R_ct:-2*s*d_ct+Y2_ct*s*s*s*2/3;
/**/
R_de:-2*s*d_de+Y2_de*s*s*s*2/3; R_df:-2*s*d_df+Y2_df*s*s*s*2/3; R_dh:-2*s*d_dh+Y2_dh*s*s*s*2/3;
R_dk:-2*s*d_dk+Y2_dk*s*s*s*2/3; R_dl:-2*s*d_dl+Y2_dl*s*s*s*2/3; R_dt:-2*s*d_dt+Y2_dt*s*s*s*2/3;
/**/
R_ef:-2*s*d_ef+Y2_ef*s*s*s*2/3; R_ek:-2*s*d_ek+Y2_ek*s*s*s*2/3; R_el:-2*s*d_el+Y2_el*s*s*s*2/3;
R_et:-2*s*d_et+Y2_et*s*s*s*2/3; R_eh:-2*s*d_eh+Y2_eh*s*s*s*2/3; R_ex:-2*s*d_ex+Y2_ex*s*s*s*2/3;
/**/
R_fg:-2*s*d_fg+Y2_fg*s*s*s*2/3; R_fl:-2*s*d_fl+Y2_fl*s*s*s*2/3; R_fk:-2*s*d_fk+Y2_fk*s*s*s*2/3;
R_fh:-2*s*d_fh+Y2_fh*s*s*s*2/3; R_ft:-2*s*d_ft+Y2_ft*s*s*s*2/3;
/**/
Q_a: +(%i)*s*D_aX; Q_b: +(%i)*s*D_bX; Q_c: +(%i)*s*D_cX; Q_d: +(%i)*s*D_dX; Q_e: +(%i)*s*D_eX;
Q_f: +(%i)*s*D_fX;
\end{verbatim}
\end{scriptsize}

\section{$G_{\emptyset/\mu\nu\rho\ldots}$ in terms of TSE $Q(s)$ and $R(s)$ in Maxima code}

These are then substituted into (\ref{Gpa})-(\ref{Gpabc}),and (\ref{Gpabcd})-(\ref{Gpabcdef}) in Maxima code:
\begin{scriptsize}
\begin{verbatim}
Gb:+R_bl*p;
Ga:+R_al*p;
Gbc:p* p* R_bl* R_ct + R_bc;
Gab:p* p* R_al* R_bt + R_ab;
Gbcd: +  R_bl* R_ct* R_dk*   p* p* p
 + (R_bc* R_dl +R_bd* R_cl + R_bl* R_cd)*   p;
Gabc: +  R_al* R_bt* R_ck*   p* p* p
 + (R_ab* R_cl +R_ac* R_bl + R_al* R_bc)*   p;
Gabcd: + R_al* R_bt* R_ck* R_dh* p* p* p* p
 + (+ R_ab* R_cl* R_dt + R_ac* R_bl* R_dt + R_al* R_bc* R_dt + R_al* R_bd* R_ct
    + R_ad* R_bt* R_cl + R_al* R_bt* R_cd)* p* p
 + R_ab* R_cd + R_ac* R_bd + R_ad* R_bc;
Gabcde: +  R_al* R_bt* R_ck* R_dh* R_ex* p* p* p* p* p
 + (R_ab* R_cl* R_dt* R_ek + R_ac* R_bl* R_dt* R_ek + R_al* R_bc* R_dt* R_ek
  + R_al* R_bd* R_ct* R_ek + R_ad* R_bt* R_cl* R_ek + R_al* R_bt* R_cd* R_ek
  + R_al* R_be* R_ck* R_dt + R_ae* R_bt* R_ck* R_dl + R_al* R_bt* R_ce* R_dk
  + R_al* R_bt* R_ck* R_de)* p* p* p
 + (+ R_ab* R_cd* R_el + R_ac* R_bd* R_el + R_ad* R_bc* R_el + R_ab* R_ce* R_dl
    + R_ac* R_be* R_dl + R_ae* R_bc* R_dl + R_ab* R_cl* R_de + R_ac* R_bl* R_de
    + R_al* R_bc* R_de + R_ad* R_be* R_cl + R_ae* R_bd* R_cl + R_ad* R_bl* R_ce
    + R_al* R_bd* R_ce + R_ae* R_bl* R_cd + R_al* R_be* R_cd)* p;
Gabcdef: + R_al* R_bt* R_ck* R_dh* R_ex* R_fg* p* p* p* p* p* p
 + (+ R_ab* R_cl* R_dt* R_ek* R_fh + R_ac* R_bl* R_dt* R_ek* R_fh + R_al* R_bc* R_dt* R_ek* R_fh
    + R_al* R_bd* R_ct* R_ek* R_fh + R_ad* R_bt* R_cl* R_ek* R_fh + R_al* R_bt* R_cd* R_ek* R_fh
    + R_al* R_be* R_ck* R_dt* R_fh + R_ae* R_bt* R_ck* R_dl* R_fh + R_al* R_bt* R_ce* R_dk* R_fh
    + R_al* R_bt* R_ck* R_de* R_fh + R_al* R_bf* R_ck* R_dh* R_et + R_af* R_bt* R_ck* R_dh* R_el
    + R_al* R_bt* R_cf* R_dk* R_eh
                      + R_al* R_bt* R_ck* R_df* R_eh + R_al* R_bt* R_ck* R_dh* R_ef)* p* p* p* p
 + (R_ab* R_cd* R_el* R_ft + R_ac* R_bd* R_el* R_ft + R_ad* R_bc* R_el* R_ft
  + R_ab* R_ce* R_dl* R_ft + R_ae* R_bc* R_dl* R_ft + R_ab* R_cl* R_de* R_ft
  + R_ac* R_be* R_dl* R_ft + R_ac* R_bl* R_de* R_ft + R_al* R_bc* R_de* R_ft
  + R_ad* R_be* R_cl* R_ft + R_ae* R_bd* R_cl* R_ft + R_ad* R_bl* R_ce* R_ft
  + R_al* R_bd* R_ce* R_ft + R_ae* R_bl* R_cd* R_ft + R_al* R_be* R_cd* R_ft
  + R_al* R_bc* R_df* R_et + R_ab* R_cl* R_df* R_et + R_ac* R_bl* R_df* R_et
  + R_ad* R_bf* R_cl* R_et + R_al* R_bd* R_cf* R_et + R_al* R_bf* R_cd* R_et
  + R_ab* R_cf* R_dt* R_el + R_af* R_be* R_cl* R_dt + R_al* R_be* R_cf* R_dt
  + R_al* R_bf* R_ce* R_dt + R_ae* R_bf* R_ct* R_dl + R_ae* R_bt* R_cf* R_dl
  + R_af* R_bt* R_ce* R_dl + R_al* R_be* R_ct* R_df + R_ae* R_bt* R_cl* R_df
  + R_al* R_bt* R_ce* R_df + R_al* R_bf* R_ct* R_de + R_af* R_bt* R_cl* R_de
  + R_al* R_bt* R_cf* R_de + R_ac* R_bf* R_dt* R_el + R_af* R_bc* R_dt* R_el
  + R_af* R_bd* R_ct* R_el + R_ad* R_bt* R_cf* R_el + R_af* R_bt* R_cd* R_el
  + R_ab* R_cl* R_dt* R_ef + R_ac* R_bl* R_dt* R_ef + R_al* R_bc* R_dt* R_ef
  + R_al* R_bd* R_ct* R_ef + R_ad* R_bt* R_cl* R_ef + R_al* R_bt* R_cd* R_ef)* p* p
 + R_ae* R_bf* R_cd + R_ad* R_bf* R_ce + R_af* R_bd* R_ce + R_af* R_be* R_cd
 + R_ab* R_cd* R_ef + R_ac* R_bd* R_ef + R_ad* R_bc* R_ef + R_ab* R_ce* R_df
 + R_ac* R_be* R_df + R_ae* R_bc* R_df + R_ab* R_cf* R_de + R_ac* R_bf* R_de
 + R_af* R_bc* R_de + R_ad* R_be* R_cf + R_ae* R_bd* R_cf;
\end{verbatim}
\end{scriptsize}

\section{Sixth-order Green function Equation in Maxima Code}

These are substituted into the sixth-order Green function equation:
\begin{scriptsize}
\begin{verbatim}
del:
-i*D_bY_ab*Ga +(2*i/3)*D_cY_ab*p*Gbc -((1/2)*D_abX+(3/8)*(D_beY_ae-D_ebY_ea))*Gab
+(1/4)*D_dcY_ab*p*Gbcd +((i/6)*D_cbaX+(i/6)*(Y_da*D_cY_db+D_bY_da*Y_dc))*Gabc
+((1/9)*D_bY_fa*D_dY_fc+(1/8)*Y_fa*D_dcY_fb)*Gabcd -((i/12)*D_bY_fa*D_edY_fc)*Gabcde
-((1/64)*D_cbY_ja*D_feY_jd)*Gabcdef;
expand(del);
\end{verbatim}
\end{scriptsize}

\section{Textual Output Simplification Algorithm}

To rewrite the exponents for easy text output handling, we implement the following substitutions:

\begin{scriptsize}
\begin{verbatim}
subst(0,s*s*s*s*s*s*s*s*s*s*s*s*s*s*s*s*s*s,%); subst(0,s*s*s*s*s*s*s*s*s*s*s*s*s*s*s*s*s,%);
subst(0,s*s*s*s*s*s*s*s*s*s*s*s*s*s*s*s,%); subst(0,s*s*s*s*s*s*s*s*s*s*s*s*s*s*s,%);
subst(0,s*s*s*s*s*s*s*s*s*s*s*s*s*s,%); subst(0,s*s*s*s*s*s*s*s*s*s*s*s*s,%);
subst(0,s*s*s*s*s*s*s*s*s*s*s*s,%); subst(0,s*s*s*s*s*s*s*s*s*s*s,%); subst(0,s*s*s*s*s*s*s*s*s*s,%);
subst(0,s*s*s*s*s*s*s*s*s,%); subst(0,s*s*s*s*s*s*s*s,%); subst(s7,s*s*s*s*s*s*s,%);
subst(s6,s*s*s*s*s*s,%); subst(s5,s*s*s*s*s,%); subst(s4,s*s*s*s,%); subst(s3,s*s*s,%);
subst(s2,s*s,%); subst(s1,s,%);
subst(p18,p*p*p*p*p*p*p*p*p*p*p*p*p*p*p*p*p*p,%); subst(p17,p*p*p*p*p*p*p*p*p*p*p*p*p*p*p*p*p,%);
subst(p16,p*p*p*p*p*p*p*p*p*p*p*p*p*p*p*p,%); subst(p15,p*p*p*p*p*p*p*p*p*p*p*p*p*p*p,%);
subst(p14,p*p*p*p*p*p*p*p*p*p*p*p*p*p,%); subst(p13,p*p*p*p*p*p*p*p*p*p*p*p*p,%);
subst(p12,p*p*p*p*p*p*p*p*p*p*p*p,%); subst(p11,p*p*p*p*p*p*p*p*p*p*p,%);
subst(p10,p*p*p*p*p*p*p*p*p*p,%); subst(p9,p*p*p*p*p*p*p*p*p,%); subst(p8,p*p*p*p*p*p*p*p,%);
subst(p7,p*p*p*p*p*p*p,%); subst(p6,p*p*p*p*p*p,%); subst(p5,p*p*p*p*p,%); subst(p4,p*p*p*p,%);
subst(p3,p*p*p,%); subst(p2,p*p,%); subst(p1,p,%);
\end{verbatim}
\end{scriptsize}

Note that one can freely replace the variables and replace with an $\ell$-subscripted variable by introducing a variable z.

\section{$G_{0/\mu\nu\rho\ldots}$ in terms of $P$, $Q$, $R$}

\begin{scriptsize}

\end{scriptsize}

\chapter{Maxima Code of Partial Derivatives of $G_0(p)$}

\section{System of Index Replacements
}
In this chapter, we temporarily make the following replacements in the indices to facilitate us in implementing our algorithm with a Maxima\footnote{This is formerly called Macsyma.} symbolic software.
\[
\{{\mu}_{\ell},{\nu}_{\ell},{\rho}_{\ell},{\sigma}_{\ell},{\alpha}_{\ell},{\beta}_{\ell}\}\to\{{\mathtt{az}},{\mathtt{bz}},{\mathtt{cz}},{\mathtt{dz}},{\mathtt{ez}},{\mathtt{fz}}\}
\]

\[
\{{\lambda}_{\ell},{\tau}_{\ell},{\kappa}_{\ell},{\eta}_{\ell},{\xi}_{\ell},{\varrho}_{\ell}\}\to\{{\mathtt{lz}},{\mathtt{tz}},{\mathtt{kz}},{\mathtt{hz}},{\mathtt{xz}},{\mathtt{gz}}\}
\]

and
\[
 s_\ell \to sz.
\]

In the following illustration, we depict how the ensuing integrations (with respect to the matrix potential $X$ and momentum $p$ integrations found in (\ref{L1G}) and proper-time $s$ integration found in (\ref{LG}) will be compressed as we formulate instructions for Maxima to be implemented in a Linux enviroment.

We identify the following partial momentum-space differentiations expressed in Maxima environment:
\begin{eqnarray}
    \mathrm{Ga}
    &\equiv &\frac{\partial G_0(p)}{\partial p_{\mu}}
    =\int^\infty_0\,\,ds
    (-R_{{\mu}{\lambda}}p_{\lambda})e^{-\Theta}
    \nonumber\\ \label{gga}
    &\equiv & \mathrm{-Q\_a-R\_al*p\_l;}
\end{eqnarray}
where
\begin{eqnarray}
 \Theta=-Xs+\frac{1}{2}\mbox{tr}Y^2s^2-\frac{1}{2}p\cdot \left(-2s+\frac{2}{3}Y^2s^3\right)\cdot p
\end{eqnarray}
\begin{eqnarray}
    \mathrm{Gb}
    &\equiv &\frac{\partial G_0(p)}{\partial p_{\nu}}
    =\int^\infty_0\,\,ds
    (-R_{{\nu}{\tau}}p_{\tau})e^{-\Theta}
    \nonumber\\
    &\equiv & \mathrm{-R\_bt*p\_t;}.
\end{eqnarray}

Differentiating (\ref{gga}) with respect to $p_{\nu}$
\begin{eqnarray}
    \mathrm{Gab}
    \!\!\!\!\!&\equiv &\!\!\!\!\!\frac{\partial^2 G_0(p)}{\partial p_{\mu}\partial p_{\nu}}
    =\int^\infty_0\,\,ds
    (-R_{{\mu}{\nu}}
        +R_{{\mu}{\tau}}R_{{\nu}{\lambda}}p_{\tau}p_{\lambda})e^{-\Theta}
    \nonumber\\ \label{Ga}
    \!\!\!\!\!&\equiv & \!\!\!\!\!
    \mathrm{- R\_ab
        + p\_l* p\_t* R\_al* R\_bt;}
\end{eqnarray}

For convenience, $\mathrm{p\_l* p\_t}$  were compressed further into $\mathrm{p* p}$. Its indices were easily rebooked by looking at the subindices ($\mathrm{l}$ and $\mathrm{l}$) of $\mathrm{R\_al* R\_bt}$.

\section{Maxima Instruction for $G_{0/\mu\nu\rho\ldots}$}

We present below the equivalent expressions of (\ref{Gpa})-(\ref{Gpabc}), and (\ref{Gpabcd})-(\ref{Gpabcdef}) in Maxima environment. The integration
\begin{eqnarray*}
  G_0(p) = \int_0^\infty ds\,\,\, e^{\Theta}
\end{eqnarray*}
with respect to the proper-time variable $s$ is suppressed in the transcription.

\begin{scriptsize}

\end{scriptsize}

\section{Mass-Dimensional Basis in Maxima code}

\begin{scriptsize}
\begin{verbatim}
thr: +(2*i/3)*D_aY_la*p1*s1+(8*i/3)*D_kY_lt*p3*s2;

for:
 + s*D_aaX
 + p2*s2*(- 2*D_ltX + D_taY_la + D_atY_la + D_aaY_lt + (3/2)*(D_atY_al - D_taY_la))
 - 2*p4*s3*D_hkY_lt;

fiv:
+(2*i/3)*p1*s2* (D_laaX + D_alaX + D_aalX + 2*D_bY_ab*Y_al + 2*(D_lY_ab + D_bY_al)*Y_ab)
-(4*i/3)*p3*s3* (D_ktlX + D_tY_al*Y_ak + D_kY_at*Y_al);

six:
+(1/2)*s2*Y_ab*(D_ccY_ab + D_cbY_ac + D_bcY_ac)
+(4/9)*s2*(D_bY_ab*D_cY_ac + D_cY_ab*D_cY_ab + D_cY_ab*D_bY_ac)
+(4/3)*p2*s3*D_aX*(D_tY_la+D_aY_lt)+D_aX*D_bY_ab
- p2*s3*((D_tlY_ab + D_tbY_al + D_blY_at)*Y_ab + (D_tbY_ab + D_btY_ab + D_bbY_at)*Y_al)
-(8/9)*p2*s3*(2*(D_bY_ab*D_tY_al + D_bY_al*D_tY_ab) + D_lY_ab*D_tY_ab  + D_bY_al*D_bY_at)
+ 2*p4*s4*D_hkY_at*Y_al +(16/9)*p4*s4*D_tY_al*D_hY_ak;

svn:
+(2*i/3)*p1*s3*(D_bY_ab*(D_lcY_ac + D_clY_ac + D_ccY_al)
             + D_cY_ab*(D_lcY_ab + D_clY_ab + D_lbY_ac + D_cbY_al + D_blY_ac + D_bcY_al)
             + D_lY_ab*(D_ccY_ab + D_cbY_ac + D_bcY_ac)
             + D_bY_al*(D_ccY_ab + D_cbY_ac + D_bcY_ac)
             - D_bY_ab*Y2_al + (2/3)*D_cY_ab*Y2_bc)
+i*p1*s3*D_aX*(D_alX+D_laX)
-(i/2)*p1*s3*D_aX*(D_bbY_la+D_abY_lb+D_baY_lb)
+(3*i/4)*p1*s3*D_aX*((D_lbY_ab+D_abY_lb)-(D_blY_ba+D_baY_bl))
-(4*i/3)*p3*s4*
  (D_bY_ab*D_ktY_al + D_lY_ab*D_ktY_ab + D_bY_al*(D_ktY_ab + D_kbY_at + D_btY_ak)
   + D_tY_ab*(D_kbY_al + D_blY_ak) + D_tY_al*(D_kbY_ab + D_bkY_ab + D_bbY_ak))
-(8*i/9)*p3*s4*(D_bY_al*Y2_bt + D_tY_ab*Y2_bl)
+i*p3*s4*D_aX*(D_ktY_la+D_aX*D_kaY_lt+D_akY_lt)
+(8*i/3)*p5*s5*D_tY_al*D_xhY_ak;


eit:
-(1/3)*s3*D_abX*Y2_ab
+(1/4)*s3*Y2_ab*(D_cbY_ca - D_bcY_ac)
+(1/8)*s3*(D_cbY_ab*(D_ddY_ac + D_dcY_ad + D_cdY_ad)
        + D_bcY_ab*(D_ddY_ac + D_dcY_ad + D_cdY_ad)
        + D_ccY_ab*(D_ddY_ab + D_dbY_ad + D_bdY_ad)
        + D_dcY_ab*((D_dcY_ab + D_cdY_ab)
          + (D_dbY_ac + D_bdY_ac) + (D_cbY_ad + D_bcY_ad)))
+(1/3)*s3*D_aX*(D_bbaX+D_abbX+D_babX)
+(1/3)*s3*D_aX*(2*D_cY_bc*Y_ba+2*(D_cY_ba+D_aY_bc)*Y_bc)
+(2/3)*p2*s4*(D_laX*Y2_at+D_atX*Y2_al)
+(1/2)*p2*s4 *((D_baY_la - D_abY_al)*Y2_bt + (D_tbY_ab - D_btY_ba)*Y2_al)
-(1/3)*p2*s4*((D_baY_la + D_abY_la + D_aaY_lb)*Y2_bt
              + (D_baY_lt + D_btY_la + D_tbY_la)*Y2_ab)
-(1/4)*p2*s4*
  (+(D_alY_bt + D_tlY_ba)*D_ccY_ba
   +(D_ccY_ba + D_caY_bc)*D_taY_bl
   +(D_lcY_ba + D_clY_ba)*D_taY_bc
   +(D_ctY_ba + D_ctY_ba)*D_caY_bl
   +(D_lcY_ba + D_caY_bl)*D_acY_bt
   +(D_alY_bc + D_acY_bl)*D_ctY_ba
   +(D_tcY_bl + D_ctY_bl)*D_aaY_bc
   +(D_caY_ba + D_acY_ba + D_caY_ba)*D_tcY_bl
   +(D_ccY_ba + D_caY_bc + D_acY_bc)*D_ltY_ba
   +(D_tcY_bc + D_ctY_bc + D_ccY_bt)*D_aaY_bl
   +(D_caY_bl + D_alY_bc + D_caY_bl)*D_tcY_ba
   +(D_taY_bc + D_caY_bt + D_atY_bc)*D_caY_bl
   +(D_lcY_ba + D_clY_ba)*(D_tcY_ba + D_ctY_ba)
   +(D_caY_ba + D_acY_ba)*(D_tlY_bc + D_ctY_bl + D_clY_bt)
   +(D_laY_ba + D_alY_ba)*(D_tcY_bc + D_ctY_bc + D_ccY_bt))
-(2/3)*p2*s4*D_aX*((D_tlaX+D_talX+D_atlX)+2*D_tY_bl*Y_ba
   +(D_lY_ba+D_aY_bl)*Y_bt+(D_tY_ba+D_aY_bt)*Y_bl)
+(2/3)*p4*s5*
   (D_akY_lt*Y2_ah + D_haY_lt*Y2_ak + D_hkY_la*Y2_at)
+(1/2)*p4*s5*
  ((D_haY_bl + D_aeY_bh)*D_ktY_ba
 + (D_haY_bt + D_ftY_ah)*D_kaY_bl
 + (D_haY_bk + D_ahY_bk)*D_atY_bl
 + (D_taY_bl + D_atY_bl + D_ltY_ba)*D_hkY_ba
 + (D_laY_ba + D_alY_ba + D_aaY_bl)*D_hkY_bt
 + (D_aaY_bh + D_ahY_ba + D_haY_ba)*D_ktY_bl)
- p6*s6*D_ktY_al*D_gxY_ah;

nin:
-(2*i/9)*p1*s4*((D_abbX + D_babX + D_bbaX + 2*D_cY_bc*Y_ba
                + (2*D_aY_bc + 2*D_cY_ba)*Y_bc)*Y2_al
              + (D_balX + D_blaX + D_lbaX + (D_bY_cl + D_lY_cb)*Y_ca
                + (D_aY_cl + D_lY_ca)*Y_cb + 2*D_bY_ca*Y_cl)*Y2_ab)
-(2*i/3)*p1*s4*D_aX*D_bX*D_bY_la
+(i/2)*p1*s4*D_aX*((D_ccY_bl+D_clY_bc+D_lcY_bc)*Y_ba
	+((D_laY_bc+D_alY_bc)+(D_lcY_ba+D_clY_ba)+(D_caY_bl+D_acY_bl))*Y_bc
	+(D_ccY_ba+D_caY_bc+D_acY_bc)*Y_bl)
+(4*i/9)*p1*s4*D_aX*(2*(D_lY_ba+D_aY_bl)*D_cY_bc
	+2*(D_cY_ba+D_aY_bc)*D_cY_bl+2*(D_cY_ba+D_aY_bc)*D_lY_bc)
+(4*i/9)*p3*s5*((D_atlX + D_tY_bl*Y_ba + D_aY_bt*Y_bl)*Y2_ak
               + (D_kalX + D_aY_bl*Y_bk + D_kY_ba*Y_bl)*Y2_at
               + (D_ktaX + D_tY_ba*Y_bk + D_kY_bt*Y_ba)*Y2_al)
-i*p3*s5*D_aX*(D_ktY_bl*Y_ba+(D_ktY_ba+D_kaY_bt+D_akY_bt)*Y_bl)
-(8*i/9)*p3*s5*D_aX*((D_lY_ba+D_aY_bl)*D_kY_bt+D_tY_bl*(D_kY_ba+D_aY_bk));

ten:
-(1/6)*s4*((D_dcY_ab + D_dbY_ac + D_bdY_ac)*Y2_cd
           + (D_dcY_ac + D_cdY_ac + D_ccY_ad)*Y2_bd)*Y_ab
-(4/27)*s4*(2*D_dY_cd*D_bY_ca + (D_aY_cd + D_dY_ca)*D_bY_cd
           + (D_aY_cd+D_dY_ca)*D_dY_cb)*Y2_ab
+(1/2)*s4*D_aX*D_bX*(D_abX+(3/4)*(D_bcY_ac+D_cbY_ac)
+(1/3)*s4*D_aX*((D_cY_ba+D_aY_bc)*(D_cdY_bd+D_dcY_bd+D_ddY_bc)
	+D_cY_bc*(D_ddY_ba+D_daY_bd+D_adY_bd)
	+D_cY_bd*((D_caY_bd+D_acY_bd)+(D_cdY_ba+D_dcY_ba)+(D_adY_bc+D_daY_bc)))
+(1/3)*p2*s5*(((D_clY_ab + D_cbY_al + D_blY_ac)*Y2_ct
              + (D_tcY_ab + D_tbY_ac + D_bcY_at)*Y2_cl
              + (D_tcY_ac + D_ctY_ac + D_ccY_at)*Y2_bl
              + (D_tcY_al + D_clY_at + D_tlY_ac)*Y2_bc)*Y_ab
              + ((D_cbY_ab + D_bcY_ab + D_bbY_ac)*Y2_ct
                +(D_bcY_at + D_btY_ac + D_tbY_ac)*Y2_cb)*Y_al)
+(8/27)*p2*s5*((2*D_bY_cb*D_aY_cl  + D_aY_cb*D_lY_cb
                + 2*(D_bY_ca + D_aY_cb)*D_bY_cl)*Y2_at
              +(2*D_bY_cb*D_tY_ca  + D_bY_ca*D_bY_ct
                + (D_aY_cb + 2*D_bY_ca)*D_tY_cb)*Y2_al
              +(2*D_bY_ca*D_tY_cl + D_tY_ca*D_bY_cl
                + D_lY_ca*D_tY_cb + D_aY_cl*(D_bY_ct + D_tY_cb))*Y2_ab)
+(1/2)*p2*s5*D_aX*D_bX*(D_tbY_la+D_btY_la+*D_baY_lt)
-(2/3)*p2*s5*D_aX*((D_lY_ba+D_aY_bl)*(D_ccY_bt+D_ctY_bc+D_tcY_bc)
	+(D_cY_ba+D_aY_bc)*(D_clY_bt+D_tcY_bl+D_tlY_bc)
	+(D_lY_bc+D_cY_bl)*(D_taY_bc+D_atY_bc+D_tcY_ba)
	+D_cY_bc*(D_tlY_ba+D_taY_bl+D_atY_bl)
	+D_cY_bl*(D_caY_bt+D_acY_bt+D_ctY_ba)
	+D_tY_bc*(D_acY_bl+D_clY_ba+D_caY_bl)
	+D_tY_bl*(D_ccY_ba+D_caY_bc+D_acY_bc))
-(2/3)*p4*s6*((D_bkY_at*Y2_bh + D_hbY_at*Y2_bk + D_hkY_ab*Y2_bt)*Y_al
              + D_hkY_at*Y2_bl*Y_ab)
-(16/27)*p4*s6*(D_tY_bl (D_aY_bk*Y2_ah + D_hY_ba*Y2_ak)
               +(D_aY_bl*Y2_at + D_tY_ba*Y2_al)*D_hY_bk)
+(4/3)*p4*s6*D_aX*(D_lY_fa*D_hkY_ft+D_aY_bl*D_hkY_bt
	+D_tY_bl*(D_hkY_ba+D_haY_bk+D_ahY_bk));
twl:
-(1/24)*s5*((D_deY_be + D_edY_be)*D_caY_bd
            + (D_deY_be + D_edY_be)*D_cdY_ba + (D_deY_be + D_edY_be)*D_dcY_ba
 + (D_aeY_be + D_eaY_be)*D_cdY_bd
  + (D_aeY_be + D_eaY_be)*(D_dcY_bd + D_ddY_bc)
  + 2*D_caY_be*D_ddY_be + (D_caY_be + D_caY_be)*D_edY_bd
 + (D_aeY_bd + D_eaY_bd)*(D_ceY_bd + D_ecY_bd)
 + (D_aeY_bd + D_eaY_bd)*(D_cdY_be + D_dcY_be)
 + (D_aeY_bd + D_eaY_bd)*(D_edY_bc + D_deY_bc)
 + (D_ceY_ba + D_ecY_ba)*((D_ddY_be + D_edY_bd)+(D_ddY_be + D_deY_bd))
 + (D_ceY_be + D_ecY_be)*D_ddY_ba + D_eeY_bc*D_ddY_ba
 + (D_edY_ba + D_deY_ba)*(D_ceY_bd + D_ecY_bd + D_edY_bc)
 + (D_edY_ba + D_deY_ba) + (D_edY_ba + D_deY_ba))*Y2_ac
+(1/4)*s5*D_aX*D_cX*((D_ddY_bc+D_dcY_bd+D_cdY_bd)Y_ba
  			+(D_caY_bd+D_cdY_ba+D_dcY_ba)*Y_bd)
+(2/9)*s5*D_aX*D_cX*(2*D_cY_ba*D_dY_bd
			+(D_dY_ba+D_aY_bd)*D_dY_bc+(D_dY_ba+D_aY_bd)*D_cY_bd)
-(1/9)*s5*D_aX*Y2_bc(D_cbaX+D_cabX+D_acbX)
+(1/9)*p2*s6*((D_cbY_la + D_bcY_la)*Y2_ab*Y2_ct
              +(- 2*D_abX + (1/6)*(D_cbY_ca - D_bcY_ac))*Y2_al*Y2_bt)
+(1/12)*p2*s6*(
  ( (D_cdY_bd + D_dcY_bd + D_ddY_bc)*(D_alY_bc+D_laY_bc)
   +(D_cdY_bd + D_dcY_bd + D_ddY_bc)*(D_acY_bl+D_caY_bl)
   +(D_ddY_bc + D_dcY_bd + D_cdY_bd)*(D_acY_bl+D_clY_ba)
   +(D_adY_bc + D_daY_bd + D_ddY_ba)*(D_lcY_bc+D_clY_bc)
   +(D_adY_bc + D_daY_bc + D_acY_bd)*(D_dlY_bc+D_ldY_bc)
   +(D_adY_bd + D_daY_bd + D_ddY_ba)* D_ccY_bl
   +(D_acY_bd + D_caY_bd)* D_dlY_bc + D_cdY_ba* D_ldY_bc
   +(D_adY_bc + D_caY_bd)* D_dcY_bl
   +(2*(D_acY_bd +D_caY_bd) + D_caY_bd + (D_cdY_ba+D_dcY_ba)
     + 2* D_daY_bc+D_adY_bc)*D_cdY_bl)* Y2_at
   +((D_cdY_bd + D_dcY_bd + D_ddY_bc)*(D_taY_bc+D_atY_bc)
   +(D_cdY_bd + D_dcY_bd + D_ddY_bc)*(D_tcY_ba+D_ctY_ba)
   +(D_dcY_bd + D_ddY_bc + D_cdY_bd)*(D_tcY_ba+D_caY_bt)
   +(D_adY_bd + D_daY_bd + D_ddY_ba)*(D_tcY_bc+D_ctY_bc)
   +(D_adY_bd + D_daY_bd + D_ddY_ba)*D_ccY_bt
   +(D_acY_bd + D_cdY_ba + D_dcY_ba)*D_dcY_bt
   +(D_adY_bc+D_daY_bc + 2*D_caY_bd+D_acY_bd
     + 2*D_dcY_ba+D_cdY_ba)*(D_tdY_bc+D_dtY_bc))* Y2_al
 + ((D_cdY_bl + D_dlY_bc)*D_taY_bd +(D_dcY_bl + D_clY_bd)*D_tdY_ba
   +(D_ldY_bd + D_dlY_bd  + D_ddY_bl)*D_caY_bt
   +(D_adY_bd + D_daY_bd + D_ddY_ba)*D_tcY_bl
   +(D_adY_bd + D_daY_bd + D_ddY_ba)*D_clY_bt
   +(D_adY_bd + D_daY_bd + D_ddY_ba)*D_tlY_bc
   +(D_laY_bd + D_daY_bl + D_adY_bl)*D_ctY_bd
   +(D_caY_bd + D_cdY_ba + D_dcY_ba)*D_ltY_bd
   +(D_cdY_bd + D_dcY_bd + D_ddY_bc)*D_ltY_ba
   +(D_laY_bd + D_adY_bl + D_daY_bl)* D_dcY_bt
   +(D_adY_bl + D_daY_bl + D_ldY_ba)*D_cdY_bt
   +(D_adY_bl + D_daY_bl + D_ldY_ba)*D_dtY_bc
   +(D_caY_bd + D_cdY_ba + D_dcY_ba)*D_dtY_bl
   +(D_dcY_ba + D_cdY_ba + D_caY_bd)*D_tlY_bd
   +(D_lcY_ba + D_clY_ba + D_caY_bl)*D_tdY_bd
   +(D_lcY_ba + D_clY_ba + D_caY_bl)*D_dtY_bd
   +(D_lcY_ba + D_clY_ba + D_caY_bl)*D_ddY_bt
   +(D_ldY_bd + D_dlY_bd + D_ddY_bl)*(D_ctY_ba+D_tcY_ba)
   +(D_cdY_bd + D_dcY_bd + D_ddY_bc)*(D_atY_bl+D_taY_bl)
   +(D_dcY_ba + D_cdY_ba + D_cdY_ba + D_caY_bd)*D_dlY_bt
   +(D_cdY_bl + D_dlY_bc + D_dcY_bl + D_clY_bd)*D_dtY_ba
   +(D_clY_bd + D_cdY_bl + D_dcY_bl + D_dlY_bc)*D_atY_bd
   +(D_alY_bd + D_laY_bd + D_daY_bl + D_adY_bl + D_dlY_ba + D_ldY_ba)*D_tcY_bd
   +(D_alY_bd + D_daY_bl + D_laY_bd + D_adY_bl + D_dlY_ba + D_ldY_ba)*D_tdY_bc
   +(D_caY_bd + D_caY_bd + D_cdY_ba + D_dcY_ba + D_dcY_ba + D_cdY_ba)*D_tdY_bl)*Y2_ac)
-(1/2)*p2*s6*D_aX*D_cX*((D_tlY_bc+D_tcY_bl+D_ctY_bl)*Y_ba
			+(D_tcY_ba+D_ctY_ba+D_caY_bt)*Y_bl)
-(4/9)*p2*s6*D_aX*D_cX*((D_cY_ba+D_aY_bc)*D_tY_bl
			+D_lY_ba*(D_tY_bc+D_cY_bt)+D_aY_bl*(D_tY_bc+D_cY_bt))
+(2/9)*p2*s6*D_aX*((D_tbaX+D_atbX+D_tabX)*Y2_bl+(D_blaX+D_ablX+D_balX)*Y2_bt)
-(2/9)* p4*s7*(D_baY_lt*Y2_ak*Y2_bh + D_bkY_la*Y2_at*Y2_bh + D_hbY_la*Y2_at*Y2_bk)
-(1/6)*p4*s7*
 ((D_kcY_bl*(D_atY_bh + D_haY_bt) + D_hkY_ba*(D_tcY_bl + D_ltY_bc)
 + D_ktY_bc*(D_alY_bh + D_haY_bl) + D_ktY_bl*(D_acY_bh + D_ahY_bc + D_haY_bc)
 + D_ctY_bl*(D_ahY_bk + D_haY_bk + D_hkY_ba)
 + D_hkY_bt*(D_alY_bc + D_acY_bl + D_laY_bc))*Y2_ca
 +(D_hkY_ba*(D_taY_bc + D_atY_bc) + D_kaY_bc*(D_haY_bt + D_atY_bh)
 + D_atY_bc*(D_haY_bk + D_ahY_bk) + D_ktY_ba*(D_haY_bc + D_acY_bh + D_ctY_ba)
 + D_hkY_bt*(D_caY_ba + D_acY_ba + D_aaY_bc)
 + D_ktY_bc*(D_haY_ba + D_dhY_bd + D_aaY_bh))*Y2_cl
 +(D_hkY_ba*(D_lcY_ba + D_caY_bl) + D_kcY_ba*(D_haY_bl + D_alY_bh)
 + D_kaY_bl*(D_haY_bc + D_acY_bh) + D_hkY_bc*(D_laY_ba + D_alY_ba + D_aaY_bl)
 + D_acY_bl*(D_hkY_ba + D_haY_bk + D_ahY_bk)
 + D_kcY_bl*(D_haY_ba + D_dhY_bd + D_aaY_bh))*Y2_ct
 +(D_hcY_ba*(D_ltY_ba + D_taY_bl) + D_ctY_ba*(D_haY_bl + D_aeY_bh)
 + D_caY_bl*(D_haY_bt + D_atY_bh) + D_hcY_bt*(D_laY_ba + D_alY_ba + D_aaY_bl)
 + D_atY_bl*(D_haY_bc + D_ahY_bc + D_heY_ba)
 + D_ctY_bl*(D_haY_ba + D_ahY_ba + D_aaY_bh))*Y2_ck
 +(D_ckY_ba*(D_ltY_ba + D_taY_bl) + D_ktY_ba*(D_caY_bl + D_alY_bc)
 + D_kaY_bl*(D_caY_bt + D_atY_bc) + D_ckY_bt*(D_aaY_bl + D_laY_ba + D_alY_ba)
 + D_atY_bl*(D_ckY_ba + D_caY_bk + D_acY_bk)
 + D_ktY_bl*(D_cdY_bd + D_dcY_bd + D_aaY_bc))*Y2_ch)
 +(1/3)*p6*s8 (D_ktY_al*(D_bxY_ah*Y2_bg + Y2_bh*D_gxY_ab)
     + D_gxY_ah*(D_btY_al*Y2_bk + D_kbY_al*Y2_bt + D_ktY_ab*Y2_bl + D_ktY_al*Y2_bx));
\end{verbatim}
\end{scriptsize}


\chapter{Momentum Integration in $D$-dimensions}
In the calculation of the one-loop effective Lagrangian prescribed by\footnote{Here,
\begin{eqnarray}
    G_0
        =\!\!\int_0^\infty\!\!\!ds\,\,
     e^{+\Theta(s)}
        =\!\!\int_0^\infty\!\!\!ds\,\,
     e^{+(m^2+X)s+P(s)+Q(s)\cdot p+\frac{1}{2}p\cdot R(s)\cdot p}
\end{eqnarray}
is expressed in terms of exact trancedental functions $P(s)$, $Q(s)$, and $R(s)$.}
\begin{eqnarray}
    {\cal L}^{(1)}=\sum^\infty_{\ell=0}{\cal L}^{(1)}_\ell
    =\frac{\hbar}{2(2\pi)^D}
    \mbox{Tr}
    \int dX\int d^Dp \,G_0\sum^{\infty}_{\ell=0}
    \left(\Delta_1 G_0\right)^\ell,
\end{eqnarray}
up to $\ell$-th order corrections, there is an ensuing $X$ integration, momentum $p$ integration, and $\ell$-th fold proper-time integrations\footnote{See Appendix D on how to handle $\ell$-th fold proper time integrals}. In this chapter, we present the technique of handling (Gaussian) momentum $p$ integrals in $D$-dimensions needed in the quasilocal background method.

In our derivation, we present (without proof) the identity
\begin{equation}\label{p0}
    \int d^Dp\,\,\,
        e^{-\frac{1}{2}p\cdot R \cdot p}
        =
        (2\pi)^{D/2}(\det R)^{-1/2}
\end{equation}
and
\begin{equation}\label{det}
     \frac{\partial}{\partial R_{{\lambda}{\tau}}}(\det R)
     =R^{-1}_{{\lambda}{\tau}}(\det R),
\end{equation}
for any symmetric non-singular matrix $R$.

We present first on how to handle integrals involving odd-powers of $p$. Since these kind of integrals vanish, a theorem can be stated (without proof) that the momentum-space derivative of such integrals also vanish. Without ambiguity in the order of differentiation and integration, the technique in handling integrals involving even-powers is derived from this theorem.

\section{Odd-powers of $p$ Integrals}
The integrals involving odd-powers of $p$ vanish:
\begin{eqnarray}\label{p1}
    \int d^Dp\,\,\,
        p_{\lambda}\,\,\,
        e^{-\frac{1}{2}p\cdot R\cdot p}
        = 0
\\ \label{p3}
    \int d^Dp\,\,\,
        p_{\lambda}p_{\tau}p_{\kappa}\,\,\,
        e^{-\frac{1}{2}p\cdot R\cdot p}
        = 0
\\ \label{p5}
    \int d^Dp\,\,\,
        p_{\lambda}p_{\tau}p_{\kappa}p_{\eta}p_{\xi}\,\,\,
        e^{-\frac{1}{2}p\cdot R\cdot p}
        = 0
\\ \label{p7}
    \int d^Dp\,\,\,
        p_{\lambda}p_{\tau}p_{\kappa}p_{\eta}p_{\xi}p_{\varrho}p_{\vartheta}\,\,\,
        e^{-\frac{1}{2}p\cdot R\cdot p}
        = 0
\\  \vdots \,\,\,\,\,\,\,\,\,\,\,\,\,\,\,\,\,\,\,\,\nonumber
\end{eqnarray}
We show odd-powers of $p$ that may be relevant in the calculation
of higher mass-dimensional effective Lagrangians up from 4 up to
8 mass-dimensions. They are the integrals that involves $p$,
$p^3$, up to $p^7$.

\section{Even-powers of $p$ Integrals}
Integrals involving even powers of $p$ are obtained by
differentiating (\ref{p0}) with respect to $R_{{\lambda}{\tau}}$
\begin{equation}
    \frac{\partial}{\partial R_{{\lambda}{\tau}}}
    \int d^Dp\,\,\,
        e^{-\frac{1}{2}p\cdot R \cdot p}
    =
    -\frac{1}{2}(2\pi)^{D/2}
        \left[\frac{\partial}{\partial R_{{\lambda}{\tau}}}(\det R)\right]
        (\det R)^{-3/2}
\end{equation}
and using (\ref{det})
\begin{equation}\label{dpp2}
    \int d^Dp\,\,\,
        p_{\lambda}p_{\tau}\,\,\,
        e^{-\frac{1}{2}p\cdot R\cdot p}
        =(2\pi)^{D/2}R^{-1}_{{\lambda}{\tau}}(\det R)^{-1/2}.
\end{equation}

Integrals involving higher even powers of $p$ can be obtained by
assuming the identity\footnote{Note this identity can be seen from (\ref{p1}).} \cite{GM}
\begin{equation}\label{dp1}
    \int d^Dp\,\,\,
        \frac{\partial}{\partial p_{\lambda}}
    \left(
        p_{\tau}\,\,\,
        e^{-\frac{1}{2}p\cdot R\cdot p}
    \right)
    =0.
\end{equation}
We use this identity to alternatively obtain (\ref{dpp2}). Having done so for the case of integral involving $p^2$, we generate similar identities to evaluate integrals involving $p^4$, $p^6$, and $p^8$ using (\ref{p3}), (\ref{p5}), and (\ref{p7}), respectively. That is, we partially differentiate each integrand of the odd-power-in-$p$ integrals ((\ref{p3}), (\ref{p5}), and (\ref{p7})) once with respect to momentum-space:
\begin{equation}\label{dp3}
    \int d^Dp\,\,\,
        \frac{\partial}{\partial p_{\eta}}
    \left(
        p_{\lambda}p_{\tau}p_{\kappa}\,\,\,
        e^{-\frac{1}{2}p\cdot R\cdot p}
    \right)
    =0
\end{equation}
\begin{equation}\label{dp5}
    \int d^Dp\,\,\,
        \frac{\partial}{\partial p_{\varrho}}
    \left(
        p_{\lambda}p_{\tau}p_{\kappa}p_{\eta}p_{\xi}\,\,\,
        e^{-\frac{1}{2}p\cdot R\cdot p}
    \right)
    =0
\end{equation}
\begin{equation}\label{dp7}
    \int d^Dp\,\,\,
        \frac{\partial}{\partial p_{\omega}}
    \left(
        p_{\lambda}p_{\tau}p_{\kappa}p_{\eta}p_{\xi}p_{\varrho}p_{\vartheta}\,\,\,
        e^{-\frac{1}{2}p\cdot R\cdot p}
    \right)
    =0.
\end{equation}

Using $p^2$-integral identity integrals involving $p^4$ is evaluated. Similarly, $p^4$-integrals will simplify integrals involving $p^6$. Similarly, $p^6$ for $p^8$ as it will be shown in the following subsections of this chapter. We limit our derivation of integrals involving even-powers from 2 up to 8 only as they will be relevant in our calculation of one-loop effective Lagrangian and collecting only products of invariants whose total mass-dimensions is eight or less.

\section{$p^2$-integral Identity}
From (\ref{dp1}), we have
\begin{eqnarray}\label{Rdpp2}
    \int d^Dp\,\,\,
    \left(
        -R_{{\lambda}{\kappa}}p_{\kappa}p_{\tau}\,\,\,
        +\delta_{{\lambda}{\tau}}
    \right)
        e^{-\frac{1}{2}p\cdot R\cdot p}
    =0.
\end{eqnarray}
Simplifying in the following manner:
\begin{small}
\begin{eqnarray*}
    R_{{\lambda}{\kappa}}\int d^Dp\,\,\,
        p_{\kappa}p_{\tau}\,\,\,
    e^{-\frac{1}{2}p\cdot R\cdot p}
    =
    \delta_{{\lambda}{\tau}}
    \int d^Dp\,\,\,
    e^{-\frac{1}{2}p\cdot R\cdot p}
    =
    \delta_{{\lambda}{\tau}}
    (2\pi)^{D/2}(\det R)^{-1/2}
\end{eqnarray*}
\vspace{-5mm}
\begin{eqnarray*}
    \int d^Dp\,\,\,
        p_{\kappa}p_{\tau}\,\,\,
    e^{-\frac{1}{2}p\cdot R\cdot p}
    =
    \delta_{{\lambda}{\tau}}
    R^{-1}_{{\lambda}{\kappa}}
    (2\pi)^{D/2}(\det R)^{-1/2}
    =
     R^{-1}_{{\tau}{\kappa}}
    (2\pi)^{D/2}(\det R)^{-1/2}.
 \end{eqnarray*}
\end{small}
Using (\ref{p0}) and since $R_{{\tau}{\kappa}}$ is symmetric,
\begin{small}
\begin{eqnarray}\label{dpp2b}
    \int d^Dp\,\,\,
        p_{\kappa}p_{\tau}\,\,\,
    e^{-\frac{1}{2}p\cdot R\cdot p}
    &=&
     (2\pi)^{D/2}
     R^{-1}_{{\kappa}{\tau}}
    (\det R)^{-1/2}.
 \end{eqnarray}
\end{small}
Renaming a dummy index $\kappa\to\lambda$:
\begin{small}
\begin{eqnarray}\label{p2lt}
    \int d^Dp\,\,\,
        p_{\lambda}p_{\tau}\,\,\,
    e^{-\frac{1}{2}p\cdot R\cdot p}
    &=&
     (2\pi)^{D/2}
     R^{-1}_{{\lambda}{\tau}}
    (\det R)^{-1/2}.
 \end{eqnarray}
\end{small}
This is exactly (\ref{dpp2}).

\section{$p^4$-integral Identity.}
From (\ref{dp3})
\begin{small}
\begin{eqnarray*}
    \int d^Dp\,\,\,
    \left(
        \delta_{{\lambda}{\eta}}p_{\tau}p_{\kappa}
        +p_{\lambda}\delta_{{\tau}{\eta}}p_{\kappa}
        +p_{\lambda}p_{\tau}\delta_{{\kappa}{\eta}}
        -R_{{\eta}{\xi}}p_{\xi}p_{\lambda}p_{\tau}p_{\kappa}
    \right)
        e^{-\frac{1}{2}p\cdot R\cdot p}
    =0
\end{eqnarray*}
\end{small}
Simplifying in the following manner:
\begin{small}
\begin{eqnarray*}
    R_{{\eta}{\xi}}
    \int d^Dp\!\!\!&&\!\!\!
        p_{\xi}p_{\lambda}p_{\tau}p_{\kappa}\,\,\,
        e^{-\frac{1}{2}p\cdot R\cdot p}
\\ \,\,\,\,\,\,\,\,\,\,\,\,\,\,\,
&&=
    \int d^Dp\,\,\,
    \left(
        \delta_{{\lambda}{\eta}}p_{\tau}p_{\kappa}
        +p_{\lambda}\delta_{{\tau}{\eta}}p_{\kappa}
        +p_{\lambda}p_{\tau}\delta_{{\kappa}{\eta}}
    \right)
    e^{-\frac{1}{2}p\cdot R\cdot p}.
    \\ \int d^Dp\!\!\!&&\!\!\!
        p_{\xi}p_{\lambda}p_{\tau}p_{\kappa}\,\,\,
        e^{-\frac{1}{2}p\cdot R\cdot p}
\\ \,\,\,\,\,\,\,\,\,\,\,\,\,\,\,
&&=
    \int d^Dp\,\,\,
    \left(
        R^{-1}_{{\eta}{\xi}}
            \delta_{{\lambda}{\eta}}p_{\tau}p_{\kappa}
        +R^{-1}_{{\eta}{\xi}}
            p_{\lambda}\delta_{{\tau}{\eta}}p_{\kappa}
        +R^{-1}_{{\eta}{\xi}}
            p_{\lambda}p_{\tau}\delta_{{\kappa}{\eta}}
    \right)
    e^{-\frac{1}{2}p\cdot R\cdot p}.
    \\ \int d^Dp\!\!\!&&\!\!\!
        p_{\xi}p_{\lambda}p_{\tau}p_{\kappa}\,\,\,
        e^{-\frac{1}{2}p\cdot R\cdot p}
\\ \,\,\,\,\,\,\,\,\,\,\,\,\,\,\,
&&=
    \int d^Dp\,\,\,
    \left(
        R^{-1}_{{\lambda}{\xi}}
            p_{\tau}p_{\kappa}
        +R^{-1}_{{\tau}{\xi}}
            p_{\lambda}p_{\kappa}
        +R^{-1}_{{\kappa}{\xi}}
            p_{\lambda}p_{\tau}
    \right)
    e^{-\frac{1}{2}p\cdot R\cdot p}.
\end{eqnarray*}
\end{small}

Repeated use of (\ref{dpp2}) or (\ref{dpp2b}),
\begin{small}
\begin{eqnarray*}
    \int d^Dp\,\,\,
        p_{\xi}p_{\lambda}p_{\tau}p_{\kappa}\,\,\,
        e^{-\frac{1}{2}p\cdot R\cdot p}
    =
    (2\pi)^{D/2}
    \left(
        R^{-1}_{{\lambda}{\xi}}
            R^{-1}_{{\tau}{\kappa}}
        +R^{-1}_{{\tau}{\xi}}
            R^{-1}_{{\lambda}{\kappa}}
        +R^{-1}_{{\kappa}{\xi}}
            R^{-1}_{{\lambda}{\tau}}
    \right)
        (\det R)^{-1/2}
\end{eqnarray*}
\end{small}
Changing dummy indices $\{\xi,\lambda,\tau,\kappa\}\to\{\lambda,\tau,\kappa,\xi\}$ and $\xi\to\eta$,
\begin{small}
\begin{equation}
    \int d^Dp\,\,\,
        p_{\lambda}p_{\tau}p_{\kappa}p_{\eta}\,\,\,
        e^{-\frac{1}{2}p\cdot R\cdot p}
    =
    (2\pi)^{D/2}
    \left(
        R^{-1}_{{\tau}{\lambda}}
            R^{-1}_{{\kappa}{\eta}}
        +R^{-1}_{{\kappa}{\lambda}}
            R^{-1}_{{\tau}{\eta}}
        +R^{-1}_{{\eta}{\lambda}}
            R^{-1}_{{\tau}{\kappa}}
    \right)
        (\det R)^{-1/2}.
\nonumber\\ \label{dpp4}
\end{equation}
\end{small}
This is the identity needed to simplify (Gaussian) integrals involving $p^4$.

\section{$p^6$-integral Identity}
From (\ref{dp5})
\begin{small}
\begin{eqnarray*}
    \int d^Dp\!\!\!&&\!\!\!
    \left(
        \delta_{{\varrho}{\lambda}}p_{\tau}p_{\kappa}p_{\eta}p_{\xi}
        +p_{\lambda}\delta_{{\varrho}{\tau}}p_{\kappa}p_{\eta}p_{\xi}
        +p_{\lambda}p_{\tau}\delta_{{\varrho}{\kappa}}p_{\eta}p_{\xi}
   \right.
    \\ &&\,\,\,\,\,\,\,\,\,\,\,\,\,\,
        +p_{\lambda}p_{\tau}p_{\kappa}\delta_{{\varrho}{\eta}}p_{\xi}
        +p_{\lambda}p_{\tau}p_{\kappa}p_{\eta}\delta_{{\varrho}{\xi}}
    \left.
        -R_{{\varrho}{\vartheta}}p_{\vartheta}p_{\lambda}p_{\tau}p_{\kappa}p_{\eta}p_{\xi}\,\,\,
        e^{-\frac{1}{2}p\cdot R\cdot p}
    \right)
    =0
\end{eqnarray*}
\end{small}
Simplifying in the following manner:
\begin{small}
\begin{eqnarray*}
    R_{{\varrho}{\vartheta}}
    \int d^Dp\!\!\!&&\!\!\!
        p_{\vartheta}p_{\lambda}p_{\tau}p_{\kappa}p_{\eta}p_{\xi}\,\,\,
     e^{-\frac{1}{2}p\cdot R\cdot p}
    = \int d^Dp\,\,\,
    \left(
        \delta_{{\varrho}{\lambda}}p_{\tau}p_{\kappa}p_{\eta}p_{\xi}
        +p_{\lambda}\delta_{{\varrho}{\tau}}p_{\kappa}p_{\eta}p_{\xi}
\right.
\\ &&\,\,\,\,\,\,\,\,\,\,\,
\left.
        +p_{\lambda}p_{\tau}\delta_{{\varrho}{\kappa}}p_{\eta}p_{\xi}
        +p_{\lambda}p_{\tau}p_{\kappa}\delta_{{\varrho}{\eta}}p_{\xi}
        +p_{\lambda}p_{\tau}p_{\kappa}p_{\eta}\delta_{{\varrho}{\xi}}
    \right)
        e^{-\frac{1}{2}p\cdot R\cdot p}.
\\    \int d^Dp\!\!\!&&\!\!\!
        p_{\vartheta}p_{\lambda}p_{\tau}p_{\kappa}p_{\eta}p_{\xi}\,\,\,
     e^{-\frac{1}{2}p\cdot R\cdot p}
     =
    \int d^Dp\,\,\,
    \left(
        R^{-1}_{{\varrho}{\vartheta}}\delta_{{\varrho}{\lambda}}p_{\tau}p_{\kappa}p_{\eta}p_{\xi}
        +R^{-1}_{{\varrho}{\vartheta}}p_{\lambda}\delta_{{\varrho}{\tau}}p_{\kappa}p_{\eta}p_{\xi}
    \right.
    \\ &&\,\,\,\,\,\,\,\,\,\,\,
    \left.
        +R^{-1}_{{\varrho}{\vartheta}}p_{\lambda}p_{\tau}\delta_{{\varrho}{\kappa}}p_{\eta}p_{\xi}
        +R^{-1}_{{\varrho}{\vartheta}}p_{\lambda}p_{\tau}p_{\kappa}\delta_{{\varrho}{\eta}}p_{\xi}
        +R^{-1}_{{\varrho}{\vartheta}}p_{\lambda}p_{\tau}p_{\kappa}p_{\eta}\delta_{{\varrho}{\xi}}
    \right)
        e^{-\frac{1}{2}p\cdot R\cdot p}.
\\    \int d^Dp\!\!\!&&\!\!\!
        p_{\kappa}p_{\lambda}p_{\tau}p_{\kappa}p_{\eta}p_{\xi}\,\,\,
     e^{-\frac{1}{2}p\cdot R\cdot p}
     =
    \int d^Dp\,\,\,
    \left(
        R^{-1}_{{\lambda}{\vartheta}}p_{\tau}p_{\kappa}p_{\eta}p_{\xi}
        +R^{-1}_{{\tau}{\vartheta}}p_{\lambda}p_{\kappa}p_{\eta}p_{\xi}
    \right.
    \\ &&\,\,\,\,\,\,\,\,\,\,\,
        +R^{-1}_{{\kappa}{\vartheta}}p_{\lambda}p_{\tau}p_{\eta}p_{\xi}
    \left.
        +R^{-1}_{{\eta}{\vartheta}}p_{\lambda}p_{\tau}p_{\kappa}p_{\xi}
        +R^{-1}_{{\xi}{\vartheta}}p_{\lambda}p_{\tau}p_{\kappa}p_{\eta}
    \right)
        e^{-\frac{1}{2}p\cdot R\cdot p}.
\end{eqnarray*}
\end{small}

Repeated use of (\ref{dpp4})
\begin{small}
\begin{eqnarray*}
    &&\!\!\!\!\!\!\!\!\!\!\!\!\!\!\!\!\!\!\!\!\!\!\!\!\!\!\!\!\!\!\!\!\!\!\!\!\!\!\!\!\!\!\!\!\!
    \int d^Dp\,\,\,
        p_{\vartheta}p_{\lambda}p_{\tau}p_{\kappa}p_{\eta}p_{\xi}\,\,\,
     e^{-\frac{1}{2}p\cdot R\cdot p}
     =
    (2\pi)^{D/2}
    \\&&\!\!\!\!\!\!\!\!\!\times\left[
        R^{-1}_{{\lambda}{\vartheta}}
    \left(
        R^{-1}_{{\xi}{\tau}}
            R^{-1}_{{\kappa}{\eta}}
        +R^{-1}_{{\tau}{\eta}}
            R^{-1}_{{\kappa}{\xi}}
        +R^{-1}_{{\tau}{\kappa}}
            R^{-1}_{{\eta}{\xi}}
    \right)\right.
    \\&&
        +R^{-1}_{{\tau}{\vartheta}}
    \left(
        R^{-1}_{{\xi}{\lambda}}
            R^{-1}_{{\kappa}{\eta}}
        +R^{-1}_{{\lambda}{\eta}}
            R^{-1}_{{\kappa}{\xi}}
        +R^{-1}_{{\lambda}{\kappa}}
            R^{-1}_{{\eta}{\xi}}
    \right)
    \\&&
        +R^{-1}_{{\kappa}{\vartheta}}
    \left(
        R^{-1}_{{\xi}{\lambda}}
            R^{-1}_{{\tau}{\eta}}
        +R^{-1}_{{\lambda}{\eta}}
            R^{-1}_{{\tau}{\xi}}
        +R^{-1}_{{\lambda}{\tau}}
            R^{-1}_{{\eta}{\xi}}
    \right)
    \\ &&
        +R^{-1}_{{\eta}{\vartheta}}
    \left(
        R^{-1}_{{\xi}{\lambda}}
            R^{-1}_{{\tau}{\kappa}}
        +R^{-1}_{{\lambda}{\kappa}}
            R^{-1}_{{\tau}{\xi}}
        +R^{-1}_{{\lambda}{\tau}}
            R^{-1}_{{\kappa}{\xi}}
    \right)
    \\ &&
    \left.
        +R^{-1}_{{\xi}{\vartheta}}
    \left(
        R^{-1}_{{\eta}{\lambda}}
            R^{-1}_{{\tau}{\kappa}}
        +R^{-1}_{{\lambda}{\kappa}}
            R^{-1}_{{\tau}{\eta}}
        +R^{-1}_{{\lambda}{\tau}}
            R^{-1}_{{\kappa}{\eta}}
    \right)
    \right]
        (\det R)^{-1/2}.
\end{eqnarray*}
\end{small}
By distributive property of multiplication and addition,
\begin{small}
\begin{eqnarray*}
    &&\!\!\!\!\!\!\!\!\!\!\!\!\!\!\!\!\!\!\!\!\!\!\!\!\!\!\!\!\!\!\!\!\!\!\!\!\!\!\!\!\!\!\!\!\!
    \int d^Dp\,\,\,
        p_{\vartheta}p_{\lambda}p_{\tau}p_{\kappa}p_{\eta}p_{\xi}\,\,\,
     e^{-\frac{1}{2}p\cdot R\cdot p}
     =
    (2\pi)^{D/2}
    \\&&\!\!\!\!\!\!\!\!\!
\times\left[
        R^{-1}_{{\lambda}{\vartheta}}R^{-1}_{{\xi}{\tau}}
            R^{-1}_{{\kappa}{\eta}}
        +R^{-1}_{{\lambda}{\vartheta}}R^{-1}_{{\tau}{\eta}}
            R^{-1}_{{\kappa}{\xi}}
        +R^{-1}_{{\lambda}{\vartheta}}R^{-1}_{{\tau}{\kappa}}
            R^{-1}_{{\eta}{\xi}}
        +
        R^{-1}_{{\tau}{\vartheta}}R^{-1}_{{\xi}{\lambda}}
            R^{-1}_{{\kappa}{\eta}}
\right.
    \\&&
        +R^{-1}_{{\tau}{\vartheta}}R^{-1}_{{\lambda}{\eta}}
            R^{-1}_{{\kappa}{\xi}}
        +R^{-1}_{{\tau}{\vartheta}}R^{-1}_{{\lambda}{\kappa}}
            R^{-1}_{{\eta}{\xi}}
        +R^{-1}_{{\kappa}{\vartheta}}R^{-1}_{{\xi}{\lambda}}
            R^{-1}_{{\tau}{\eta}}
        +R^{-1}_{{\kappa}{\vartheta}}R^{-1}_{{\lambda}{\eta}}
            R^{-1}_{{\tau}{\xi}}
    \\&&
        +R^{-1}_{{\kappa}{\vartheta}}R^{-1}_{{\lambda}{\tau}}
            R^{-1}_{{\eta}{\xi}}
        +R^{-1}_{{\eta}{\vartheta}}R^{-1}_{{\xi}{\lambda}}
            R^{-1}_{{\tau}{\kappa}}
        +R^{-1}_{{\eta}{\vartheta}}R^{-1}_{{\lambda}{\kappa}}
            R^{-1}_{{\tau}{\xi}}
        +R^{-1}_{{\eta}{\vartheta}}R^{-1}_{{\lambda}{\tau}}
            R^{-1}_{{\kappa}{\xi}}
    \\ &&
    \left.
        +
        R^{-1}_{{\xi}{\vartheta}}R^{-1}_{{\eta}{\lambda}}
            R^{-1}_{{\tau}{\kappa}}
        +R^{-1}_{{\xi}{\vartheta}}R^{-1}_{{\lambda}{\kappa}}
            R^{-1}_{{\tau}{\eta}}
        +R^{-1}_{{\xi}{\vartheta}}R^{-1}_{{\lambda}{\tau}}
            R^{-1}_{{\kappa}{\eta}}
    \right]
        (\det R)^{-1/2}.
\end{eqnarray*}
\end{small}
Changing dummy indices $\{\vartheta,\lambda,\tau,\kappa,\eta,\xi\}\to\{\lambda,\tau,\kappa,\eta,\xi,\vartheta\}$ and $\vartheta\to\varrho$,
\begin{small}
\begin{eqnarray}
    &&\!\!\!\!\!\!\!\!\!\!\!\!\!\!\!\!\!\!\!\!\!\!\!\!\!\!\!\!\!\!\!\!\!\!\!\!\!\!\!\!\!\!\!\!\!
    \int d^Dp\,\,\,
        p_{\lambda}p_{\tau}p_{\kappa}p_{\eta}p_{\xi}p_{\varrho}\,\,\,
     e^{-\frac{1}{2}p\cdot R\cdot p}
     =
    (2\pi)^{D/2}
\nonumber
    \\&&\!\!\!\!\!\!\!\!\!
\times
\left[
        R^{-1}_{{\tau}{\lambda}}R^{-1}_{{\varrho}{\kappa}}
            R^{-1}_{{\eta}{\xi}}
        +R^{-1}_{{\tau}{\lambda}}R^{-1}_{{\kappa}{\xi}}
            R^{-1}_{{\eta}{\varrho}}
        +R^{-1}_{{\tau}{\lambda}}R^{-1}_{{\kappa}{\eta}}
            R^{-1}_{{\xi}{\varrho}}
\right.
\nonumber
    \\&&
        +
        R^{-1}_{{\kappa}{\lambda}}R^{-1}_{{\varrho}{\tau}}
            R^{-1}_{{\eta}{\xi}}
        +R^{-1}_{{\kappa}{\lambda}}R^{-1}_{{\tau}{\xi}}
            R^{-1}_{{\eta}{\varrho}}
        +R^{-1}_{{\kappa}{\lambda}}R^{-1}_{{\tau}{\eta}}
            R^{-1}_{{\xi}{\varrho}}
\nonumber
    \\&&
        +R^{-1}_{{\eta}{\lambda}}R^{-1}_{{\varrho}{\tau}}
            R^{-1}_{{\kappa}{\xi}}
        +R^{-1}_{{\eta}{\lambda}}R^{-1}_{{\tau}{\xi}}
            R^{-1}_{{\kappa}{\varrho}}
        +R^{-1}_{{\eta}{\lambda}}R^{-1}_{{\tau}{\kappa}}
            R^{-1}_{{\xi}{\varrho}}
\nonumber
    \\ &&
        +R^{-1}_{{\xi}{\lambda}}R^{-1}_{{\varrho}{\tau}}
            R^{-1}_{{\kappa}{\eta}}
        +R^{-1}_{{\xi}{\lambda}}R^{-1}_{{\tau}{\eta}}
            R^{-1}_{{\kappa}{\varrho}}
        +R^{-1}_{{\xi}{\lambda}}R^{-1}_{{\tau}{\kappa}}
            R^{-1}_{{\eta}{\varrho}}
\nonumber
    \\ &&
    \left.
        +
        R^{-1}_{{\varrho}{\lambda}}R^{-1}_{{\xi}{\tau}}
            R^{-1}_{{\kappa}{\eta}}
        +R^{-1}_{{\varrho}{\lambda}}R^{-1}_{{\tau}{\eta}}
            R^{-1}_{{\kappa}{\xi}}
        +R^{-1}_{{\varrho}{\lambda}}R^{-1}_{{\tau}{\kappa}}
            R^{-1}_{{\eta}{\xi}}
    \right]
        (\det R)^{-1/2}\label{dpp6}
\end{eqnarray}
\end{small}

\section{$p^8$-integral Identity}
From (\ref{dp7})
\begin{small}
\begin{eqnarray*}
    \int d^Dp\!\!\!&&\!\!\!
    \left(
        \delta_{\lambda\omega}p_{\tau}p_{\kappa}p_{\eta}p_{\xi}p_{\varrho}p_{\vartheta}
        +p_{\lambda}\delta_{\tau\omega}p_{\kappa}p_{\eta}p_{\xi}p_{\varrho}p_{\vartheta}
\right.
\\ &&
        +p_{\lambda}p_{\tau}\delta_{\kappa\omega}p_{\eta}p_{\xi}p_{\varrho}p_{\vartheta}
        +p_{\lambda}p_{\tau}p_{\kappa}\delta_{\eta\omega}p_{\xi}p_{\varrho}p_{\vartheta}
\\ &&
        +p_{\lambda}p_{\tau}p_{\kappa}p_{\eta}\delta_{\xi\omega}p_{\varrho}p_{\vartheta}
        +p_{\lambda}p_{\tau}p_{\kappa}p_{\eta}p_{\xi}\delta_{\varrho\omega}p_{\vartheta}
\\ &&
\left.
        +p_{\lambda}p_{\tau}p_{\kappa}p_{\eta}p_{\xi}p_{\varrho}\delta_{\vartheta\omega}
        -R_{\omega\epsilon}p_{\epsilon}p_{\tau}p_{\kappa}p_{\eta}p_{\xi}p_{\varrho}p_{\vartheta}
    \right)
        e^{-\frac{1}{2}p\cdot R\cdot p}
    =0
\end{eqnarray*}
\end{small}
Transposing and simplifying in the following manner:
\begin{small}
\begin{eqnarray*}
    \int d^Dp\!\!\!&&\!\!\!
        p_{\epsilon}p_{\lambda}p_{\tau}p_{\kappa}p_{\eta}p_{\xi}p_{\varrho}p_{\vartheta}
        e^{-\frac{1}{2}p\cdot R\cdot p}
     =\int d^Dp
    \left(
        R_{\omega\epsilon}^{-1}\delta_{\lambda\omega}p_{\tau}p_{\kappa}p_{\eta}p_{\xi}p_{\varrho}p_{\vartheta}
\right.
\\ &&
        +R_{\omega\epsilon}^{-1}p_{\lambda}\delta_{\tau\omega}p_{\kappa}p_{\eta}p_{\xi}p_{\varrho}p_{\vartheta}
        +R_{\omega\epsilon}^{-1}p_{\lambda}p_{\tau}\delta_{\kappa\omega}p_{\eta}p_{\xi}p_{\varrho}p_{\vartheta}
\\ &&
        +R_{\omega\epsilon}^{-1}p_{\lambda}p_{\tau}p_{\kappa}\delta_{\eta\omega}p_{\xi}p_{\varrho}p_{\vartheta}
        +R_{\omega\epsilon}^{-1}p_{\lambda}p_{\tau}p_{\kappa}p_{\eta}\delta_{\xi\omega}p_{\varrho}p_{\vartheta}
\\ &&
\left.
        +R_{\omega\epsilon}^{-1}p_{\lambda}p_{\tau}p_{\kappa}p_{\eta}p_{\xi}\delta_{\varrho\omega}p_{\vartheta}
        +R_{\omega\epsilon}^{-1}p_{\lambda}p_{\tau}p_{\kappa}p_{\eta}p_{\xi}p_{\varrho}\delta_{\vartheta\omega}
    \right)
        e^{-\frac{1}{2}p\cdot R\cdot p}
\end{eqnarray*}
\begin{eqnarray*}
    \int d^Dp\!\!\!&&\!\!\!
        p_{\epsilon}p_{\lambda}p_{\tau}p_{\kappa}p_{\eta}p_{\xi}p_{\varrho}p_{\vartheta}
        e^{-\frac{1}{2}p\cdot R\cdot p}
     =\int d^Dp
    \left(
        R_{\lambda\epsilon}^{-1}p_{\tau}p_{\kappa}p_{\eta}p_{\xi}p_{\varrho}p_{\vartheta}
\right.
\\ &&
        +R_{\tau\epsilon}^{-1}p_{\lambda}p_{\kappa}p_{\eta}p_{\xi}p_{\varrho}p_{\vartheta}
        +R_{\kappa\epsilon}^{-1}p_{\lambda}p_{\tau}p_{\eta}p_{\xi}p_{\varrho}p_{\vartheta}
\\ &&
        +R_{\eta\epsilon}^{-1}p_{\lambda}p_{\tau}p_{\kappa}p_{\xi}p_{\varrho}p_{\vartheta}
        +R_{\xi\epsilon}^{-1}p_{\lambda}p_{\tau}p_{\kappa}p_{\eta}p_{\varrho}p_{\vartheta}
\\ &&
\left.
        +R_{\varrho\epsilon}^{-1}p_{\lambda}p_{\tau}p_{\kappa}p_{\eta}p_{\xi}p_{\vartheta}
        +R_{\vartheta\epsilon}^{-1}p_{\lambda}p_{\tau}p_{\kappa}p_{\eta}p_{\xi}p_{\varrho}
    \right)
        e^{-\frac{1}{2}p\cdot R\cdot p}
\end{eqnarray*}
\end{small}
Repeated use of (\ref{dpp6})
\begin{small}
\begin{eqnarray*}
    \int d^Dp\!\!\!&&\!\!\!
        p_{\epsilon}p_{\lambda}p_{\tau}p_{\kappa}p_{\eta}p_{\xi}p_{\varrho}p_{\vartheta}
        e^{-\frac{1}{2}p\cdot R\cdot p}
     =(2\pi)^{D/2}
    \\&&\!\!\!\!\!\!\!\!\!\!\!\!\!\!\!\!\!\!
    \left(
        R_{\lambda\epsilon}^{-1}
\right.\left[
        R^{-1}_{{\kappa}{\tau}}R^{-1}_{{\vartheta}{\eta}}
            R^{-1}_{{\xi}{\varrho}}
        +R^{-1}_{{\kappa}{\tau}}R^{-1}_{{\eta}{\varrho}}
            R^{-1}_{{\xi}{\vartheta}}
        +R^{-1}_{{\kappa}{\tau}}R^{-1}_{{\eta}{\xi}}
            R^{-1}_{{\varrho}{\vartheta}}
        +
        R^{-1}_{{\eta}{\tau}}R^{-1}_{{\vartheta}{\kappa}}
            R^{-1}_{{\xi}{\varrho}}
\right.
    \\&&
        +R^{-1}_{{\eta}{\tau}}R^{-1}_{{\kappa}{\varrho}}
            R^{-1}_{{\xi}{\vartheta}}
        +R^{-1}_{{\eta}{\tau}}R^{-1}_{{\kappa}{\xi}}
            R^{-1}_{{\varrho}{\vartheta}}
        +R^{-1}_{{\xi}{\tau}}R^{-1}_{{\vartheta}{\kappa}}
            R^{-1}_{{\eta}{\varrho}}
        +R^{-1}_{{\xi}{\tau}}R^{-1}_{{\kappa}{\varrho}}
            R^{-1}_{{\eta}{\vartheta}}
    \\&&
        +R^{-1}_{{\xi}{\tau}}R^{-1}_{{\kappa}{\eta}}
            R^{-1}_{{\varrho}{\vartheta}}
        +R^{-1}_{{\varrho}{\tau}}R^{-1}_{{\vartheta}{\kappa}}
            R^{-1}_{{\eta}{\xi}}
        +R^{-1}_{{\varrho}{\tau}}R^{-1}_{{\kappa}{\xi}}
            R^{-1}_{{\eta}{\vartheta}}
        +R^{-1}_{{\varrho}{\tau}}R^{-1}_{{\kappa}{\eta}}
            R^{-1}_{{\xi}{\vartheta}}
    \\ &&
    \left.
        +
        R^{-1}_{{\vartheta}{\tau}}R^{-1}_{{\varrho}{\kappa}}
            R^{-1}_{{\eta}{\xi}}
        +R^{-1}_{{\vartheta}{\tau}}R^{-1}_{{\kappa}{\xi}}
            R^{-1}_{{\eta}{\varrho}}
        +R^{-1}_{{\vartheta}{\tau}}R^{-1}_{{\kappa}{\eta}}
            R^{-1}_{{\xi}{\varrho}}
    \right]
    \\&&\!\!\!\!\!\!\!\!\!\!\!\!\!\!\!\!\!\!
        +R_{\tau\epsilon}^{-1}
\left[
        R^{-1}_{{\kappa}{\lambda}}R^{-1}_{{\vartheta}{\eta}}
            R^{-1}_{{\xi}{\varrho}}
        +R^{-1}_{{\kappa}{\lambda}}R^{-1}_{{\eta}{\varrho}}
            R^{-1}_{{\xi}{\vartheta}}
        +R^{-1}_{{\kappa}{\lambda}}R^{-1}_{{\eta}{\xi}}
            R^{-1}_{{\varrho}{\vartheta}}
        +
        R^{-1}_{{\eta}{\lambda}}R^{-1}_{{\vartheta}{\kappa}}
            R^{-1}_{{\xi}{\varrho}}
\right.
    \\&&
        +R^{-1}_{{\eta}{\lambda}}R^{-1}_{{\kappa}{\varrho}}
            R^{-1}_{{\xi}{\vartheta}}
        +R^{-1}_{{\eta}{\lambda}}R^{-1}_{{\kappa}{\xi}}
            R^{-1}_{{\varrho}{\vartheta}}
        +R^{-1}_{{\xi}{\lambda}}R^{-1}_{{\vartheta}{\kappa}}
            R^{-1}_{{\eta}{\varrho}}
        +R^{-1}_{{\xi}{\lambda}}R^{-1}_{{\kappa}{\varrho}}
            R^{-1}_{{\eta}{\vartheta}}
    \\&&
        +R^{-1}_{{\xi}{\lambda}}R^{-1}_{{\kappa}{\eta}}
            R^{-1}_{{\varrho}{\vartheta}}
        +R^{-1}_{{\varrho}{\lambda}}R^{-1}_{{\vartheta}{\kappa}}
            R^{-1}_{{\eta}{\xi}}
        +R^{-1}_{{\varrho}{\lambda}}R^{-1}_{{\kappa}{\xi}}
            R^{-1}_{{\eta}{\vartheta}}
        +R^{-1}_{{\varrho}{\lambda}}R^{-1}_{{\kappa}{\eta}}
            R^{-1}_{{\xi}{\vartheta}}
    \\ &&
    \left.
        +
        R^{-1}_{{\vartheta}{\lambda}}R^{-1}_{{\varrho}{\kappa}}
            R^{-1}_{{\eta}{\xi}}
        +R^{-1}_{{\vartheta}{\lambda}}R^{-1}_{{\kappa}{\xi}}
            R^{-1}_{{\eta}{\varrho}}
        +R^{-1}_{{\vartheta}{\lambda}}R^{-1}_{{\kappa}{\eta}}
            R^{-1}_{{\xi}{\varrho}}
    \right]
    \\&&\!\!\!\!\!\!\!\!\!\!\!\!\!\!\!\!\!\!
        +R_{\kappa\epsilon}^{-1}
\left[
        R^{-1}_{{\tau}{\lambda}}R^{-1}_{{\vartheta}{\eta}}
            R^{-1}_{{\xi}{\varrho}}
        +R^{-1}_{{\tau}{\lambda}}R^{-1}_{{\eta}{\varrho}}
            R^{-1}_{{\xi}{\vartheta}}
        +R^{-1}_{{\tau}{\lambda}}R^{-1}_{{\eta}{\xi}}
            R^{-1}_{{\varrho}{\vartheta}}
        +
        R^{-1}_{{\eta}{\lambda}}R^{-1}_{{\vartheta}{\tau}}
            R^{-1}_{{\xi}{\varrho}}
\right.
    \\&&
        +R^{-1}_{{\eta}{\lambda}}R^{-1}_{{\tau}{\varrho}}
            R^{-1}_{{\xi}{\vartheta}}
        +R^{-1}_{{\eta}{\lambda}}R^{-1}_{{\tau}{\xi}}
            R^{-1}_{{\varrho}{\vartheta}}
        +R^{-1}_{{\xi}{\lambda}}R^{-1}_{{\vartheta}{\tau}}
            R^{-1}_{{\eta}{\varrho}}
        +R^{-1}_{{\xi}{\lambda}}R^{-1}_{{\tau}{\varrho}}
            R^{-1}_{{\eta}{\vartheta}}
    \\&&
        +R^{-1}_{{\xi}{\lambda}}R^{-1}_{{\tau}{\eta}}
            R^{-1}_{{\varrho}{\vartheta}}
        +R^{-1}_{{\varrho}{\lambda}}R^{-1}_{{\vartheta}{\tau}}
            R^{-1}_{{\eta}{\xi}}
        +R^{-1}_{{\varrho}{\lambda}}R^{-1}_{{\tau}{\xi}}
            R^{-1}_{{\eta}{\vartheta}}
        +R^{-1}_{{\varrho}{\lambda}}R^{-1}_{{\tau}{\eta}}
            R^{-1}_{{\xi}{\vartheta}}
    \\ &&
    \left.
        +
        R^{-1}_{{\vartheta}{\lambda}}R^{-1}_{{\varrho}{\tau}}
            R^{-1}_{{\eta}{\xi}}
        +R^{-1}_{{\vartheta}{\lambda}}R^{-1}_{{\tau}{\xi}}
            R^{-1}_{{\eta}{\varrho}}
        +R^{-1}_{{\vartheta}{\lambda}}R^{-1}_{{\tau}{\eta}}
            R^{-1}_{{\xi}{\varrho}}
    \right]
    \\&&\!\!\!\!\!\!\!\!\!\!\!\!\!\!\!\!\!\!
        +R_{\eta\epsilon}^{-1}
\left[
        R^{-1}_{{\tau}{\lambda}}R^{-1}_{{\vartheta}{\kappa}}
            R^{-1}_{{\xi}{\varrho}}
        +R^{-1}_{{\tau}{\lambda}}R^{-1}_{{\kappa}{\varrho}}
            R^{-1}_{{\xi}{\vartheta}}
        +R^{-1}_{{\tau}{\lambda}}R^{-1}_{{\kappa}{\xi}}
            R^{-1}_{{\varrho}{\vartheta}}
        +
        R^{-1}_{{\kappa}{\lambda}}R^{-1}_{{\vartheta}{\tau}}
            R^{-1}_{{\xi}{\varrho}}
\right.
    \\&&
        +R^{-1}_{{\kappa}{\lambda}}R^{-1}_{{\tau}{\varrho}}
            R^{-1}_{{\xi}{\vartheta}}
        +R^{-1}_{{\kappa}{\lambda}}R^{-1}_{{\tau}{\xi}}
            R^{-1}_{{\varrho}{\vartheta}}
        +R^{-1}_{{\xi}{\lambda}}R^{-1}_{{\vartheta}{\tau}}
            R^{-1}_{{\kappa}{\varrho}}
        +R^{-1}_{{\xi}{\lambda}}R^{-1}_{{\tau}{\varrho}}
            R^{-1}_{{\kappa}{\vartheta}}
    \\&&
        +R^{-1}_{{\xi}{\lambda}}R^{-1}_{{\tau}{\kappa}}
            R^{-1}_{{\varrho}{\vartheta}}
        +R^{-1}_{{\varrho}{\lambda}}R^{-1}_{{\vartheta}{\tau}}
            R^{-1}_{{\kappa}{\xi}}
        +R^{-1}_{{\varrho}{\lambda}}R^{-1}_{{\tau}{\xi}}
            R^{-1}_{{\kappa}{\vartheta}}
        +R^{-1}_{{\varrho}{\lambda}}R^{-1}_{{\tau}{\kappa}}
            R^{-1}_{{\xi}{\vartheta}}
    \\ &&
    \left.
        +
        R^{-1}_{{\vartheta}{\lambda}}R^{-1}_{{\varrho}{\tau}}
            R^{-1}_{{\kappa}{\xi}}
        +R^{-1}_{{\vartheta}{\lambda}}R^{-1}_{{\tau}{\xi}}
            R^{-1}_{{\kappa}{\varrho}}
        +R^{-1}_{{\vartheta}{\lambda}}R^{-1}_{{\tau}{\kappa}}
            R^{-1}_{{\xi}{\varrho}}
    \right]
    \\&&\!\!\!\!\!\!\!\!\!\!\!\!\!\!\!\!\!\!
        +R_{\xi\epsilon}^{-1}
\left[
        R^{-1}_{{\tau}{\lambda}}R^{-1}_{{\vartheta}{\kappa}}
            R^{-1}_{{\eta}{\varrho}}
        +R^{-1}_{{\tau}{\lambda}}R^{-1}_{{\kappa}{\varrho}}
            R^{-1}_{{\eta}{\vartheta}}
        +R^{-1}_{{\tau}{\lambda}}R^{-1}_{{\kappa}{\eta}}
            R^{-1}_{{\varrho}{\vartheta}}
        +
        R^{-1}_{{\kappa}{\lambda}}R^{-1}_{{\vartheta}{\tau}}
            R^{-1}_{{\eta}{\varrho}}
\right.
    \\&&
        +R^{-1}_{{\kappa}{\lambda}}R^{-1}_{{\tau}{\varrho}}
            R^{-1}_{{\eta}{\vartheta}}
        +R^{-1}_{{\kappa}{\lambda}}R^{-1}_{{\tau}{\eta}}
            R^{-1}_{{\varrho}{\vartheta}}
        +R^{-1}_{{\eta}{\lambda}}R^{-1}_{{\vartheta}{\tau}}
            R^{-1}_{{\kappa}{\varrho}}
        +R^{-1}_{{\eta}{\lambda}}R^{-1}_{{\tau}{\varrho}}
            R^{-1}_{{\kappa}{\vartheta}}
    \\&&
        +R^{-1}_{{\eta}{\lambda}}R^{-1}_{{\tau}{\kappa}}
            R^{-1}_{{\varrho}{\vartheta}}
        +R^{-1}_{{\varrho}{\lambda}}R^{-1}_{{\vartheta}{\tau}}
            R^{-1}_{{\kappa}{\eta}}
        +R^{-1}_{{\varrho}{\lambda}}R^{-1}_{{\tau}{\eta}}
            R^{-1}_{{\kappa}{\vartheta}}
        +R^{-1}_{{\varrho}{\lambda}}R^{-1}_{{\tau}{\kappa}}
            R^{-1}_{{\eta}{\vartheta}}
    \\ &&
    \left.
        +
        R^{-1}_{{\vartheta}{\lambda}}R^{-1}_{{\varrho}{\tau}}
            R^{-1}_{{\kappa}{\eta}}
        +R^{-1}_{{\vartheta}{\lambda}}R^{-1}_{{\tau}{\eta}}
            R^{-1}_{{\kappa}{\varrho}}
        +R^{-1}_{{\vartheta}{\lambda}}R^{-1}_{{\tau}{\kappa}}
            R^{-1}_{{\eta}{\varrho}}
    \right]
    \\&&\!\!\!\!\!\!\!\!\!\!\!\!\!\!\!\!\!\!
        +R_{\varrho\epsilon}^{-1}
\left[
        R^{-1}_{{\tau}{\lambda}}R^{-1}_{{\vartheta}{\kappa}}
            R^{-1}_{{\eta}{\xi}}
        +R^{-1}_{{\tau}{\lambda}}R^{-1}_{{\kappa}{\xi}}
            R^{-1}_{{\eta}{\vartheta}}
        +R^{-1}_{{\tau}{\lambda}}R^{-1}_{{\kappa}{\eta}}
            R^{-1}_{{\xi}{\vartheta}}
        +
        R^{-1}_{{\kappa}{\lambda}}R^{-1}_{{\vartheta}{\tau}}
            R^{-1}_{{\eta}{\xi}}
\right.
    \\&&
        +R^{-1}_{{\kappa}{\lambda}}R^{-1}_{{\tau}{\xi}}
            R^{-1}_{{\eta}{\vartheta}}
        +R^{-1}_{{\kappa}{\lambda}}R^{-1}_{{\tau}{\eta}}
            R^{-1}_{{\xi}{\vartheta}}
        +R^{-1}_{{\eta}{\lambda}}R^{-1}_{{\vartheta}{\tau}}
            R^{-1}_{{\kappa}{\xi}}
        +R^{-1}_{{\eta}{\lambda}}R^{-1}_{{\tau}{\xi}}
            R^{-1}_{{\kappa}{\vartheta}}
    \\&&
        +R^{-1}_{{\eta}{\lambda}}R^{-1}_{{\tau}{\kappa}}
            R^{-1}_{{\xi}{\vartheta}}
        +R^{-1}_{{\xi}{\lambda}}R^{-1}_{{\vartheta}{\tau}}
            R^{-1}_{{\kappa}{\eta}}
        +R^{-1}_{{\xi}{\lambda}}R^{-1}_{{\tau}{\eta}}
            R^{-1}_{{\kappa}{\vartheta}}
        +R^{-1}_{{\xi}{\lambda}}R^{-1}_{{\tau}{\kappa}}
            R^{-1}_{{\eta}{\vartheta}}
    \\ &&
    \left.
        +
        R^{-1}_{{\vartheta}{\lambda}}R^{-1}_{{\xi}{\tau}}
            R^{-1}_{{\kappa}{\eta}}
        +R^{-1}_{{\vartheta}{\lambda}}R^{-1}_{{\tau}{\eta}}
            R^{-1}_{{\kappa}{\xi}}
        +R^{-1}_{{\vartheta}{\lambda}}R^{-1}_{{\tau}{\kappa}}
            R^{-1}_{{\eta}{\xi}}
    \right]
    \\&&\!\!\!\!\!\!\!\!\!\!\!\!\!\!\!\!\!\!
        +R_{\vartheta\epsilon}^{-1}
     \left[
        R^{-1}_{{\tau}{\lambda}}R^{-1}_{{\varrho}{\kappa}}
            R^{-1}_{{\eta}{\xi}}
        +R^{-1}_{{\tau}{\lambda}}R^{-1}_{{\kappa}{\xi}}
            R^{-1}_{{\eta}{\varrho}}
        +R^{-1}_{{\tau}{\lambda}}R^{-1}_{{\kappa}{\eta}}
            R^{-1}_{{\xi}{\varrho}}
        +
        R^{-1}_{{\kappa}{\lambda}}R^{-1}_{{\varrho}{\tau}}
            R^{-1}_{{\eta}{\xi}}
\right.
    \\&&
        +R^{-1}_{{\kappa}{\lambda}}R^{-1}_{{\tau}{\xi}}
            R^{-1}_{{\eta}{\varrho}}
        +R^{-1}_{{\kappa}{\lambda}}R^{-1}_{{\tau}{\eta}}
            R^{-1}_{{\xi}{\varrho}}
        +R^{-1}_{{\eta}{\lambda}}R^{-1}_{{\varrho}{\tau}}
            R^{-1}_{{\kappa}{\xi}}
        +R^{-1}_{{\eta}{\lambda}}R^{-1}_{{\tau}{\xi}}
            R^{-1}_{{\kappa}{\varrho}}
    \\&&
        +R^{-1}_{{\eta}{\lambda}}R^{-1}_{{\tau}{\kappa}}
            R^{-1}_{{\xi}{\varrho}}
        +R^{-1}_{{\xi}{\lambda}}R^{-1}_{{\varrho}{\tau}}
            R^{-1}_{{\kappa}{\eta}}
        +R^{-1}_{{\xi}{\lambda}}R^{-1}_{{\tau}{\eta}}
            R^{-1}_{{\kappa}{\varrho}}
        +R^{-1}_{{\xi}{\lambda}}R^{-1}_{{\tau}{\kappa}}
            R^{-1}_{{\eta}{\varrho}}
    \\ &&
    \left.\left.
        +
        R^{-1}_{{\varrho}{\lambda}}R^{-1}_{{\xi}{\tau}}
            R^{-1}_{{\kappa}{\eta}}
        +R^{-1}_{{\varrho}{\lambda}}R^{-1}_{{\tau}{\eta}}
            R^{-1}_{{\kappa}{\xi}}
        +R^{-1}_{{\varrho}{\lambda}}R^{-1}_{{\tau}{\kappa}}
            R^{-1}_{{\eta}{\xi}}
    \right]
\right)(\det R)^{-1/2}
\end{eqnarray*}
\end{small}
By distributive property of multiplication and addition,
\begin{small}
\begin{eqnarray*}
    \int d^Dp\!\!\!&&\!\!\!
        p_{\epsilon}p_{\lambda}p_{\tau}p_{\kappa}p_{\eta}p_{\xi}p_{\varrho}p_{\vartheta}
        e^{-\frac{1}{2}p\cdot R\cdot p}
     =(2\pi)^{D/2}
    \\&&
\left[\right.
        R_{\lambda\epsilon}^{-1}R^{-1}_{{\kappa}{\tau}}R^{-1}_{{\vartheta}{\eta}}
            R^{-1}_{{\xi}{\varrho}}
        +R_{\lambda\epsilon}^{-1}R^{-1}_{{\kappa}{\tau}}R^{-1}_{{\eta}{\varrho}}
            R^{-1}_{{\xi}{\vartheta}}
        +R_{\lambda\epsilon}^{-1}R^{-1}_{{\kappa}{\tau}}R^{-1}_{{\eta}{\xi}}
            R^{-1}_{{\varrho}{\vartheta}}
        +
        R_{\lambda\epsilon}^{-1}R^{-1}_{{\eta}{\tau}}R^{-1}_{{\vartheta}{\kappa}}
            R^{-1}_{{\xi}{\varrho}}
    \\&&
        +R_{\lambda\epsilon}^{-1}R^{-1}_{{\eta}{\tau}}R^{-1}_{{\kappa}{\varrho}}
            R^{-1}_{{\xi}{\vartheta}}
        +R_{\lambda\epsilon}^{-1}R^{-1}_{{\eta}{\tau}}R^{-1}_{{\kappa}{\xi}}
            R^{-1}_{{\varrho}{\vartheta}}
        +R_{\lambda\epsilon}^{-1}R^{-1}_{{\xi}{\tau}}R^{-1}_{{\vartheta}{\kappa}}
            R^{-1}_{{\eta}{\varrho}}
        +R_{\lambda\epsilon}^{-1}R^{-1}_{{\xi}{\tau}}R^{-1}_{{\kappa}{\varrho}}
            R^{-1}_{{\eta}{\vartheta}}
    \\&&
        +R_{\lambda\epsilon}^{-1}R^{-1}_{{\xi}{\tau}}R^{-1}_{{\kappa}{\eta}}
            R^{-1}_{{\varrho}{\vartheta}}
        +R_{\lambda\epsilon}^{-1}R^{-1}_{{\varrho}{\tau}}R^{-1}_{{\vartheta}{\kappa}}
            R^{-1}_{{\eta}{\xi}}
        +R_{\lambda\epsilon}^{-1}R^{-1}_{{\varrho}{\tau}}R^{-1}_{{\kappa}{\xi}}
            R^{-1}_{{\eta}{\vartheta}}
        +R_{\lambda\epsilon}^{-1}R^{-1}_{{\varrho}{\tau}}R^{-1}_{{\kappa}{\eta}}
            R^{-1}_{{\xi}{\vartheta}}
    \\ &&
        +
        R_{\lambda\epsilon}^{-1}R^{-1}_{{\vartheta}{\tau}}R^{-1}_{{\varrho}{\kappa}}
            R^{-1}_{{\eta}{\xi}}
        +R_{\lambda\epsilon}^{-1}R^{-1}_{{\vartheta}{\tau}}R^{-1}_{{\kappa}{\xi}}
            R^{-1}_{{\eta}{\varrho}}
        +R_{\lambda\epsilon}^{-1}R^{-1}_{{\vartheta}{\tau}}R^{-1}_{{\kappa}{\eta}}
            R^{-1}_{{\xi}{\varrho}}
        +
        R_{\tau\epsilon}^{-1}R^{-1}_{{\kappa}{\lambda}}R^{-1}_{{\vartheta}{\eta}}
            R^{-1}_{{\xi}{\varrho}}
\\ &&
        +R_{\tau\epsilon}^{-1}R^{-1}_{{\kappa}{\lambda}}R^{-1}_{{\eta}{\varrho}}
            R^{-1}_{{\xi}{\vartheta}}
        +R_{\tau\epsilon}^{-1}R^{-1}_{{\kappa}{\lambda}}R^{-1}_{{\eta}{\xi}}
            R^{-1}_{{\varrho}{\vartheta}}
        +
        R_{\tau\epsilon}^{-1}R^{-1}_{{\eta}{\lambda}}R^{-1}_{{\vartheta}{\kappa}}
            R^{-1}_{{\xi}{\varrho}}
        +R_{\tau\epsilon}^{-1}R^{-1}_{{\eta}{\lambda}}R^{-1}_{{\kappa}{\varrho}}
            R^{-1}_{{\xi}{\vartheta}}
    \\&&
        +R_{\tau\epsilon}^{-1}R^{-1}_{{\eta}{\lambda}}R^{-1}_{{\kappa}{\xi}}
            R^{-1}_{{\varrho}{\vartheta}}
        +R_{\tau\epsilon}^{-1}R^{-1}_{{\xi}{\lambda}}R^{-1}_{{\vartheta}{\kappa}}
            R^{-1}_{{\eta}{\varrho}}
        +R_{\tau\epsilon}^{-1}R^{-1}_{{\xi}{\lambda}}R^{-1}_{{\kappa}{\varrho}}
            R^{-1}_{{\eta}{\vartheta}}
        +R_{\tau\epsilon}^{-1}R^{-1}_{{\xi}{\lambda}}R^{-1}_{{\kappa}{\eta}}
            R^{-1}_{{\varrho}{\vartheta}}
    \\ &&
        +R_{\tau\epsilon}^{-1}R^{-1}_{{\varrho}{\lambda}}R^{-1}_{{\vartheta}{\kappa}}
            R^{-1}_{{\eta}{\xi}}
        +R_{\tau\epsilon}^{-1}R^{-1}_{{\varrho}{\lambda}}R^{-1}_{{\kappa}{\xi}}
            R^{-1}_{{\eta}{\vartheta}}
        +R_{\tau\epsilon}^{-1}R^{-1}_{{\varrho}{\lambda}}R^{-1}_{{\kappa}{\eta}}
            R^{-1}_{{\xi}{\vartheta}}
        +
        R_{\tau\epsilon}^{-1}R^{-1}_{{\vartheta}{\lambda}}R^{-1}_{{\varrho}{\kappa}}
            R^{-1}_{{\eta}{\xi}}
    \\ &&
        +R_{\tau\epsilon}^{-1}R^{-1}_{{\vartheta}{\lambda}}R^{-1}_{{\kappa}{\xi}}
            R^{-1}_{{\eta}{\varrho}}
        +R_{\tau\epsilon}^{-1}R^{-1}_{{\vartheta}{\lambda}}R^{-1}_{{\kappa}{\eta}}
            R^{-1}_{{\xi}{\varrho}}
        +
        R_{\kappa\epsilon}^{-1}R^{-1}_{{\tau}{\lambda}}R^{-1}_{{\vartheta}{\eta}}
            R^{-1}_{{\xi}{\varrho}}
        +R_{\kappa\epsilon}^{-1}R^{-1}_{{\tau}{\lambda}}R^{-1}_{{\eta}{\varrho}}
            R^{-1}_{{\xi}{\vartheta}}
    \\ &&
        +R_{\kappa\epsilon}^{-1}R^{-1}_{{\tau}{\lambda}}R^{-1}_{{\eta}{\xi}}
            R^{-1}_{{\varrho}{\vartheta}}
        +
        R_{\kappa\epsilon}^{-1}R^{-1}_{{\eta}{\lambda}}R^{-1}_{{\vartheta}{\tau}}
            R^{-1}_{{\xi}{\varrho}}
        +R_{\kappa\epsilon}^{-1}R^{-1}_{{\eta}{\lambda}}R^{-1}_{{\tau}{\varrho}}
            R^{-1}_{{\xi}{\vartheta}}
        +R_{\kappa\epsilon}^{-1}R^{-1}_{{\eta}{\lambda}}R^{-1}_{{\tau}{\xi}}
            R^{-1}_{{\varrho}{\vartheta}}
    \\&&
        +R_{\kappa\epsilon}^{-1}R^{-1}_{{\xi}{\lambda}}R^{-1}_{{\vartheta}{\tau}}
            R^{-1}_{{\eta}{\varrho}}
        +R_{\kappa\epsilon}^{-1}R^{-1}_{{\xi}{\lambda}}R^{-1}_{{\tau}{\varrho}}
            R^{-1}_{{\eta}{\vartheta}}
        +R_{\kappa\epsilon}^{-1}R^{-1}_{{\xi}{\lambda}}R^{-1}_{{\tau}{\eta}}
            R^{-1}_{{\varrho}{\vartheta}}
        +R_{\kappa\epsilon}^{-1}R^{-1}_{{\varrho}{\lambda}}R^{-1}_{{\vartheta}{\tau}}
            R^{-1}_{{\eta}{\xi}}
    \\ &&
        +R_{\kappa\epsilon}^{-1}R^{-1}_{{\varrho}{\lambda}}R^{-1}_{{\tau}{\xi}}
            R^{-1}_{{\eta}{\vartheta}}
        +R_{\kappa\epsilon}^{-1}R^{-1}_{{\varrho}{\lambda}}R^{-1}_{{\tau}{\eta}}
            R^{-1}_{{\xi}{\vartheta}}
        +
        R_{\kappa\epsilon}^{-1}R^{-1}_{{\vartheta}{\lambda}}R^{-1}_{{\varrho}{\tau}}
            R^{-1}_{{\eta}{\xi}}
        +R_{\kappa\epsilon}^{-1}R^{-1}_{{\vartheta}{\lambda}}R^{-1}_{{\tau}{\xi}}
            R^{-1}_{{\eta}{\varrho}}
    \\ &&
        +R_{\kappa\epsilon}^{-1}R^{-1}_{{\vartheta}{\lambda}}R^{-1}_{{\tau}{\eta}}
            R^{-1}_{{\xi}{\varrho}}
        +
        R_{\eta\epsilon}^{-1}R^{-1}_{{\tau}{\lambda}}R^{-1}_{{\vartheta}{\kappa}}
            R^{-1}_{{\xi}{\varrho}}
        +R_{\eta\epsilon}^{-1}R^{-1}_{{\tau}{\lambda}}R^{-1}_{{\kappa}{\varrho}}
            R^{-1}_{{\xi}{\vartheta}}
        +R_{\eta\epsilon}^{-1}R^{-1}_{{\tau}{\lambda}}R^{-1}_{{\kappa}{\xi}}
            R^{-1}_{{\varrho}{\vartheta}}
    \\&&
        +
        R_{\eta\epsilon}^{-1}R^{-1}_{{\kappa}{\lambda}}R^{-1}_{{\vartheta}{\tau}}
            R^{-1}_{{\xi}{\varrho}}
        +R_{\eta\epsilon}^{-1}R^{-1}_{{\kappa}{\lambda}}R^{-1}_{{\tau}{\varrho}}
            R^{-1}_{{\xi}{\vartheta}}
        +R_{\eta\epsilon}^{-1}R^{-1}_{{\kappa}{\lambda}}R^{-1}_{{\tau}{\xi}}
            R^{-1}_{{\varrho}{\vartheta}}
        +R_{\eta\epsilon}^{-1}R^{-1}_{{\xi}{\lambda}}R^{-1}_{{\vartheta}{\tau}}
            R^{-1}_{{\kappa}{\varrho}}
    \\ &&
        +R_{\eta\epsilon}^{-1}R^{-1}_{{\xi}{\lambda}}R^{-1}_{{\tau}{\varrho}}
            R^{-1}_{{\kappa}{\vartheta}}
        +R_{\eta\epsilon}^{-1}R^{-1}_{{\xi}{\lambda}}R^{-1}_{{\tau}{\kappa}}
            R^{-1}_{{\varrho}{\vartheta}}
        +R_{\eta\epsilon}^{-1}R^{-1}_{{\varrho}{\lambda}}R^{-1}_{{\vartheta}{\tau}}
            R^{-1}_{{\kappa}{\xi}}
        +R_{\eta\epsilon}^{-1}R^{-1}_{{\varrho}{\lambda}}R^{-1}_{{\tau}{\xi}}
            R^{-1}_{{\kappa}{\vartheta}}
    \\ &&
        +R_{\eta\epsilon}^{-1}R^{-1}_{{\varrho}{\lambda}}R^{-1}_{{\tau}{\kappa}}
            R^{-1}_{{\xi}{\vartheta}}
        +
        R_{\eta\epsilon}^{-1}R^{-1}_{{\vartheta}{\lambda}}R^{-1}_{{\varrho}{\tau}}
            R^{-1}_{{\kappa}{\xi}}
        +R_{\eta\epsilon}^{-1}R^{-1}_{{\vartheta}{\lambda}}R^{-1}_{{\tau}{\xi}}
            R^{-1}_{{\kappa}{\varrho}}
        +R_{\eta\epsilon}^{-1}R^{-1}_{{\vartheta}{\lambda}}R^{-1}_{{\tau}{\kappa}}
            R^{-1}_{{\xi}{\varrho}}
\\ &&
        +
        R_{\xi\epsilon}^{-1}R^{-1}_{{\tau}{\lambda}}R^{-1}_{{\vartheta}{\kappa}}
            R^{-1}_{{\eta}{\varrho}}
        +R_{\xi\epsilon}^{-1}R^{-1}_{{\tau}{\lambda}}R^{-1}_{{\kappa}{\varrho}}
            R^{-1}_{{\eta}{\vartheta}}
        +R_{\xi\epsilon}^{-1}R^{-1}_{{\tau}{\lambda}}R^{-1}_{{\kappa}{\eta}}
            R^{-1}_{{\varrho}{\vartheta}}
        +
        R_{\xi\epsilon}^{-1}R^{-1}_{{\kappa}{\lambda}}R^{-1}_{{\vartheta}{\tau}}
            R^{-1}_{{\eta}{\varrho}}
    \\ &&
        +R_{\xi\epsilon}^{-1}R^{-1}_{{\kappa}{\lambda}}R^{-1}_{{\tau}{\varrho}}
            R^{-1}_{{\eta}{\vartheta}}
        +R_{\xi\epsilon}^{-1}R^{-1}_{{\kappa}{\lambda}}R^{-1}_{{\tau}{\eta}}
            R^{-1}_{{\varrho}{\vartheta}}
        +R_{\xi\epsilon}^{-1}R^{-1}_{{\eta}{\lambda}}R^{-1}_{{\vartheta}{\tau}}
            R^{-1}_{{\kappa}{\varrho}}
        +R_{\xi\epsilon}^{-1}R^{-1}_{{\eta}{\lambda}}R^{-1}_{{\tau}{\varrho}}
            R^{-1}_{{\kappa}{\vartheta}}
    \\ &&
        +R_{\xi\epsilon}^{-1}R^{-1}_{{\eta}{\lambda}}R^{-1}_{{\tau}{\kappa}}
            R^{-1}_{{\varrho}{\vartheta}}
        +R_{\xi\epsilon}^{-1}R^{-1}_{{\varrho}{\lambda}}R^{-1}_{{\vartheta}{\tau}}
            R^{-1}_{{\kappa}{\eta}}
        +R_{\xi\epsilon}^{-1}R^{-1}_{{\varrho}{\lambda}}R^{-1}_{{\tau}{\eta}}
            R^{-1}_{{\kappa}{\vartheta}}
        +R_{\xi\epsilon}^{-1}R^{-1}_{{\varrho}{\lambda}}R^{-1}_{{\tau}{\kappa}}
            R^{-1}_{{\eta}{\vartheta}}
    \\ &&
        +
        R_{\xi\epsilon}^{-1}R^{-1}_{{\vartheta}{\lambda}}R^{-1}_{{\varrho}{\tau}}
            R^{-1}_{{\kappa}{\eta}}
        +R_{\xi\epsilon}^{-1}R^{-1}_{{\vartheta}{\lambda}}R^{-1}_{{\tau}{\eta}}
            R^{-1}_{{\kappa}{\varrho}}
        +R_{\xi\epsilon}^{-1}R^{-1}_{{\vartheta}{\lambda}}R^{-1}_{{\tau}{\kappa}}
            R^{-1}_{{\eta}{\varrho}}
        +
        R_{\varrho\epsilon}^{-1}R^{-1}_{{\tau}{\lambda}}R^{-1}_{{\vartheta}{\kappa}}
            R^{-1}_{{\eta}{\xi}}
    \\ &&
        +R_{\varrho\epsilon}^{-1}R^{-1}_{{\tau}{\lambda}}R^{-1}_{{\kappa}{\xi}}
            R^{-1}_{{\eta}{\vartheta}}
        +R_{\varrho\epsilon}^{-1}R^{-1}_{{\tau}{\lambda}}R^{-1}_{{\kappa}{\eta}}
            R^{-1}_{{\xi}{\vartheta}}
        +
        R_{\varrho\epsilon}^{-1}R^{-1}_{{\kappa}{\lambda}}R^{-1}_{{\vartheta}{\tau}}
            R^{-1}_{{\eta}{\xi}}
        +R_{\varrho\epsilon}^{-1}R^{-1}_{{\kappa}{\lambda}}R^{-1}_{{\tau}{\xi}}
            R^{-1}_{{\eta}{\vartheta}}
    \\ &&
        +R_{\varrho\epsilon}^{-1}R^{-1}_{{\kappa}{\lambda}}R^{-1}_{{\tau}{\eta}}
            R^{-1}_{{\xi}{\vartheta}}
        +R_{\varrho\epsilon}^{-1}R^{-1}_{{\eta}{\lambda}}R^{-1}_{{\vartheta}{\tau}}
            R^{-1}_{{\kappa}{\xi}}
        +R_{\varrho\epsilon}^{-1}R^{-1}_{{\eta}{\lambda}}R^{-1}_{{\tau}{\xi}}
            R^{-1}_{{\kappa}{\vartheta}}
        +R_{\varrho\epsilon}^{-1}R^{-1}_{{\eta}{\lambda}}R^{-1}_{{\tau}{\kappa}}
            R^{-1}_{{\xi}{\vartheta}}
    \\ &&
        +R_{\varrho\epsilon}^{-1}R^{-1}_{{\xi}{\lambda}}R^{-1}_{{\vartheta}{\tau}}
            R^{-1}_{{\kappa}{\eta}}
        +R_{\varrho\epsilon}^{-1}R^{-1}_{{\xi}{\lambda}}R^{-1}_{{\tau}{\eta}}
            R^{-1}_{{\kappa}{\vartheta}}
        +R_{\varrho\epsilon}^{-1}R^{-1}_{{\xi}{\lambda}}R^{-1}_{{\tau}{\kappa}}
            R^{-1}_{{\eta}{\vartheta}}
        +
        R_{\varrho\epsilon}^{-1}R^{-1}_{{\vartheta}{\lambda}}R^{-1}_{{\xi}{\tau}}
            R^{-1}_{{\kappa}{\eta}}
    \\ &&
        +R_{\varrho\epsilon}^{-1}R^{-1}_{{\vartheta}{\lambda}}R^{-1}_{{\tau}{\eta}}
            R^{-1}_{{\kappa}{\xi}}
        +R_{\varrho\epsilon}^{-1}R^{-1}_{{\vartheta}{\lambda}}R^{-1}_{{\tau}{\kappa}}
            R^{-1}_{{\eta}{\xi}}
        +
        R_{\vartheta\epsilon}^{-1}R^{-1}_{{\tau}{\lambda}}R^{-1}_{{\varrho}{\kappa}}
            R^{-1}_{{\eta}{\xi}}
        +R_{\vartheta\epsilon}^{-1}R^{-1}_{{\tau}{\lambda}}R^{-1}_{{\kappa}{\xi}}
            R^{-1}_{{\eta}{\varrho}}
    \\ &&
        +R_{\vartheta\epsilon}^{-1}R^{-1}_{{\tau}{\lambda}}R^{-1}_{{\kappa}{\eta}}
            R^{-1}_{{\xi}{\varrho}}
        +
        R_{\vartheta\epsilon}^{-1}R^{-1}_{{\kappa}{\lambda}}R^{-1}_{{\varrho}{\tau}}
            R^{-1}_{{\eta}{\xi}}
        +R_{\vartheta\epsilon}^{-1}R^{-1}_{{\kappa}{\lambda}}R^{-1}_{{\tau}{\xi}}
            R^{-1}_{{\eta}{\varrho}}
        +R_{\vartheta\epsilon}^{-1}R^{-1}_{{\kappa}{\lambda}}R^{-1}_{{\tau}{\eta}}
            R^{-1}_{{\xi}{\varrho}}
    \\&&
        +R_{\vartheta\epsilon}^{-1}R^{-1}_{{\eta}{\lambda}}R^{-1}_{{\varrho}{\tau}}
            R^{-1}_{{\kappa}{\xi}}
        +R_{\vartheta\epsilon}^{-1}R^{-1}_{{\eta}{\lambda}}R^{-1}_{{\tau}{\xi}}
            R^{-1}_{{\kappa}{\varrho}}
        +R_{\vartheta\epsilon}^{-1}R^{-1}_{{\eta}{\lambda}}R^{-1}_{{\tau}{\kappa}}
            R^{-1}_{{\xi}{\varrho}}
        +R_{\vartheta\epsilon}^{-1}R^{-1}_{{\xi}{\lambda}}R^{-1}_{{\varrho}{\tau}}
            R^{-1}_{{\kappa}{\eta}}
    \\ &&
        +R_{\vartheta\epsilon}^{-1}R^{-1}_{{\xi}{\lambda}}R^{-1}_{{\tau}{\eta}}
            R^{-1}_{{\kappa}{\varrho}}
        +R_{\vartheta\epsilon}^{-1}R^{-1}_{{\xi}{\lambda}}R^{-1}_{{\tau}{\kappa}}
            R^{-1}_{{\eta}{\varrho}}
        +
        R_{\vartheta\epsilon}^{-1}R^{-1}_{{\varrho}{\lambda}}R^{-1}_{{\xi}{\tau}}
            R^{-1}_{{\kappa}{\eta}}
        +R_{\vartheta\epsilon}^{-1}R^{-1}_{{\varrho}{\lambda}}R^{-1}_{{\tau}{\eta}}
            R^{-1}_{{\kappa}{\xi}}
    \\ &&
        +R_{\vartheta\epsilon}^{-1}R^{-1}_{{\varrho}{\lambda}}R^{-1}_{{\tau}{\kappa}}
            R^{-1}_{{\eta}{\xi}}
\left.\right)\left(\det R\right)^{-1/2}
\end{eqnarray*}
\end{small}
Changing dummy indices $\{\epsilon,\lambda,\tau,\kappa,\eta,\xi,\varrho,\vartheta,\}\to\{\lambda,\tau,\kappa,\eta,\xi,\varrho,\vartheta,\epsilon\}$ and $\epsilon\to\omega$,
\begin{small}
\begin{eqnarray}
    \int d^Dp\!\!\!&&\!\!\!
        p_{\lambda}p_{\tau}p_{\kappa}p_{\eta}p_{\xi}p_{\varrho}p_{\vartheta}p_{\omega}
        e^{-\frac{1}{2}p\cdot R\cdot p}
     =(2\pi)^{D/2}
\nonumber
    \\&&
\left[\right.
        R_{\tau\lambda}^{-1}R^{-1}_{{\eta}{\kappa}}R^{-1}_{{\omega}{\xi}}
            R^{-1}_{{\varrho}{\vartheta}}
        +R_{\tau\lambda}^{-1}R^{-1}_{{\eta}{\kappa}}R^{-1}_{{\xi}{\vartheta}}
            R^{-1}_{{\varrho}{\omega}}
        +R_{\tau\lambda}^{-1}R^{-1}_{{\eta}{\kappa}}R^{-1}_{{\xi}{\varrho}}
            R^{-1}_{{\vartheta}{\omega}}
        +
        R_{\tau\lambda}^{-1}R^{-1}_{{\xi}{\kappa}}R^{-1}_{{\omega}{\eta}}
            R^{-1}_{{\varrho}{\vartheta}}
\nonumber
    \\&&
        +R_{\tau\lambda}^{-1}R^{-1}_{{\xi}{\kappa}}R^{-1}_{{\eta}{\vartheta}}
            R^{-1}_{{\varrho}{\omega}}
        +R_{\tau\lambda}^{-1}R^{-1}_{{\xi}{\kappa}}R^{-1}_{{\eta}{\varrho}}
            R^{-1}_{{\vartheta}{\omega}}
        +R_{\tau\lambda}^{-1}R^{-1}_{{\varrho}{\kappa}}R^{-1}_{{\omega}{\eta}}
            R^{-1}_{{\xi}{\vartheta}}
        +R_{\tau\lambda}^{-1}R^{-1}_{{\varrho}{\kappa}}R^{-1}_{{\eta}{\vartheta}}
            R^{-1}_{{\xi}{\omega}}
\nonumber
    \\&&
        +R_{\tau\lambda}^{-1}R^{-1}_{{\varrho}{\kappa}}R^{-1}_{{\eta}{\xi}}
            R^{-1}_{{\vartheta}{\omega}}
        +R_{\tau\lambda}^{-1}R^{-1}_{{\vartheta}{\kappa}}R^{-1}_{{\omega}{\eta}}
            R^{-1}_{{\xi}{\varrho}}
        +R_{\tau\lambda}^{-1}R^{-1}_{{\vartheta}{\kappa}}R^{-1}_{{\eta}{\varrho}}
            R^{-1}_{{\xi}{\omega}}
        +R_{\tau\lambda}^{-1}R^{-1}_{{\vartheta}{\kappa}}R^{-1}_{{\eta}{\xi}}
            R^{-1}_{{\varrho}{\omega}}
\nonumber
    \\ &&
        +
        R_{\tau\lambda}^{-1}R^{-1}_{{\omega}{\kappa}}R^{-1}_{{\vartheta}{\eta}}
            R^{-1}_{{\xi}{\varrho}}
        +R_{\tau\lambda}^{-1}R^{-1}_{{\omega}{\kappa}}R^{-1}_{{\eta}{\varrho}}
            R^{-1}_{{\xi}{\vartheta}}
        +R_{\tau\lambda}^{-1}R^{-1}_{{\omega}{\kappa}}R^{-1}_{{\eta}{\xi}}
            R^{-1}_{{\varrho}{\vartheta}}
        +
        R_{\kappa\lambda}^{-1}R^{-1}_{{\eta}{\tau}}R^{-1}_{{\omega}{\xi}}
            R^{-1}_{{\varrho}{\vartheta}}
\nonumber
\\ &&
        +R_{\kappa\lambda}^{-1}R^{-1}_{{\eta}{\tau}}R^{-1}_{{\xi}{\vartheta}}
            R^{-1}_{{\varrho}{\omega}}
        +R_{\kappa\lambda}^{-1}R^{-1}_{{\eta}{\tau}}R^{-1}_{{\xi}{\varrho}}
            R^{-1}_{{\vartheta}{\omega}}
        +
        R_{\kappa\lambda}^{-1}R^{-1}_{{\xi}{\tau}}R^{-1}_{{\omega}{\eta}}
            R^{-1}_{{\varrho}{\vartheta}}
        +R_{\kappa\lambda}^{-1}R^{-1}_{{\xi}{\tau}}R^{-1}_{{\eta}{\vartheta}}
            R^{-1}_{{\varrho}{\omega}}
\nonumber
    \\&&
        +R_{\kappa\lambda}^{-1}R^{-1}_{{\xi}{\tau}}R^{-1}_{{\eta}{\varrho}}
            R^{-1}_{{\vartheta}{\omega}}
        +R_{\kappa\lambda}^{-1}R^{-1}_{{\varrho}{\tau}}R^{-1}_{{\omega}{\eta}}
            R^{-1}_{{\xi}{\vartheta}}
        +R_{\kappa\lambda}^{-1}R^{-1}_{{\varrho}{\tau}}R^{-1}_{{\eta}{\vartheta}}
            R^{-1}_{{\xi}{\omega}}
        +R_{\kappa\lambda}^{-1}R^{-1}_{{\varrho}{\tau}}R^{-1}_{{\eta}{\xi}}
            R^{-1}_{{\vartheta}{\omega}}
\nonumber
    \\ &&
        +R_{\kappa\lambda}^{-1}R^{-1}_{{\vartheta}{\tau}}R^{-1}_{{\omega}{\eta}}
            R^{-1}_{{\xi}{\varrho}}
        +R_{\kappa\lambda}^{-1}R^{-1}_{{\vartheta}{\tau}}R^{-1}_{{\eta}{\varrho}}
            R^{-1}_{{\xi}{\omega}}
        +R_{\kappa\lambda}^{-1}R^{-1}_{{\vartheta}{\tau}}R^{-1}_{{\eta}{\xi}}
            R^{-1}_{{\varrho}{\omega}}
        +
        R_{\kappa\lambda}^{-1}R^{-1}_{{\omega}{\tau}}R^{-1}_{{\vartheta}{\eta}}
            R^{-1}_{{\xi}{\varrho}}
\nonumber
    \\ &&
        +R_{\kappa\lambda}^{-1}R^{-1}_{{\omega}{\tau}}R^{-1}_{{\eta}{\varrho}}
            R^{-1}_{{\xi}{\vartheta}}
        +R_{\kappa\lambda}^{-1}R^{-1}_{{\omega}{\tau}}R^{-1}_{{\eta}{\xi}}
            R^{-1}_{{\varrho}{\vartheta}}
        +
        R_{\eta\lambda}^{-1}R^{-1}_{{\kappa}{\tau}}R^{-1}_{{\omega}{\xi}}
            R^{-1}_{{\varrho}{\vartheta}}
        +R_{\eta\lambda}^{-1}R^{-1}_{{\kappa}{\tau}}R^{-1}_{{\xi}{\vartheta}}
            R^{-1}_{{\varrho}{\omega}}
\nonumber
    \\ &&
        +R_{\eta\lambda}^{-1}R^{-1}_{{\kappa}{\tau}}R^{-1}_{{\xi}{\varrho}}
            R^{-1}_{{\vartheta}{\omega}}
        +
        R_{\eta\lambda}^{-1}R^{-1}_{{\xi}{\tau}}R^{-1}_{{\omega}{\kappa}}
            R^{-1}_{{\varrho}{\vartheta}}
        +R_{\eta\lambda}^{-1}R^{-1}_{{\xi}{\tau}}R^{-1}_{{\kappa}{\vartheta}}
            R^{-1}_{{\varrho}{\omega}}
        +R_{\eta\lambda}^{-1}R^{-1}_{{\xi}{\tau}}R^{-1}_{{\kappa}{\varrho}}
            R^{-1}_{{\vartheta}{\omega}}
\nonumber
    \\&&
        +R_{\eta\lambda}^{-1}R^{-1}_{{\varrho}{\tau}}R^{-1}_{{\omega}{\kappa}}
            R^{-1}_{{\xi}{\vartheta}}
        +R_{\eta\lambda}^{-1}R^{-1}_{{\varrho}{\tau}}R^{-1}_{{\kappa}{\vartheta}}
            R^{-1}_{{\xi}{\omega}}
        +R_{\eta\lambda}^{-1}R^{-1}_{{\varrho}{\tau}}R^{-1}_{{\kappa}{\xi}}
            R^{-1}_{{\vartheta}{\omega}}
        +R_{\eta\lambda}^{-1}R^{-1}_{{\vartheta}{\tau}}R^{-1}_{{\omega}{\kappa}}
            R^{-1}_{{\xi}{\varrho}}
\nonumber
    \\&&
        +R_{\eta\lambda}^{-1}R^{-1}_{{\vartheta}{\tau}}R^{-1}_{{\kappa}{\varrho}}
            R^{-1}_{{\xi}{\omega}}
        +R_{\eta\lambda}^{-1}R^{-1}_{{\vartheta}{\tau}}R^{-1}_{{\kappa}{\xi}}
            R^{-1}_{{\varrho}{\omega}}
        +
        R_{\eta\lambda}^{-1}R^{-1}_{{\omega}{\tau}}R^{-1}_{{\vartheta}{\kappa}}
            R^{-1}_{{\xi}{\varrho}}
        +R_{\eta\lambda}^{-1}R^{-1}_{{\omega}{\tau}}R^{-1}_{{\kappa}{\varrho}}
            R^{-1}_{{\xi}{\vartheta}}
\nonumber
    \\&&
        +R_{\eta\lambda}^{-1}R^{-1}_{{\omega}{\tau}}R^{-1}_{{\kappa}{\xi}}
            R^{-1}_{{\varrho}{\vartheta}}
        +
        R_{\xi\lambda}^{-1}R^{-1}_{{\kappa}{\tau}}R^{-1}_{{\omega}{\eta}}
            R^{-1}_{{\varrho}{\vartheta}}
        +R_{\xi\lambda}^{-1}R^{-1}_{{\kappa}{\tau}}R^{-1}_{{\eta}{\vartheta}}
            R^{-1}_{{\varrho}{\omega}}
        +R_{\xi\lambda}^{-1}R^{-1}_{{\kappa}{\tau}}R^{-1}_{{\eta}{\varrho}}
            R^{-1}_{{\vartheta}{\omega}}
\nonumber
    \\&&
        +
        R_{\xi\lambda}^{-1}R^{-1}_{{\eta}{\tau}}R^{-1}_{{\omega}{\kappa}}
            R^{-1}_{{\varrho}{\vartheta}}
        +R_{\xi\lambda}^{-1}R^{-1}_{{\eta}{\tau}}R^{-1}_{{\kappa}{\vartheta}}
            R^{-1}_{{\varrho}{\omega}}
        +R_{\xi\lambda}^{-1}R^{-1}_{{\eta}{\tau}}R^{-1}_{{\kappa}{\varrho}}
            R^{-1}_{{\vartheta}{\omega}}
        +R_{\xi\lambda}^{-1}R^{-1}_{{\varrho}{\tau}}R^{-1}_{{\omega}{\kappa}}
            R^{-1}_{{\eta}{\vartheta}}
\nonumber
    \\&&
        +R_{\xi\lambda}^{-1}R^{-1}_{{\varrho}{\tau}}R^{-1}_{{\kappa}{\vartheta}}
            R^{-1}_{{\eta}{\omega}}
        +R_{\xi\lambda}^{-1}R^{-1}_{{\varrho}{\tau}}R^{-1}_{{\kappa}{\eta}}
            R^{-1}_{{\vartheta}{\omega}}
        +R_{\xi\lambda}^{-1}R^{-1}_{{\vartheta}{\tau}}R^{-1}_{{\omega}{\kappa}}
            R^{-1}_{{\eta}{\varrho}}
        +R_{\xi\lambda}^{-1}R^{-1}_{{\vartheta}{\tau}}R^{-1}_{{\kappa}{\varrho}}
            R^{-1}_{{\eta}{\omega}}
\nonumber
    \\&&
        +R_{\xi\lambda}^{-1}R^{-1}_{{\vartheta}{\tau}}R^{-1}_{{\kappa}{\eta}}
            R^{-1}_{{\varrho}{\omega}}
        +
        R_{\xi\lambda}^{-1}R^{-1}_{{\omega}{\tau}}R^{-1}_{{\vartheta}{\kappa}}
            R^{-1}_{{\eta}{\varrho}}
        +R_{\xi\lambda}^{-1}R^{-1}_{{\omega}{\tau}}R^{-1}_{{\kappa}{\varrho}}
            R^{-1}_{{\eta}{\vartheta}}
        +R_{\xi\lambda}^{-1}R^{-1}_{{\omega}{\tau}}R^{-1}_{{\kappa}{\eta}}
            R^{-1}_{{\varrho}{\vartheta}}
\nonumber
\\ &&
        +
        R_{\varrho\lambda}^{-1}R^{-1}_{{\kappa}{\tau}}R^{-1}_{{\omega}{\eta}}
            R^{-1}_{{\xi}{\vartheta}}
        +R_{\varrho\lambda}^{-1}R^{-1}_{{\kappa}{\tau}}R^{-1}_{{\eta}{\vartheta}}
            R^{-1}_{{\xi}{\omega}}
        +R_{\varrho\lambda}^{-1}R^{-1}_{{\kappa}{\tau}}R^{-1}_{{\eta}{\xi}}
            R^{-1}_{{\vartheta}{\omega}}
        +
        R_{\varrho\lambda}^{-1}R^{-1}_{{\eta}{\tau}}R^{-1}_{{\omega}{\kappa}}
            R^{-1}_{{\xi}{\vartheta}}
\nonumber
    \\&&
        +R_{\varrho\lambda}^{-1}R^{-1}_{{\eta}{\tau}}R^{-1}_{{\kappa}{\vartheta}}
            R^{-1}_{{\xi}{\omega}}
        +R_{\varrho\lambda}^{-1}R^{-1}_{{\eta}{\tau}}R^{-1}_{{\kappa}{\xi}}
            R^{-1}_{{\vartheta}{\omega}}
        +R_{\varrho\lambda}^{-1}R^{-1}_{{\xi}{\tau}}R^{-1}_{{\omega}{\kappa}}
            R^{-1}_{{\eta}{\vartheta}}
        +R_{\varrho\lambda}^{-1}R^{-1}_{{\xi}{\tau}}R^{-1}_{{\kappa}{\vartheta}}
            R^{-1}_{{\eta}{\omega}}
\nonumber
    \\&&
        +R_{\varrho\lambda}^{-1}R^{-1}_{{\xi}{\tau}}R^{-1}_{{\kappa}{\eta}}
            R^{-1}_{{\vartheta}{\omega}}
        +R_{\varrho\lambda}^{-1}R^{-1}_{{\vartheta}{\tau}}R^{-1}_{{\omega}{\kappa}}
            R^{-1}_{{\eta}{\xi}}
        +R_{\varrho\lambda}^{-1}R^{-1}_{{\vartheta}{\tau}}R^{-1}_{{\kappa}{\xi}}
            R^{-1}_{{\eta}{\omega}}
        +R_{\varrho\lambda}^{-1}R^{-1}_{{\vartheta}{\tau}}R^{-1}_{{\kappa}{\eta}}
            R^{-1}_{{\xi}{\omega}}
\nonumber
    \\ &&
        +
        R_{\varrho\lambda}^{-1}R^{-1}_{{\omega}{\tau}}R^{-1}_{{\vartheta}{\kappa}}
            R^{-1}_{{\eta}{\xi}}
        +R_{\varrho\lambda}^{-1}R^{-1}_{{\omega}{\tau}}R^{-1}_{{\kappa}{\xi}}
            R^{-1}_{{\eta}{\vartheta}}
        +R_{\varrho\lambda}^{-1}R^{-1}_{{\omega}{\tau}}R^{-1}_{{\kappa}{\eta}}
            R^{-1}_{{\xi}{\vartheta}}
        +
        R_{\vartheta\lambda}^{-1}R^{-1}_{{\kappa}{\tau}}R^{-1}_{{\omega}{\eta}}
            R^{-1}_{{\xi}{\varrho}}
\nonumber
    \\&&
        +R_{\vartheta\lambda}^{-1}R^{-1}_{{\kappa}{\tau}}R^{-1}_{{\eta}{\varrho}}
            R^{-1}_{{\xi}{\omega}}
        +R_{\vartheta\lambda}^{-1}R^{-1}_{{\kappa}{\tau}}R^{-1}_{{\eta}{\xi}}
            R^{-1}_{{\varrho}{\omega}}
        +
        R_{\vartheta\lambda}^{-1}R^{-1}_{{\eta}{\tau}}R^{-1}_{{\omega}{\kappa}}
            R^{-1}_{{\xi}{\varrho}}
        +R_{\vartheta\lambda}^{-1}R^{-1}_{{\eta}{\tau}}R^{-1}_{{\kappa}{\varrho}}
            R^{-1}_{{\xi}{\omega}}
\nonumber
    \\&&
        +R_{\vartheta\lambda}^{-1}R^{-1}_{{\eta}{\tau}}R^{-1}_{{\kappa}{\xi}}
            R^{-1}_{{\varrho}{\omega}}
        +R_{\vartheta\lambda}^{-1}R^{-1}_{{\xi}{\tau}}R^{-1}_{{\omega}{\kappa}}
            R^{-1}_{{\eta}{\varrho}}
        +R_{\vartheta\lambda}^{-1}R^{-1}_{{\xi}{\tau}}R^{-1}_{{\kappa}{\varrho}}
            R^{-1}_{{\eta}{\omega}}
        +R_{\vartheta\lambda}^{-1}R^{-1}_{{\xi}{\tau}}R^{-1}_{{\kappa}{\eta}}
            R^{-1}_{{\varrho}{\omega}}
\nonumber
    \\ &&
        +R_{\vartheta\lambda}^{-1}R^{-1}_{{\varrho}{\tau}}R^{-1}_{{\omega}{\kappa}}
            R^{-1}_{{\eta}{\xi}}
        +R_{\vartheta\lambda}^{-1}R^{-1}_{{\varrho}{\tau}}R^{-1}_{{\kappa}{\xi}}
            R^{-1}_{{\eta}{\omega}}
        +R_{\vartheta\lambda}^{-1}R^{-1}_{{\varrho}{\tau}}R^{-1}_{{\kappa}{\eta}}
            R^{-1}_{{\xi}{\omega}}
        +
        R_{\vartheta\lambda}^{-1}R^{-1}_{{\omega}{\tau}}R^{-1}_{{\varrho}{\kappa}}
            R^{-1}_{{\eta}{\xi}}
\nonumber
    \\&&
        +R_{\vartheta\lambda}^{-1}R^{-1}_{{\omega}{\tau}}R^{-1}_{{\kappa}{\xi}}
            R^{-1}_{{\eta}{\varrho}}
        +R_{\vartheta\lambda}^{-1}R^{-1}_{{\omega}{\tau}}R^{-1}_{{\kappa}{\eta}}
            R^{-1}_{{\xi}{\varrho}}
        +
        R_{\omega\lambda}^{-1}R^{-1}_{{\kappa}{\tau}}R^{-1}_{{\vartheta}{\eta}}
            R^{-1}_{{\xi}{\varrho}}
        +R_{\omega\lambda}^{-1}R^{-1}_{{\kappa}{\tau}}R^{-1}_{{\eta}{\varrho}}
            R^{-1}_{{\xi}{\vartheta}}
\nonumber
    \\&&
        +R_{\omega\lambda}^{-1}R^{-1}_{{\kappa}{\tau}}R^{-1}_{{\eta}{\xi}}
            R^{-1}_{{\varrho}{\vartheta}}
        +
        R_{\omega\lambda}^{-1}R^{-1}_{{\eta}{\tau}}R^{-1}_{{\vartheta}{\kappa}}
            R^{-1}_{{\xi}{\varrho}}
        +R_{\omega\lambda}^{-1}R^{-1}_{{\eta}{\tau}}R^{-1}_{{\kappa}{\varrho}}
            R^{-1}_{{\xi}{\vartheta}}
        +R_{\omega\lambda}^{-1}R^{-1}_{{\eta}{\tau}}R^{-1}_{{\kappa}{\xi}}
            R^{-1}_{{\varrho}{\vartheta}}
\nonumber
    \\&&
        +R_{\omega\lambda}^{-1}R^{-1}_{{\xi}{\tau}}R^{-1}_{{\vartheta}{\kappa}}
            R^{-1}_{{\eta}{\varrho}}
        +R_{\omega\lambda}^{-1}R^{-1}_{{\xi}{\tau}}R^{-1}_{{\kappa}{\varrho}}
            R^{-1}_{{\eta}{\vartheta}}
        +R_{\omega\lambda}^{-1}R^{-1}_{{\xi}{\tau}}R^{-1}_{{\kappa}{\eta}}
            R^{-1}_{{\varrho}{\vartheta}}
        +R_{\omega\lambda}^{-1}R^{-1}_{{\varrho}{\tau}}R^{-1}_{{\vartheta}{\kappa}}
            R^{-1}_{{\eta}{\xi}}
\nonumber
    \\&&
        +R_{\omega\lambda}^{-1}R^{-1}_{{\varrho}{\tau}}R^{-1}_{{\kappa}{\xi}}
            R^{-1}_{{\eta}{\vartheta}}
        +R_{\omega\lambda}^{-1}R^{-1}_{{\varrho}{\tau}}R^{-1}_{{\kappa}{\eta}}
            R^{-1}_{{\xi}{\vartheta}}
        +
        R_{\omega\lambda}^{-1}R^{-1}_{{\vartheta}{\tau}}R^{-1}_{{\varrho}{\kappa}}
            R^{-1}_{{\eta}{\xi}}
        +R_{\omega\lambda}^{-1}R^{-1}_{{\vartheta}{\tau}}R^{-1}_{{\kappa}{\xi}}
            R^{-1}_{{\eta}{\varrho}}
\nonumber
    \\&&
        +R_{\omega\lambda}^{-1}R^{-1}_{{\vartheta}{\tau}}R^{-1}_{{\kappa}{\eta}}
            R^{-1}_{{\xi}{\varrho}}
\left.\right]\left(\det R\right)^{-1/2}
\label{dpp8}
\end{eqnarray}
\end{small}

\section{Flat-space momentum integrals}

\begin{small}
\begin{eqnarray}
\int d^Dp\,\,e^{-p^2\left(\sum_{i=0}^n s_i\right)} &=& \frac{\pi^{D/2}}{2^0\left(\sum_{i=0}^n s_i\right)^{0+D/2}}
\label{pdzer}
\end{eqnarray}
\end{small}
\begin{small}
\begin{eqnarray}
\int d^Dp\,\,p_\lambda\,p_\tau\,e^{-p^2\left(\sum_{i=0}^n s_i\right)}&=&\frac{\pi^{D/2}}{2^1\left(\sum_{i=0}^n s_i\right)^{1+D/2}}\delta_{\lambda\tau}
\label{pdtwo}
\end{eqnarray}
\end{small}
\begin{small}
\begin{eqnarray}
&&\int d^Dp\,\,p_\lambda\,p_\tau\,p_\kappa\,p_\eta\,e^{-p^2\left(\sum_{i=0}^n s_i\right)}
=\frac{\pi^{D/2}}{2^2\left(\sum_{i=0}^n s_i\right)^{2+D/2}}
(\delta_{\lambda\tau}\delta_{\kappa\eta}
+\delta_{\lambda\kappa}\delta_{\tau\eta}
+\delta_{\lambda\eta}\delta_{\tau\kappa})
\nonumber\\
\label{pdfor}
\end{eqnarray}
\vspace{-0.5cm}
\end{small}
\begin{small}
\begin{eqnarray}
&&\!\!\!\!\!\!\!\!\!\!\!\!\!\!\!\!\!\!\!\!\!\!\!\!\!\!\!\!\!\!\!\!\!\!\!\!\!\!\!\!\!\!\!\!\!\!\!\!\!\!\!\!\!\!\!\!\!\!\!\!\!
\int d^Dp\,\,p_\lambda\,p_\tau\,p_\kappa\,p_\eta\,p_\xi\,p_\varrho\,e^{-p^2\left(\sum_{i=0}^n s_i\right)}
\nonumber
\\=&&\!\!\!\!\!\!\!\!
\frac{\pi^{D/2}}{2^3\left(\sum_{i=0}^n s_i\right)^{3+D/2}}
\nonumber\\\,\,\,\,\,\,&&\times
             [{\delta}_{\lambda\tau} {\delta}_{\kappa\eta} {\delta}_{\xi\varrho}
              + {\delta}_{\lambda\tau} {\delta}_{\kappa\xi} {\delta}_{\eta\varrho}
              + {\delta}_{\lambda\tau} {\delta}_{\kappa\varrho} {\delta}_{\eta\xi}
              + {\delta}_{\lambda\kappa} {\delta}_{\tau\eta} {\delta}_{\xi\varrho}
              + {\delta}_{\lambda\kappa} {\delta}_{\tau\xi} {\delta}_{\eta\varrho}
\nonumber\\&&\,\,\,\,\,\,\,\,
              + {\delta}_{\lambda\kappa} {\delta}_{\tau\varrho} {\delta}_{\eta\xi}
              + {\delta}_{\lambda\eta} {\delta}_{\tau\kappa} {\delta}_{\xi\varrho}
              + {\delta}_{\lambda\eta} {\delta}_{\tau\xi} {\delta}_{\kappa\varrho}
              + {\delta}_{\lambda\eta} {\delta}_{\tau\varrho} {\delta}_{\kappa\xi}
              + {\delta}_{\lambda\xi} {\delta}_{\tau\kappa} {\delta}_{\eta\varrho}
\nonumber\\&&\,\,\,\,\,\,\,\,
              + {\delta}_{\lambda\xi} {\delta}_{\tau\eta} {\delta}_{\kappa\varrho}
              + {\delta}_{\lambda\xi} {\delta}_{\tau\varrho} {\delta}_{\kappa\eta}
              + {\delta}_{\lambda\varrho} {\delta}_{\tau\kappa} {\delta}_{\eta\xi}
              + {\delta}_{\lambda\varrho} {\delta}_{\tau\eta} {\delta}_{\kappa\xi}
              + {\delta}_{\lambda\varrho} {\delta}_{\tau\xi} {\delta}_{\kappa\eta}]
\label{pdsix}
\end{eqnarray}
\end{small}
\vspace{-0.5cm}
\begin{small}
\begin{eqnarray}
&&\!\!\!\!\!\!\!\!\!\!\!\!\!\!\!\!\!\!\!\!\!\!\!\!\!\!\!\!\!\!\!\!\!\!
\int d^Dp
       \,p_{\lambda}\,p_{\tau}\,p_{\kappa}\,p_{\eta}\,p_{\xi}\,p_{\varrho}\,p_{\vartheta}\,p_{\omega}
       \,e^{-p^2\left(\sum_{i=0}^n s_i\right)}
\nonumber
    \\&&
\!\!\!\!\!\!\!\!\!\!\!\!\!\!\!\!\!\!\!\!\!\!\!\!\!\!\!\!\!\!\!\!
     =\frac{\pi^{D/2}}{2^4\left(\sum_{i=0}^n s_i\right)^{4+D/2}}
\left[\right.
        \delta_{\tau\lambda}\delta_{{\eta}{\kappa}}\delta_{{\omega}{\xi}}
            \delta_{{\varrho}{\vartheta}}
        +\delta_{\tau\lambda}\delta_{{\eta}{\kappa}}\delta_{{\xi}{\vartheta}}
            \delta_{{\varrho}{\omega}}
        +\delta_{\tau\lambda}\delta_{{\eta}{\kappa}}\delta_{{\xi}{\varrho}}
            \delta_{{\vartheta}{\omega}}
\nonumber
    \\&&
\!\!\!\!\!\!\!\!\!\!\!\!\!\!\!\!\!\!\!\!\!\!\!\!\!\!
        +
        \delta_{\tau\lambda}\delta_{{\xi}{\kappa}}\delta_{{\omega}{\eta}}
            \delta_{{\varrho}{\vartheta}}
        +\delta_{\tau\lambda}\delta_{{\xi}{\kappa}}\delta_{{\eta}{\vartheta}}
            \delta_{{\varrho}{\omega}}
        +\delta_{\tau\lambda}\delta_{{\xi}{\kappa}}\delta_{{\eta}{\varrho}}
            \delta_{{\vartheta}{\omega}}
        +\delta_{\tau\lambda}\delta_{{\varrho}{\kappa}}\delta_{{\omega}{\eta}}
            \delta_{{\xi}{\vartheta}}
        +\delta_{\tau\lambda}\delta_{{\varrho}{\kappa}}\delta_{{\eta}{\vartheta}}
            \delta_{{\xi}{\omega}}
        +\delta_{\tau\lambda}\delta_{{\varrho}{\kappa}}\delta_{{\eta}{\xi}}
            \delta_{{\vartheta}{\omega}}
\nonumber
    \\&&
\!\!\!\!\!\!\!\!\!\!\!\!\!\!\!\!\!\!\!\!\!\!\!\!\!\!
        +\delta_{\tau\lambda}\delta_{{\vartheta}{\kappa}}\delta_{{\omega}{\eta}}
            \delta_{{\xi}{\varrho}}
        +\delta_{\tau\lambda}\delta_{{\vartheta}{\kappa}}\delta_{{\eta}{\varrho}}
            \delta_{{\xi}{\omega}}
        +\delta_{\tau\lambda}\delta_{{\vartheta}{\kappa}}\delta_{{\eta}{\xi}}
            \delta_{{\varrho}{\omega}}
        +
        \delta_{\tau\lambda}\delta_{{\omega}{\kappa}}\delta_{{\vartheta}{\eta}}
            \delta_{{\xi}{\varrho}}
        +\delta_{\tau\lambda}\delta_{{\omega}{\kappa}}\delta_{{\eta}{\varrho}}
            \delta_{{\xi}{\vartheta}}
        +\delta_{\tau\lambda}\delta_{{\omega}{\kappa}}\delta_{{\eta}{\xi}}
            \delta_{{\varrho}{\vartheta}}
\nonumber
    \\&&
\!\!\!\!\!\!\!\!\!\!\!\!\!\!\!\!\!\!\!\!\!\!\!\!\!\!
        +
        \delta_{\kappa\lambda}\delta_{{\eta}{\tau}}\delta_{{\omega}{\xi}}
            \delta_{{\varrho}{\vartheta}}
        +\delta_{\kappa\lambda}\delta_{{\eta}{\tau}}\delta_{{\xi}{\vartheta}}
            \delta_{{\varrho}{\omega}}
        +\delta_{\kappa\lambda}\delta_{{\eta}{\tau}}\delta_{{\xi}{\varrho}}
            \delta_{{\vartheta}{\omega}}
        +
        \delta_{\kappa\lambda}\delta_{{\xi}{\tau}}\delta_{{\omega}{\eta}}
            \delta_{{\varrho}{\vartheta}}
        +\delta_{\kappa\lambda}\delta_{{\xi}{\tau}}\delta_{{\eta}{\vartheta}}
            \delta_{{\varrho}{\omega}}
        +\delta_{\kappa\lambda}\delta_{{\xi}{\tau}}\delta_{{\eta}{\varrho}}
            \delta_{{\vartheta}{\omega}}
\nonumber
    \\&&
\!\!\!\!\!\!\!\!\!\!\!\!\!\!\!\!\!\!\!\!\!\!\!\!\!\!
        +\delta_{\kappa\lambda}\delta_{{\varrho}{\tau}}\delta_{{\omega}{\eta}}
            \delta_{{\xi}{\vartheta}}
        +\delta_{\kappa\lambda}\delta_{{\varrho}{\tau}}\delta_{{\eta}{\vartheta}}
            \delta_{{\xi}{\omega}}
        +\delta_{\kappa\lambda}\delta_{{\varrho}{\tau}}\delta_{{\eta}{\xi}}
            \delta_{{\vartheta}{\omega}}
        +\delta_{\kappa\lambda}\delta_{{\vartheta}{\tau}}\delta_{{\omega}{\eta}}
            \delta_{{\xi}{\varrho}}
        +\delta_{\kappa\lambda}\delta_{{\vartheta}{\tau}}\delta_{{\eta}{\varrho}}
            \delta_{{\xi}{\omega}}
        +\delta_{\kappa\lambda}\delta_{{\vartheta}{\tau}}\delta_{{\eta}{\xi}}
            \delta_{{\varrho}{\omega}}
\nonumber
    \\&&
\!\!\!\!\!\!\!\!\!\!\!\!\!\!\!\!\!\!\!\!\!\!\!\!\!\!
        +
        \delta_{\kappa\lambda}\delta_{{\omega}{\tau}}\delta_{{\vartheta}{\eta}}
            \delta_{{\xi}{\varrho}}
        +\delta_{\kappa\lambda}\delta_{{\omega}{\tau}}\delta_{{\eta}{\varrho}}
            \delta_{{\xi}{\vartheta}}
        +\delta_{\kappa\lambda}\delta_{{\omega}{\tau}}\delta_{{\eta}{\xi}}
            \delta_{{\varrho}{\vartheta}}
        +
        \delta_{\eta\lambda}\delta_{{\kappa}{\tau}}\delta_{{\omega}{\xi}}
            \delta_{{\varrho}{\vartheta}}
        +\delta_{\eta\lambda}\delta_{{\kappa}{\tau}}\delta_{{\xi}{\vartheta}}
            \delta_{{\varrho}{\omega}}
        +\delta_{\eta\lambda}\delta_{{\kappa}{\tau}}\delta_{{\xi}{\varrho}}
            \delta_{{\vartheta}{\omega}}
\nonumber
    \\&&
\!\!\!\!\!\!\!\!\!\!\!\!\!\!\!\!\!\!\!\!\!\!\!\!\!\!
        +
        \delta_{\eta\lambda}\delta_{{\xi}{\tau}}\delta_{{\omega}{\kappa}}
            \delta_{{\varrho}{\vartheta}}
        +\delta_{\eta\lambda}\delta_{{\xi}{\tau}}\delta_{{\kappa}{\vartheta}}
            \delta_{{\varrho}{\omega}}
        +\delta_{\eta\lambda}\delta_{{\xi}{\tau}}\delta_{{\kappa}{\varrho}}
            \delta_{{\vartheta}{\omega}}
        +\delta_{\eta\lambda}\delta_{{\varrho}{\tau}}\delta_{{\omega}{\kappa}}
            \delta_{{\xi}{\vartheta}}
        +\delta_{\eta\lambda}\delta_{{\varrho}{\tau}}\delta_{{\kappa}{\vartheta}}
            \delta_{{\xi}{\omega}}
        +\delta_{\eta\lambda}\delta_{{\varrho}{\tau}}\delta_{{\kappa}{\xi}}
            \delta_{{\vartheta}{\omega}}
\nonumber
    \\&&
\!\!\!\!\!\!\!\!\!\!\!\!\!\!\!\!\!\!\!\!\!\!\!\!\!\!
        +\delta_{\eta\lambda}\delta_{{\vartheta}{\tau}}\delta_{{\omega}{\kappa}}
            \delta_{{\xi}{\varrho}}
        +\delta_{\eta\lambda}\delta_{{\vartheta}{\tau}}\delta_{{\kappa}{\varrho}}
            \delta_{{\xi}{\omega}}
        +\delta_{\eta\lambda}\delta_{{\vartheta}{\tau}}\delta_{{\kappa}{\xi}}
            \delta_{{\varrho}{\omega}}
        +
        \delta_{\eta\lambda}\delta_{{\omega}{\tau}}\delta_{{\vartheta}{\kappa}}
            \delta_{{\xi}{\varrho}}
        +\delta_{\eta\lambda}\delta_{{\omega}{\tau}}\delta_{{\kappa}{\varrho}}
            \delta_{{\xi}{\vartheta}}
        +\delta_{\eta\lambda}\delta_{{\omega}{\tau}}\delta_{{\kappa}{\xi}}
            \delta_{{\varrho}{\vartheta}}
\nonumber
    \\&&
\!\!\!\!\!\!\!\!\!\!\!\!\!\!\!\!\!\!\!\!\!\!\!\!\!\!
        +
        \delta_{\xi\lambda}\delta_{{\kappa}{\tau}}\delta_{{\omega}{\eta}}
            \delta_{{\varrho}{\vartheta}}
        +\delta_{\xi\lambda}\delta_{{\kappa}{\tau}}\delta_{{\eta}{\vartheta}}
            \delta_{{\varrho}{\omega}}
        +\delta_{\xi\lambda}\delta_{{\kappa}{\tau}}\delta_{{\eta}{\varrho}}
            \delta_{{\vartheta}{\omega}}
        +
        \delta_{\xi\lambda}\delta_{{\eta}{\tau}}\delta_{{\omega}{\kappa}}
            \delta_{{\varrho}{\vartheta}}
        +\delta_{\xi\lambda}\delta_{{\eta}{\tau}}\delta_{{\kappa}{\vartheta}}
            \delta_{{\varrho}{\omega}}
        +\delta_{\xi\lambda}\delta_{{\eta}{\tau}}\delta_{{\kappa}{\varrho}}
            \delta_{{\vartheta}{\omega}}
\nonumber
    \\&&
\!\!\!\!\!\!\!\!\!\!\!\!\!\!\!\!\!\!\!\!\!\!\!\!\!\!
        +\delta_{\xi\lambda}\delta_{{\varrho}{\tau}}\delta_{{\omega}{\kappa}}
            \delta_{{\eta}{\vartheta}}
        +\delta_{\xi\lambda}\delta_{{\varrho}{\tau}}\delta_{{\kappa}{\vartheta}}
            \delta_{{\eta}{\omega}}
        +\delta_{\xi\lambda}\delta_{{\varrho}{\tau}}\delta_{{\kappa}{\eta}}
            \delta_{{\vartheta}{\omega}}
        +\delta_{\xi\lambda}\delta_{{\vartheta}{\tau}}\delta_{{\omega}{\kappa}}
            \delta_{{\eta}{\varrho}}
        +\delta_{\xi\lambda}\delta_{{\vartheta}{\tau}}\delta_{{\kappa}{\varrho}}
            \delta_{{\eta}{\omega}}
        +\delta_{\xi\lambda}\delta_{{\vartheta}{\tau}}\delta_{{\kappa}{\eta}}
            \delta_{{\varrho}{\omega}}
\nonumber
    \\&&
\!\!\!\!\!\!\!\!\!\!\!\!\!\!\!\!\!\!\!\!\!\!\!\!\!\!
        +
        \delta_{\xi\lambda}\delta_{{\omega}{\tau}}\delta_{{\vartheta}{\kappa}}
            \delta_{{\eta}{\varrho}}
        +\delta_{\xi\lambda}\delta_{{\omega}{\tau}}\delta_{{\kappa}{\varrho}}
            \delta_{{\eta}{\vartheta}}
        +\delta_{\xi\lambda}\delta_{{\omega}{\tau}}\delta_{{\kappa}{\eta}}
            \delta_{{\varrho}{\vartheta}}
        +
        \delta_{\varrho\lambda}\delta_{{\kappa}{\tau}}\delta_{{\omega}{\eta}}
            \delta_{{\xi}{\vartheta}}
        +\delta_{\varrho\lambda}\delta_{{\kappa}{\tau}}\delta_{{\eta}{\vartheta}}
            \delta_{{\xi}{\omega}}
        +\delta_{\varrho\lambda}\delta_{{\kappa}{\tau}}\delta_{{\eta}{\xi}}
            \delta_{{\vartheta}{\omega}}
\nonumber
    \\&&
\!\!\!\!\!\!\!\!\!\!\!\!\!\!\!\!\!\!\!\!\!\!\!\!\!\!
        +
        \delta_{\varrho\lambda}\delta_{{\eta}{\tau}}\delta_{{\omega}{\kappa}}
            \delta_{{\xi}{\vartheta}}
        +\delta_{\varrho\lambda}\delta_{{\eta}{\tau}}\delta_{{\kappa}{\vartheta}}
            \delta_{{\xi}{\omega}}
        +\delta_{\varrho\lambda}\delta_{{\eta}{\tau}}\delta_{{\kappa}{\xi}}
            \delta_{{\vartheta}{\omega}}
        +\delta_{\varrho\lambda}\delta_{{\xi}{\tau}}\delta_{{\omega}{\kappa}}
            \delta_{{\eta}{\vartheta}}
        +\delta_{\varrho\lambda}\delta_{{\xi}{\tau}}\delta_{{\kappa}{\vartheta}}
            \delta_{{\eta}{\omega}}
        +\delta_{\varrho\lambda}\delta_{{\xi}{\tau}}\delta_{{\kappa}{\eta}}
            \delta_{{\vartheta}{\omega}}
\nonumber
    \\&&
\!\!\!\!\!\!\!\!\!\!\!\!\!\!\!\!\!\!\!\!\!\!\!\!\!\!
        +\delta_{\varrho\lambda}\delta_{{\vartheta}{\tau}}\delta_{{\omega}{\kappa}}
            \delta_{{\eta}{\xi}}
        +\delta_{\varrho\lambda}\delta_{{\vartheta}{\tau}}\delta_{{\kappa}{\xi}}
            \delta_{{\eta}{\omega}}
        +\delta_{\varrho\lambda}\delta_{{\vartheta}{\tau}}\delta_{{\kappa}{\eta}}
            \delta_{{\xi}{\omega}}
        +
        \delta_{\varrho\lambda}\delta_{{\omega}{\tau}}\delta_{{\vartheta}{\kappa}}
            \delta_{{\eta}{\xi}}
        +\delta_{\varrho\lambda}\delta_{{\omega}{\tau}}\delta_{{\kappa}{\xi}}
            \delta_{{\eta}{\vartheta}}
        +\delta_{\varrho\lambda}\delta_{{\omega}{\tau}}\delta_{{\kappa}{\eta}}
            \delta_{{\xi}{\vartheta}}
\nonumber
    \\&&
\!\!\!\!\!\!\!\!\!\!\!\!\!\!\!\!\!\!\!\!\!\!\!\!\!\!
        +
        \delta_{\vartheta\lambda}\delta_{{\kappa}{\tau}}\delta_{{\omega}{\eta}}
            \delta_{{\xi}{\varrho}}
        +\delta_{\vartheta\lambda}\delta_{{\kappa}{\tau}}\delta_{{\eta}{\varrho}}
            \delta_{{\xi}{\omega}}
        +\delta_{\vartheta\lambda}\delta_{{\kappa}{\tau}}\delta_{{\eta}{\xi}}
            \delta_{{\varrho}{\omega}}
        +
        \delta_{\vartheta\lambda}\delta_{{\eta}{\tau}}\delta_{{\omega}{\kappa}}
            \delta_{{\xi}{\varrho}}
        +\delta_{\vartheta\lambda}\delta_{{\eta}{\tau}}\delta_{{\kappa}{\varrho}}
            \delta_{{\xi}{\omega}}
        +\delta_{\vartheta\lambda}\delta_{{\eta}{\tau}}\delta_{{\kappa}{\xi}}
            \delta_{{\varrho}{\omega}}
\nonumber
    \\&&
\!\!\!\!\!\!\!\!\!\!\!\!\!\!\!\!\!\!\!\!\!\!\!\!\!\!
        +\delta_{\vartheta\lambda}\delta_{{\xi}{\tau}}\delta_{{\omega}{\kappa}}
            \delta_{{\eta}{\varrho}}
        +\delta_{\vartheta\lambda}\delta_{{\xi}{\tau}}\delta_{{\kappa}{\varrho}}
            \delta_{{\eta}{\omega}}
        +\delta_{\vartheta\lambda}\delta_{{\xi}{\tau}}\delta_{{\kappa}{\eta}}
            \delta_{{\varrho}{\omega}}
        +\delta_{\vartheta\lambda}\delta_{{\varrho}{\tau}}\delta_{{\omega}{\kappa}}
            \delta_{{\eta}{\xi}}
        +\delta_{\vartheta\lambda}\delta_{{\varrho}{\tau}}\delta_{{\kappa}{\xi}}
            \delta_{{\eta}{\omega}}
        +\delta_{\vartheta\lambda}\delta_{{\varrho}{\tau}}\delta_{{\kappa}{\eta}}
            \delta_{{\xi}{\omega}}
\nonumber
    \\&&
\!\!\!\!\!\!\!\!\!\!\!\!\!\!\!\!\!\!\!\!\!\!\!\!\!\!
        +
        \delta_{\vartheta\lambda}\delta_{{\omega}{\tau}}\delta_{{\varrho}{\kappa}}
            \delta_{{\eta}{\xi}}
        +\delta_{\vartheta\lambda}\delta_{{\omega}{\tau}}\delta_{{\kappa}{\xi}}
            \delta_{{\eta}{\varrho}}
        +\delta_{\vartheta\lambda}\delta_{{\omega}{\tau}}\delta_{{\kappa}{\eta}}
            \delta_{{\xi}{\varrho}}
        +
        \delta_{\omega\lambda}\delta_{{\kappa}{\tau}}\delta_{{\vartheta}{\eta}}
            \delta_{{\xi}{\varrho}}
        +\delta_{\omega\lambda}\delta_{{\kappa}{\tau}}\delta_{{\eta}{\varrho}}
            \delta_{{\xi}{\vartheta}}
        +\delta_{\omega\lambda}\delta_{{\kappa}{\tau}}\delta_{{\eta}{\xi}}
            \delta_{{\varrho}{\vartheta}}
\nonumber
    \\&&
\!\!\!\!\!\!\!\!\!\!\!\!\!\!\!\!\!\!\!\!\!\!\!\!\!\!
        +
        \delta_{\omega\lambda}\delta_{{\eta}{\tau}}\delta_{{\vartheta}{\kappa}}
            \delta_{{\xi}{\varrho}}
        +\delta_{\omega\lambda}\delta_{{\eta}{\tau}}\delta_{{\kappa}{\varrho}}
            \delta_{{\xi}{\vartheta}}
        +\delta_{\omega\lambda}\delta_{{\eta}{\tau}}\delta_{{\kappa}{\xi}}
            \delta_{{\varrho}{\vartheta}}
        +\delta_{\omega\lambda}\delta_{{\xi}{\tau}}\delta_{{\vartheta}{\kappa}}
            \delta_{{\eta}{\varrho}}
        +\delta_{\omega\lambda}\delta_{{\xi}{\tau}}\delta_{{\kappa}{\varrho}}
            \delta_{{\eta}{\vartheta}}
        +\delta_{\omega\lambda}\delta_{{\xi}{\tau}}\delta_{{\kappa}{\eta}}
            \delta_{{\varrho}{\vartheta}}
\nonumber
    \\&&
\!\!\!\!\!\!\!\!\!\!\!\!\!\!\!\!\!\!\!\!\!\!\!\!\!\!
        +\delta_{\omega\lambda}\delta_{{\varrho}{\tau}}\delta_{{\vartheta}{\kappa}}
            \delta_{{\eta}{\xi}}
        +\delta_{\omega\lambda}\delta_{{\varrho}{\tau}}\delta_{{\kappa}{\xi}}
            \delta_{{\eta}{\vartheta}}
        +\delta_{\omega\lambda}\delta_{{\varrho}{\tau}}\delta_{{\kappa}{\eta}}
            \delta_{{\xi}{\vartheta}}
\left.
        +
        \delta_{\omega\lambda}\delta_{{\vartheta}{\tau}}\delta_{{\varrho}{\kappa}}
            \delta_{{\eta}{\xi}}
        +\delta_{\omega\lambda}\delta_{{\vartheta}{\tau}}\delta_{{\kappa}{\xi}}
            \delta_{{\eta}{\varrho}}
        +\delta_{\omega\lambda}\delta_{{\vartheta}{\tau}}\delta_{{\kappa}{\eta}}
            \delta_{{\xi}{\varrho}}
\right]
\nonumber\\
\label{pdeight}
\end{eqnarray}
\end{small}

\chapter{$N$-fold Proper Time Integrations}
In the calculation of the one-loop effective Lagrangian prescribed by\footnote{Here,
\begin{eqnarray}
    G_0
        =\!\!\int_0^\infty\!\!\!ds\,\,
     e^{-\Theta(s)}
        =\!\!\int_0^\infty\!\!\!ds\,\,
     e^{-(m^2+X)s-P(s)-Q(s)\cdot p-\frac{1}{2}p\cdot R(s)\cdot p}
\end{eqnarray}
is expressed in terms of exact trancedental functions $P(s)$, $Q(s)$, and $R(s)$.}

\begin{eqnarray}
    {\cal L}^{(1)}=\sum^\infty_{\ell=0}{\cal L}^{(1)}_\ell
    =\frac{\hbar}{2(2\pi)^D}
    \mbox{Tr}
    \int dX\int d^Dp \,G_0\sum^{\infty}_{\ell=0}
    \left(\Delta_1 G_0\right)^\ell,
\end{eqnarray}
up to $\ell$-th order corrections, there is an ensuing $X$ integration, $p$ integration\footnote{See Appendix C on how to handle momentum integrations in $D$-dimensions.}, and $\ell$-th fold proper-time integrations:
\begin{eqnarray}\label{1pt}
    {\cal L}^{(1)}_0
    =\frac{\hbar}{2(2\pi)^D}
    \mbox{Tr}\!\!
    \int\!\!\!dX\!\!\int\!\!\!d^Dp
    \!\!\int_0^\infty\!\!\!ds_1\,\,
         f\left(s_1\right)
\end{eqnarray}
\begin{eqnarray}\label{2pt}
    {\cal L}^{(1)}_1
    =\frac{\hbar}{2(2\pi)^D}
    \mbox{Tr}\!\!
    \int\!\!\!dX\!\!\int\!\!\!d^Dp
    \!\!\int_0^\infty\!\!\!ds_2\,\,
       \!\!\int_0^\infty\!\!\!ds_1\,\, f\left(s_1,s_2\right)
\end{eqnarray}
\begin{eqnarray}\label{3pt}
    {\cal L}^{(1)}_2
    =\frac{\hbar}{2(2\pi)^D}
    \mbox{Tr}\!\!
    \int\!\!\!dX\!\!\int\!\!\!d^Dp
    \!\!\int_0^\infty\!\!\!ds_3\,\,
      \!\!\int_0^\infty\!\!\!ds_2\,\,
      \!\!\int_0^\infty\!\!\!ds_1\,\,f\left(s_1,s_2,s_3\right)
\end{eqnarray}
for $\ell=1$ to $\ell=3$, respectively.

In this chapter, we present a method of handling two-fold\footnote{The one-fold proper time integration (\ref{1pt}) is handled by an ordinary technique in integral calculus, which we choose to exclude in our presentation here.} (\ref{2pt}) up to three-fold (\ref{3pt}) proper-time integrations which this thesis extensively used.

For purposes of emphasis, we choose to isolate the ensuing proper-time integration. We opt to express the $(\ell+1)$st-fold proper-time integration in the following suppressed notation:
\begin{eqnarray}
f(s_{q_0},\ldots,s_{q_\ell})&\equiv &(q_0,q_1,\ldots,q_{\ell}\,\mathrm{;}\,{\cal A},{\cal B})
\nonumber\\
 &\equiv & \frac{1}{\ell+1}
\prod^\ell_{k=0}
   \int^\infty_0 ds_k
   \frac{\prod^{\ell}_{r=0} {s_{r}}^{q_{r}}}{{\cal A}\left(\sum^{\ell}_{r=0} s_r\right)^{{\cal B}+D/2}}\,e^{-m^2\left(\sum^{\ell}_{r=0}{s_{r}}\right)}
\label{psupp}
\end{eqnarray}
with $\ell+1$ is the number of proper-time variables $s_0,\ldots,s_{\ell}$ and also the number of folds of integration. The $q$'s stand for the powers of the proper-time variables. A semicolon separates these powers of proper-time variables from the resulting $X$-integration number $A$ in the denominator and power $B$ of $(\sum^n_{r=0} s_r)$. With this notation, the proper-time integration will not be written explicitly from then on.

In our technique of integration, we will be using repeatedly for every fold the Euler integral representation of the Gamma function
\begin{equation}\label{GAMMA}
  \Gamma(z+1)
   =\int^\infty_0\,\,dx\,\,e^{-x}x^z.
\end{equation}

Note: To remove the ambiguity of the variable $D$ either as a variable of integration or as dimensions of integration space, we denote (in this chapter only) the latter meaning as $D'$. Otherwise in the rest of the chapters, $D$ denotes dimensions.

\section{Two-fold Proper Time Integration}
The two-fold proper time integral of the form
\begin{equation}
  \int^\infty_0\,\,ds_1
    \int^\infty_0\,\,ds_0\,\,
    f(s_0,s_1)e^{-m^2(s_0+s_1)}
\end{equation}
can be evaluated through the simultaneous substitution
\begin{equation}
\begin{array}{l}
    B=s_1+s_0 \\
    A=-s_1+s_0.
  \end{array}
\end{equation}
Eliminating $s_0$ and $s_1$ in favor of $A$ and $B$ through
\begin{equation}
\begin{array}{l}
    s_1=(1/2)(B-A) \\
    s_0=(1/2)(B+A).
  \end{array}
\end{equation}
Its Jacobian is given by
\begin{equation}
  \frac{\partial (s_0,s_1)}{\partial (A,B)}
    =\frac{1}{2}.
\end{equation}
Changing variables entails a change in the limits of integration. These are
\begin{equation}
\begin{array}{c}
    -B \leq A \leq B\\
    0 \leq B < \infty.
  \end{array}
\end{equation}
Thus\footnote{The limits of the variable of integration $A$ are obtained simply by setting $s_1=0$ for the upper limit and $s_0=0$ for the lower limit. The limits of last variable of integration (for this case $B$) are $\infty$ and $0$ for upper and lower limit, respectively.},
\begin{eqnarray}
  \int^\infty_0\,\,ds_1
    \int^\infty_0\,\,ds_0\,\,
    f(s_0,s_1)e^{-m^2(s_0+s_1)}
=
   \frac{1}{2}
   \int^\infty_0\!\!dB
     \int^{B}_{-B}\!\!dA\,\,
   f(A,B)e^{-m^2B}\label{2ptime}
\end{eqnarray}

The following are some of the two-fold proper-time integrations used in this thesis:
\begin{eqnarray}
\!\!\!\!\!\!\!\!\!\!\!\!\!\!\!\!\!\!\!\!\!\!\!\!\!\!\!
(1,0;1,1)
\!\!\!\!&\equiv &\!\!\!\!
 \int^\infty_0\!\!\!ds_1\!\!
 \int^\infty_0\!\!\! ds_0\,
  \frac{s_0 e^{-m^2(s_0+s_1)}}{1(s_0+s_1)^{1+D/2}}
 \!=\!\frac{1}{2}\!\int^\infty_0\!\!\!ds\frac{e^{-m^2s}}{s^{-1+D/2}}
 \!=\!\frac{1}{2}\frac{\Gamma[2-D/2]}{m^{4-D/2}}
\end{eqnarray}
\begin{eqnarray}
\!\!\!\!\!\!\!\!\!\!\!\!\!\!\!\!\!\!\!\!\!\!\!\!\!\!\!
(2,0;2,2)
\!\!\!\!&\equiv &\!\!\!\!
 \int^\infty_0\!\!\!ds_1\!\!
 \int^\infty_0\!\!\!ds_0\,
  \frac{{s_0}^2 e^{-m^2(s_0+s_1)}}{2(s_0+s_1)^{2+D/2}}
  \!=\!\frac{1}{2}\!\cdot\!\frac{1}{3}\!\int^\infty_0\!\!\!ds\frac{e^{-m^2s}}{s^{-1+D/2}}
 \!=\!\frac{1}{6}\frac{\Gamma[2-D/2]}{m^{4-D/2}}
\end{eqnarray}
\begin{eqnarray}
\!\!\!\!\!\!\!\!\!\!\!\!\!\!\!\!\!\!\!\!\!\!\!\!\!\!\!
(3,0;4,3)
\!\!\!\!&\equiv &\!\!\!\!
 \int^\infty_0\!\!\!ds_1\!\!
 \int^\infty_0\!\!\!ds_0\,
  \frac{{s_0}^3 e^{-m^2(s_0+s_1)}}{4(s_0+s_1)^{3+D/2}}
 \!=\!\frac{1}{2}\!\cdot\!\frac{1}{8}\!\int^\infty_0\!\!\!ds\frac{e^{-m^2s}}{s^{-1+D/2}}
 \!=\!\frac{1}{16}\!\frac{\Gamma[2-D/2]}{m^{4-D/2}}
\end{eqnarray}
where
\begin{equation}\label{2fss}
  f(s_0,s_1)
   =\frac{{s_0}^p}{2^{p-1}(s_0+s_1)^{p+D'/2}}\,\,\,\,\,\,\,\mbox{for}\,\,p=1,2,3
\end{equation}
respectively.
In all cases,
\begin{eqnarray}
\int^\infty_0\!\!\!ds\frac{e^{-m^2s}}{s^{-1+D/2}}
 =\frac{\Gamma[2-D/2]}{m^{4-D/2}}.
\end{eqnarray}
This uses the Euler integral representation of the Gamma function (\ref{GAMMA}).

\section{Example: Calculating (3,0;4,3) using Mathematica}

\begin{small}$In[1]:=$\end{small}\vspace{-0.75cm}
\begin{verbatim}
     (1/2)*Integrate[(((1/2)*(b+a))^3)*(((1/2)*(b-a))^0)*
                       Exp[-b*m^2]/(4*b^(3+D/2)),{a,-b,b}]
\end{verbatim}
\begin{small}$Out[1]:=$\end{small}\vspace{-0.5cm}
\begin{eqnarray*}
\frac{1}{16\,{}}{{\mathrm{b}}^{1 - {\frac{{\mathrm{D}}}{2}}}}\,{\mathrm{E}}^{-{\mathrm{b}}\,{{\mathrm{m}}^2}}
\end{eqnarray*}
\begin{small}$In[1]:=$\end{small}\vspace{-0.8cm}
\begin{verbatim}
     Integrate[%,{b,0,Infinity}]
\end{verbatim}
\begin{small}$Out[1]:=$\end{small}\vspace{-0.3cm}
\begin{eqnarray*}
\mbox{If}\left[\mbox{Re}[{{\mathrm{m}}^2}] > 0 \,\,\,\&\&\,\,\, \mbox{Re}[{\mathrm{D}}] < 4,\,\,\,\,
  {\frac{{{{({\mathrm{m}}^2)}}^{{\frac{\mathrm{D}}{2}}}}\,\mbox{Gamma}[2 - {\frac{{\mathrm{D}}}{2}}]}{16
\,{{\mathrm{m}}^4}}},\,\,\,\,
  \int _{0}^{\infty}{\frac{{{\mathrm{b}}^{1 - {\frac{{\mathrm{D}}}{2}}}}\,{\mathrm{E}}^{-{\mathrm{b}}\,{{\mathrm{m}}^2}}}{16\,{}}}\,\mathrm{db}\right]
\end{eqnarray*}

\section{Three-fold Proper Time Integration}
The three-fold proper time integral of the form
\begin{equation}
  \int^\infty_0\,\,ds_2
  \int^\infty_0\,\,ds_1
    \int^\infty_0\,\,ds_0\,\,
    f(s_0,s_1,s_2)e^{-m^2(s_0+s_1+s_2)}
\end{equation}
can be evaluated through the simultaneous substitution
\begin{equation}
\begin{array}{l}
    C=s_2+s_1+s_0 \\
    B=-s_2+(1/2)(s_1+s_0) \\
    A=-s_1+s_0.
  \end{array}
\end{equation}
Eliminating $s_0$, $s_1$ and $s_2$ in favor of $A$, $B$ and $C$ through
\begin{equation}
\begin{array}{l}
    s_2=(1/3)(C-2B)\\
    s_1=(1/3)(C+B)-(1/2)A \\
    s_0=(1/3)(C+B)+(1/2)A.
  \end{array}
\end{equation}
Its Jacobian is given by
\begin{equation}
  \frac{\partial (s_0,s_1,s_2)}{\partial (A,B,C)}
    =\frac{1}{3}.
\end{equation}
Changing variables entails a change in the limits of integration. These are
\begin{equation}
\begin{array}{c}
    -\frac{2}{3}(C+B) \leq A \leq \frac{2}{3}(C+B)\\
    -C \leq B \leq C/2\\
    0 \leq C < \infty.
  \end{array}
\end{equation}
Thus\footnote{The limits of the variable of integration $A$ are obtained simply by setting $s_1=0$ for the upper limit and $s_0=0$ for the lower limit. For the variable of integration $B$, these are obtained by setting $s_2=0$ and $s_0+s_1=0$, respectively for upper and lower limit of $\int\,dB$. The limits of last variable of integration (for this case $C$) are $\infty$ and $0$ for upper and lower limit, respectively.},
\begin{eqnarray}
  &&
  \int^\infty_0\,\,ds_2
  \int^\infty_0\,\,ds_1
    \int^\infty_0\,\,ds_0\,\,
    f(s_0,s_1,s_2)e^{-m^2(s_0+s_1+s_2)}
 \nonumber\\
  &=&
   \frac{1}{3}
   \int^\infty_0\!\!dC
   \int^{C/2}_{-C}\!\!dB
     \int^{\frac{2}{3}(C+B)}_{\frac{2}{3}(-C-B)}\!\!dA\,\,
   f(A,B,C)e^{-m^2C}\label{3ptime}
\end{eqnarray}

The following are some of the three-fold proper-time integration used in this thesis:
\begin{scriptsize}
\begin{eqnarray}
(1,1,0;2,2)
\!\!\!\!&\equiv &\!\!\!\!
 \int^\infty_0\!\!\!\!\!\!\!ds_2\!\!
 \int^\infty_0\!\!\!\!\!\!\!ds_1\!\!
 \int^\infty_0\!\!\!\!\!\!\! ds_0\,
  \frac{{s_0}{s_1} e^{-m^2(s_0+s_1+s_2)}}{2(s_0+s_1+s_2)^{2+D/2}}
 \!=\!\frac{1}{3}\cdot\frac{1}{16}\!\int^\infty_0\!\!\!ds\frac{e^{-m^2s}}{s^{-2+D/2}}
 \!=\!\frac{1}{48}\frac{\Gamma[3-D/2]}{m^{6-D/2}}
\\
(2,1,0;4,3)
\!\!\!\!&\equiv &\!\!\!\!
 \int^\infty_0\!\!\!\!\!\!\!ds_2\!\!
 \int^\infty_0\!\!\!\!\!\!\!ds_1\!\!
 \int^\infty_0\!\!\!\!\!\!\!ds_0\,
  \frac{{s_0}^2{s_1} e^{-m^2(s_0+s_1+s_2)}}{4(s_0+s_1+s_2)^{3+D/2}}
 \!=\!\frac{1}{3}\cdot\frac{1}{80}\!\int^\infty_0\!\!\!ds\frac{e^{-m^2s}}{s^{-2+D/2}}
 \!=\!\frac{1}{240}\frac{\Gamma[3-D/2]}{m^{6-D/2}}
\\
(1,2,0;4,3)
\!\!\!\!&\equiv &\!\!\!\!
 \int^\infty_0\!\!\!\!\!\!\!ds_2\!\!
 \int^\infty_0\!\!\!\!\!\!\!ds_1\!\!
 \int^\infty_0\!\!\!\!\!\!\!ds_0\,
  \frac{{s_0}{s_1}^2 e^{-m^2(s_0+s_1+s_2)}}{4(s_0+s_1+s_2)^{3+D/2}}
 \!=\!\frac{1}{3}\cdot\frac{1}{80}\!\int^\infty_0\!\!\!ds\frac{e^{-m^2s}}{s^{-2+D/2}}
 \!=\!\frac{1}{240}\frac{\Gamma[3-D/2]}{m^{6-D/2}}
\\
(2,2,0;8,4)
\!\!\!\!&\equiv &\!\!\!\!
 \int^\infty_0\!\!\!\!\!\!\!ds_2\!\!
 \int^\infty_0\!\!\!\!\!\!\!ds_1\!\!
 \int^\infty_0\!\!\!\!\!\!\!ds_0\,\!
  \frac{{s_0}^2{s_1}^2 e^{-m^2(s_0+s_1+s_2)}}{8(s_0+s_1+s_2)^{4+D/2}}
 \!=\!\frac{1}{3}\cdot\frac{1}{480}\!\int^\infty_0\!\!\!\!ds\frac{e^{-m^2s}}{s^{-2+D/2}}
 \!=\!\frac{1}{1440}\frac{\Gamma[3-D/2]}{m^{6-D/2}}
\end{eqnarray}
\end{scriptsize}
where
\begin{eqnarray}
  f(s_0,s_1,s_2)=
  \frac{{s_0}^p{s_1}^q e^{-m^2(s_0+s_1+s_2)}}{2^{(p+q)-1}(s_0+s_1+s_2)^{4+D/2}}
  \,\,\,\,\,\,\,\mbox{for}\,\,p+q=2,3,4
\end{eqnarray}
In all cases,
\begin{eqnarray}
 \int^\infty_0\!\!\!ds\frac{e^{-m^2s}}{s^{-2+D/2}}
 =\frac{\Gamma[3-D/2]}{m^{6-D/2}}
\end{eqnarray}
respectively.

This again uses the Euler integral representation of the Gamma function (\ref{GAMMA}).

\section{Example: Calculating (2,2,0;8,4) using Mathematica}

\begin{small}$In[1]:=$\end{small}\vspace{-0.73cm}
\begin{verbatim}
     (1/3)*Integrate[((((c+b)/3+a/2)^2)*(((c+b)/3-a/2)^2)
                     *((c/3-(2/3)*b)^0)*Exp[-c*m^2])/(8*c^(4+D/2)),
                                        {a,-(2/3)*(c+b),(2/3)*(c+b)}]
\end{verbatim}
\begin{small}$Out[1]:=$\end{small}\vspace{-0.4cm}
\begin{eqnarray*}
{\frac{4}{10935\,{}}\,{{\mathrm{c}}^{-4 - {\frac{{\mathrm{D}}}{2}}}}\,{{\left( {\mathrm{b}} + {\mathrm{c}} \right) }^5}\,{\mathrm{E}}^{-\mathrm{c}\,{{\mathrm{m}}^2}}}
\end{eqnarray*}

\newpage
\hspace{-.8cm}
\begin{small}$In[2]:=$\end{small}\vspace{-0.8cm}
\begin{verbatim}
    Integrate[%,{b,-c,c/2}]
\end{verbatim}
\begin{small}$Out[2]:=$\end{small}\vspace{-0.5cm}
\begin{eqnarray*}
\frac{1}{1440\,{}}\,{\mathrm{c}^{2 - {\frac{\mathrm{D}}{2}}}}\,\mathrm{E}^{-{\mathrm{c}}\,{\mathrm{m}^2}}
\end{eqnarray*}
\begin{small}$In[3]:=$\end{small}\vspace{-0.8cm}
\begin{verbatim}
    Integrate[%,{c,0,Infinity}]
\end{verbatim}
\begin{small}$Out[3]:=$\end{small}\vspace{-0.4cm}
\begin{eqnarray*}
\mbox{If}\left[\mbox{Re}[{{\mathrm{m}}^2}] > 0 \,\,\&\&\,\, \mbox{Re}[{\mathrm{D}}] < 6,\,\,\,\,
  {\frac{{{{({\mathrm{m}}^2)}}^{{\frac{{\mathrm{D}}}{2}}}}\,\mbox{Gamma}[3 - {\frac{{\mathrm{D}}}{2}}]}{1440
\,{{\mathrm{m}}^6}}},\,\,\,\,
  \int _{0}^{\infty}{\frac{{{\mathrm{c}}^{2 - {\frac{{\mathrm{D}}}{2}}}}\,{\mathrm{E}}^{-{\mathrm{c}}\,{{\mathrm{m}}^2}}}{1440\,{}}}
\,\mathrm{dc}\right]
\end{eqnarray*}

\section{Proper-time Integral and Gamma Function Equivalence}

\begin{eqnarray}
\int^\infty_0\!\!\!ds\frac{e^{-m^2s}}{s^{-k+D/2}}
 \equiv\frac{\Gamma[(k+1)-D/2]}{m^{2(k+1)-D/2}}.
\end{eqnarray}
with $k=0,1,2,3,4,5,6,7$.


\chapter{Scratchwork in the Calculation of ${\cal L}^{(1)[8]}$}

\section{Maxima Code Implemented}

\subsection{First-order Corrections}
In this subsection, we will solve for (\ref{1[8]}):
\begin{eqnarray}
{\cal L}^{(1)[8]}_1
    = \frac{\hbar}{2(2\pi)^D}\mbox{Tr}\int dX
    \int d^Dp 
    \,\,\, G_{\emptyset_{\mathrm{t}}}[8]
\end{eqnarray}
with $[8]$ given in (\ref{eit}).

\begin{scriptsize}
\begin{verbatim}
/*Momentum and Proper-time Integration*/
/*s3=(3,0;1,1)*/ s3: 1/4;
/*p2s4=(4,0;2,2)*/ p2: d_lt; s4: 1/10;
/*p4s5=(5,0;4,3)*/ p4: (d_lt*d_kh + d_lk*d_th + d_lh*d_tk); s5: 1/24;
/*p6s6=(6,0;8,4)*/
p6: (d_lt*d_kh*d_gx +d_lt*d_kg*d_hx +d_lt*d_kx*d_hg +d_lk*d_th*d_gx +d_lk*d_tg*d_hx
    +d_lk*d_tx*d_hg +d_lh*d_tk*d_gx +d_lh*d_tg*d_kx +d_lh*d_tx*d_kg +d_lg*d_tk*d_hx
    +d_lg*d_th*d_kx +d_lg*d_tx*d_kh +d_lx*d_tk*d_hg +d_lx*d_th*d_kg +d_lx*d_tg*d_kh);
s6: 1/56;

/*eit:*/
expand(
-(1/3)*s3*D_abX*Y2_ab
+(1/4)*s3*Y2_ab*(D_cbY_ca - D_bcY_ac)
+(1/8)*s3*(D_cbY_ab*(D_ddY_ac + D_dcY_ad + D_cdY_ad)
        + D_bcY_ab*(D_ddY_ac + D_dcY_ad + D_cdY_ad)+ D_ccY_ab*(D_ddY_ab + D_dbY_ad + D_bdY_ad)
        + D_dcY_ab*((D_dcY_ab + D_cdY_ab) + (D_dbY_ac + D_bdY_ac) + (D_cbY_ad + D_bcY_ad)))
+(1/3)*s3*D_aX*(D_bbaX+D_abbX+D_babX) +(1/3)*s3*D_aX*(2*D_cY_bc*Y_ba+2*(D_cY_ba+D_aY_bc)*Y_bc)
+(2/3)*p2*s4*(D_laX*Y2_at+D_atX*Y2_al)
+(1/2)*p2*s4 *((D_baY_la - D_abY_al)*Y2_bt + (D_tbY_ab - D_btY_ba)*Y2_al)
-(1/3)*p2*s4*((D_baY_la + D_abY_la + D_aaY_lb)*Y2_bt + (D_baY_lt + D_btY_la + D_tbY_la)*Y2_ab)
-(1/4)*p2*s4*
  (+(D_alY_bt + D_tlY_ba)*D_ccY_ba +(D_ccY_ba + D_caY_bc)*D_taY_bl+(D_lcY_ba + D_clY_ba)*D_taY_bc
   +(D_ctY_ba + D_ctY_ba)*D_caY_bl +(D_lcY_ba + D_caY_bl)*D_acY_bt+(D_alY_bc + D_acY_bl)*D_ctY_ba
   +(D_tcY_bl + D_ctY_bl)*D_aaY_bc +(D_caY_ba + D_acY_ba + D_caY_ba)*D_tcY_bl
   +(D_ccY_ba + D_caY_bc + D_acY_bc)*D_ltY_ba +(D_tcY_bc + D_ctY_bc + D_ccY_bt)*D_aaY_bl
   +(D_caY_bl + D_alY_bc + D_caY_bl)*D_tcY_ba +(D_taY_bc + D_caY_bt + D_atY_bc)*D_caY_bl
   +(D_lcY_ba + D_clY_ba)*(D_tcY_ba + D_ctY_ba)
   +(D_caY_ba + D_acY_ba)*(D_tlY_bc + D_ctY_bl + D_clY_bt)
   +(D_laY_ba + D_alY_ba)*(D_tcY_bc + D_ctY_bc + D_ccY_bt))
-(2/3)*p2*s4*D_aX*((D_tlaX+D_talX+D_atlX)+2*D_tY_bl*Y_ba
                  +(D_lY_ba+D_aY_bl)*Y_bt+(D_tY_ba+D_aY_bt)*Y_bl)
+(2/3)*p4*s5*(D_akY_lt*Y2_ah + D_haY_lt*Y2_ak + D_hkY_la*Y2_at)
+(1/2)*p4*s5*((D_haY_bl + D_alY_bh)*D_ktY_ba + (D_haY_bt + D_btY_ah)*D_kaY_bl
            + (D_haY_bk + D_ahY_bk)*D_atY_bl + (D_taY_bl + D_atY_bl + D_ltY_ba)*D_hkY_ba
            + (D_laY_ba + D_alY_ba + D_aaY_bl)*D_hkY_bt + (D_aaY_bh + D_ahY_ba + D_haY_ba)*D_ktY_bl)
- p6*s6*D_ktY_al*D_gxY_ah);
/*save file in Maxima output*/
save(eit1st8,%);
\end{verbatim}
\end{scriptsize}

Manipulating the Maxima generated output like sorting (using MS Excel) according to string length and labeling each eight mass-dimensional invariant accordingly for later reference, we have:

\begin{scriptsize}
\begin{verbatim}
a1: ((1 12) D_abbX D_aX
a2: ((1 12) D_aX D_babX
a3: ((1 12) D_aX D_bbaX
a4: ((-1 12) D_abX Y2_ab
a5: ((1 16) D_cbY_ca Y2_ab
a6: ((-1 16) D_bcY_ac Y2_ab
a7: ((1 32) D_bcY_ab D_cdY_ad
a8: ((1 32) D_bcY_ab D_dcY_ad
a9: ((1 32) D_bcY_ab D_ddY_ac
a10: ((1 32) D_bcY_ad D_dcY_ab
a11: ((1 32) D_bdY_ac D_dcY_ab
a12: ((1 32) D_bdY_ad D_ccY_ab
a13: ((1 32) D_cbY_ab D_cdY_ad
a14: ((1 32) D_cbY_ab D_dcY_ad
a15: ((1 32) D_cbY_ab D_ddY_ac
a16: ((1 32) D_cbY_ad D_dcY_ab
a17: ((1 32) D_ccY_ab D_dbY_ad
a18: ((1 32) D_ccY_ab D_ddY_ab
a19: ((1 32) D_cdY_ab D_dcY_ab
a20: ((1 32) D_dbY_ac D_dcY_ab
a21: ((1 32) D_dcY_ab D_dcY_ab
a22: ((1 15) D_atX d_lt Y2_al
a23: ((1 15) D_laX d_lt Y2_at
a24: ((1 6) D_aX D_aY_bc Y_bc
a25: ((1 6) D_aX D_cY_ba Y_bc
a26: ((1 6) D_aX D_cY_bc Y_ba
a27: ((-1 15) D_atlX D_aX d_lt
a28: ((-1 15) D_aX d_lt D_talX
a29: ((-1 15) D_aX d_lt D_tlaX
a30: ((1 20) d_lt D_tbY_ab Y2_al
a31: ((1 60) D_baY_la d_lt Y2_bt
a32: ((-1 20) D_abY_al d_lt Y2_bt
a33: ((-1 20) D_btY_ba d_lt Y2_al
a34: ((-1 30) D_aaY_lb d_lt Y2_bt
a35: ((-1 30) D_abY_la d_lt Y2_bt
a36: ((-1 30) D_baY_lt d_lt Y2_ab
a37: ((-1 30) D_btY_la d_lt Y2_ab
a38: ((-1 30) d_lt D_tbY_la Y2_ab
a39: ((-1 20) D_caY_ba d_lt D_tcY_bl
a40: ((-1 20) D_caY_bl D_ctY_ba d_lt
a41: ((-1 20) D_caY_bl d_lt D_tcY_ba
a42: ((-1 40) D_aaY_bc D_ctY_bl d_lt
a43: ((-1 40) D_aaY_bc d_lt D_tcY_bl
a44: ((-1 40) D_aaY_bl D_ccY_bt d_lt
a45: ((-1 40) D_aaY_bl D_ctY_bc d_lt
a46: ((-1 40) D_aaY_bl d_lt D_tcY_bc
a47: ((-1 40) D_acY_ba D_clY_bt d_lt
a48: ((-1 40) D_acY_ba D_ctY_bl d_lt
a49: ((-1 40) D_acY_ba d_lt D_tcY_bl
a50: ((-1 40) D_acY_ba d_lt D_tlY_bc
a51: ((-1 40) D_acY_bc d_lt D_ltY_ba
a52: ((-1 40) D_acY_bl D_ctY_ba d_lt
a53: ((-1 40) D_acY_bt D_caY_bl d_lt
a54: ((-1 40) D_acY_bt D_lcY_ba d_lt
a55: ((-1 40) D_alY_ba D_ccY_bt d_lt
a56: ((-1 40) D_alY_ba D_ctY_bc d_lt
a57: ((-1 40) D_alY_ba d_lt D_tcY_bc
a58: ((-1 40) D_alY_bc D_ctY_ba d_lt
a59: ((-1 40) D_alY_bc d_lt D_tcY_ba
a60: ((-1 40) D_alY_bt D_ccY_ba d_lt
a61: ((-1 40) D_atY_bc D_caY_bl d_lt
a62: ((-1 40) D_caY_ba D_clY_bt d_lt
a63: ((-1 40) D_caY_ba D_ctY_bl d_lt
a64: ((-1 40) D_caY_ba d_lt D_tlY_bc
a65: ((-1 40) D_caY_bc d_lt D_ltY_ba
a66: ((-1 40) D_caY_bc d_lt D_taY_bl
a67: ((-1 40) D_caY_bl D_caY_bt d_lt
a68: ((-1 40) D_caY_bl d_lt D_taY_bc
a69: ((-1 40) D_ccY_ba d_lt D_ltY_ba
a70: ((-1 40) D_ccY_ba d_lt D_taY_bl
a71: ((-1 40) D_ccY_ba d_lt D_tlY_ba
a72: ((-1 40) D_ccY_bt D_laY_ba d_lt
a73: ((-1 40) D_clY_ba D_ctY_ba d_lt
a74: ((-1 40) D_clY_ba d_lt D_taY_bc
a75: ((-1 40) D_clY_ba d_lt D_tcY_ba
a76: ((-1 40) D_ctY_ba D_lcY_ba d_lt
a77: ((-1 40) D_ctY_bc D_laY_ba d_lt
a78: ((-1 40) D_laY_ba d_lt D_tcY_bc
a79: ((-1 40) D_lcY_ba d_lt D_taY_bc
a80: ((-1 40) D_lcY_ba d_lt D_tcY_ba
a81: ((-1 15) D_aX D_aY_bl d_lt Y_bt
a82: ((-1 15) D_aX D_aY_bt d_lt Y_bl
a83: ((-1 15) D_aX d_lt D_lY_ba Y_bt
a84: ((-1 15) D_aX d_lt D_tY_ba Y_bl
a85: ((-2 15) D_aX d_lt D_tY_bl Y_ba
a86: ((1 36) D_akY_lt d_kh d_lt Y2_ah
a87: ((1 36) D_akY_lt d_lh d_tk Y2_ah
a88: ((1 36) D_akY_lt d_lk d_th Y2_ah
a89: ((1 36) D_haY_lt d_kh d_lt Y2_ak
a90: ((1 36) D_haY_lt d_lh d_tk Y2_ak
a91: ((1 36) D_haY_lt d_lk d_th Y2_ak
a92: ((1 36) D_hkY_la d_kh d_lt Y2_at
a93: ((1 36) D_hkY_la d_lh d_tk Y2_at
a94: ((1 36) D_hkY_la d_lk d_th Y2_at
a95: ((1 48) D_aaY_bh d_kh D_ktY_bl d_lt
a96: ((1 48) D_aaY_bh D_ktY_bl d_lh d_tk
a97: ((1 48) D_aaY_bh D_ktY_bl d_lk d_th
a98: ((1 48) D_aaY_bl D_hkY_bt d_kh d_lt
a99: ((1 48) D_aaY_bl D_hkY_bt d_lh d_tk
a100: ((1 48) D_aaY_bl D_hkY_bt d_lk d_th
a101: ((1 48) D_alY_bh d_kh D_ktY_ba d_lt
a102: ((1 48) D_alY_bh D_ktY_ba d_lh d_tk
a103: ((1 48) D_alY_bh D_ktY_ba d_lk d_th
a104: ((1 48) D_ahY_ba d_kh D_ktY_bl d_lt
a105: ((1 48) D_ahY_ba D_ktY_bl d_lh d_tk
a106: ((1 48) D_ahY_ba D_ktY_bl d_lk d_th
a107: ((1 48) D_ahY_bk D_atY_bl d_kh d_lt
a108: ((1 48) D_ahY_bk D_atY_bl d_lh d_tk
a109: ((1 48) D_ahY_bk D_atY_bl d_lk d_th
a110: ((1 48) D_alY_ba D_hkY_bt d_kh d_lt
a111: ((1 48) D_alY_ba D_hkY_bt d_lh d_tk
a112: ((1 48) D_alY_ba D_hkY_bt d_lk d_th
a113: ((1 48) D_atY_bl D_haY_bk d_kh d_lt
a114: ((1 48) D_atY_bl D_haY_bk d_lh d_tk
a115: ((1 48) D_atY_bl D_haY_bk d_lk d_th
a116: ((1 48) D_atY_bl D_hkY_ba d_kh d_lt
a117: ((1 48) D_atY_bl D_hkY_ba d_lh d_tk
a118: ((1 48) D_atY_bl D_hkY_ba d_lk d_th
a119: ((1 48) D_btY_ah D_kaY_bl d_kh d_lt
a120: ((1 48) D_btY_ah D_kaY_bl d_lh d_tk
a121: ((1 48) D_btY_ah D_kaY_bl d_lk d_th
a122: ((1 48) D_haY_ba d_kh D_ktY_bl d_lt
a123: ((1 48) D_haY_ba D_ktY_bl d_lh d_tk
a124: ((1 48) D_haY_ba D_ktY_bl d_lk d_th
a125: ((1 48) D_haY_bl d_kh D_ktY_ba d_lt
a126: ((1 48) D_haY_bl D_ktY_ba d_lh d_tk
a127: ((1 48) D_haY_bl D_ktY_ba d_lk d_th
a128: ((1 48) D_haY_bt D_kaY_bl d_kh d_lt
a129: ((1 48) D_haY_bt D_kaY_bl d_lh d_tk
a130: ((1 48) D_haY_bt D_kaY_bl d_lk d_th
a131: ((1 48) D_hkY_ba d_kh d_lt D_ltY_ba
a132: ((1 48) D_hkY_ba d_kh d_lt D_taY_bl
a133: ((1 48) D_hkY_ba d_lh D_ltY_ba d_tk
a134: ((1 48) D_hkY_ba d_lh D_taY_bl d_tk
a135: ((1 48) D_hkY_ba d_lk D_ltY_ba d_th
a136: ((1 48) D_hkY_ba d_lk D_taY_bl d_th
a137: ((1 48) D_hkY_bt d_kh D_laY_ba d_lt
a138: ((1 48) D_hkY_bt D_laY_ba d_lh d_tk
a139: ((1 48) D_hkY_bt D_laY_ba d_lk d_th
a140: ((-1 56) d_gx D_gxY_ah d_kh D_ktY_al d_lt
a141: ((-1 56) d_gx D_gxY_ah D_ktY_al d_lh d_tk
a142: ((-1 56) d_gx D_gxY_ah D_ktY_al d_lk d_th
a143: ((-1 56) D_gxY_ah d_hg D_ktY_al d_kx d_lt
a144: ((-1 56) D_gxY_ah d_hg D_ktY_al d_lk d_tx
a145: ((-1 56) D_gxY_ah d_hg D_ktY_al d_lx d_tk
a146: ((-1 56) D_gxY_ah d_hx d_kg D_ktY_al d_lt
a147: ((-1 56) D_gxY_ah d_hx D_ktY_al d_lg d_tk
a148: ((-1 56) D_gxY_ah d_hx D_ktY_al d_lk d_tg
a149: ((-1 56) D_gxY_ah d_kg D_ktY_al d_lh d_tx
a150: ((-1 56) D_gxY_ah d_kg D_ktY_al d_lx d_th
a151: ((-1 56) D_gxY_ah d_kh D_ktY_al d_lg d_tx
a152: ((-1 56) D_gxY_ah d_kh D_ktY_al d_lx d_tg
a153: ((-1 56) D_gxY_ah D_ktY_al d_kx d_lg d_th
a154: ((-1 56) D_gxY_ah D_ktY_al d_kx d_lh d_tg
\end{verbatim}
\end{scriptsize}

Renaming dummy indices following an arbitrary pattern so as to combine similar type of invariants, we have:

\begin{scriptsize}
\begin{verbatim}
DX DY Y type of invariant
a24:+(1/6)*D_aX*D_aY_bc*Y_bc
a25:+(1/6)*D_aX*D_cY_ba*Y_bc
a26:+(1/6)*D_aX*D_cY_bc*Y_ba
a81:-(1/15)*D_aX*D_aY_bc*Y_bc
a82:-(1/15)*D_aX*D_aY_bc*Y_bc
a83:-(1/15)*D_aX*D_cY_ba*Y_bc
a84:-(1/15)*D_aX*D_cY_ba*Y_bc
a85:-(2/15)*D_aX*D_cY_bc*Y_ba
Output:=========================
+(1/30) D_aX D_cY_bc Y_ba +(1/30) D_aX D_aY_bc Y_bc +(1/30) D_aX D_cY_ba Y_bc
================================
DDX YY type of invariant
a4:-(1/12)*D_abX*Y2_ab
a22:+(1/15)*D_abX*Y2_ab
a23:+(1/15)*D_baX*Y2_ab=+(1/15)*D_abX*Y2_ba=+(1/15)*D_abX*Y2_ab
Output:=========================
+(1/20) D_abX Y2_ab
================================
DDY YY
a5: +(1/16)*D_cbY_ca*Y2_ab
a6: -(1/16)*D_bcY_ac*Y2_ab
a30: +(1/20)*D_bcY_ac*Y2_ab
a31: +(1/60)*D_caY_ba*Y2_cb
a32: -(1/20)*D_acY_ab*Y2_cb
a33: -(1/20)*D_cbY_ca*Y2_ab
a34: -(1/30)*D_aaY_bc*Y2_cb
a35: -(1/30)*D_acY_ba*Y2_cb
a36: -(1/30)*D_baY_cc*Y2_ab
a37: -(1/30)*D_bcY_ca*Y2_ab
a38: -(1/30)*D_cbY_ca*Y2_ab
a86: +(1/36)*D_abY_cc*Y2_ab
a87: +(1/36)*D_acY_bc*Y2_ab
a88: +(1/36)*D_acY_cb*Y2_ab
a89: +(1/36)*D_baY_cc*Y2_ab
a90: +(1/36)*D_caY_cb*Y2_ab
a91: +(1/36)*D_caY_bc*Y2_ab
a92: +(1/36)*D_ccY_ba*Y2_ab
a93: +(1/36)*D_cbY_ca*Y2_ab
a94: +(1/36)*D_bcY_ca*Y2_ab
Output:=========================
+(1/36) D_abY_cc Y2_ab +(2/45) D_acY_bc Y2_ab +(1/36) D_acY_cb Y2_ab -(1/180) D_baY_cc Y2_ab
-(1/80) D_bcY_ac Y2_ab -(1/180) D_bcY_ca Y2_ab -(1/180) D_caY_bc Y2_ab -(1/45) D_caY_cb Y2_ab
+(1/144) D_cbY_ca Y2_ab -(1/180) D_ccY_ba Y2_ab
================================
DDX DDX type of invariant
a1:+(1/12) D_abbX D_aX = -(1/12) D_abX D_abX
a2:+(1/12) D_babX D_aX = -(1/12) D_baX D_abX
a3:+(1/12) D_bbaX D_aX = -(1/12) D_bbX D_aaX
a27:-(1/15) D_abbX D_aX = +(1/15) D_abX D_abX
a28:-(1/15) D_babX D_aX = +(1/15) D_baX D_abX
a29:-(1/15) D_bbaX D_aX = +(1/15) D_bbX D_aaX
Output:=========================
-(1/60) D_abX D_abX-(1/60) D_abX D_baX-(1/60) D_aaX D_bbX
================================
 type of invariant
DDY DDY
a7:+(1/32) D_bcY_ab D_cdY_ad
a8:+(1/32) D_bcY_ab D_dcY_ad
a9:+(1/32) D_bcY_ab D_ddY_ac
a10:+(1/32) D_bcY_ad D_dcY_ab
a11:+(1/32) D_cdY_ab D_dbY_ac
a12:+(1/32) D_bdY_ad D_ccY_ab
a13:+(1/32) D_cbY_ab D_cdY_ad
a14:+(1/32) D_cbY_ab D_dcY_ad
a15:+(1/32) D_cbY_ab D_ddY_ac
a16:+(1/32) D_cbY_ad D_dcY_ab
a17:+(1/32) D_ccY_ab D_dbY_ad
a18:+(1/32) D_ccY_ab D_ddY_ab
a19:+(1/32) D_dcY_ab D_cdY_ab
a20:+(1/32) D_cdY_ab D_cbY_ad
a21:+(1/32) D_cdY_ab D_cdY_ab
a39:-(1/20) D_dbY_ab D_cdY_ac
a40:-(1/20) D_dcY_ab D_dbY_ac
a41:-(1/20) D_dcY_ab D_bdY_ac
a42:-(1/40) D_ccY_ab D_bdY_ad
a43:-(1/40) D_ccY_ab D_dbY_ad
a44:-(1/40) D_ddY_ab D_ccY_ab
a45:-(1/40) D_ddY_ab D_cbY_ac
a46:-(1/40) D_ddY_ab D_bcY_ac
a47:-(1/40) D_bcY_ab D_cdY_ad
a48:-(1/40) D_bcY_ab D_cdY_ad
a49:-(1/40) D_bcY_ab D_dcY_ad
a50:-(1/40) D_bcY_ab D_ddY_ac
a51:-(1/40) D_cbY_ab D_ddY_ac
a52:-(1/40) D_dcY_ab D_cbY_ad
a53:-(1/40) D_dcY_ab D_cdY_ab
a54:-(1/40) D_dcY_ab D_bcY_ad
a55:-(1/40) D_bdY_ab D_ccY_ad
a56:-(1/40) D_bdY_ab D_cdY_ac
a57:-(1/40) D_bdY_ab D_dcY_ac
a58:-(1/40) D_cdY_ab D_bdY_ac
a59:-(1/40) D_cdY_ab D_dbY_ac
a60:-(1/40) D_cbY_ab D_ddY_ac
a61:-(1/40) D_cdY_ab D_bcY_ad
a62:-(1/40) D_cbY_ab D_cdY_ad
a63:-(1/40) D_cbY_ab D_cdY_ad
a64:-(1/40) D_cbY_ab D_ddY_ac
a65:-(1/40) D_bcY_ab D_ddY_ac
a66:-(1/40) D_bcY_ab D_dcY_ad
a67:-(1/40) D_cdY_ab D_cdY_ab
a68:-(1/40) D_cdY_ab D_bdY_ac
a69:-(1/40) D_ccY_ab D_ddY_ab
a70:-(1/40) D_ccY_ab D_dbY_ad
a71:-(1/40) D_ccY_ab D_ddY_ab
a72:-(1/40) D_ccY_ab D_bdY_ad
a73:-(1/40) D_cdY_ab D_cdY_ab
a74:-(1/40) D_cdY_ab D_dbY_ac
a75:-(1/40) D_cdY_ab D_dcY_ab
a76:-(1/40) D_cdY_ab D_dcY_ab
a77:-(1/40) D_bdY_ab D_dcY_ac
a78:-(1/40) D_cbY_ab D_cdY_ad
a79:-(1/40) D_dcY_ab D_dbY_ac
a80:-(1/40) D_dcY_ab D_dcY_ab
a95:+(1/48) D_ccY_ab D_bdY_ad
a96:+(1/48) D_ccY_ab D_ddY_ab
a97:+(1/48) D_ccY_ab D_dbY_ad
a98:+(1/48) D_ccY_ab D_ddY_ab
a99:+(1/48) D_ccY_ab D_bdY_ad
a100:+(1/48) D_ccY_ab D_dbY_ad
a101:+(1/48) D_cdY_ab D_bdY_ac
a102:+(1/48) D_cbY_ab D_ddY_ac
a103:+(1/48) D_cdY_ab D_dbY_ac
a104:+(1/48) D_bcY_ab D_cdY_ad
a105:+(1/48) D_bcY_ab D_ddY_ac
a106:+(1/48) D_bcY_ab D_dcY_ad
a107:+(1/48) D_cbY_ab D_cdY_ad
a108:+(1/48) D_cdY_ab D_cbY_ad
a109:+(1/48) D_cdY_ab D_cdY_ab
a110:+(1/48) D_bcY_ab D_ddY_ac
a111:+(1/48) D_bcY_ab D_cdY_ad
a112:+(1/48) D_bcY_ab D_dcY_ad
a113:+(1/48) D_cbY_ab D_dcY_ad
a114:+(1/48) D_cdY_ab D_bcY_ad
a115:+(1/48) D_cdY_ab D_dcY_ab
a116:+(1/48) D_cbY_ab D_ddY_ac
a117:+(1/48) D_cdY_ab D_bdY_ac
a118:+(1/48) D_cdY_ab D_dbY_ac
a119:+(1/48) D_bcY_ad D_daY_bc
a120:+(1/48) D_cdY_ab D_daY_cb
a121:+(1/48) D_cbY_ab D_daY_cd
a122:+(1/48) D_cbY_ab D_cdY_ad
a123:+(1/48) D_cbY_ab D_ddY_ac
a124:+(1/48) D_cbY_ab D_dcY_ad
a125:+(1/48) D_dcY_ab D_dbY_ac
a126:+(1/48) D_bcY_ab D_ddY_ac
a127:+(1/48) D_dcY_ab D_bdY_ac
a128:+(1/48) D_dcY_ab D_dcY_ab
a129:+(1/48) D_dcY_ab D_bcY_ad
a130:+(1/48) D_bcY_ab D_dcY_ad
a131:+(1/48) D_ccY_ab D_ddY_ab
a132:+(1/48) D_ccY_ab D_dbY_ad
a133:+(1/48) D_dcY_ab D_dcY_ab
a134:+(1/48) D_dcY_ab D_cbY_ad
a135:+(1/48) D_dcY_ab D_cdY_ab
a136:+(1/48) D_dcY_ab D_dbY_ac
a137:+(1/48) D_ccY_ab D_bdY_ad
a138:+(1/48) D_cbY_ab D_cdY_ad
a139:+(1/48) D_bcY_ab D_cdY_ad
a140:-(1/56) D_ccY_ab D_bdY_ad
a141:-(1/56) D_ccY_ab D_ddY_ab
a142:-(1/56) D_ccY_ab D_dbY_ad
a143:-(1/56) D_bcY_ab D_cdY_ad
a144:-(1/56) D_bcY_ab D_dcY_ad
a145:-(1/56) D_bcY_ab D_ddY_ac
a146:-(1/56) D_cbY_ab D_cdY_ad
a147:-(1/56) D_cbY_ab D_ddY_ac
a148:-(1/56) D_cbY_ab D_dcY_ad
a149:-(1/56) D_dcY_ab D_dcY_ab
a150:-(1/56) D_dcY_ab D_dbY_ac
a151:-(1/56) D_dcY_ab D_bcY_ad
a152:-(1/56) D_dcY_ab D_bdY_ac
a153:-(1/56) D_dcY_ab D_cbY_ad
a154:-(1/56) D_dcY_ab D_cdY_ab
Output:=========================
 +(1/48) (D_daY_bc) (D_bcY_ad) +(29/1120) (D_bdY_ad) (D_ccY_ab) +(-1/40) (D_bdY_ab) (D_ccY_ad)
 +(-1/240) (D_bcY_ad) (D_cdY_ab) +(-1/120) (D_bdY_ac) (D_cdY_ab) +(5/96) (D_cbY_ad) (D_cdY_ab)
 +(1/480) (D_cdY_ab) (D_cdY_ab) +(-1/40) (D_bdY_ab) (D_cdY_ac) +(29/1120) (D_bcY_ab) (D_cdY_ad)
 +(1/1120) (D_cbY_ab) (D_cdY_ad) +(1/48) (D_cdY_ab) (D_daY_cb) +(1/48) (D_cbY_ab) (D_daY_cd)
 +(-1/20) (D_cdY_ac) (D_dbY_ab) +(11/480) (D_cdY_ab) (D_dbY_ac) +(29/1120) (D_ccY_ab) (D_dbY_ad)
 +(31/3360) (D_bcY_ad) (D_dcY_ab) +(-79/1680) (D_bdY_ac) (D_dcY_ab) +(31/3360) (D_cbY_ad) (D_dcY_ab)
 +(-67/3360) (D_cdY_ab) (D_dcY_ab) +(-43/840) (D_dbY_ac) (D_dcY_ab) +(-1/840) (D_dcY_ab) (D_dcY_ab)
 +(-1/20) (D_bdY_ab) (D_dcY_ac) +(29/1120) (D_bcY_ab) (D_dcY_ad) +(37/672) (D_cbY_ab) (D_dcY_ad)
 +(-1/40) (D_bcY_ac) (D_ddY_ab) +(-1/40) (D_cbY_ac) (D_ddY_ab) +(1/1120) (D_ccY_ab) (D_ddY_ab)
 +(29/1120) (D_bcY_ab) (D_ddY_ac) +(1/1120) (D_cbY_ab) (D_ddY_ac)
================================
\end{verbatim}
\end{scriptsize}

Combining all Maxima generated output will give ${\cal L}^{(1)[8]}_{1}$ as given in (\ref{L18}).

\subsection{Second-order Corrections}

In this subsection, we will solve for (\ref{2[8]})
\begin{eqnarray}
{\cal L}^{(1)[8]}_2
    = \frac{\hbar}{2(2\pi)^D}\mbox{Tr}\int dX
    \int d^Dp 
    \,\,\, G_{\emptyset_{\mathrm{t}}}\left([3][5]+[4][4]+[5][3]\right)
\end{eqnarray}

Let us first solve for $[3][5]$. The following are $[3]_\ell$ with $\ell=1$ and $[5]_\ell$ with $\ell=2$ transcribed for Maxima environment implementation:

\begin{scriptsize}
\begin{verbatim}
thr: +(2*i/3)*D_aY_la*p1*s1+(8*i/3)*D_kY_lt*p3*s2;
thrz: +(2*(%i)/3)*D_azY_lzaz*p*sz1+(8*i/3)*D_kzY_lztz*p*p*p*sz2;
thr1: +(2*(%i)/3)*D_a1Y_l1a1*p*s11+(8*i/3)*D_k1Y_l1t1*p*p*p*s12;

fiv:
+(2*i/3)*p1*s2* (D_laaX + D_alaX + D_aalX + 2*D_bY_ab*Y_al + 2*(D_lY_ab + D_bY_al)*Y_ab)
-(4*i/3)*p3*s3* (D_ktlX + D_tY_al*Y_ak + D_kY_at*Y_al);

fivz:
+(2*(%i)/3)*p*sz2* (D_lzazazX + D_azlzazX + D_azazlzX + 2*D_bzY_azbz*Y_azlz
                    + 2*(D_lzY_azbz + D_bzY_azlz)*Y_azbz)
-(4*(%i)/3)*p*p*p*sz3* (D_kztzlzX + D_tzY_azlz*Y_azkz + D_kzY_aztz*Y_azlz);
fiv2:
+(2*(%i)/3)*p21*s22* (D_l2a2a2X + D_a2l2a2X + D_a2a2l2X + 2*D_b2Y_a2b2*Y_a2l2
                    + 2*(D_l2Y_a2b2 + D_b2Y_a2l2)*Y_a2b2)
-(4*(%i)/3)*p23*s23* (D_k2t2l2X + D_t2Y_a2l2*Y_a2k2 + D_k2Y_a2t2*Y_a2l2);

thr1: +(2*(%i)/3)*D_a1Y_l1a1*p11*s11+(8*(%i)/3)*D_k1Y_l1t1*p13*s12;
fiv2:
+(2*(%i)/3)*p21*s22* (D_l2a2a2X + D_a2l2a2X + D_a2a2l2X + 2*D_b2Y_a2b2*Y_a2l2
                    + 2*(D_l2Y_a2b2 + D_b2Y_a2l2)*Y_a2b2)
-(4*(%i)/3)*p23*s23* (D_k2t2l2X + D_t2Y_a2l2*Y_a2k2 + D_k2Y_a2t2*Y_a2l2);
expand(thr1*fiv2);

Output:=========================
+D_a1a1X D_a2a2X s11 s21 -2 D_a2a2X D_l1t1X p12 s12 s21 -2 D_a1a1X D_l2t2X p22 s11 s22
+D_a1a1X D_a2a2Y_l2t2 p22 s11 s22 +D_a1a1X D_a2t2Y_l2a2 p22 s11 s22
+D_a1a1Y_l1t1 D_a2a2X p12 s12 s21 +D_a1t1Y_l1a1 D_a2a2X p12 s12 s21
+D_t1a1Y_l1a1 D_a2a2X p12 s12 s21 +D_t2a2Y_l2a2 D_a1a1X p22 s11 s22
+(4 D_l1t1X D_l2t2X p12 p22 s12 s22 -2 D_a2a2X D_h1k1Y_l1t1 p14 s13 s21
-2 D_a1a1X D_h2k2Y_l2t2 p24 s11 s23 +(3 2) D_a1a1X D_a2t2Y_a2l2 p22 s11 s22
+(3 2) D_a1t1Y_a1l1 D_a2a2X p12 s12 s21 +(-3 2) D_a1a1X D_t2a2Y_l2a2 p22 s11 s22
+(-3 2) D_a2a2X D_t1a1Y_l1a1 p12 s12 s21 +3 D_l1t1X D_t2a2Y_l2a2 p12 p22 s12 s22
+3 D_l2t2X D_t1a1Y_l1a1 p12 p22 s12 s22 +4 D_h1k1Y_l1t1 D_l2t2X p14 p22 s13 s22
+4 D_h2k2Y_l2t2 D_l1t1X p12 p24 s12 s23 -2 D_a1a1Y_l1t1 D_l2t2X p12 p22 s12 s22
-2 D_a1t1Y_l1a1 D_l2t2X p12 p22 s12 s22 -2 D_a2a2Y_l2t2 D_l1t1X p12 p22 s12 s22
-2 D_a2t2Y_l2a2 D_l1t1X p12 p22 s12 s22 -2 D_t1a1Y_l1a1 D_l2t2X p12 p22 s12 s22
-2 D_t2a2Y_l2a2 D_l1t1X p12 p22 s12 s22 -3 D_a1t1Y_a1l1 D_l2t2X p12 p22 s12 s22
-3 D_a2t2Y_a2l2 D_l1t1X p12 p22 s12 s22 +D_a1a1Y_l1t1 D_a2a2Y_l2t2 p12 p22 s12 s22
+D_a1a1Y_l1t1 D_a2t2Y_l2a2 p12 p22 s12 s22 +D_a1t1Y_l1a1 D_a2a2Y_l2t2 p12 p22 s12 s22
+D_a1t1Y_l1a1 D_a2t2Y_l2a2 p12 p22 s12 s22 +D_t1a1Y_l1a1 D_a2a2Y_l2t2 p12 p22 s12 s22
+D_t1a1Y_l1a1 D_a2t2Y_l2a2 p12 p22 s12 s22 +D_t1a1Y_l1a1 D_t2a2Y_l2a2 p12 p22 s12 s22
+D_t2a2Y_l2a2 D_a1a1Y_l1t1 p12 p22 s12 s22 +D_t2a2Y_l2a2 D_a1t1Y_l1a1 p12 p22 s12 s22
+3 D_h1k1Y_l1t1 D_t2a2Y_l2a2 p14 p22 s13 s22 +3 D_h2k2Y_l2t2 D_t1a1Y_l1a1 p12 p24 s12 s23
+4 D_h1k1Y_l1t1 D_h2k2Y_l2t2 p14 p24 s13 s23 -2 D_a1a1Y_l1t1 D_h2k2Y_l2t2 p12 p24 s12 s23
-2 D_a1t1Y_l1a1 D_h2k2Y_l2t2 p12 p24 s12 s23 -2 D_a2a2Y_l2t2 D_h1k1Y_l1t1 p14 p22 s13 s22
-2 D_a2t2Y_l2a2 D_h1k1Y_l1t1 p14 p22 s13 s22 -2 D_t1a1Y_l1a1 D_h2k2Y_l2t2 p12 p24 s12 s23
-2 D_t2a2Y_l2a2 D_h1k1Y_l1t1 p14 p22 s13 s22 -3 D_a1t1Y_a1l1 D_h2k2Y_l2t2 p12 p24 s12 s23
-3 D_a2t2Y_a2l2 D_h1k1Y_l1t1 p14 p22 s13 s22 +(3 2) D_a1a1Y_l1t1 D_a2t2Y_a2l2 p12 p22 s12 s22
+(3 2) D_a1t1Y_a1l1 D_a2a2Y_l2t2 p12 p22 s12 s22 +(3 2) D_a1t1Y_a1l1 D_a2t2Y_l2a2 p12 p22 s12 s22
+(3 2) D_a1t1Y_l1a1 D_a2t2Y_a2l2 p12 p22 s12 s22 +(3 2) D_t1a1Y_l1a1 D_a2t2Y_a2l2 p12 p22 s12 s22
+(3 2) D_t2a2Y_l2a2 D_a1t1Y_a1l1 p12 p22 s12 s22 +(9 4) D_a1t1Y_a1l1 D_a2t2Y_a2l2 p12 p22 s12 s22
+(9 4) D_t1a1Y_l1a1 D_t2a2Y_l2a2 p12 p22 s12 s22 +(-3 2) D_a1a1Y_l1t1 D_t2a2Y_l2a2 p12 p22 s12 s22
+(-3 2) D_a1t1Y_l1a1 D_t2a2Y_l2a2 p12 p22 s12 s22 +(-3 2) D_a2a2Y_l2t2 D_t1a1Y_l1a1 p12 p22 s12 s22
+(-3 2) D_a2t2Y_l2a2 D_t1a1Y_l1a1 p12 p22 s12 s22 +(-3 2) D_t1a1Y_l1a1 D_t2a2Y_l2a2 p12 p22 s12 s22
+(-3 2) D_t2a2Y_l2a2 D_t1a1Y_l1a1 p12 p22 s12 s22 +(-9 4) D_a1t1Y_a1l1 D_t2a2Y_l2a2 p12 p22 s12 s22
+(-9 4) D_a2t2Y_a2l2 D_t1a1Y_l1a1 p12 p22 s12 s22
================================
\end{verbatim}
\end{scriptsize}

The impending momentum and proper-time integrations are performed by implementing the following Maxima code:

\begin{scriptsize}
\begin{verbatim}
/*three-fold proper time integrations*/
/*(1,1,0;1,1)*/ s11s21: (1/3)*(1/8);
/*(2,1,0;2,2)*/ s12s21: (1/3)*(1/40);
/*(1,2,0;2,2)*/ s11s22: (1/3)*(1/40);
/*(2,2,0;4,3)*/ s12s22: (1/3)*(1/240);
/*(1,3,0;4,3)*/ s11s23: (1/3)*(1/160);
/*(3,1,0;4,3)*/ s13s21: (1/3)*(1/160);
/*(2,3,0;8,4)*/ s12s23: (1/3)*(1/1120);
/*(3,2,0;8,4)*/ s13s22: (1/3)*(1/1120);
/*(3,3,0;16,4)*/ s13s23: (1/3)*(1/17920);

/*p2, p4, p6 momentum integrations*/
/*p2*/
p11p21: d_l1l2;
/*p4*/
p13p21:(d_l2l1*d_t1k1 +d_l2t1*d_l1k1 +d_l2k1*d_l1t1);
p11p23:(d_l1l2*d_t2k2 +d_l1t2*d_l2k2 +d_l1k2*d_l2t2);
/*p6*/
p13p23:
(d_l1t1*d_k1l2*d_t2k2 +d_l1t1*d_k1t2*d_l2k2 +d_l1t1*d_k1k2*d_l2t2
 +d_l1k1*d_t1l2*d_t2k2 +d_l1k1*d_t1t2*d_l2k2 +d_l1k1*d_t1k2*d_l2t2
 +d_l1l2*d_t1k1*d_t2k2 +d_l1l2*d_t1t2*d_k1k2 +d_l1l2*d_t1k2*d_k1t2
 +d_l1t2*d_t1k1*d_l2k2 +d_l1t2*d_t1l2*d_k1k2 +d_l1t2*d_t1k2*d_k1l2
 +d_l1k2*d_t1k1*d_l2t2 +d_l1k2*d_t1l2*d_k1t2 +d_l1k2*d_t1t2*d_k1l2);

/*[3][5]*/
expand(
-(4/9)*D_a1Y_l1a1*D_a2a2l2X*p11p21*s11s22 -(4/9)*D_a1Y_l1a1*D_a2l2a2X*p11p21*s11s22
-(4/9)*D_a1Y_l1a1*D_l2a2a2X*p11p21*s11s22 -(16/9)*D_a2a2l2X*D_k1Y_l1t1*p13p21*s12s22
-(16/9)*D_a2l2a2X*D_k1Y_l1t1*p13p21*s12s22 -(16/9)*D_k1Y_l1t1*D_l2a2a2X*p13p21*s12s22
+(8/9)*D_a1Y_l1a1*D_k2t2l2X*p11p23*s11s23 +(32/9)*D_k1Y_l1t1*D_k2t2l2X*p13p23*s12s23
-(8/9)*D_a1Y_l1a1*D_b2Y_a2l2*p11p21*s11s22*Y_a2b2 -(8/9)*D_a1Y_l1a1*D_l2Y_a2b2*p11p21*s11s22*Y_a2b2
-(32/9)*D_b2Y_a2l2*D_k1Y_l1t1*p13p21*s12s22*Y_a2b2 -(32/9)*D_k1Y_l1t1*D_l2Y_a2b2*p13p21*s12s22*Y_a2b2
+(8/9)*D_a1Y_l1a1*D_t2Y_a2l2*p11p23*s11s23*Y_a2k2 +(32/9)*D_k1Y_l1t1*D_t2Y_a2l2*p13p23*s12s23*Y_a2k2
-(8/9)*D_a1Y_l1a1*D_b2Y_a2b2*p11p21*s11s22*Y_a2l2 -(32/9)*D_b2Y_a2b2*D_k1Y_l1t1*p13p21*s12s22*Y_a2l2
+(8/9)*D_a1Y_l1a1*D_k2Y_a2t2*p11p23*s11s23*Y_a2l2 +(32/9)*D_k1Y_l1t1*D_k2Y_a2t2*p13p23*s12s23*Y_a2l2);
save(eit2nd35,%);
\end{verbatim}
\end{scriptsize}

Sorting according to string length, labeling each term, contracting indices, grouping to common type of invariant, and
renaming dummy indices, we have:

\begin{scriptsize}
\begin{verbatim}
a1->a
a2->b
l1->c
l2->d
t1->e
t2->f
k1->g
b2->h
\end{verbatim}
\end{scriptsize}

we have

\begin{scriptsize}
\begin{verbatim}
Output:=========================
DDDX DY type of invariant
b1: +(-1 270) D_bbcX D_aY_ca
b2: +(-1 270) D_bcbX D_aY_ca
b3: +(-1 270) D_cbbX D_aY_ca
b4: +(1 540) D_cddX D_aY_ca
b27: +(1 945) D_cddX D_eY_ce
b5: +(-1 405) D_bbcX D_eY_ce
b11: +(-1 405) D_bcbX D_eY_ce
b17: +(-1 405) D_cbbX D_eY_ce
b29: +(1 945) D_ffcX D_eY_ce
b9: +(1 540) D_ffcX D_aY_ca
b19: +(1 945) D_ffgX D_gY_cc
b8: +(1 945) D_gddX D_gY_cc
b58: +(1 945) D_gecX D_gY_ce
b15: +(1 945) D_cegX D_gY_ce
b14: +(1 945) D_gceX D_gY_ce
b26: +(1 945) D_ffeX D_cY_ce
b25: +(1 945) D_eddX D_cY_ce
b20: +(1 945) D_ecgX D_gY_ce
b22: +(1 945) D_egcX D_gY_ce
b21: +(1 945) D_cgeX D_gY_ce
b23: +(1 945) D_dgdX D_gY_cc
b13: +(-1 405) D_bbeX D_cY_ce
b6: +(-1 405) D_bebX D_cY_ce
b12: +(-1 405) D_ebbX D_cY_ce
b24: +(1 945) D_dedX D_cY_ce
b30: +(1 945) D_dcdX D_eY_ce
b10: +(1 540) D_dcdX D_aY_ca
b28: +(-1 405) D_bbgX D_gY_cc
b7: +(-1 405) D_bgbX D_gY_cc
b16: +(-1 405) D_gbbX D_gY_cc

DY DY Y  type of invariant
b18: +(-1 135) D_aY_ca D_hY_bh Y_bc
b32: +(-1 135) D_aY_ca D_hY_bc Y_bh
b35: +(-1 135) D_aY_ca D_cY_bh Y_bh
b31: +(1 540) D_aY_ca D_cY_bd Y_bd
b54: +(1 540) D_aY_ca D_dY_bd Y_bc
b69: +(1 945) D_eY_ce D_cY_bd Y_bd
b75: +(1 945) D_eY_ce D_dY_bd Y_bc
b34: +(-2 405) D_hY_bh D_eY_ce Y_bc
b38: +(-2 405) D_hY_bc D_eY_ce Y_bh
b62: +(-2 405) D_eY_ce D_cY_bh Y_bh
b70: +(1 945) D_eY_ce D_fY_bf Y_bc
b77: +(1 945) D_eY_ce D_fY_bc Y_bf
b41: +(1 540) D_aY_ca D_fY_bf Y_bc
b61: +(1 540) D_aY_ca D_fY_bc Y_bf
b48: +(1 945) D_gY_cc D_fY_bf Y_bg
b51: +(1 945) D_gY_cc D_fY_bg Y_bf
b40: +(1 945) D_gY_cc D_gY_bd Y_bd
b45: +(1 945) D_gY_cc D_dY_bd Y_bg
b66: +(1 945) D_gY_ce D_gY_be Y_bc
b44: +(1 945) D_gY_ce D_eY_bc Y_bg
b47: +(1 945) D_gY_ce D_cY_be Y_bg
b50: +(1 945) D_gY_ce D_eY_bg Y_bc
b65: +(1 945) D_cY_ce D_dY_be Y_bd
b72: +(1 945) D_cY_ce D_eY_bd Y_bd
b56: +(1 945) D_gY_cc D_dY_bg Y_bd
b64: +(1 945) D_gY_cc D_gY_bd Y_bd
b55: +(1 945) D_gY_ce D_eY_bg Y_bc
b60: +(1 945) D_gY_ce D_gY_bc Y_be
b67: +(1 945) D_cY_ce D_eY_bd Y_bd
b73: +(1 945) D_cY_ce D_dY_bd Y_be
b49: +(1 945) D_gY_ce D_eY_bc Y_bg
b52: +(1 945) D_gY_ce D_cY_bg Y_be
b71: +(1 945) D_eY_ce D_dY_bc Y_bd
b78: +(1 945) D_eY_ce D_cY_bd Y_bd
b42: +(1 540) D_aY_ca D_dY_bc Y_bd
b63: +(1 540) D_aY_ca D_cY_bd Y_bd
b43: +(1 945) D_gY_ce D_gY_bc Y_be
b46: +(1 945) D_gY_ce D_cY_be Y_bg
b53: +(1 945) D_gY_ce D_cY_bg Y_be
b57: +(1 945) D_gY_ce D_gY_be Y_bc
b68: +(1 945) D_cY_ce D_fY_bf Y_be
b74: +(1 945) D_cY_ce D_fY_be Y_bf
b76: +(-2 405) D_hY_bh D_cY_ce Y_be
b36: +(-2 405) D_hY_be D_cY_ce Y_bh
b39: +(-2 405) D_cY_ce D_eY_bh Y_bh
b33: +(-2 405) D_hY_bh D_gY_cc Y_bg
b37: +(-2 405) D_hY_bg D_gY_cc Y_bh
b59: +(-2 405) D_gY_cc D_gY_bh Y_bh
================================
\end{verbatim}
\end{scriptsize}

\begin{scriptsize}
\begin{verbatim}
/*[5][3]*/
/*a1->a3,a2->a1,a3->a2*/
/*b1->b2*/
/*l1->l3,l2->l1,l3->l2*/
/*t1->t3,t2->t1,t3->t2*/
/*k1->k3,k2->k1,k3->k2*/

/*p2*/
p11p21: d_l2l1;
/*p4*/
p13p21:(d_l1l2*d_t2k2 +d_l1t2*d_l2k2 +d_l1k2*d_l2t2);
p11p23:(d_l2l1*d_t1k1 +d_l2t1*d_l1k1 +d_l2k1*d_l1t1);
/*p6*/
p13p23:
(d_l2t2*d_k2l1*d_t1k1 +d_l2t2*d_k2t1*d_l1k1 +d_l2t2*d_k2k1*d_l1t1
 +d_l2k2*d_t2l1*d_t1k1 +d_l2k2*d_t2t1*d_l1k1 +d_l2k2*d_t2k1*d_l1t1
 +d_l2l1*d_t2k2*d_t1k1 +d_l2l1*d_t2t1*d_k2k1 +d_l2l1*d_t2k1*d_k2t1
 +d_l2t1*d_t2k2*d_l1k1 +d_l2t1*d_t2l1*d_k2k1 +d_l2t1*d_t2k1*d_k2l1
 +d_l2k1*d_t2k2*d_l1t1 +d_l2k1*d_t2l1*d_k2t1 +d_l2k1*d_t2t1*d_k2l1);

/*[5][3]*/
expand(
-(4/9)*D_a2Y_l2a2*D_a1a1l1X*p11p21*s11s22
-(4/9)*D_a2Y_l2a2*D_a1l1a1X*p11p21*s11s22
-(4/9)*D_a2Y_l2a2*D_l1a1a1X*p11p21*s11s22
-(16/9)*D_a1a1l1X*D_k2Y_l2t2*p13p21*s12s22
-(16/9)*D_a1l1a1X*D_k2Y_l2t2*p13p21*s12s22
-(16/9)*D_k2Y_l2t2*D_l1a1a1X*p13p21*s12s22
+(8/9)*D_a2Y_l2a2*D_k1t1l1X*p11p23*s11s23
+(32/9)*D_k2Y_l2t2*D_k1t1l1X*p13p23*s12s23
-(8/9)*D_a2Y_l2a2*D_b1Y_a1l1*p11p21*s11s22*Y_a1b1
-(8/9)*D_a2Y_l2a2*D_l1Y_a1b1*p11p21*s11s22*Y_a1b1
-(32/9)*D_b1Y_a1l1*D_k2Y_l2t2*p13p21*s12s22*Y_a1b1
-(32/9)*D_k2Y_l2t2*D_l1Y_a1b1*p13p21*s12s22*Y_a1b1
+(8/9)*D_a2Y_l2a2*D_t1Y_a1l1*p11p23*s11s23*Y_a1k1
+(32/9)*D_k2Y_l2t2*D_t1Y_a1l1*p13p23*s12s23*Y_a1k1
-(8/9)*D_a2Y_l2a2*D_b1Y_a1b1*p11p21*s11s22*Y_a1l1
-(32/9)*D_b1Y_a1b1*D_k2Y_l2t2*p13p21*s12s22*Y_a1l1
+(8/9)*D_a2Y_l2a2*D_k1Y_a1t1*p11p23*s11s23*Y_a1l1
+(32/9)*D_k2Y_l2t2*D_k1Y_a1t1*p13p23*s12s23*Y_a1l1);
save(eit2nd53,%);
\end{verbatim}
\end{scriptsize}

The result is identical to $[3][5]$ (with 1 and 2 subsubscripts swapped) if one implements the following (renaming of dummy indices)

\begin{scriptsize}
\begin{verbatim}
a1<->a2 a2->a3, a1->a2, a3->a1
b2->b1
l1<->l2   l2->l3, l1->l2, l3->l1
t1<->t2   t1->t3, t2->t1, t3->t2
k1->k2

a2->a
a1->b
l2->c
l1->d
t2->e
t1->f
k2->g
b1->h
\end{verbatim}
\end{scriptsize}
after performing kronecker delta index contractions.

Because the result $[3][5]$ is identical to $[5][3]$ in a chosen system of dummy-index replacements (to be able to combine similar types). We can work out with either one of them and multiply the result by 2. Choosing $[5][3]$, we have the following (written in Maxima code) grouped into similar type of invariant

\begin{scriptsize}
\begin{verbatim}
DDDX DY type of invariant
-(1/270)*D_ccaX*D_bY_ab -(1/270)*D_cacX*D_bY_ab -(1/270)*D_accX*D_bY_ab +(1/540)*D_accX*D_bY_ab
+(1/945)*D_accX*D_bY_ab -(1/405)*D_ccaX*D_bY_ab  -(1/405)*D_cacX*D_bY_ab-(1/405)*D_accX*D_bY_ab
+(1/945)*D_ccaX*D_bY_ab +(1/540)*D_ccaX*D_bY_ab +(1/945)*D_cbaX*D_cY_ab +(1/945)*D_abcX*D_cY_ab
+(1/945)*D_cabX*D_cY_ab +(1/945)*D_ccbX*D_aY_ab +(1/945)*D_bccX*D_aY_ab +(1/945)*D_bacX*D_cY_ab
+(1/945)*D_bcaX*D_cY_ab +(1/945)*D_acbX*D_cY_ab -(1/405)*D_ccbX*D_aY_ab -(1/405)*D_cbcX*D_aY_ab
-(1/405)*D_bccX*D_aY_ab +(1/945)*D_cbcX*D_aY_ab +(1/945)*D_cacX*D_bY_ab +(1/540)*D_cacX*D_bY_ab;
save(ddxdy,%);
Output:=========================
+(1 945) D_abcX D_cY_ab +(1 945) D_acbX D_cY_ab +(1 945) D_bacX D_cY_ab +(1 945) D_bcaX D_cY_ab
+(1 945) D_cabX D_cY_ab +(1 945) D_cbaX D_cY_ab +(-4 2835) D_aY_ab D_ccbX +(-4 2835) D_aY_ab D_bccX
+(-4 2835) D_aY_ab D_cbcX +(-37 11340) D_accX D_bY_ab +(-37 11340) D_bY_ab D_cacX
+(-37 11340) D_bY_ab D_ccaX
================================
DY DY Y type of invariant
-(1/135)*D_cY_bc*D_dY_ad*Y_ab -(1/135)*D_cY_dc*D_bY_ad*Y_ab -(1/135)*D_cY_dc*D_dY_ab*Y_ab
+(1/540)*D_cY_dc*D_dY_ab*Y_ab +(1/540)*D_cY_bc*D_dY_ad*Y_ab +(1/945)*D_cY_dc*D_dY_ab*Y_ab
+(1/945)*D_cY_bc*D_dY_ad*Y_ab -(2/405)*D_cY_ac*D_dY_bd*Y_ab -(2/405)*D_bY_ac*D_dY_cd*Y_ab
-(2/405)*D_cY_dc*D_dY_ab*Y_ab +(1/945)*D_cY_bc*D_dY_ad*Y_ab +(1/945)*D_cY_dc*D_bY_ad*Y_ab
+(1/540)*D_cY_bc*D_dY_ad*Y_ab +(1/540)*D_cY_dc*D_bY_ad*Y_ab +(1/945)*D_bY_cc*D_dY_ad*Y_ab
+(1/945)*D_bY_cc*D_dY_ad*Y_ab +(1/945)*D_cY_bd*D_cY_ad*Y_ab +(1/945)*D_bY_cd*D_dY_ac*Y_ab
+(1/945)*D_bY_cd*D_cY_ad*Y_ab +(1/945)*D_cY_bd*D_dY_ac*Y_ab +(1/945)*D_cY_cd*D_bY_ad*Y_ab
+(1/945)*D_cY_cd*D_dY_ab*Y_ab +(1/945)*D_cY_bd*D_dY_ac*Y_ab +(1/945)*D_cY_db*D_cY_ad*Y_ab
+(1/945)*D_cY_cd*D_dY_ab*Y_ab +(1/945)*D_cY_cb*D_dY_ad*Y_ab +(1/945)*D_bY_cd*D_dY_ac*Y_ab
+(1/945)*D_cY_db*D_dY_ac*Y_ab +(1/945)*D_cY_dc*D_bY_ad*Y_ab +(1/945)*D_cY_dc*D_dY_ab*Y_ab
+(1/540)*D_cY_dc*D_bY_ad*Y_ab +(1/540)*D_cY_dc*D_dY_ab*Y_ab +(1/945)*D_cY_db*D_cY_ad*Y_ab
+(1/945)*D_bY_cd*D_cY_ad*Y_ab +(1/945)*D_cY_db*D_dY_ac*Y_ab +(1/945)*D_cY_bd*D_cY_ad*Y_ab
+(1/945)*D_cY_cb*D_dY_ad*Y_ab +(1/945)*D_cY_cd*D_bY_ad*Y_ab -(2/405)*D_cY_ac*D_dY_db*Y_ab
-(2/405)*D_bY_ac*D_dY_dc*Y_ab -(2/405)*D_cY_cd*D_dY_ab*Y_ab;
save(dydyy,%);
Output:=========================
+(2 945) D_bY_cd D_cY_ad Y_ab+(2 945) D_cY_ad D_cY_bd Y_ab +(2 945) D_bY_ad D_cY_cd Y_ab
+(2 945) D_cY_ad D_cY_db Y_ab +(2 945) D_bY_cd D_dY_ac Y_ab +(2 945) D_cY_bd D_dY_ac Y_ab
+(2 945) D_cY_db D_dY_ac Y_ab +(2 945) D_bY_cc D_dY_ad Y_ab +(2 945) D_cY_cb D_dY_ad Y_ab
+(-1 630) D_cY_bc D_dY_ad Y_ab +(-2 405) D_cY_ac D_dY_bd Y_ab +(-2 405) D_bY_ac D_dY_cd Y_ab
+(-2 405) D_cY_ac D_dY_db Y_ab +(-2 405) D_bY_ac D_dY_dc Y_ab+(-1 630) D_bY_ad D_cY_dc Y_ab
+(-8 2835) D_cY_cd D_dY_ab Y_ab +(-37 5670) D_cY_dc D_dY_ab Y_ab
================================
\end{verbatim}
\end{scriptsize}

Combining all Maxima generated output and then multiplying by a factor of 2 will give ${\cal L}^{(1)[8]}_{2\,\,\,\,[3][5]+[5][3]}$ as given in (\ref{L28_35+53}).

Finally, let us solve for $[4][4]$. The following are $[4]_\ell$ with $\ell=1$ and $\ell=2$ transcribed for Maxima environment implementation:

\begin{scriptsize}
\begin{verbatim}
for:
 + s*D_aaX
 + p2*s2*(- 2*D_ltX + D_taY_la + D_atY_la + D_aaY_lt + (3/2)*(D_atY_al - D_taY_la))
 - 2*p4*s3*D_hkY_lt;
forz:
 + sz1*D_azazX
 + pz2*sz2*(- 2*D_lztzX + D_tzazY_lzaz + D_aztzY_lzaz + D_azazY_lztz
            + (3/2)*(D_aztzY_azlz - D_tzazY_lzaz))
 - 2*pz4*sz3*D_hzkzY_lztz;

for1:
 + s11*D_a1a1X
 + p12*s12*(- 2*D_l1t1X + D_t1a1Y_l1a1 + D_a1t1Y_l1a1 + D_a1a1Y_l1t1
            + (3/2)*(D_a1t1Y_a1l1 - D_t1a1Y_l1a1))
 - 2*p14*s13*D_h1k1Y_l1t1;
for2:
 + s21*D_a2a2X
 + p22*s22*(- 2*D_l2t2X + D_t2a2Y_l2a2 + D_a2t2Y_l2a2 + D_a2a2Y_l2t2
            + (3/2)*(D_a2t2Y_a2l2 - D_t2a2Y_l2a2))
 - 2*p24*s23*D_h2k2Y_l2t2;
expand(for1*for2);
\end{verbatim}
\end{scriptsize}

The result will be expanded:

\begin{scriptsize}
\begin{verbatim}
expand(+D_a1a1X*D_a2a2X*s11s21 +D_t1a1Y_l1a1*D_a2a2X*p12*s12s21 +D_a1a1Y_l1t1*D_a2a2X*p12*s12s21
+D_a1t1Y_l1a1*D_a2a2X*p12*s12s21 -2*D_a2a2X*D_l1t1X*p12*s12s21 +D_a1a1X*D_a2a2Y_l2t2*p22*s11s22
+D_a1a1X*D_a2t2Y_l2a2*p22*s11s22 -2*D_a1a1X*D_l2t2X*p22*s11s22 +D_t2a2Y_l2a2*D_a1a1X*p22*s11s22
-2*D_a2a2X*D_h1k1Y_l1t1*p14*s13s21 +4*D_l1t1X*D_l2t2X*p12p22*s12s22
-2*D_a1a1X*D_h2k2Y_l2t2*p24*s11s23 +(3/2)*D_a1t1Y_a1l1*D_a2a2X*p12*s12s21
+(3/2)*D_a1a1X*D_a2t2Y_a2l2*p22*s11s22 -(3/2)*D_a2a2X*D_t1a1Y_l1a1*p12*s12s21
-(3/2)*D_a1a1X*D_t2a2Y_l2a2*p22*s11s22 +3*D_l2t2X*D_t1a1Y_l1a1*p12p22*s12s22
+3*D_l1t1X*D_t2a2Y_l2a2*p12p22*s12s22 +4*D_h1k1Y_l1t1*D_l2t2X*p14p22*s13s22
+4*D_h2k2Y_l2t2*D_l1t1X*p12p24*s12s23 +D_t1a1Y_l1a1*D_a2a2Y_l2t2*p12p22*s12s22
+D_a1a1Y_l1t1*D_a2a2Y_l2t2*p12p22*s12s22 +D_a1t1Y_l1a1*D_a2a2Y_l2t2*p12p22*s12s22
+D_t1a1Y_l1a1*D_a2t2Y_l2a2*p12p22*s12s22 +D_a1a1Y_l1t1*D_a2t2Y_l2a2*p12p22*s12s22
+D_a1t1Y_l1a1*D_a2t2Y_l2a2*p12p22*s12s22 -2*D_a2a2Y_l2t2*D_l1t1X*p12p22*s12s22
-3*D_a2t2Y_a2l2*D_l1t1X*p12p22*s12s22 -2*D_a2t2Y_l2a2*D_l1t1X*p12p22*s12s22
-2*D_t1a1Y_l1a1*D_l2t2X*p12p22*s12s22 -2*D_a1a1Y_l1t1*D_l2t2X*p12p22*s12s22
-3*D_a1t1Y_a1l1*D_l2t2X*p12p22*s12s22 -2*D_a1t1Y_l1a1*D_l2t2X*p12p22*s12s22
+D_t1a1Y_l1a1*D_t2a2Y_l2a2*p12p22*s12s22 +D_t2a2Y_l2a2*D_a1a1Y_l1t1*p12p22*s12s22
+D_t2a2Y_l2a2*D_a1t1Y_l1a1*p12p22*s12s22 -2*D_t2a2Y_l2a2*D_l1t1X*p12p22*s12s22
+3*D_h1k1Y_l1t1*D_t2a2Y_l2a2*p14p22*s13s22 +3*D_h2k2Y_l2t2*D_t1a1Y_l1a1*p12p24*s12s23
+4*D_h1k1Y_l1t1*D_h2k2Y_l2t2*p14*p24*s13s23 -2*D_a2a2Y_l2t2*D_h1k1Y_l1t1*p14p22*s13s22
-3*D_a2t2Y_a2l2*D_h1k1Y_l1t1*p14p22*s13s22 -2*D_a2t2Y_l2a2*D_h1k1Y_l1t1*p14p22*s13s22
-2*D_t1a1Y_l1a1*D_h2k2Y_l2t2*p12p24*s12s23 -2*D_a1a1Y_l1t1*D_h2k2Y_l2t2*p12p24*s12s23
-3*D_a1t1Y_a1l1*D_h2k2Y_l2t2*p12p24*s12s23 -2*D_a1t1Y_l1a1*D_h2k2Y_l2t2*p12p24*s12s23
-2*D_t2a2Y_l2a2*D_h1k1Y_l1t1*p14p22*s13s22 +(3/2)*D_a1t1Y_a1l1*D_a2a2Y_l2t2*p12p22*s12s22
+(3/2)*D_t1a1Y_l1a1*D_a2t2Y_a2l2*p12p22*s12s22 +(3/2)*D_a1a1Y_l1t1*D_a2t2Y_a2l2*p12p22*s12s22
+(9/4)*D_a1t1Y_a1l1*D_a2t2Y_a2l2*p12p22*s12s22 +(3/2)*D_a1t1Y_l1a1*D_a2t2Y_a2l2*p12p22*s12s22
+(3/2)*D_a1t1Y_a1l1*D_a2t2Y_l2a2*p12p22*s12s22 +(9/4)*D_t1a1Y_l1a1*D_t2a2Y_l2a2*p12p22*s12s22
+(3/2)*D_t2a2Y_l2a2*D_a1t1Y_a1l1*p12p22*s12s22 -(3/2)*D_a2a2Y_l2t2*D_t1a1Y_l1a1*p12p22*s12s22
-(9/4)*D_a2t2Y_a2l2*D_t1a1Y_l1a1*p12p22*s12s22 -(3/2)*D_a2t2Y_l2a2*D_t1a1Y_l1a1*p12p22*s12s22
-(3/2)*D_t1a1Y_l1a1*D_t2a2Y_l2a2*p12p22*s12s22 -(3/2)*D_a1a1Y_l1t1*D_t2a2Y_l2a2*p12p22*s12s22
-(9/4)*D_a1t1Y_a1l1*D_t2a2Y_l2a2*p12p22*s12s22 -(3/2)*D_a1t1Y_l1a1*D_t2a2Y_l2a2*p12p22*s12s22
-(3/2)*D_t2a2Y_l2a2*D_t1a1Y_l1a1*p12p22*s12s22);
\end{verbatim}
\end{scriptsize}

This is further expanded as one performs up to $p^6$ momentum and three-fold proper-time integration by implementing the following substitutions:

\begin{scriptsize}
\begin{verbatim}
/*(1,1,0;1,1)*/ s11s21: (1/3)*(1/8);
/*(2,1,0;2,2)*/ s12s21: (1/3)*(1/40);
/*(1,2,0;2,2)*/ s11s22: (1/3)*(1/40);
/*(2,2,0;4,3)*/ s12s22: (1/3)*(1/240);
/*(1,3,0;4,3)*/ s11s23: (1/3)*(1/160);
/*(3,1,0;4,3)*/ s13s21: (1/3)*(1/160);
/*(2,3,0;8,4)*/ s12s23: (1/3)*(1/1120);
/*(3,2,0;8,4)*/ s13s22: (1/3)*(1/1120);
/*(3,3,0;16,4)*/ s13s23: (1/3)*(1/17920);
/*l1t1*/ p12: d_l1t1;
/*l2t2*/ p22: d_l2t2;
/*l1t1k1h1*/ p14: (d_l1t1*d_k1h1 +d_l1k1*d_t1h1 +d_l1h1*d_t1k1);
/*l2t2k2h2*/ p24: (d_l2t2*d_k2h2 +d_l2k2*d_t2h2 +d_l2h2*d_t2k2);
/*l1t1l2t2*/ p12p22: (d_l1t1*d_l2t2 +d_l1l2*d_t1t2 +d_l1t2*d_t1l2);
/*p4*/
p13p21:(d_l2l1*d_t1k1 +d_l2t1*d_l1k1 +d_l2k1*d_l1t1);
p11p23:(d_l1l2*d_t2k2 +d_l1t2*d_l2k2 +d_l1k2*d_l2t2);
/*p6*/
 p14p22:
(d_l1t1*d_k1h1*d_l2t2 +d_l1t1*d_k1l2*d_h1t2 +d_l1t1*d_k1t2*d_h1l2
 +d_l1k1*d_t1h1*d_l2t2 +d_l1k1*d_t1l2*d_h1t2 +d_l1k1*d_t1t2*d_h1l2
 +d_l1h1*d_t1k1*d_l2t2 +d_l1h1*d_t1l2*d_k1t2 +d_l1h1*d_t1t2*d_k1l2
 +d_l1l2*d_t1k1*d_h1t2 +d_l1l2*d_t1h1*d_k1t2 +d_l1l2*d_t1t2*d_k1h1
 +d_l1t2*d_t1k1*d_h1l2 +d_l1t2*d_t1h1*d_k1l2 +d_l1t2*d_t1l2*d_k1h1);

 p12p24:
(d_l2t2*d_k2h2*d_l1t1 +d_l2t2*d_k2l1*d_h2t1 +d_l2t2*d_k2t1*d_h2l1
 +d_l2k2*d_t2h2*d_l1t1 +d_l2k2*d_t2l1*d_h2t1 +d_l2k2*d_t2t1*d_h2l1
 +d_l2h2*d_t2k2*d_l1t1 +d_l2h2*d_t2l1*d_k2t1 +d_l2h2*d_t2t1*d_k2l1
 +d_l2l1*d_t2k2*d_h2t1 +d_l2l1*d_t2h2*d_k2t1 +d_l2l1*d_t2t1*d_k2h2
 +d_l2t1*d_t2k2*d_h2l1 +d_l2t1*d_t2h2*d_k2l1 +d_l2t1*d_t2l1*d_k2h2);
\end{verbatim}
\end{scriptsize}

Such implementation will result to a 285-term which can be grouped into three types of invariant:
\begin{scriptsize}
\begin{verbatim}
DDX DDX type of invariant
f1: +(1/24) D_a1a1X D_a2a2X
f2: +(-1/60) D_a2a2X D_l1t1X d_l1t1
f3: +(-1/60) D_a1a1X D_l2t2X d_l2t2
f4: +(1/180) D_l1t1X D_l2t2X d_l1t1 d_l2t2
f5: +(1/180) D_l1t1X D_l2t2X d_l1t2 d_t1l2
f6: +(1/180) D_l1t1X D_l2t2X d_l1l2 d_t1t2
DDX DDY type of invariant
f7: +(1/80) D_a2a2X D_a1t1Y_a1l1 d_l1t1
f8: +(1/80) D_a1a1X D_a2t2Y_a2l2 d_l2t2
f9: +(1/120) D_a2a2X D_a1a1Y_l1t1 d_l1t1
f10: +(1/120) D_a2a2X D_a1t1Y_l1a1 d_l1t1
f11: +(1/120) D_a1a1X D_a2a2Y_l2t2 d_l2t2
f12: +(1/120) D_a1a1X D_a2t2Y_l2a2 d_l2t2
f13: +(-1/80) D_a1a1X D_t2a2Y_l2a2 d_l2t2
f14: +(1/120) D_a1a1X D_t2a2Y_l2a2 d_l2t2
f15: +(-1/240) D_a2a2X D_t1a1Y_l1a1 d_l1t1
f16: +(-1/240) D_a2a2X D_h1k1Y_l1t1 d_k1h1 d_l1t1
f17: +(-1/240) D_a2a2X D_h1k1Y_l1t1 d_l1k1 d_t1h1
f18: +(-1/240) D_a2a2X D_h1k1Y_l1t1 d_l1h1 d_t1k1
f19: +(1/720) D_l2t2X D_t1a1Y_l1a1 d_l1t1 d_l2t2
f20: +(1/720) D_l2t2X D_t1a1Y_l1a1 d_l1t2 d_t1l2
f21: +(1/720) D_l2t2X D_t1a1Y_l1a1 d_l1l2 d_t1t2
f22: +(1/240) D_l1t1X D_t2a2Y_l2a2 d_l1t1 d_l2t2
f23: +(1/240) D_l1t1X D_t2a2Y_l2a2 d_l1t2 d_t1l2
f24: +(1/240) D_l1t1X D_t2a2Y_l2a2 d_l1l2 d_t1t2
f25: +(-1/240) D_a1a1X D_h2k2Y_l2t2 d_k2h2 d_l2t2
f26: +(-1/360) D_l1t1X D_a2a2Y_l2t2 d_l1t1 d_l2t2
f27: +(-1/240) D_l1t1X D_a2t2Y_a2l2 d_l1t1 d_l2t2
f28: +(-1/360) D_l1t1X D_a2t2Y_l2a2 d_l1t1 d_l2t2
f29: +(-1/360) D_l2t2X D_a1a1Y_l1t1 d_l1t1 d_l2t2
f30: +(-1/240) D_l2t2X D_a1t1Y_a1l1 d_l1t1 d_l2t2
f31: +(-1/360) D_l2t2X D_a1t1Y_l1a1 d_l1t1 d_l2t2
f32: +(-1/360) D_l1t1X D_a2a2Y_l2t2 d_l1t2 d_t1l2
f33: +(-1/240) D_l1t1X D_a2t2Y_a2l2 d_l1t2 d_t1l2
f34: +(-1/360) D_l1t1X D_a2t2Y_l2a2 d_l1t2 d_t1l2
f35: +(-1/360) D_l2t2X D_a1a1Y_l1t1 d_l1t2 d_t1l2
f36: +(-1/240) D_l2t2X D_a1t1Y_a1l1 d_l1t2 d_t1l2
f37: +(-1/360) D_l2t2X D_a1t1Y_l1a1 d_l1t2 d_t1l2
f38: +(-1/360) D_l1t1X D_a2a2Y_l2t2 d_l1l2 d_t1t2
f39: +(-1/240) D_l1t1X D_a2t2Y_a2l2 d_l1l2 d_t1t2
f40: +(-1/360) D_l1t1X D_a2t2Y_l2a2 d_l1l2 d_t1t2
f41: +(-1/360) D_l2t2X D_a1a1Y_l1t1 d_l1l2 d_t1t2
f42: +(-1/240) D_l2t2X D_a1t1Y_a1l1 d_l1l2 d_t1t2
f43: +(-1/360) D_l2t2X D_a1t1Y_l1a1 d_l1l2 d_t1t2
f44: +(-1/240) D_a1a1X D_h2k2Y_l2t2 d_l2k2 d_t2h2
f45: +(-1/240) D_a1a1X D_h2k2Y_l2t2 d_l2h2 d_t2k2
f46: +(-1/360) D_l1t1X D_t2a2Y_l2a2 d_l1t1 d_l2t2
f47: +(-1/360) D_l1t1X D_t2a2Y_l2a2 d_l1t2 d_t1l2
f48: +(-1/360) D_l1t1X D_t2a2Y_l2a2 d_l1l2 d_t1t2
f49: +(1/840) D_l1t1X D_h2k2Y_l2t2 d_h2t1 d_k2l1 d_l2t2
f50: +(1/840) D_l1t1X D_h2k2Y_l2t2 d_h2l1 d_k2t1 d_l2t2
f51: +(1/840) D_l1t1X D_h2k2Y_l2t2 d_k2h2 d_l1t1 d_l2t2
f52: +(1/840) D_l2t2X D_h1k1Y_l1t1 d_h1t2 d_k1l2 d_l1t1
f53: +(1/840) D_l2t2X D_h1k1Y_l1t1 d_h1l2 d_k1t2 d_l1t1
f54: +(1/840) D_l2t2X D_h1k1Y_l1t1 d_k1h1 d_l1t1 d_l2t2
f55: +(1/840) D_l2t2X D_h1k1Y_l1t1 d_k1t2 d_l1l2 d_t1h1
f56: +(1/840) D_l2t2X D_h1k1Y_l1t1 d_k1l2 d_l1t2 d_t1h1
f57: +(1/840) D_l2t2X D_h1k1Y_l1t1 d_l1k1 d_l2t2 d_t1h1
f58: +(1/840) D_l2t2X D_h1k1Y_l1t1 d_h1t2 d_l1l2 d_t1k1
f59: +(1/840) D_l2t2X D_h1k1Y_l1t1 d_h1l2 d_l1t2 d_t1k1
f60: +(1/840) D_l2t2X D_h1k1Y_l1t1 d_l1h1 d_l2t2 d_t1k1
f61: +(1/840) D_l2t2X D_h1k1Y_l1t1 d_k1t2 d_l1h1 d_t1l2
f62: +(1/840) D_l2t2X D_h1k1Y_l1t1 d_h1t2 d_l1k1 d_t1l2
f63: +(1/840) D_l2t2X D_h1k1Y_l1t1 d_k1h1 d_l1t2 d_t1l2
f64: +(1/840) D_l2t2X D_h1k1Y_l1t1 d_k1l2 d_l1h1 d_t1t2
f65: +(1/840) D_l2t2X D_h1k1Y_l1t1 d_h1l2 d_l1k1 d_t1t2
f66: +(1/840) D_l2t2X D_h1k1Y_l1t1 d_k1h1 d_l1l2 d_t1t2
f67: +(1/840) D_l1t1X D_h2k2Y_l2t2 d_l1t1 d_l2k2 d_t2h2
f68: +(1/840) D_l1t1X D_h2k2Y_l2t2 d_k2t1 d_l2l1 d_t2h2
f69: +(1/840) D_l1t1X D_h2k2Y_l2t2 d_k2l1 d_l2t1 d_t2h2
f70: +(1/840) D_l1t1X D_h2k2Y_l2t2 d_l1t1 d_l2h2 d_t2k2
f71: +(1/840) D_l1t1X D_h2k2Y_l2t2 d_h2t1 d_l2l1 d_t2k2
f72: +(1/840) D_l1t1X D_h2k2Y_l2t2 d_h2l1 d_l2t1 d_t2k2
f73: +(1/840) D_l1t1X D_h2k2Y_l2t2 d_k2t1 d_l2h2 d_t2l1
f74: +(1/840) D_l1t1X D_h2k2Y_l2t2 d_h2t1 d_l2k2 d_t2l1
f75: +(1/840) D_l1t1X D_h2k2Y_l2t2 d_k2h2 d_l2t1 d_t2l1
f76: +(1/840) D_l1t1X D_h2k2Y_l2t2 d_k2l1 d_l2h2 d_t2t1
f77: +(1/840) D_l1t1X D_h2k2Y_l2t2 d_h2l1 d_l2k2 d_t2t1
f78: +(1/840) D_l1t1X D_h2k2Y_l2t2 d_k2h2 d_l2l1 d_t2t1
f79: +(-1/240) D_a2a2X D_h1k1Y_l1t1 d_k1h1 d_l1t1
f80: +(-1/240) D_a2a2X D_h1k1Y_l1t1 d_l1k1 d_t1h1
f81: +(-1/240) D_a2a2X D_h1k1Y_l1t1 d_l1h1 d_t1k1
DDY DDY type of invariant
f82: +(1/720) D_a1a1Y_l1t1 D_a2a2Y_l2t2 d_l1t1 d_l2t2
f83: +(1/480) D_a1t1Y_a1l1 D_a2a2Y_l2t2 d_l1t1 d_l2t2
f84: +(1/720) D_a1t1Y_l1a1 D_a2a2Y_l2t2 d_l1t1 d_l2t2
f85: +(1/480) D_a1a1Y_l1t1 D_a2t2Y_a2l2 d_l1t1 d_l2t2
f86: +(1/320) D_a1t1Y_a1l1 D_a2t2Y_a2l2 d_l1t1 d_l2t2
f87: +(1/480) D_a1t1Y_l1a1 D_a2t2Y_a2l2 d_l1t1 d_l2t2
f88: +(1/720) D_a1a1Y_l1t1 D_a2t2Y_l2a2 d_l1t1 d_l2t2
f89: +(1/480) D_a1t1Y_a1l1 D_a2t2Y_l2a2 d_l1t1 d_l2t2
f90: +(1/720) D_a1t1Y_l1a1 D_a2t2Y_l2a2 d_l1t1 d_l2t2
f91: +(1/720) D_a1a1Y_l1t1 D_a2a2Y_l2t2 d_l1t2 d_t1l2
f92: +(1/480) D_a1t1Y_a1l1 D_a2a2Y_l2t2 d_l1t2 d_t1l2
f93: +(1/720) D_a1t1Y_l1a1 D_a2a2Y_l2t2 d_l1t2 d_t1l2
f94: +(1/480) D_a1a1Y_l1t1 D_a2t2Y_a2l2 d_l1t2 d_t1l2
f95: +(1/320) D_a1t1Y_a1l1 D_a2t2Y_a2l2 d_l1t2 d_t1l2
f96: +(1/480) D_a1t1Y_l1a1 D_a2t2Y_a2l2 d_l1t2 d_t1l2
f97: +(1/720) D_a1a1Y_l1t1 D_a2t2Y_l2a2 d_l1t2 d_t1l2
f98: +(1/480) D_a1t1Y_a1l1 D_a2t2Y_l2a2 d_l1t2 d_t1l2
f99: +(1/720) D_a1t1Y_l1a1 D_a2t2Y_l2a2 d_l1t2 d_t1l2
f100: +(1/720) D_a1a1Y_l1t1 D_a2a2Y_l2t2 d_l1l2 d_t1t2
f101: +(1/480) D_a1t1Y_a1l1 D_a2a2Y_l2t2 d_l1l2 d_t1t2
f102: +(1/720) D_a1t1Y_l1a1 D_a2a2Y_l2t2 d_l1l2 d_t1t2
f103: +(1/480) D_a1a1Y_l1t1 D_a2t2Y_a2l2 d_l1l2 d_t1t2
f104: +(1/320) D_a1t1Y_a1l1 D_a2t2Y_a2l2 d_l1l2 d_t1t2
f105: +(1/480) D_a1t1Y_l1a1 D_a2t2Y_a2l2 d_l1l2 d_t1t2
f106: +(1/720) D_a1a1Y_l1t1 D_a2t2Y_l2a2 d_l1l2 d_t1t2
f107: +(1/480) D_a1t1Y_a1l1 D_a2t2Y_l2a2 d_l1l2 d_t1t2
f108: +(1/720) D_a1t1Y_l1a1 D_a2t2Y_l2a2 d_l1l2 d_t1t2
f109: +(1/960) D_t1a1Y_l1a1 D_t2a2Y_l2a2 d_l1t1 d_l2t2
f110: +(1/960) D_t1a1Y_l1a1 D_t2a2Y_l2a2 d_l1t2 d_t1l2
f111: +(1/960) D_t1a1Y_l1a1 D_t2a2Y_l2a2 d_l1l2 d_t1t2
f112: +(1/720) D_t2a2Y_l2a2 D_a1a1Y_l1t1 d_l1t1 d_l2t2
f113: +(1/480) D_t2a2Y_l2a2 D_a1t1Y_a1l1 d_l1t1 d_l2t2
f114: +(1/720) D_t2a2Y_l2a2 D_a1t1Y_l1a1 d_l1t1 d_l2t2
f115: +(-1/960) D_a2t2Y_a2l2 D_t1a1Y_l1a1 d_l1t1 d_l2t2
f116: +(1/720) D_t2a2Y_l2a2 D_a1a1Y_l1t1 d_l1t2 d_t1l2
f117: +(1/480) D_t2a2Y_l2a2 D_a1t1Y_a1l1 d_l1t2 d_t1l2
f118: +(1/720) D_t2a2Y_l2a2 D_a1t1Y_l1a1 d_l1t2 d_t1l2
f119: +(-1/960) D_a2t2Y_a2l2 D_t1a1Y_l1a1 d_l1t2 d_t1l2
f120: +(1/720) D_t2a2Y_l2a2 D_a1a1Y_l1t1 d_l1l2 d_t1t2
f121: +(1/480) D_t2a2Y_l2a2 D_a1t1Y_a1l1 d_l1l2 d_t1t2
f122: +(1/720) D_t2a2Y_l2a2 D_a1t1Y_l1a1 d_l1l2 d_t1t2
f123: +(-1/960) D_a2t2Y_a2l2 D_t1a1Y_l1a1 d_l1l2 d_t1t2
f124: +(-1/480) D_a1a1Y_l1t1 D_t2a2Y_l2a2 d_l1t1 d_l2t2
f125: +(-1/320) D_a1t1Y_a1l1 D_t2a2Y_l2a2 d_l1t1 d_l2t2
f126: +(-1/480) D_a1t1Y_l1a1 D_t2a2Y_l2a2 d_l1t1 d_l2t2
f127: +(-1/480) D_a1a1Y_l1t1 D_t2a2Y_l2a2 d_l1t2 d_t1l2
f128: +(-1/320) D_a1t1Y_a1l1 D_t2a2Y_l2a2 d_l1t2 d_t1l2
f129: +(-1/480) D_a1t1Y_l1a1 D_t2a2Y_l2a2 d_l1t2 d_t1l2
f130: +(-1/480) D_a1a1Y_l1t1 D_t2a2Y_l2a2 d_l1l2 d_t1t2
f131: +(-1/320) D_a1t1Y_a1l1 D_t2a2Y_l2a2 d_l1l2 d_t1t2
f132: +(-1/480) D_a1t1Y_l1a1 D_t2a2Y_l2a2 d_l1l2 d_t1t2
f133: +(-1/1440) D_a2a2Y_l2t2 D_t1a1Y_l1a1 d_l1t1 d_l2t2
f134: +(-1/1440) D_a2t2Y_l2a2 D_t1a1Y_l1a1 d_l1t1 d_l2t2
f135: +(-1/1440) D_a2a2Y_l2t2 D_t1a1Y_l1a1 d_l1t2 d_t1l2
f136: +(-1/1440) D_a2t2Y_l2a2 D_t1a1Y_l1a1 d_l1t2 d_t1l2
f137: +(-1/1440) D_a2a2Y_l2t2 D_t1a1Y_l1a1 d_l1l2 d_t1t2
f138: +(-1/1440) D_a2t2Y_l2a2 D_t1a1Y_l1a1 d_l1l2 d_t1t2
f139: +(-1/1440) D_t2a2Y_l2a2 D_t1a1Y_l1a1 d_l1t1 d_l2t2
f140: +(-1/1440) D_t2a2Y_l2a2 D_t1a1Y_l1a1 d_l1t2 d_t1l2
f141: +(-1/1440) D_t2a2Y_l2a2 D_t1a1Y_l1a1 d_l1l2 d_t1t2
f142: +(1/3360) D_h2k2Y_l2t2 D_t1a1Y_l1a1 d_h2t1 d_k2l1 d_l2t2
f143: +(1/3360) D_h2k2Y_l2t2 D_t1a1Y_l1a1 d_h2l1 d_k2t1 d_l2t2
f144: +(1/3360) D_h2k2Y_l2t2 D_t1a1Y_l1a1 d_k2h2 d_l1t1 d_l2t2
f145: +(1/1120) D_h1k1Y_l1t1 D_t2a2Y_l2a2 d_h1t2 d_k1l2 d_l1t1
f146: +(1/1120) D_h1k1Y_l1t1 D_t2a2Y_l2a2 d_h1l2 d_k1t2 d_l1t1
f147: +(1/1120) D_h1k1Y_l1t1 D_t2a2Y_l2a2 d_k1h1 d_l1t1 d_l2t2
f148: +(1/1120) D_h1k1Y_l1t1 D_t2a2Y_l2a2 d_k1t2 d_l1l2 d_t1h1
f149: +(1/1120) D_h1k1Y_l1t1 D_t2a2Y_l2a2 d_k1l2 d_l1t2 d_t1h1
f150: +(1/1120) D_h1k1Y_l1t1 D_t2a2Y_l2a2 d_l1k1 d_l2t2 d_t1h1
f151: +(1/1120) D_h1k1Y_l1t1 D_t2a2Y_l2a2 d_h1t2 d_l1l2 d_t1k1
f152: +(1/1120) D_h1k1Y_l1t1 D_t2a2Y_l2a2 d_h1l2 d_l1t2 d_t1k1
f153: +(1/1120) D_h1k1Y_l1t1 D_t2a2Y_l2a2 d_l1h1 d_l2t2 d_t1k1
f154: +(1/1120) D_h1k1Y_l1t1 D_t2a2Y_l2a2 d_k1t2 d_l1h1 d_t1l2
f155: +(1/1120) D_h1k1Y_l1t1 D_t2a2Y_l2a2 d_h1t2 d_l1k1 d_t1l2
f156: +(1/1120) D_h1k1Y_l1t1 D_t2a2Y_l2a2 d_k1h1 d_l1t2 d_t1l2
f157: +(1/1120) D_h1k1Y_l1t1 D_t2a2Y_l2a2 d_k1l2 d_l1h1 d_t1t2
f158: +(1/1120) D_h1k1Y_l1t1 D_t2a2Y_l2a2 d_h1l2 d_l1k1 d_t1t2
f159: +(1/1120) D_h1k1Y_l1t1 D_t2a2Y_l2a2 d_k1h1 d_l1l2 d_t1t2
f160: +(1/3360) D_h2k2Y_l2t2 D_t1a1Y_l1a1 d_l1t1 d_l2k2 d_t2h2
f161: +(1/3360) D_h2k2Y_l2t2 D_t1a1Y_l1a1 d_k2t1 d_l2l1 d_t2h2
f162: +(1/3360) D_h2k2Y_l2t2 D_t1a1Y_l1a1 d_k2l1 d_l2t1 d_t2h2
f163: +(1/3360) D_h2k2Y_l2t2 D_t1a1Y_l1a1 d_l1t1 d_l2h2 d_t2k2
f164: +(1/3360) D_h2k2Y_l2t2 D_t1a1Y_l1a1 d_h2t1 d_l2l1 d_t2k2
f165: +(1/3360) D_h2k2Y_l2t2 D_t1a1Y_l1a1 d_h2l1 d_l2t1 d_t2k2
f166: +(1/3360) D_h2k2Y_l2t2 D_t1a1Y_l1a1 d_k2t1 d_l2h2 d_t2l1
f167: +(1/3360) D_h2k2Y_l2t2 D_t1a1Y_l1a1 d_h2t1 d_l2k2 d_t2l1
f168: +(1/3360) D_h2k2Y_l2t2 D_t1a1Y_l1a1 d_k2h2 d_l2t1 d_t2l1
f169: +(1/3360) D_h2k2Y_l2t2 D_t1a1Y_l1a1 d_k2l1 d_l2h2 d_t2t1
f170: +(1/3360) D_h2k2Y_l2t2 D_t1a1Y_l1a1 d_h2l1 d_l2k2 d_t2t1
f171: +(1/3360) D_h2k2Y_l2t2 D_t1a1Y_l1a1 d_k2h2 d_l2l1 d_t2t1
f172: +(-1/1680) D_a2a2Y_l2t2 D_h1k1Y_l1t1 d_h1t2 d_k1l2 d_l1t1
f173: +(-1/1120) D_a2t2Y_a2l2 D_h1k1Y_l1t1 d_h1t2 d_k1l2 d_l1t1
f174: +(-1/1680) D_a2t2Y_l2a2 D_h1k1Y_l1t1 d_h1t2 d_k1l2 d_l1t1
f175: +(-1/1680) D_a2a2Y_l2t2 D_h1k1Y_l1t1 d_h1l2 d_k1t2 d_l1t1
f176: +(-1/1120) D_a2t2Y_a2l2 D_h1k1Y_l1t1 d_h1l2 d_k1t2 d_l1t1
f177: +(-1/1680) D_a2t2Y_l2a2 D_h1k1Y_l1t1 d_h1l2 d_k1t2 d_l1t1
f178: +(-1/1680) D_a1a1Y_l1t1 D_h2k2Y_l2t2 d_h2t1 d_k2l1 d_l2t2
f179: +(-1/1120) D_a1t1Y_a1l1 D_h2k2Y_l2t2 d_h2t1 d_k2l1 d_l2t2
f180: +(-1/1680) D_a1t1Y_l1a1 D_h2k2Y_l2t2 d_h2t1 d_k2l1 d_l2t2
f181: +(-1/1680) D_a1a1Y_l1t1 D_h2k2Y_l2t2 d_h2l1 d_k2t1 d_l2t2
f182: +(-1/1120) D_a1t1Y_a1l1 D_h2k2Y_l2t2 d_h2l1 d_k2t1 d_l2t2
f183: +(-1/1680) D_a1t1Y_l1a1 D_h2k2Y_l2t2 d_h2l1 d_k2t1 d_l2t2
f184: +(-1/1680) D_a2a2Y_l2t2 D_h1k1Y_l1t1 d_k1h1 d_l1t1 d_l2t2
f185: +(-1/1120) D_a2t2Y_a2l2 D_h1k1Y_l1t1 d_k1h1 d_l1t1 d_l2t2
f186: +(-1/1680) D_a2t2Y_l2a2 D_h1k1Y_l1t1 d_k1h1 d_l1t1 d_l2t2
f187: +(-1/1680) D_a1a1Y_l1t1 D_h2k2Y_l2t2 d_k2h2 d_l1t1 d_l2t2
f188: +(-1/1120) D_a1t1Y_a1l1 D_h2k2Y_l2t2 d_k2h2 d_l1t1 d_l2t2
f189: +(-1/1680) D_a1t1Y_l1a1 D_h2k2Y_l2t2 d_k2h2 d_l1t1 d_l2t2
f190: +(-1/1680) D_a2a2Y_l2t2 D_h1k1Y_l1t1 d_k1t2 d_l1l2 d_t1h1
f191: +(-1/1120) D_a2t2Y_a2l2 D_h1k1Y_l1t1 d_k1t2 d_l1l2 d_t1h1
f192: +(-1/1680) D_a2t2Y_l2a2 D_h1k1Y_l1t1 d_k1t2 d_l1l2 d_t1h1
f193: +(-1/1680) D_a2a2Y_l2t2 D_h1k1Y_l1t1 d_k1l2 d_l1t2 d_t1h1
f194: +(-1/1120) D_a2t2Y_a2l2 D_h1k1Y_l1t1 d_k1l2 d_l1t2 d_t1h1
f195: +(-1/1680) D_a2t2Y_l2a2 D_h1k1Y_l1t1 d_k1l2 d_l1t2 d_t1h1
f196: +(-1/1680) D_a2a2Y_l2t2 D_h1k1Y_l1t1 d_l1k1 d_l2t2 d_t1h1
f197: +(-1/1120) D_a2t2Y_a2l2 D_h1k1Y_l1t1 d_l1k1 d_l2t2 d_t1h1
f198: +(-1/1680) D_a2t2Y_l2a2 D_h1k1Y_l1t1 d_l1k1 d_l2t2 d_t1h1
f199: +(-1/1680) D_a2a2Y_l2t2 D_h1k1Y_l1t1 d_h1t2 d_l1l2 d_t1k1
f200: +(-1/1120) D_a2t2Y_a2l2 D_h1k1Y_l1t1 d_h1t2 d_l1l2 d_t1k1
f201: +(-1/1680) D_a2t2Y_l2a2 D_h1k1Y_l1t1 d_h1t2 d_l1l2 d_t1k1
f202: +(-1/1680) D_a2a2Y_l2t2 D_h1k1Y_l1t1 d_h1l2 d_l1t2 d_t1k1
f203: +(-1/1120) D_a2t2Y_a2l2 D_h1k1Y_l1t1 d_h1l2 d_l1t2 d_t1k1
f204: +(-1/1680) D_a2t2Y_l2a2 D_h1k1Y_l1t1 d_h1l2 d_l1t2 d_t1k1
f205: +(-1/1680) D_a2a2Y_l2t2 D_h1k1Y_l1t1 d_l1h1 d_l2t2 d_t1k1
f206: +(-1/1120) D_a2t2Y_a2l2 D_h1k1Y_l1t1 d_l1h1 d_l2t2 d_t1k1
f207: +(-1/1680) D_a2t2Y_l2a2 D_h1k1Y_l1t1 d_l1h1 d_l2t2 d_t1k1
f208: +(-1/1680) D_a2a2Y_l2t2 D_h1k1Y_l1t1 d_k1t2 d_l1h1 d_t1l2
f209: +(-1/1120) D_a2t2Y_a2l2 D_h1k1Y_l1t1 d_k1t2 d_l1h1 d_t1l2
f210: +(-1/1680) D_a2t2Y_l2a2 D_h1k1Y_l1t1 d_k1t2 d_l1h1 d_t1l2
f211: +(-1/1680) D_a2a2Y_l2t2 D_h1k1Y_l1t1 d_h1t2 d_l1k1 d_t1l2
f212: +(-1/1120) D_a2t2Y_a2l2 D_h1k1Y_l1t1 d_h1t2 d_l1k1 d_t1l2
f213: +(-1/1680) D_a2t2Y_l2a2 D_h1k1Y_l1t1 d_h1t2 d_l1k1 d_t1l2
f214: +(-1/1680) D_a2a2Y_l2t2 D_h1k1Y_l1t1 d_k1h1 d_l1t2 d_t1l2
f215: +(-1/1120) D_a2t2Y_a2l2 D_h1k1Y_l1t1 d_k1h1 d_l1t2 d_t1l2
f216: +(-1/1680) D_a2t2Y_l2a2 D_h1k1Y_l1t1 d_k1h1 d_l1t2 d_t1l2
f217: +(-1/1680) D_a2a2Y_l2t2 D_h1k1Y_l1t1 d_k1l2 d_l1h1 d_t1t2
f218: +(-1/1120) D_a2t2Y_a2l2 D_h1k1Y_l1t1 d_k1l2 d_l1h1 d_t1t2
f219: +(-1/1680) D_a2t2Y_l2a2 D_h1k1Y_l1t1 d_k1l2 d_l1h1 d_t1t2
f220: +(-1/1680) D_a2a2Y_l2t2 D_h1k1Y_l1t1 d_h1l2 d_l1k1 d_t1t2
f221: +(-1/1120) D_a2t2Y_a2l2 D_h1k1Y_l1t1 d_h1l2 d_l1k1 d_t1t2
f222: +(-1/1680) D_a2t2Y_l2a2 D_h1k1Y_l1t1 d_h1l2 d_l1k1 d_t1t2
f223: +(-1/1680) D_a2a2Y_l2t2 D_h1k1Y_l1t1 d_k1h1 d_l1l2 d_t1t2
f224: +(-1/1120) D_a2t2Y_a2l2 D_h1k1Y_l1t1 d_k1h1 d_l1l2 d_t1t2
f225: +(-1/1680) D_a2t2Y_l2a2 D_h1k1Y_l1t1 d_k1h1 d_l1l2 d_t1t2
f226: +(-1/1680) D_a1a1Y_l1t1 D_h2k2Y_l2t2 d_l1t1 d_l2k2 d_t2h2
f227: +(-1/1120) D_a1t1Y_a1l1 D_h2k2Y_l2t2 d_l1t1 d_l2k2 d_t2h2
f228: +(-1/1680) D_a1t1Y_l1a1 D_h2k2Y_l2t2 d_l1t1 d_l2k2 d_t2h2
f229: +(-1/1680) D_a1a1Y_l1t1 D_h2k2Y_l2t2 d_k2t1 d_l2l1 d_t2h2
f230: +(-1/1120) D_a1t1Y_a1l1 D_h2k2Y_l2t2 d_k2t1 d_l2l1 d_t2h2
f231: +(-1/1680) D_a1t1Y_l1a1 D_h2k2Y_l2t2 d_k2t1 d_l2l1 d_t2h2
f232: +(-1/1680) D_a1a1Y_l1t1 D_h2k2Y_l2t2 d_k2l1 d_l2t1 d_t2h2
f233: +(-1/1120) D_a1t1Y_a1l1 D_h2k2Y_l2t2 d_k2l1 d_l2t1 d_t2h2
f234: +(-1/1680) D_a1t1Y_l1a1 D_h2k2Y_l2t2 d_k2l1 d_l2t1 d_t2h2
f235: +(-1/1680) D_a1a1Y_l1t1 D_h2k2Y_l2t2 d_l1t1 d_l2h2 d_t2k2
f236: +(-1/1120) D_a1t1Y_a1l1 D_h2k2Y_l2t2 d_l1t1 d_l2h2 d_t2k2
f237: +(-1/1680) D_a1t1Y_l1a1 D_h2k2Y_l2t2 d_l1t1 d_l2h2 d_t2k2
f238: +(-1/1680) D_a1a1Y_l1t1 D_h2k2Y_l2t2 d_h2t1 d_l2l1 d_t2k2
f239: +(-1/1120) D_a1t1Y_a1l1 D_h2k2Y_l2t2 d_h2t1 d_l2l1 d_t2k2
f240: +(-1/1680) D_a1t1Y_l1a1 D_h2k2Y_l2t2 d_h2t1 d_l2l1 d_t2k2
f241: +(-1/1680) D_a1a1Y_l1t1 D_h2k2Y_l2t2 d_h2l1 d_l2t1 d_t2k2
f242: +(-1/1120) D_a1t1Y_a1l1 D_h2k2Y_l2t2 d_h2l1 d_l2t1 d_t2k2
f243: +(-1/1680) D_a1t1Y_l1a1 D_h2k2Y_l2t2 d_h2l1 d_l2t1 d_t2k2
f244: +(-1/1680) D_a1a1Y_l1t1 D_h2k2Y_l2t2 d_k2t1 d_l2h2 d_t2l1
f245: +(-1/1120) D_a1t1Y_a1l1 D_h2k2Y_l2t2 d_k2t1 d_l2h2 d_t2l1
f246: +(-1/1680) D_a1t1Y_l1a1 D_h2k2Y_l2t2 d_k2t1 d_l2h2 d_t2l1
f247: +(-1/1680) D_a1a1Y_l1t1 D_h2k2Y_l2t2 d_h2t1 d_l2k2 d_t2l1
f248: +(-1/1120) D_a1t1Y_a1l1 D_h2k2Y_l2t2 d_h2t1 d_l2k2 d_t2l1
f249: +(-1/1680) D_a1t1Y_l1a1 D_h2k2Y_l2t2 d_h2t1 d_l2k2 d_t2l1
f250: +(-1/1680) D_a1a1Y_l1t1 D_h2k2Y_l2t2 d_k2h2 d_l2t1 d_t2l1
f251: +(-1/1120) D_a1t1Y_a1l1 D_h2k2Y_l2t2 d_k2h2 d_l2t1 d_t2l1
f252: +(-1/1680) D_a1t1Y_l1a1 D_h2k2Y_l2t2 d_k2h2 d_l2t1 d_t2l1
f253: +(-1/1680) D_a1a1Y_l1t1 D_h2k2Y_l2t2 d_k2l1 d_l2h2 d_t2t1
f254: +(-1/1120) D_a1t1Y_a1l1 D_h2k2Y_l2t2 d_k2l1 d_l2h2 d_t2t1
f255: +(-1/1680) D_a1t1Y_l1a1 D_h2k2Y_l2t2 d_k2l1 d_l2h2 d_t2t1
f256: +(-1/1680) D_a1a1Y_l1t1 D_h2k2Y_l2t2 d_h2l1 d_l2k2 d_t2t1
f257: +(-1/1120) D_a1t1Y_a1l1 D_h2k2Y_l2t2 d_h2l1 d_l2k2 d_t2t1
f258: +(-1/1680) D_a1t1Y_l1a1 D_h2k2Y_l2t2 d_h2l1 d_l2k2 d_t2t1
f259: +(-1/1680) D_a1a1Y_l1t1 D_h2k2Y_l2t2 d_k2h2 d_l2l1 d_t2t1
f260: +(-1/1120) D_a1t1Y_a1l1 D_h2k2Y_l2t2 d_k2h2 d_l2l1 d_t2t1
f261: +(-1/1680) D_a1t1Y_l1a1 D_h2k2Y_l2t2 d_k2h2 d_l2l1 d_t2t1
f262: +(-1/1680) D_t2a2Y_l2a2 D_h1k1Y_l1t1 d_h1t2 d_k1l2 d_l1t1
f263: +(-1/1680) D_t2a2Y_l2a2 D_h1k1Y_l1t1 d_h1l2 d_k1t2 d_l1t1
f264: +(-1/1680) D_t2a2Y_l2a2 D_h1k1Y_l1t1 d_k1h1 d_l1t1 d_l2t2
f265: +(-1/1680) D_t2a2Y_l2a2 D_h1k1Y_l1t1 d_k1t2 d_l1l2 d_t1h1
f266: +(-1/1680) D_t2a2Y_l2a2 D_h1k1Y_l1t1 d_k1l2 d_l1t2 d_t1h1
f267: +(-1/1680) D_t2a2Y_l2a2 D_h1k1Y_l1t1 d_l1k1 d_l2t2 d_t1h1
f268: +(-1/1680) D_t2a2Y_l2a2 D_h1k1Y_l1t1 d_h1t2 d_l1l2 d_t1k1
f269: +(-1/1680) D_t2a2Y_l2a2 D_h1k1Y_l1t1 d_h1l2 d_l1t2 d_t1k1
f270: +(-1/1680) D_t2a2Y_l2a2 D_h1k1Y_l1t1 d_l1h1 d_l2t2 d_t1k1
f271: +(-1/1680) D_t2a2Y_l2a2 D_h1k1Y_l1t1 d_k1t2 d_l1h1 d_t1l2
f272: +(-1/1680) D_t2a2Y_l2a2 D_h1k1Y_l1t1 d_h1t2 d_l1k1 d_t1l2
f273: +(-1/1680) D_t2a2Y_l2a2 D_h1k1Y_l1t1 d_k1h1 d_l1t2 d_t1l2
f274: +(-1/1680) D_t2a2Y_l2a2 D_h1k1Y_l1t1 d_k1l2 d_l1h1 d_t1t2
f275: +(-1/1680) D_t2a2Y_l2a2 D_h1k1Y_l1t1 d_h1l2 d_l1k1 d_t1t2
f276: +(-1/1680) D_t2a2Y_l2a2 D_h1k1Y_l1t1 d_k1h1 d_l1l2 d_t1t2
f277: +(1/13440) D_h1k1Y_l1t1 D_h2k2Y_l2t2 d_k1h1 d_k2h2 d_l1t1 d_l2t2
f278: +(1/13440) D_h1k1Y_l1t1 D_h2k2Y_l2t2 d_k2h2 d_l1k1 d_l2t2 d_t1h1
f279: +(1/13440) D_h1k1Y_l1t1 D_h2k2Y_l2t2 d_k2h2 d_l1h1 d_l2t2 d_t1k1
f280: +(1/13440) D_h1k1Y_l1t1 D_h2k2Y_l2t2 d_k1h1 d_l1t1 d_l2k2 d_t2h2
f281: +(1/13440) D_h1k1Y_l1t1 D_h2k2Y_l2t2 d_l1k1 d_l2k2 d_t1h1 d_t2h2
f282: +(1/13440) D_h1k1Y_l1t1 D_h2k2Y_l2t2 d_l1h1 d_l2k2 d_t1k1 d_t2h2
f283: +(1/13440) D_h1k1Y_l1t1 D_h2k2Y_l2t2 d_k1h1 d_l1t1 d_l2h2 d_t2k2
f284: +(1/13440) D_h1k1Y_l1t1 D_h2k2Y_l2t2 d_l1k1 d_l2h2 d_t1h1 d_t2k2
f285: +(1/13440) D_h1k1Y_l1t1 D_h2k2Y_l2t2 d_l1h1 d_l2h2 d_t1k1 d_t2k2
\end{verbatim}
\end{scriptsize}

Using Mathematica to contract kronecker delta indices, we have:

\begin{scriptsize}
\begin{verbatim}
f1: {a1,a1,a2,a2}
f2: {a2,a2,l1,t1}/.{l1->t1}
f3: {a1,a1,l2,t2}/.{l2->t2}
f4: {l1,t1,l2,t2}/.{l1->t1,l2->t2}
f5: {l1,t1,l2,t2}/.{l1->t2,t1->l2}
f6{l1,t1,l2,t2}/.{l1->l2,t1->t2}
:
f7: {a2,a2,a1,t1,a1,l1}/.{l1->t1}
f8: {a1,a1,a2,t2,a2,l2}/.{l2->t2}
f9: {a2,a2,a1,a1,l1,t1}/.{l1->t1}
f10: {a2,a2,a1,t1,l1,a1}/.{l1->t1}
f11: {a1,a1,a2,a2,l2,t2}/.{l2->t2}
f12: {a1,a1,a2,t2,l2,a2}/.{l2->t2}
f13: {a1,a1,t2,a2,l2,a2}/.{l2->t2}
f14: {a1,a1,t2,a2,l2,a2}/.{l2->t2}
f15: {a2,a2,t1,a1,l1,a1}/.{l1->t1}
f16: {a2,a2,h1,k1,l1,t1}/.{k1->h1,l1->t1}
f17: {a2,a2,h1,k1,l1,t1}/.{l1->k1,t1->h1}
f18: {a2,a2,h1,k1,l1,t1}/.{l1->h1,t1->k1}
f19: {l2,t2,t1,a1,l1,a1}/.{l1->t1,l2->t2}
f20: {l2,t2,t1,a1,l1,a1}/.{l1->t2,t1->l2}
f21: {l2,t2,t1,a1,l1,a1}/.{l1->l2,t1->t2}
f22: {l1,t1,t2,a2,l2,a2}/.{l1->t1,l2->t2}
f23: {l1,t1,t2,a2,l2,a2}/.{l1->t2,t1->l2}
f24: {l1,t1,t2,a2,l2,a2}/.{l1->l2,t1->t2}
f25: {a1,a1,h2,k2,l2,t2}/.{k2->h2,l2->t2}
f26: {l1,t1,a2,a2,l2,t2}/.{l1->t1,l2->t2}
f27: {l1,t1,a2,t2,a2,l2}/.{l1->t1,l2->t2}
f28: {l1,t1,a2,t2,l2,a2}/.{l1->t1,l2->t2}
f29: {l2,t2,a1,a1,l1,t1}/.{l1->t1,l2->t2}
f30: {l2,t2,a1,t1,a1,l1}/.{l1->t1,l2->t2}
f31: {l2,t2,a1,t1,l1,a1}/.{l1->t1,l2->t2}
f32: {l1,t1,a2,a2,l2,t2}/.{l1->t2,t1->l2}
f33: {l1,t1,a2,t2,a2,l2}/.{l1->t2,t1->l2}
f34: {l1,t1,a2,t2,l2,a2}/.{l1->t2,t1->l2}
f35: {l2,t2,a1,a1,l1,t1}/.{l1->t2,t1->l2}
f36: {l2,t2,a1,t1,a1,l1}/.{l1->t2,t1->l2}
f37: {l2,t2,a1,t1,l1,a1}/.{l1->t2,t1->l2}
f38: {l1,t1,a2,a2,l2,t2}/.{l1->l2,t1->t2}
f39: {l1,t1,a2,t2,a2,l2}/.{l1->l2,t1->t2}
f40: {l1,t1,a2,t2,l2,a2}/.{l1->l2,t1->t2}
f41: {l2,t2,a1,a1,l1,t1}/.{l1->l2,t1->t2}
f42: {l2,t2,a1,t1,a1,l1}/.{l1->l2,t1->t2}
f43: {l2,t2,a1,t1,l1,a1}/.{l1->l2,t1->t2}
f44: {a1,a1,h2,k2,l2,t2}/.{l2->k2,t2->h2}
f45: {a1,a1,h2,k2,l2,t2}/.{l2->h2,t2->k2}
f46: {l1,t1,t2,a2,l2,a2}/.{l1->t1,l2->t2}
f47: {l1,t1,t2,a2,l2,a2}/.{l1->t2,t1->l2}
f48: {l1,t1,t2,a2,l2,a2}/.{l1->l2,t1->t2}
f49: {a2,a2,h1,k1,l1,t1}/.{k1->h1,l1->t1}
f50: {a2,a2,h1,k1,l1,t1}/.{l1->k1,t1->h1}
f51: {a2,a2,h1,k1,l1,t1}/.{l1->h1,t1->k1}
f52: {l1,t1,h2,k2,l2,t2}/.{h2->t1,k2->l1,l2->t2}
f53: {l1,t1,h2,k2,l2,t2}/.{h2->l1,k2->t1,l2->t2}
f54: {l1,t1,h2,k2,l2,t2}/.{k2->h2,l1->t1,l2->t2}
f55: {l2,t2,h1,k1,l1,t1}/.{h1->t2,k1->l2,l1->t1}
f56: {l2,t2,h1,k1,l1,t1}/.{h1->l2,k1->t2,l1->t1}
f57: {l2,t2,h1,k1,l1,t1}/.{k1->h1,l1->t1,l2->t2}
f58: {l2,t2,h1,k1,l1,t1}/.{k1->t2,l1->l2,t1->h1}
f59: {l2,t2,h1,k1,l1,t1}/.{k1->l2,l1->t2,t1->h1}
f60: {l2,t2,h1,k1,l1,t1}/.{l1->k1,l2->t2,t1->h1}
f61: {l2,t2,h1,k1,l1,t1}/.{h1->t2,l1->l2,t1->k1}
f62: {l2,t2,h1,k1,l1,t1}/.{h1->l2,l1->t2,t1->k1}
f63: {l2,t2,h1,k1,l1,t1}/.{l1->h1,l2->t2,t1->k1}
f64: {l2,t2,h1,k1,l1,t1}/.{k1->t2,l1->h1,t1->l2}
f65: {l2,t2,h1,k1,l1,t1}/.{h1->t2,l1->k1,t1->l2}
f66: {l2,t2,h1,k1,l1,t1}/.{k1->h1,l1->t2,t1->l2}
f67: {l2,t2,h1,k1,l1,t1}/.{k1->l2,l1->h1,t1->t2}
f68: {l2,t2,h1,k1,l1,t1}/.{h1->l2,l1->k1,t1->t2}
f69: {l2,t2,h1,k1,l1,t1}/.{k1->h1,l1->l2,t1->t2}
f70: {l1,t1,h2,k2,l2,t2}/.{l1->t1,l2->k2,t2->h2}
f71: {l1,t1,h2,k2,l2,t2}/.{k2->t1,l2->l1,t2->h2}
f72: {l1,t1,h2,k2,l2,t2}/.{k2->l1,l2->t1,t2->h2}
f73: {l1,t1,h2,k2,l2,t2}/.{l1->t1,l2->h2,t2->k2}
f74: {l1,t1,h2,k2,l2,t2}/.{h2->t1,l2->l1,t2->k2}
f75: {l1,t1,h2,k2,l2,t2}/.{h2->l1,l2->t1,t2->k2}
f76: {l1,t1,h2,k2,l2,t2}/.{k2->t1,l2->h2,t2->l1}
f77: {l1,t1,h2,k2,l2,t2}/.{h2->t1,l2->k2,t2->l1}
f78: {l1,t1,h2,k2,l2,t2}/.{k2->h2,l2->t1,t2->l1}
f79: {l1,t1,h2,k2,l2,t2}/.{k2->l1,l2->h2,t2->t1}
f80: {l1,t1,h2,k2,l2,t2}/.{h2->l1,l2->k2,t2->t1}
f81{l1,t1,h2,k2,l2,t2}/.{k2->h2,l2->l1,t2->t1}
:
f82: {a1,a1,l1,t1,a2,a2,l2,t2}/.{l1->t1,l2->t2}
f83: {a1,t1,a1,l1,a2,a2,l2,t2}/.{l1->t1,l2->t2}
f84: {a1,t1,l1,a1,a2,a2,l2,t2}/.{l1->t1,l2->t2}
f85: {a1,a1,l1,t1,a2,t2,a2,l2}/.{l1->t1,l2->t2}
f86: {a1,t1,a1,l1,a2,t2,a2,l2}/.{l1->t1,l2->t2}
f87: {a1,t1,l1,a1,a2,t2,a2,l2}/.{l1->t1,l2->t2}
f88: {a1,a1,l1,t1,a2,t2,l2,a2}/.{l1->t1,l2->t2}
f89: {a1,t1,a1,l1,a2,t2,l2,a2}/.{l1->t1,l2->t2}
f90: {a1,t1,l1,a1,a2,t2,l2,a2}/.{l1->t1,l2->t2}
f91: {a1,a1,l1,t1,a2,a2,l2,t2}/.{l1->t2,t1->l2}
f92: {a1,t1,a1,l1,a2,a2,l2,t2}/.{l1->t2,t1->l2}
f93: {a1,t1,l1,a1,a2,a2,l2,t2}/.{l1->t2,t1->l2}
f94: {a1,a1,l1,t1,a2,t2,a2,l2}/.{l1->t2,t1->l2}
f95: {a1,t1,a1,l1,a2,t2,a2,l2}/.{l1->t2,t1->l2}
f96: {a1,t1,l1,a1,a2,t2,a2,l2}/.{l1->t2,t1->l2}
f97: {a1,a1,l1,t1,a2,t2,l2,a2}/.{l1->t2,t1->l2}
f98: {a1,t1,a1,l1,a2,t2,l2,a2}/.{l1->t2,t1->l2}
f99: {a1,t1,l1,a1,a2,t2,l2,a2}/.{l1->t2,t1->l2}
f100: {a1,a1,l1,t1,a2,a2,l2,t2}/.{l1->l2,t1->t2}
f101: {a1,t1,a1,l1,a2,a2,l2,t2}/.{l1->l2,t1->t2}
f102: {a1,t1,l1,a1,a2,a2,l2,t2}/.{l1->l2,t1->t2}
f103: {a1,a1,l1,t1,a2,t2,a2,l2}/.{l1->l2,t1->t2}
f104: {a1,t1,a1,l1,a2,t2,a2,l2}/.{l1->l2,t1->t2}
f105: {a1,t1,l1,a1,a2,t2,a2,l2}/.{l1->l2,t1->t2}
f106: {a1,a1,l1,t1,a2,t2,l2,a2}/.{l1->l2,t1->t2}
f107: {a1,t1,a1,l1,a2,t2,l2,a2}/.{l1->l2,t1->t2}
f108: {a1,t1,l1,a1,a2,t2,l2,a2}/.{l1->l2,t1->t2}
f109: {t1,a1,l1,a1,t2,a2,l2,a2}/.{l1->t1,l2->t2}
f110: {t1,a1,l1,a1,t2,a2,l2,a2}/.{l1->t2,t1->l2}
f111: {t1,a1,l1,a1,t2,a2,l2,a2}/.{l1->l2,t1->t2}
f112: {t2,a2,l2,a2,a1,a1,l1,t1}/.{l1->t1,l2->t2}
f113: {t2,a2,l2,a2,a1,t1,a1,l1}/.{l1->t1,l2->t2}
f114: {t2,a2,l2,a2,a1,t1,l1,a1}/.{l1->t1,l2->t2}
f115: {a2,t2,a2,l2,t1,a1,l1,a1}/.{l1->t1,l2->t2}
f116: {t2,a2,l2,a2,a1,a1,l1,t1}/.{l1->t2,t1->l2}
f117: {t2,a2,l2,a2,a1,t1,a1,l1}/.{l1->t2,t1->l2}
f118: {t2,a2,l2,a2,a1,t1,l1,a1}/.{l1->t2,t1->l2}
f119: {a2,t2,a2,l2,t1,a1,l1,a1}/.{l1->t2,t1->l2}
f120: {t2,a2,l2,a2,a1,a1,l1,t1}/.{l1->l2,t1->t2}
f121: {t2,a2,l2,a2,a1,t1,a1,l1}/.{l1->l2,t1->t2}
f122: {t2,a2,l2,a2,a1,t1,l1,a1}/.{l1->l2,t1->t2}
f123: {a2,t2,a2,l2,t1,a1,l1,a1}/.{l1->l2,t1->t2}
f124: {a1,a1,l1,t1,t2,a2,l2,a2}/.{l1->t1,l2->t2}
f125: {a1,t1,a1,l1,t2,a2,l2,a2}/.{l1->t1,l2->t2}
f126: {a1,t1,l1,a1,t2,a2,l2,a2}/.{l1->t1,l2->t2}
f127: {a1,a1,l1,t1,t2,a2,l2,a2}/.{l1->t2,t1->l2}
f128: {a1,t1,a1,l1,t2,a2,l2,a2}/.{l1->t2,t1->l2}
f129: {a1,t1,l1,a1,t2,a2,l2,a2}/.{l1->t2,t1->l2}
f130: {a1,a1,l1,t1,t2,a2,l2,a2}/.{l1->l2,t1->t2}
f131: {a1,t1,a1,l1,t2,a2,l2,a2}/.{l1->l2,t1->t2}
f132: {a1,t1,l1,a1,t2,a2,l2,a2}/.{l1->l2,t1->t2}
f133: {a2,a2,l2,t2,t1,a1,l1,a1}/.{l1->t1,l2->t2}
f134: {a2,t2,l2,a2,t1,a1,l1,a1}/.{l1->t1,l2->t2}
f135: {a2,a2,l2,t2,t1,a1,l1,a1}/.{l1->t2,t1->l2}
f136: {a2,t2,l2,a2,t1,a1,l1,a1}/.{l1->t2,t1->l2}
f137: {a2,a2,l2,t2,t1,a1,l1,a1}/.{l1->l2,t1->t2}
f138: {a2,t2,l2,a2,t1,a1,l1,a1}/.{l1->l2,t1->t2}
f139: {t2,a2,l2,a2,t1,a1,l1,a1}/.{l1->t1,l2->t2}
f140: {t2,a2,l2,a2,t1,a1,l1,a1}/.{l1->t2,t1->l2}
f141: {t2,a2,l2,a2,t1,a1,l1,a1}/.{l1->l2,t1->t2}
f142: {h2,k2,l2,t2,t1,a1,l1,a1}/.{h2->t1,k2->l1,l2->t2}
f143: {h2,k2,l2,t2,t1,a1,l1,a1}/.{h2->l1,k2->t1,l2->t2}
f144: {h2,k2,l2,t2,t1,a1,l1,a1}/.{k2->h2,l1->t1,l2->t2}
f145: {h1,k1,l1,t1,t2,a2,l2,a2}/.{h1->t2,k1->l2,l1->t1}
f146: {h1,k1,l1,t1,t2,a2,l2,a2}/.{h1->l2,k1->t2,l1->t1}
f147: {h1,k1,l1,t1,t2,a2,l2,a2}/.{k1->h1,l1->t1,l2->t2}
f148: {h1,k1,l1,t1,t2,a2,l2,a2}/.{k1->t2,l1->l2,t1->h1}
f149: {h1,k1,l1,t1,t2,a2,l2,a2}/.{k1->l2,l1->t2,t1->h1}
f150: {h1,k1,l1,t1,t2,a2,l2,a2}/.{l1->k1,l2->t2,t1->h1}
f151: {h1,k1,l1,t1,t2,a2,l2,a2}/.{h1->t2,l1->l2,t1->k1}
f152: {h1,k1,l1,t1,t2,a2,l2,a2}/.{h1->l2,l1->t2,t1->k1}
f153: {h1,k1,l1,t1,t2,a2,l2,a2}/.{l1->h1,l2->t2,t1->k1}
f154: {h1,k1,l1,t1,t2,a2,l2,a2}/.{k1->t2,l1->h1,t1->l2}
f155: {h1,k1,l1,t1,t2,a2,l2,a2}/.{h1->t2,l1->k1,t1->l2}
f156: {h1,k1,l1,t1,t2,a2,l2,a2}/.{k1->h1,l1->t2,t1->l2}
f157: {h1,k1,l1,t1,t2,a2,l2,a2}/.{k1->l2,l1->h1,t1->t2}
f158: {h1,k1,l1,t1,t2,a2,l2,a2}/.{h1->l2,l1->k1,t1->t2}
f159: {h1,k1,l1,t1,t2,a2,l2,a2}/.{k1->h1,l1->l2,t1->t2}
f160: {h2,k2,l2,t2,t1,a1,l1,a1}/.{l1->t1,l2->k2,t2->h2}
f161: {h2,k2,l2,t2,t1,a1,l1,a1}/.{k2->t1,l2->l1,t2->h2}
f162: {h2,k2,l2,t2,t1,a1,l1,a1}/.{k2->l1,l2->t1,t2->h2}
f163: {h2,k2,l2,t2,t1,a1,l1,a1}/.{l1->t1,l2->h2,t2->k2}
f164: {h2,k2,l2,t2,t1,a1,l1,a1}/.{h2->t1,l2->l1,t2->k2}
f165: {h2,k2,l2,t2,t1,a1,l1,a1}/.{h2->l1,l2->t1,t2->k2}
f166: {h2,k2,l2,t2,t1,a1,l1,a1}/.{k2->t1,l2->h2,t2->l1}
f167: {h2,k2,l2,t2,t1,a1,l1,a1}/.{h2->t1,l2->k2,t2->l1}
f168: {h2,k2,l2,t2,t1,a1,l1,a1}/.{k2->h2,l2->t1,t2->l1}
f169: {h2,k2,l2,t2,t1,a1,l1,a1}/.{k2->l1,l2->h2,t2->t1}
f170: {h2,k2,l2,t2,t1,a1,l1,a1}/.{h2->l1,l2->k2,t2->t1}
f171: {h2,k2,l2,t2,t1,a1,l1,a1}/.{k2->h2,l2->l1,t2->t1}
f172: {a2,a2,l2,t2,h1,k1,l1,t1}/.{h1->t2,k1->l2,l1->t1}
f173: {a2,t2,a2,l2,h1,k1,l1,t1}/.{h1->t2,k1->l2,l1->t1}
f174: {a2,t2,l2,a2,h1,k1,l1,t1}/.{h1->t2,k1->l2,l1->t1}
f175: {a2,a2,l2,t2,h1,k1,l1,t1}/.{h1->l2,k1->t2,l1->t1}
f176: {a2,t2,a2,l2,h1,k1,l1,t1}/.{h1->l2,k1->t2,l1->t1}
f177: {a2,t2,l2,a2,h1,k1,l1,t1}/.{h1->l2,k1->t2,l1->t1}
f178: {a1,a1,l1,t1,h2,k2,l2,t2}/.{h2->t1,k2->l1,l2->t2}
f179: {a1,t1,a1,l1,h2,k2,l2,t2}/.{h2->t1,k2->l1,l2->t2}
f180: {a1,t1,l1,a1,h2,k2,l2,t2}/.{h2->t1,k2->l1,l2->t2}
f181: {a1,a1,l1,t1,h2,k2,l2,t2}/.{h2->l1,k2->t1,l2->t2}
f182: {a1,t1,a1,l1,h2,k2,l2,t2}/.{h2->l1,k2->t1,l2->t2}
f183: {a1,t1,l1,a1,h2,k2,l2,t2}/.{h2->l1,k2->t1,l2->t2}
f184: {a2,a2,l2,t2,h1,k1,l1,t1}/.{k1->h1,l1->t1,l2->t2}
f185: {a2,t2,a2,l2,h1,k1,l1,t1}/.{k1->h1,l1->t1,l2->t2}
f186: {a2,t2,l2,a2,h1,k1,l1,t1}/.{k1->h1,l1->t1,l2->t2}
f187: {a1,a1,l1,t1,h2,k2,l2,t2}/.{k2->h2,l1->t1,l2->t2}
f188: {a1,t1,a1,l1,h2,k2,l2,t2}/.{k2->h2,l1->t1,l2->t2}
f189: {a1,t1,l1,a1,h2,k2,l2,t2}/.{k2->h2,l1->t1,l2->t2}
f190: {a2,a2,l2,t2,h1,k1,l1,t1}/.{k1->t2,l1->l2,t1->h1}
f191: {a2,t2,a2,l2,h1,k1,l1,t1}/.{k1->t2,l1->l2,t1->h1}
f192: {a2,t2,l2,a2,h1,k1,l1,t1}/.{k1->t2,l1->l2,t1->h1}
f193: {a2,a2,l2,t2,h1,k1,l1,t1}/.{k1->l2,l1->t2,t1->h1}
f194: {a2,t2,a2,l2,h1,k1,l1,t1}/.{k1->l2,l1->t2,t1->h1}
f195: {a2,t2,l2,a2,h1,k1,l1,t1}/.{k1->l2,l1->t2,t1->h1}
f196: {a2,a2,l2,t2,h1,k1,l1,t1}/.{l1->k1,l2->t2,t1->h1}
f197: {a2,t2,a2,l2,h1,k1,l1,t1}/.{l1->k1,l2->t2,t1->h1}
f198: {a2,t2,l2,a2,h1,k1,l1,t1}/.{l1->k1,l2->t2,t1->h1}
f199: {a2,a2,l2,t2,h1,k1,l1,t1}/.{h1->t2,l1->l2,t1->k1}
f200: {a2,t2,a2,l2,h1,k1,l1,t1}/.{h1->t2,l1->l2,t1->k1}
f201: {a2,t2,l2,a2,h1,k1,l1,t1}/.{h1->t2,l1->l2,t1->k1}
f202: {a2,a2,l2,t2,h1,k1,l1,t1}/.{h1->l2,l1->t2,t1->k1}
f203: {a2,t2,a2,l2,h1,k1,l1,t1}/.{h1->l2,l1->t2,t1->k1}
f204: {a2,t2,l2,a2,h1,k1,l1,t1}/.{h1->l2,l1->t2,t1->k1}
f205: {a2,a2,l2,t2,h1,k1,l1,t1}/.{l1->h1,l2->t2,t1->k1}
f206: {a2,t2,a2,l2,h1,k1,l1,t1}/.{l1->h1,l2->t2,t1->k1}
f207: {a2,t2,l2,a2,h1,k1,l1,t1}/.{l1->h1,l2->t2,t1->k1}
f208: {a2,a2,l2,t2,h1,k1,l1,t1}/.{k1->t2,l1->h1,t1->l2}
f209: {a2,t2,a2,l2,h1,k1,l1,t1}/.{k1->t2,l1->h1,t1->l2}
f210: {a2,t2,l2,a2,h1,k1,l1,t1}/.{k1->t2,l1->h1,t1->l2}
f211: {a2,a2,l2,t2,h1,k1,l1,t1}/.{h1->t2,l1->k1,t1->l2}
f212: {a2,t2,a2,l2,h1,k1,l1,t1}/.{h1->t2,l1->k1,t1->l2}
f213: {a2,t2,l2,a2,h1,k1,l1,t1}/.{h1->t2,l1->k1,t1->l2}
f214: {a2,a2,l2,t2,h1,k1,l1,t1}/.{k1->h1,l1->t2,t1->l2}
f215: {a2,t2,a2,l2,h1,k1,l1,t1}/.{k1->h1,l1->t2,t1->l2}
f216: {a2,t2,l2,a2,h1,k1,l1,t1}/.{k1->h1,l1->t2,t1->l2}
f217: {a2,a2,l2,t2,h1,k1,l1,t1}/.{k1->l2,l1->h1,t1->t2}
f218: {a2,t2,a2,l2,h1,k1,l1,t1}/.{k1->l2,l1->h1,t1->t2}
f219: {a2,t2,l2,a2,h1,k1,l1,t1}/.{k1->l2,l1->h1,t1->t2}
f220: {a2,a2,l2,t2,h1,k1,l1,t1}/.{h1->l2,l1->k1,t1->t2}
f221: {a2,t2,a2,l2,h1,k1,l1,t1}/.{h1->l2,l1->k1,t1->t2}
f222: {a2,t2,l2,a2,h1,k1,l1,t1}/.{h1->l2,l1->k1,t1->t2}
f223: {a2,a2,l2,t2,h1,k1,l1,t1}/.{k1->h1,l1->l2,t1->t2}
f224: {a2,t2,a2,l2,h1,k1,l1,t1}/.{k1->h1,l1->l2,t1->t2}
f225: {a2,t2,l2,a2,h1,k1,l1,t1}/.{k1->h1,l1->l2,t1->t2}
f226: {a1,a1,l1,t1,h2,k2,l2,t2}/.{l1->t1,l2->k2,t2->h2}
f227: {a1,t1,a1,l1,h2,k2,l2,t2}/.{l1->t1,l2->k2,t2->h2}
f228: {a1,t1,l1,a1,h2,k2,l2,t2}/.{l1->t1,l2->k2,t2->h2}
f229: {a1,a1,l1,t1,h2,k2,l2,t2}/.{k2->t1,l2->l1,t2->h2}
f230: {a1,t1,a1,l1,h2,k2,l2,t2}/.{k2->t1,l2->l1,t2->h2}
f231: {a1,t1,l1,a1,h2,k2,l2,t2}/.{k2->t1,l2->l1,t2->h2}
f232: {a1,a1,l1,t1,h2,k2,l2,t2}/.{k2->l1,l2->t1,t2->h2}
f233: {a1,t1,a1,l1,h2,k2,l2,t2}/.{k2->l1,l2->t1,t2->h2}
f234: {a1,t1,l1,a1,h2,k2,l2,t2}/.{k2->l1,l2->t1,t2->h2}
f235: {a1,a1,l1,t1,h2,k2,l2,t2}/.{l1->t1,l2->h2,t2->k2}
f236: {a1,t1,a1,l1,h2,k2,l2,t2}/.{l1->t1,l2->h2,t2->k2}
f237: {a1,t1,l1,a1,h2,k2,l2,t2}/.{l1->t1,l2->h2,t2->k2}
f238: {a1,a1,l1,t1,h2,k2,l2,t2}/.{h2->t1,l2->l1,t2->k2}
f239: {a1,t1,a1,l1,h2,k2,l2,t2}/.{h2->t1,l2->l1,t2->k2}
f240: {a1,t1,l1,a1,h2,k2,l2,t2}/.{h2->t1,l2->l1,t2->k2}
f241: {a1,a1,l1,t1,h2,k2,l2,t2}/.{h2->l1,l2->t1,t2->k2}
f242: {a1,t1,a1,l1,h2,k2,l2,t2}/.{h2->l1,l2->t1,t2->k2}
f243: {a1,t1,l1,a1,h2,k2,l2,t2}/.{h2->l1,l2->t1,t2->k2}
f244: {a1,a1,l1,t1,h2,k2,l2,t2}/.{k2->t1,l2->h2,t2->l1}
f245: {a1,t1,a1,l1,h2,k2,l2,t2}/.{k2->t1,l2->h2,t2->l1}
f246: {a1,t1,l1,a1,h2,k2,l2,t2}/.{k2->t1,l2->h2,t2->l1}
f247: {a1,a1,l1,t1,h2,k2,l2,t2}/.{h2->t1,l2->k2,t2->l1}
f248: {a1,t1,a1,l1,h2,k2,l2,t2}/.{h2->t1,l2->k2,t2->l1}
f249: {a1,t1,l1,a1,h2,k2,l2,t2}/.{h2->t1,l2->k2,t2->l1}
f250: {a1,a1,l1,t1,h2,k2,l2,t2}/.{k2->h2,l2->t1,t2->l1}
f251: {a1,t1,a1,l1,h2,k2,l2,t2}/.{k2->h2,l2->t1,t2->l1}
f252: {a1,t1,l1,a1,h2,k2,l2,t2}/.{k2->h2,l2->t1,t2->l1}
f253: {a1,a1,l1,t1,h2,k2,l2,t2}/.{k2->l1,l2->h2,t2->t1}
f254: {a1,t1,a1,l1,h2,k2,l2,t2}/.{k2->l1,l2->h2,t2->t1}
f255: {a1,t1,l1,a1,h2,k2,l2,t2}/.{k2->l1,l2->h2,t2->t1}
f256: {a1,a1,l1,t1,h2,k2,l2,t2}/.{h2->l1,l2->k2,t2->t1}
f257: {a1,t1,a1,l1,h2,k2,l2,t2}/.{h2->l1,l2->k2,t2->t1}
f258: {a1,t1,l1,a1,h2,k2,l2,t2}/.{h2->l1,l2->k2,t2->t1}
f259: {a1,a1,l1,t1,h2,k2,l2,t2}/.{k2->h2,l2->l1,t2->t1}
f260: {a1,t1,a1,l1,h2,k2,l2,t2}/.{k2->h2,l2->l1,t2->t1}
f261: {a1,t1,l1,a1,h2,k2,l2,t2}/.{k2->h2,l2->l1,t2->t1}
f262: {t2,a2,l2,a2,h1,k1,l1,t1}/.{h1->t2,k1->l2,l1->t1}
f263: {t2,a2,l2,a2,h1,k1,l1,t1}/.{h1->l2,k1->t2,l1->t1}
f264: {t2,a2,l2,a2,h1,k1,l1,t1}/.{k1->h1,l1->t1,l2->t2}
f265: {t2,a2,l2,a2,h1,k1,l1,t1}/.{k1->t2,l1->l2,t1->h1}
f266: {t2,a2,l2,a2,h1,k1,l1,t1}/.{k1->l2,l1->t2,t1->h1}
f267: {t2,a2,l2,a2,h1,k1,l1,t1}/.{l1->k1,l2->t2,t1->h1}
f268: {t2,a2,l2,a2,h1,k1,l1,t1}/.{h1->t2,l1->l2,t1->k1}
f269: {t2,a2,l2,a2,h1,k1,l1,t1}/.{h1->l2,l1->t2,t1->k1}
f270: {t2,a2,l2,a2,h1,k1,l1,t1}/.{l1->h1,l2->t2,t1->k1}
f271: {t2,a2,l2,a2,h1,k1,l1,t1}/.{k1->t2,l1->h1,t1->l2}
f272: {t2,a2,l2,a2,h1,k1,l1,t1}/.{h1->t2,l1->k1,t1->l2}
f273: {t2,a2,l2,a2,h1,k1,l1,t1}/.{k1->h1,l1->t2,t1->l2}
f274: {t2,a2,l2,a2,h1,k1,l1,t1}/.{k1->l2,l1->h1,t1->t2}
f275: {t2,a2,l2,a2,h1,k1,l1,t1}/.{h1->l2,l1->k1,t1->t2}
f276: {t2,a2,l2,a2,h1,k1,l1,t1}/.{k1->h1,l1->l2,t1->t2}
f277: {h1,k1,l1,t1,h2,k2,l2,t2}/.{k1->h1,k2->h2,l1->t1,l2->t2}
f278: {h1,k1,l1,t1,h2,k2,l2,t2}/.{k2->h2,l1->k1,l2->t2,t1->h1}
f279: {h1,k1,l1,t1,h2,k2,l2,t2}/.{k2->h2,l1->h1,l2->t2,t1->k1}
f280: {h1,k1,l1,t1,h2,k2,l2,t2}/.{k1->h1,l1->t1,l2->k2,t2->h2}
f281: {h1,k1,l1,t1,h2,k2,l2,t2}/.{l1->k1,l2->k2,t1->h1,t2->h2}
f282: {h1,k1,l1,t1,h2,k2,l2,t2}/.{l1->h1,l2->k2,t1->k1,t2->h2}
f283: {h1,k1,l1,t1,h2,k2,l2,t2}/.{k1->h1,l1->t1,l2->h2,t2->k2}
f284: {h1,k1,l1,t1,h2,k2,l2,t2}/.{l1->k1,l2->h2,t1->h1,t2->k2}
f285: {h1,k1,l1,t1,h2,k2,l2,t2}/.{l1->h1,l2->h2,t1->k1,t2->k2}
\end{verbatim}
\end{scriptsize}

Renaming dummy indices, we have:

\begin{scriptsize}
\begin{verbatim}
a1->a
a2->b
l1->c
l2->d
t1->e
t2->f
k1->g
b2->h
h1->j
h2->k
k2->m
h2->n
\end{verbatim}
\end{scriptsize}

And finally restricting the number of dummy indices to four and rearranging them to an arbitrary pattern, we have the following:

\begin{scriptsize}
\begin{verbatim}
DDX DDX type of invariant
 +(1/24)*D_aaX*D_bbX
 +(-1/60)*D_aaX*D_bbX
 +(-1/60)*D_aaX*D_bbX
 +(1/180)*D_aaX*D_bbX
 +(1/180)*D_abX*D_baX
 +(1/180)*D_abX*D_abX;
Output:=========================
+(1 180) D_abX D_abX+(1 180) D_abX| D_baX+(1 72) D_aaX| D_bbX
================================
DDX DDY type of invariant
 +(1/80)*D_aaX*D_bcY_bc +(1/80)*D_aaX*D_bcY_bc +(1/120)*D_aaX*D_bbY_cc +(1/120)*D_aaX*D_bcY_cb
 +(1/120)*D_aaX*D_bbY_cc +(1/120)*D_aaX*D_bcY_cb +(-1/80)*D_aaX*D_bcY_bc +(1/120)*D_aaX*D_bcY_bc
 +(-1/240)*D_aaX*D_bcY_bc +(-1/240)*D_aaX*D_bbY_cc +(-1/240)*D_aaX*D_bcY_cb +(-1/240)*D_aaX*D_bcY_bc
 +(1/720)*D_aaX*D_bcY_bc +(1/720)*D_abX*D_acY_bc +(1/720)*D_abX*D_bcY_ac +(1/240)*D_aaX*D_bcY_bc
 +(1/240)*D_abX*D_acY_bc +(1/240)*D_abX*D_bcY_ac +(-1/240)*D_aaX*D_bbY_cc +(-1/360)*D_aaX*D_bbY_cc
 +(-1/240)*D_aaX*D_bcY_bc +(-1/360)*D_aaX*D_bcY_cb +(-1/360)*D_aaX*D_bbY_cc +(-1/240)*D_aaX*D_bcY_bc
 +(-1/360)*D_aaX*D_bcY_cb +(-1/360)*D_abX*D_ccY_ba +(-1/240)*D_abX*D_caY_cb +(-1/360)*D_abX*D_caY_bc
 +(-1/360)*D_abX*D_ccY_ba +(-1/240)*D_abX*D_caY_cb +(-1/360)*D_abX*D_caY_bc +(-1/360)*D_abX*D_ccY_ab
 +(-1/240)*D_abX*D_cbY_ca +(-1/360)*D_abX*D_cbY_ac +(-1/360)*D_abX*D_ccY_ab +(-1/240)*D_abX*D_cbY_ca
 +(-1/360)*D_abX*D_cbY_ac +(-1/240)*D_aaX*D_bcY_cb +(-1/240)*D_aaX*D_bcY_bc +(-1/360)*D_aaX*D_bcY_bc
 +(-1/360)*D_abX*D_acY_bc +(-1/360)*D_abX*D_bcY_ac +(1/840)*D_aaX*D_bbY_cc +(1/840)*D_aaX*D_bcY_cb
 +(1/840)*D_aaX*D_bcY_bc +(1/840)*D_abX*D_baY_cc +(1/840)*D_abX*D_abY_cc +(1/840)*D_aaX*D_bbY_cc
 +(1/840)*D_abX*D_baY_cc +(1/840)*D_abX*D_abY_cc +(1/840)*D_aaX*D_bbY_cc +(1/840)*D_abX*D_cbY_ac
 +(1/840)*D_abX*D_caY_bc +(1/840)*D_aaX*D_bcY_cb +(1/840)*D_abX*D_bcY_ac +(1/840)*D_abX*D_acY_bc
 +(1/840)*D_aaX*D_bcY_bc +(1/840)*D_abX*D_cbY_ca +(1/840)*D_abX*D_bcY_ca +(1/840)*D_abX*D_ccY_ba
 +(1/840)*D_abX*D_caY_cb +(1/840)*D_abX*D_acY_cb +(1/840)*D_abX*D_ccY_ab +(1/840)*D_aaX*D_bcY_cb
 +(1/840)*D_abX*D_cbY_ac +(1/840)*D_abX*D_caY_bc +(1/840)*D_aaX*D_bcY_bc +(1/840)*D_abX*D_bcY_ac
 +(1/840)*D_abX*D_acY_bc +(1/840)*D_abX*D_cbY_ca +(1/840)*D_abX*D_bcY_ca +(1/840)*D_abX*D_ccY_ba
 +(-1/240)*D_abX*D_caY_cb +(-1/240)*D_abX*D_acY_cb +(-1/240)*D_abX*D_ccY_ab;
Output:=========================
+(1 420) D_abX D_abY_cc+(13 2520) D_abX D_acY_bc+(-1 336) D_abX D_acY_cb+(1 420) D_abX D_baY_cc
+(2 315) D_aaX D_bbY_cc+(13 2520) D_abX D_bcY_ac+(2 315) D_aaX D_bcY_bc+(1 420) D_abX D_bcY_ca
+(2 315) D_aaX D_bcY_cb+(-1 315) D_abX D_caY_bc+(-19 1680) D_abX D_caY_cb+(-1 315) D_abX D_cbY_ac
+(-1 168) D_abX D_cbY_ca+(-43 5040) D_abX D_ccY_ab+(-1 315) D_abX D_ccY_ba
================================
DDY DDY type of invariant
 +(1/320)*D_dcY_dc*D_abY_ab  +(1/480)*D_dcY_cd*D_abY_ab +(1/480)*D_dcY_dc*D_baY_ab
 +(1/720)*D_dcY_cd*D_baY_ab +(1/720)*D_ddY_ba*D_ccY_ab +(1/480)*D_daY_db*D_ccY_ab
 +(1/720)*D_daY_bd*D_ccY_ab +(1/480)*D_ddY_cb*D_acY_ab +(1/320)*D_dbY_dc*D_acY_ab
 +(1/480)*D_dbY_cd*D_acY_ab +(1/720)*D_ddY_ca*D_bcY_ab +(1/480)*D_daY_dc*D_bcY_ab
 +(1/720)*D_daY_cd*D_bcY_ab +(1/720)*D_ddY_ab*D_ccY_ab +(1/480)*D_dbY_da*D_ccY_ab
 +(1/720)*D_dbY_ad*D_ccY_ab +(1/480)*D_ddY_bc*D_acY_ab +(1/320)*D_dcY_db*D_acY_ab
 +(1/480)*D_dcY_bd*D_acY_ab +(1/720)*D_ddY_ac*D_bcY_ab +(1/480)*D_dcY_da*D_bcY_ab
 +(1/720)*D_dcY_ad*D_bcY_ab +(1/960)*D_dcY_dc*D_abY_ab +(1/960)*D_adY_cd*D_cbY_ab
 +(1/960)*D_dcY_ac*D_dbY_ab +(1/480)*D_dcY_dc*D_abY_ab +(1/720)*D_dcY_dc*D_baY_ab
 +(-1/960)*D_dcY_dc*D_abY_ab +(1/720)*D_acY_bc*D_ddY_ab +(1/480)*D_bdY_cd*D_acY_ab
 +(1/720)*D_adY_cd*D_bcY_ab +(-1/960)*D_daY_dc*D_cbY_ab +(1/720)*D_bcY_ac*D_ddY_ab
 +(1/480)*D_dcY_bc*D_adY_ab +(1/720)*D_dcY_ac*D_bdY_ab +(-1/960)*D_dcY_da*D_cbY_ab
 +(-1/320)*D_dcY_dc*D_abY_ab +(-1/480)*D_dcY_cd*D_abY_ab +(-1/480)*D_ddY_ca*D_cbY_ab
 +(-1/320)*D_daY_dc*D_cbY_ab +(-1/480)*D_daY_cd*D_cbY_ab +(-1/480)*D_ddY_ac*D_cbY_ab
 +(-1/320)*D_dcY_da*D_cbY_ab +(-1/480)*D_dcY_ad*D_cbY_ab +(-1/1440)*D_dcY_cd*D_abY_ab
 +(-1/1440)*D_ddY_ca*D_cbY_ab +(-1/1440)*D_daY_cd*D_cbY_ab +(-1/1440)*D_ddY_ac*D_cbY_ab
 +(-1/1440)*D_dcY_ad*D_cbY_ab +(-1/1440)*D_dcY_dc*D_abY_ab +(-1/1440)*D_adY_cd*D_cbY_ab
 +(-1/1440)*D_dcY_ac*D_dbY_ab +(1/1120)*D_dcY_ad*D_cbY_ab +(1/1120)*D_daY_cd*D_cbY_ab
 +(1/1120)*D_dcY_cd*D_abY_ab +(1/1120)*D_dcY_ac*D_dbY_ab +(1/1120)*D_adY_cd*D_cbY_ab
 +(1/1120)*D_dcY_dc*D_abY_ab +(1/1120)*D_dcY_da*D_cbY_ab +(1/1120)*D_dcY_ca*D_dbY_ab
 +(1/1120)*D_ddY_ca*D_cbY_ab +(1/1120)*D_daY_dc*D_cbY_ab +(1/1120)*D_acY_cd*D_dbY_ab
 +(1/1120)*D_ddY_ac*D_cbY_ab +(1/3360)*D_dcY_cd*D_abY_ab +(1/3360)*D_dcY_ad*D_cbY_ab
 +(1/3360)*D_daY_cd*D_cbY_ab +(1/3360)*D_dcY_dc*D_abY_ab +(1/3360)*D_dcY_ac*D_dbY_ab
 +(1/3360)*D_adY_cd*D_cbY_ab +(1/3360)*D_dcY_da*D_cbY_ab +(1/3360)*D_dcY_ca*D_dbY_ab
 +(1/3360)*D_ddY_ca*D_cbY_ab +(1/3360)*D_daY_dc*D_cbY_ab +(1/3360)*D_acY_cd*D_dbY_ab
 +(1/3360)*D_ddY_ac*D_cbY_ab +(-1/1680)*D_ddY_ac*D_bcY_ab +(-1/1120)*D_dcY_da*D_bcY_ab
 +(-1/1680)*D_dcY_ad*D_bcY_ab +(-1/1680)*D_ddY_ca*D_bcY_ab +(-1/1120)*D_daY_dc*D_bcY_ab
 +(-1/1680)*D_daY_cd*D_bcY_ab +(-1/1120)*D_dcY_dc*D_baY_ab +(-1/1680)*D_dcY_cd*D_baY_ab
 +(-1/1680)*D_ddY_ac*D_cbY_ab +(-1/1120)*D_dcY_da*D_cbY_ab +(-1/1680)*D_dcY_ad*D_cbY_ab
 +(-1/1680)*D_ddY_ca*D_cbY_ab +(-1/1120)*D_daY_dc*D_cbY_ab +(-1/1680)*D_daY_cd*D_cbY_ab
 +(-1/1120)*D_dcY_dc*D_abY_ab +(-1/1680)*D_dcY_cd*D_abY_ab +(-1/1680)*D_ddY_bc*D_acY_ab
 +(-1/1120)*D_dcY_db*D_acY_ab +(-1/1680)*D_dcY_bd*D_acY_ab +(-1/1680)*D_ddY_bc*D_caY_ab
 +(-1/1120)*D_dcY_db*D_caY_ab +(-1/1680)*D_dcY_bd*D_caY_ab +(-1/1680)*D_ddY_ba*D_ccY_ab
 +(-1/1120)*D_daY_db*D_ccY_ab +(-1/1680)*D_daY_bd*D_ccY_ab +(-1/1680)*D_ddY_cb*D_acY_ab
 +(-1/1120)*D_dbY_dc*D_acY_ab +(-1/1680)*D_dbY_cd*D_acY_ab +(-1/1680)*D_ddY_cb*D_caY_ab
 +(-1/1120)*D_dbY_dc*D_caY_ab +(-1/1680)*D_dbY_cd*D_caY_ab +(-1/1680)*D_ddY_ab*D_ccY_ab
 +(-1/1120)*D_dbY_da*D_ccY_ab +(-1/1680)*D_dbY_ad*D_ccY_ab +(-1/1120)*D_dcY_dc*D_baY_ab
 +(-1/1680)*D_dcY_cd*D_baY_ab +(-1/1680)*D_ddY_ac*D_bcY_ab +(-1/1120)*D_dcY_da*D_bcY_ab
 +(-1/1680)*D_dcY_ad*D_bcY_ab +(-1/1680)*D_ddY_ca*D_bcY_ab +(-1/1120)*D_daY_dc*D_bcY_ab
 +(-1/1680)*D_daY_cd*D_bcY_ab +(-1/1120)*D_dcY_dc*D_abY_ab +(-1/1680)*D_dcY_cd*D_abY_ab
 +(-1/1680)*D_ddY_ac*D_cbY_ab +(-1/1120)*D_dcY_da*D_cbY_ab +(-1/1680)*D_dcY_ad*D_cbY_ab
 +(-1/1680)*D_ddY_ca*D_cbY_ab +(-1/1120)*D_daY_dc*D_cbY_ab +(-1/1680)*D_daY_cd*D_cbY_ab
 +(-1/1680)*D_ddY_bc*D_acY_ab +(-1/1120)*D_dcY_db*D_acY_ab +(-1/1680)*D_dcY_bd*D_acY_ab
 +(-1/1680)*D_ddY_bc*D_caY_ab +(-1/1120)*D_dcY_db*D_caY_ab +(-1/1680)*D_dcY_bd*D_caY_ab
 +(-1/1680)*D_ddY_ba*D_ccY_ab +(-1/1120)*D_daY_db*D_ccY_ab +(-1/1680)*D_daY_bd*D_ccY_ab
 +(-1/1680)*D_ddY_cb*D_acY_ab +(-1/1120)*D_dbY_dc*D_acY_ab +(-1/1680)*D_dbY_cd*D_acY_ab
 +(-1/1680)*D_ddY_cb*D_caY_ab +(-1/1120)*D_dbY_dc*D_caY_ab +(-1/1680)*D_dbY_cd*D_caY_ab
 +(-1/1680)*D_ddY_ab*D_ccY_ab +(-1/1120)*D_dbY_da*D_ccY_ab +(-1/1680)*D_dbY_ad*D_ccY_ab
 +(-1/1680)*D_dcY_ac*D_bdY_ab +(-1/1680)*D_adY_cd*D_bcY_ab +(-1/1680)*D_dcY_dc*D_baY_ab
 +(-1/1680)*D_dcY_ac*D_dbY_ab +(-1/1680)*D_adY_cd*D_cbY_ab +(-1/1680)*D_dcY_dc*D_abY_ab
 +(-1/1680)*D_dcY_bc*D_adY_ab +(-1/1680)*D_dcY_bc*D_daY_ab +(-1/1680)*D_acY_bc*D_ddY_ab
 +(-1/1680)*D_bdY_cd*D_acY_ab +(-1/1680)*D_bdY_cd*D_caY_ab +(-1/1680)*D_bcY_ac*D_ddY_ab
 +(1/13440)*D_dcY_cd*D_baY_ab +(1/13440)*D_dcY_dc*D_baY_ab +(1/13440)*D_dcY_cd*D_abY_ab
 +(1/13440)*D_dcY_dc*D_abY_ab;
Output:=========================
+(1 1120) D_acY_ab D_dbY_cd +(1 1120) D_acY_ab D_dcY_bd +(1 1120) D_acY_ab D_ddY_bc
+(1 1120) D_acY_ab D_ddY_cb +(1 1260) D_acY_bc D_ddY_ab +(1 1260) D_adY_cd D_bcY_ab
+(1 1260) D_bcY_ac D_ddY_ab +(1 1260) D_bdY_ab D_dcY_ac +(-1 1680) D_bdY_cd D_caY_ab
+(-1 1680) D_daY_ab D_dcY_bc +(-1 210) D_cbY_ab D_daY_dc +(-1 210) D_cbY_ab D_dcY_da
+(1 3360) D_bcY_ab D_daY_dc +(1 3360) D_bcY_ab D_dcY_da +(1 3360) D_ccY_ab D_daY_db
+(1 3360) D_ccY_ab D_dbY_da +(-1 360) D_cbY_ab D_daY_cd +(-1 360) D_cbY_ab D_dcY_ad
+(-1 360) D_cbY_ab D_ddY_ac +(-1 360) D_cbY_ab D_ddY_ca +(1 5040) D_bcY_ab D_daY_cd
+(1 5040) D_bcY_ab D_dcY_ad +(1 5040) D_bcY_ab D_ddY_ac +(1 5040) D_bcY_ab D_ddY_ca
+(1 5040) D_ccY_ab D_daY_bd +(1 5040) D_ccY_ab D_dbY_ad +(1 5040) D_ccY_ab D_ddY_ab
+(1 5040) D_ccY_ab D_ddY_ba +(-1 560) D_caY_ab D_dbY_dc +(-1 560) D_caY_ab D_dcY_db
+(1 672) D_acY_ab D_bdY_cd +(1 672) D_adY_ab D_dcY_bc +(1 840) D_acY_cd D_dbY_ab
+(-1 840) D_caY_ab D_dbY_cd +(-1 840) D_caY_ab D_dcY_bd +(-1 840) D_caY_ab D_ddY_bc
+(-1 840) D_caY_ab D_ddY_cb +(1 840) D_dbY_ab D_dcY_ca +(11 40320) D_abY_ab D_dcY_dc
+(11 40320) D_baY_ab D_dcY_cd +(19 20160) D_adY_cd D_cbY_ab +(19 20160) D_dbY_ab D_dcY_ac
+(3 2240) D_acY_ab D_dbY_dc +(3 2240) D_acY_ab D_dcY_db +(47 40320) D_baY_ab D_dcY_dc
+(-5 8064) D_abY_ab D_dcY_cd
================================
\end{verbatim}
\end{scriptsize}

Combining all Maxima generated output will give ${\cal L}^{(1)[8]}_{2\,\,\,\,[4][4]}$ as given in (\ref{L28_44}).

\end{appendix}


\begin{thebibliography}{99}
%
\bibitem{HwHt}
   S.W.~Hawking and T. Hertog, {\it Living with Ghosts}, \textit{Phys. Rev.} \textbf{D65}, 103515 (2002). \texttt{hep-th/0107088}.
%
\bibitem{Birr}
   N.D.~Birrel and P.C.W.~Davies
   \textit{Quantum Fields in Curved Space}
   (Cambridge University Press, Cambridge, England, 1982).
%
\bibitem{Stll}
   K.S.~Stelle, \textit{Gen. Rel. Grav.} \textbf{9}, 353 (1978),
   \textit{Phys. Rev.} \textbf{D16}, 953 (1977).
%
\bibitem{Boul}
   D.G.~Boulware, S.~Deser, and K.S.~Stelle, in \textit{Quantum Field Theory and Quantum Statistics: Essays in Honor of the Sixtieth Birthday of E.S.~Fradkin} edited by I.A.~Batalin, C.J.~Isham, and C.A.~Vilkovisky (Hilger, Bristol, England, 1987).
%
\bibitem{Polc}
 J. Polchinski, {\sc TASI lectures on D-branes} \texttt{hep-th/9611050}.
%
\bibitem{John}
   C.V.  Johnson, {\sc D-brane primer} \texttt{hep-th/0007170}.
%
\bibitem{Curt}
   T.L.~Curtright, G.L.~Ghandour, and C.K.~Zachos
   \textit{Phys. Rev.} \textbf{D36} 3811 (1986).
%
\bibitem{Maed}
   K.~Maeda and D.~Turok,
   \textit{Phys. Lett.} \textbf{B202}, 376 (1988).
%
\bibitem{Dirc}
   P.A.M.~Dirac,
   \textit{Proc. Roy. Soc. London} \textbf{A167}, 148 (1938).
%
\bibitem{Buch}
    I.L. Buchbinder, S.D. Odintsov, I.L. Shapiro {\it Effective Actions in Quantum Gravity}
    (Bristol, IOP Pub. Ltd. 1992) 419pp.
%
\bibitem{Avrm}
    I.G. Avramidi, {\sc Covariant Methods for the Calculation of the Effective
    Action in Quantum Field Theory and Investigation of Higher Derivative
    Quantum Gravity} PhD Disseration, (Moscow Lomonosov State University,
    Moscow, 1986) 159 pp.
%
\bibitem{Mign}
    S.~Mignemi and D.L.~Wiltshire, {\sc Black holes in Higher Derivative
    Gravity Theories} {\it Phys. Rev.} {\bf D46} No. 4 (1992)
    1475--1506, and the references therein.
%
\bibitem{Mazz}
    F.D.~Mazzitelli, {\sc Higher Derivatives and renormalization in quantum
    Cosmology} {\it Phys. Rev.} {\bf D45} (1992) 2814--2822, and
    the references therein.
%
\bibitem{tHVt}
   G.~t'~Hooft and M.~Veltman, \textit{Ann. Inst. Poincare} \textbf{20}, 69 (1974).
%
\bibitem{Stll2}
   K.S.~Stelle, \textit{Phys. Rev.} \textbf{D16}, 953 (1977).
%
\bibitem{FrdTsy}
   E.S.~Fradkin and A.A.~Tseytlin, \textit{Nucl. Phys.} \textbf{B201}, 469 (1982).
%
\bibitem{Simn}
   J.Z.~Simon, {\it Higher-derivative Lagrangians, nonlocality, problems, and solutions}, \textit{Phys. Rev.} \textbf{D41}, 3720 (1990).
%
\bibitem{Shaf}
    Q.~Shafi and C.~Wetterich, \textit{Phys. Lett.} \textbf{129B}, 387 (1983), \textbf{152B}, 51 (1985)
%
\bibitem{Maed2}
    K.~Maeda, \textit{Phys. Rev.} \textbf{D37}, 858 (1983), \textbf{D39}, 3159 (1989).
%
\bibitem{Schw}
J. S. Schwinger, {\it On gauge invariance and vacuum
polarization}, Phys. Rev., 1951, vol. 82, No 5, pp. 664--679.
%
\bibitem{Fck}
V. A. Fock, {\it The proper time in classical and quantum
mechanics}, Izvestiya of USSR Academy of Sciences, Physics, 1937, No 4,5, pp.
551--568.
%
\bibitem{Sch} J. S. Schwinger, {\it The theory of quantized fields V.}, Phys.
Rev., 1954, vol. 93, No 3, pp. 615--628.
%
\bibitem{DeWt}
    B. S. De~Witt, {\it Quantum theory of gravity II. The manifestly
covariant theory}, Phys. Rev., 1967, vol. 162, No 5, pp. 1195--1238.
%
\bibitem{DWt-Ish} B. S. De~Witt {\it Quantum Gravity II}, Eds. C.J. Isham, R. Penrose and D.W. Sciama, (Oxford: Oxford University Press, 1981)
%
\bibitem{Brw-Dff}
    M.R.~Brown and M.J.~Duff,
    {\sc Exact Results for Effective Lagrangian}
    {\it Phys. Rev.} {\bf D11} (1975) 2124.
\bibitem{Rodf-Diss}
    E.T. Rodulfo, {\sc Low Energy Effective Lagrangian for Arbitrary
    Quasi-local Gauge Field Theory in Any Dimension}, PhD
    Dissertation, University of the Philippines, 1998.
%
\bibitem{Tiam}
    M.T.~Tiamzon,
    MS Thesis, De La Salle University, 1998.
%
\bibitem{Tiam-Rodf}
    M.T.~Tiamzon, E.T.~Rodulfo,
    {\it The Manila J. of Sci.} {\bf 2} (1999) 6.
\bibitem{bfta} B. S. De~Witt, {\it Dynamical theory of groups and fields}, (New
York: Gordon and Breach, 1965), 230 pp.
%
\bibitem{27.} B. S. De~Witt, {\it Quantum theory of gravity III. The
application of the covariant theory}, Phys. Rev., 1967, vol. 162, No 5, pp.
1239--1256.
%
\bibitem{28.} B. S. De~Witt, {\it Quantum field theory in curved spacetime},
Phys. Rep. C, 1975, vol. 19, pp. 296--357.
%
\bibitem{29.} R. E. Kallosh, {\it Renormalization in non--Abelian gauge
theories}, Nucl. Phys. B, 1974, vol. 78, No 2, pp. 293--312.
%
\bibitem{GNW} M. T. Grisaru, P. van Nieuwenhuizen and C. C. Wu, {\it
Background--field method versus normal field theory in explicit examples:
One--loop divergences in the $S$--matrix and Green's functions for Yang--Mills and
gravitational field}, Phys. Rev. D, 1975, vol. 12, No 10, pp. 3203--3213.
%
\bibitem{31.} N. K. Nielsen, {\it Ghost counting in supergravity}, Nucl. Phys.
B, 1978, vol.140, No 3, pp. 499--509.
%
\bibitem{32.} R. E. Kallosh, {\it Modified Feynman rules in supergravity},
Nucl. Phys. B, 1978, vol. 141, No 1,2, pp. 141--152.
%
\bibitem{33.} B. S. De~Witt, {\it Quantum gravity: New synthesis}, in: {\it
General relativity}, Eds. S. Hawking and W. Israel, (Cambridge: Cambridge Univ.
Press. 1979)
%
\bibitem{34.} B. S. De~Witt, {\it Gauge invariant effective action}, in: {\it
Quantum gravity II}, Second Oxford Symp. 1980, Eds. C. J. Isham, R. Penrose and
D. W. Sciama, (Oxford: Oxford Univ. Press, 1981), pp. 449--487.
%
\bibitem{35.} D. G. Boulware, {\it Gauge dependence of the effective action},
Phys. Rev. D, 1981, vol. 23, No 2, pp. 389--396.
%
\bibitem{Abbt} L. F. Abbot, {\it The background field method beyond one loop},
Nucl. Phys. B, 1981, vol. 185, No 1, pp. 189--203.
%
\bibitem{37.} I. Jack and H. Osborn, {\it Two--loop background field
calculations for arbitrary background fields}, Nucl. Phys. B, 1982, vol. 207,
No 3, pp. 474--504.
%
\bibitem{Ichi} S. Ichinose and M. Omote, {\it Renormalization using the
background field method}, Nucl. Phys. B, 1982, vol. 203, No 2, pp. 221--267.
%
\bibitem{39.} C. Lee, {\it Proper--time renormalization of multi--loop amplitudes
in the background field method (I). $\Phi^4$--theory}, Nucl. Phys. B, 1982, vol.
207, No 1, pp. 157--188.
%
\bibitem{40.} D. M. Capper and A. Mac--Lean, {\it The background field method at
two loops. A general gauge Yang--Mills calculation}, Nucl. Phys. B, 1982, vol.
203, No 3, pp. 413--422.
%
\bibitem{41.} I. Jack and H. Osborn, {\it Background field calculations in
curved space--time (I). General formalism and application to scalar fields},
Nucl. Phys. B, 1984, vol. 234, No 2, pp. 331--364.
%
\bibitem{42.} I. Jack, {\it Background field calculations in curved space--time
(II). Application to a pure gauge theory}, Nucl. Phys. B, 1984, vol. 234, No 2,
pp. 365--378.
%
\bibitem{43.} A. O. Barvinsky and G. A. Vilkovisky, {\it The generalized
Schwinger--De~Witt technique and unique effective action in quantum gravity},
in: {\it Quantum gravity}, Proc. IIIrd Sem. Quantum Gravity, Moscow 1984, Eds.
M. A. Markov, V. A. Berezin and V. P. Frolov, (Singapore: World Sci. Publ.,
1985), pp. 141--160.
%
\bibitem{44.} D. M. Capper, J. J. Dulwich and R. M. Medrano, {\it The
background field method for quantum gravity at two loops}, Nucl. Phys. B, 1985,
vol. 254, No 3,4, pp. 737--746.
%
\bibitem{45.} G. A. Vilkovisky, {\it The Gospel according to De~Witt}, in: {\it
Quantum Gravity}, Ed. S. Christensen, (Bristol: Hilger, 1983), pp. 169--209.
%
\bibitem{46.} G. A. Vilkovisky, {\it The unique effective action in quantum
field theory}, Nucl. Phys. B, 1984, vol. 234, pp. 125--137.
%
\bibitem{bftz} E. S. Fradkin and A. A. Tseytlin, {\it On the new definition of
off--shell effective action}, Nucl. Phys. B, 1984, vol. 234, No 2, pp. 509--523.
%
\bibitem{tHft}
    G. 't Hooft, {\it Nucl. Phys.} \textbf{B62} (1973) 444.
%
\bibitem{Velt}
    M.~Veltman, in {\it Methods in Field Theory}, ed. R Balian and
    J. Zinn-Justin, (North Holland Publ. Co., 1976) 267.
%
\bibitem{Bhbb}
    H.J.~Bhabha, \textit{Phys. Rev.} \textbf{70}, 759 (1946).
%
\bibitem{Pag-Rodf}
    E.T.~Rodulfo and J.A.G.~Pagaran  {\sc A Closed Integral Form of the
    Background Gauge Connection} {\it The Manila J. of Sci.} \textbf{Vol. 4 No. 2} (2001) 17-21.
    Preprint: \texttt{hep-th/0212336}
%
%
%
%
\bibitem{Avra}
    I.G.~Avramidi, {\it J. Math. Phys.} {\bf 36}, 5055 (1995), {\it Nucl. Phys.} {\bf B355}, 712 (1991).
%
\bibitem{Rodf-Delb}
    E.T.~Rodulfo and R.~Delbourgo,
        {\sc One-Loop Effective Multigluon Lagrangian in Arbitrary
            Dimensions}.
        {\it Int. J. Mod. Phys.} {\bf A14} 1999 4457--4471,
        \texttt{hep-th/9709201}.
%
\bibitem{Rodf}
    E.T.~Rodulfo,
    {\sc Quasilocal Background Field Method Applied to Gauge
    Theories},
     in {\it Proc. Workshop on Nonperturbative Methods in Quantum
     Field Theory} in Adelaide, eds A.W. Schreiber $et.al.$,
    (World  Scientific, Singapore, 1998) 231-236.
%
\bibitem{Flie}
    D.~Fliegner, P.~Haberl, M.G.~Schmidt and C.~Schubert,
    {\sc The higher Derivative Expansion of the Effective Action by
    the String Inspired Method. Part II} , \texttt{hep-th/9707189}
%
\bibitem{Flig}
    D.~Fliegner, M.~Reuter, M.G.~Schmidt and C.~Schubert,
    {\sc The Two-Loop Euler-Heisenberg Lagrangian in Dimensional
    Regularization} LANL preprint, \texttt{hep-th/9704194}
%
\bibitem{Shif}
    M.A.~Shifman, {\it Nucl. Phys.} {\bf B173} (1980) 13
%
\bibitem{Itzy}
    C.~Itzykson and J.--B.~Zuber,
    {\it Quantum field theory}, (McGraw--Hill, New York, 1980).
%
\bibitem{Oko}
    A.~Okopinska,
     {\sc Solving Schwinger-Dyson equations by trunction in zero-dimensional scalar quantum field theory.}
    \textit{Phys. Rev.} \textbf{D43}, 3561 (1991).
%
\bibitem{Rod-Pag}
    E.T.~Rodulfo and J.A.G. Pagaran, In preparation.
%
\bibitem{Wyll}
   N. Wyllard,
   {\sc Derivative corrections to D-brane actions with constant
    background fields},
    \textit{Nucl. Phys.} \textbf{B598}, 247 (2001),
    \texttt{hep-th/0008125}.
%
\bibitem{Bila}
   A. Bilal,
   {\sc Higher Derivative Corrections to the Non-Abelian Born-Infeld Action} \texttt{hep-th/010602v1}.
%
\bibitem{Vilk}
    G.A.~Vilkovisky, {\sc The Gospel According to DeWitt}
    in {\it Quantum gravity}, ed. S. Christensen
    (Adam Hilger,Bristol,1983) 169--209.
%
\bibitem{AGS}
    L.F.~Abbott, M.T.~Grisaru and R.K.~Schaeffer,
    {\it Nucl. Phys.}
    {\bf B229} (1983) 372.
%
\bibitem{Ball}
    R.D.~Ball,
    {\it Phys. Rep.} {\bf 182}, Nos. 1 \& 2 (1989) 1.
%
\bibitem{DeWt'}
    B. S. De~Witt,
        {\sc Quantum gravity: New synthesis},
        in {\it General relativity},
        eds. S. Hawking and W. Israel, (Cambridge:
        Cambridge Univ. Press. 1979).
%
\bibitem{DeWt''}
    B. S. De~Witt,
    {\sc Quantum field theory in curved spacetime},
    {\it Phys. Rep. C}, 1975, {\bf 19}, pp. 296--357.
%
\bibitem{Bogo}
    N. N. Bogolyubov and D. V. Shirkov,
    {\it Introduction to the theory of quantized fields},
    (Nauka, Moscow, 1976), 479 pp.
%
\bibitem{Ramn}
    P. Ramond, {\it Field theory: A modern primer},
    (Benjamin, Reading, Massachusetts, 1981).
%
\bibitem{Duff}
    M.J. Duff and M. Ramon-Medrano,
    {\sc Effective Lagrangian for the Yang-Mills field}
    {\it Phys. Rev.} \textbf{D12} (1975) 3357.
%
\bibitem{Delb-Ritz}
    R.~Delbourgo and A.~Ritz,
        {\sc The Low Energy Effective Lagrangian for Photon Interactions
        in Any Dimension}
        {\it Int. J. Mod. Phys.} {\bf A11} (1996)
        253, \texttt{hep-th/9503160 v2}.
%
\bibitem{GM}
 T.D.~Gargett and I.N.~McArthur,
       {\sc Derivative Expansions of One-Loop Effective Actions}
       (unpublished).
%
\bibitem{ven}
    A.E.M.~de Ven,
    {\it Nucl. Phys.} \textbf{B250} (1985) 593.
%
\bibitem{vanN}
    P.~van~Nieuwenhuizen, {\it Ann. of Phys.} {\bf 104} (1977) 197
%
\bibitem{Gilk}
    P.B. Gilkey,
    {\it J. Diff. Geom.} \textbf{10} (1975) 601.
%
\bibitem{Fradk}
    E.S. Fradkin and A. A. Tseytlin,
    {\it Nucl. Phys.} \textbf{B227} (1983) 252.
%
\bibitem{Mue}
    Uwe M$\ddot{\mathrm{u}}$ller,
    {\it in New Computing Techniques in Physics Research IV, Proceedings of the AIHENP-95 workshop, Pisa, 1995,}
    eds. B. Denby, D. Perret-Gallix (World Scientific, Singapore, 1995), p.19, \texttt{hep-th/9508031}; \texttt{hep-th/9701124}.
\end{thebibliography}
\end{document}